\documentclass[a4paper,12pt]{article}
\usepackage[utf8]{inputenc} 	% We use UTF-8 encoded files only 
\usepackage[T1]{fontenc}     % for the right font encoding
\usepackage{lmodern}         % fonts for good pdf files
\usepackage{braket}
\usepackage[text={160mm,250mm},centering]{geometry}

\usepackage{url}
\DeclareUrlCommand\email{\urlstyle{tt}}

\usepackage[british]{babel} % for proper hyphenation  (English)
\usepackage{latexsym}
\usepackage{amsmath,amssymb,amscd,amsbsy,amsthm}
\usepackage{upgreek}
\usepackage{multicol}
\usepackage{paralist} % si usas compactdesc

\usepackage{tensor}
\usepackage{graphicx}        % standard LaTeX graphics tool
\usepackage{makeidx}         % allows index generation

\usepackage{float}           % for placement of floats
\usepackage[cmtip,all,barr]{xy}   % xy plotting package for diaxy

\usepackage{authblk}
\newcommand{\autheadcr}{\authorcr}

\usepackage{epigraph}
\usepackage{rcs} \RCS $Revision: 1.16 $ \RCS $Date: 2026/01/23 22:30:36 $ \RCS $Author: hgm $ \RCS $RCSfile: 23_QC-algebra.tex,v $ \RCS $Id: 23_QC-algebra.tex,v 1.16 2026/01/23 22:30:36 hgm Exp $

\usepackage[breaklinks,colorlinks=true,plainpages=false,linkcolor=blue]{hyperref}

\usepackage{tikz}
\usetikzlibrary{arrows.meta,calc,positioning,decorations.pathreplacing}

\usepackage{refdef}
\usepackage{ifontdef}

\usepackage{amsmath}

\numberwithin{equation}{section}

\newtheorem{thm}{Theorem}[section]
\newtheorem{theorem}[thm]{Theorem}

\newtheorem{prop}[thm]{Proposition}
\newtheorem{rem}[thm]{Remark}
\newtheorem{coro}[thm]{Corollary}
\newtheorem{corollary}[thm]{Corollary}

\newtheorem{defi}[thm]{Definition}

\newtheorem{lem}[thm]{Lemma}

\newcommand{\ignore}[1]{}

\newcommand{\Lp}{\mrm{L}}
\newcommand{\Wp}{\mrm{W}}

\newcommand{\Ck}{\mrm{C}}

\newcommand{\vrho}{\varrho}

\newcommand{\vsigma}{\varsigma}

\newcommand{\vphi}{\varphi}
\newcommand{\vpi}{\varpi}

\newcommand{\vek}[1]{\mathchoice{\displaystyle\boldsymbol{#1}}
{\textstyle\boldsymbol{#1}}{\scriptstyle\boldsymbol{#1}}
{\scriptscriptstyle\boldsymbol{#1}}}

\newcommand{\mat}[1]{\mathchoice{\displaystyle\mathbf{#1}}
{\textstyle\mathbf{#1}}{\scriptstyle\mathbf{#1}}
{\scriptscriptstyle\mathbf{#1}}}

\newcommand{\opb}[1]{\vek{{\mathsf{#1}}}}

\newcommand{\ops}[1]{\mathchoice{\displaystyle\mathsf{#1}}
{\textstyle\mathsf{#1}}{\scriptstyle\mathsf{#1}}
{\scriptscriptstyle\mathsf{#1}}}

\newcommand{\tnb}[1]{\mathchoice{\displaystyle\mathboldsans{#1}}
{\textstyle\mathboldsans{#1}}{\scriptstyle\mathboldsans{#1}}
{\scriptscriptstyle\mathboldsans{#1}}}

\newcommand{\tns}[1]{\mathchoice{\displaystyle\mathsans{#1}}
{\textstyle\mathsans{#1}}{\scriptstyle\mathsans{#1}}
{\scriptscriptstyle\mathsans{#1}}}

\newcommand{\vhat}[1]{\vek{\hat{#1}}}

\DeclareMathOperator{\diag}{diag}

\DeclareMathOperator{\im}{im}

\DeclareMathOperator{\rank}{rank}
\DeclareMathOperator{\tr}{tr}

\DeclareMathOperator{\spn}{span}

\DeclareMathOperator{\supp}{supp}

\DeclareMathOperator{\cl}{cl}

\DeclareMathOperator{\co}{co}

\newcommand{\di}{\mathop{}\!\mathrm{d}}

\newcommand{\ii}{\mathchoice{\displaystyle\mathrm i}
{\textstyle\mathrm i}{\scriptstyle\mathrm i}
{\scriptscriptstyle\mathrm i}}

\newcommand{\ip}[2]{\left\langle #1 , #2 \right\rangle}
\newcommand{\bkt}[2]{\left\langle #1 \mid #2 \right\rangle}
\newcommand{\ipj}[1]{\left\langle #1 \right\rangle}

\newcommand{\ns}[1]{\left| #1 \right|}
\newcommand{\nd}[1]{\left\Vert #1 \right\Vert}

\newcommand{\trpos}{{\ops{T}}}

\newcommand{\BIGOP}[1]{\mathop{\mathchoice%
{\raise-0.22em\hbox{\huge $#1$}} {\raise-0.05em\hbox{\Large $#1$}}
{\hbox{\large $#1$}}{#1}}}
\newcommand{\BIGboxplus}{\mathop{\mathchoice%
{\raise-0.35em\hbox{\huge $\boxplus$}}%
{\raise-0.15em\hbox{\Large $\boxplus$}}{\hbox{\large
$\boxplus$}}{\boxplus}}}

\newcommand{\bbbone}{{\mathchoice {\rm 1\mskip-4mu l} {\rm 1\mskip-4mu l} {\rm 1\mskip-4.5mu l} {\rm 1\mskip-5mu l}}}

\newcommand{\frkt}[2]{{{\raise0.6ex\hbox{{\leavevmode$\textstyle #1$}}}{\raise0.25ex\hbox{\kern-0.35ex\hbox{/}}}{\raise-0.3ex\hbox{\kern-0.4ex\hbox{{\leavevmode$\textstyle #2$}}}}}}
\newcommand{\frks}[2]{{{\raise0.7ex\hbox{{\leavevmode$\scriptstyle #1$}}}{\raise0.2ex\hbox{\kern-0.4ex\hbox{\footnotesize /}}}{\raise-0.2ex\hbox{\kern-0.4ex\hbox{{\leavevmode$\scriptstyle #2$}}}}}}
\newcommand{\frkx}[2]{{{\raise0.75ex\hbox{{\leavevmode$\scriptscriptstyle #1$}}}{\raise0.17ex\hbox{\kern-0.45ex\hbox{\scriptsize /}}}{\raise-0.15ex\hbox{\kern-0.4ex\hbox{{\leavevmode$\scriptscriptstyle #2$}}}}}}
\newcommand{\frkz}[2]{{{\raise0.75ex\hbox{{\leavevmode$\scriptscriptstyle #1$}}}{\raise0.17ex\hbox{\kern-0.45ex\hbox{\tiny /}}}{\raise-0.15ex\hbox{\kern-0.4ex\hbox{{\leavevmode$\scriptscriptstyle #2$}}}}}}

\newcommand{\frk}[2]{{\mathchoice{{\frkt{#1}{#2}}}{{\frks{#1}{#2}}}{{\frkx{#1}{#2}}}{{\frkz{#1}{#2}}}}}

\newtheorem{xmpn}{Example}

\ignore{     %%% BEGIN IGNORE

\usepackage[sfdefault=cmbr,OMLmathsans]{isomath}  % isomath for slanted sans-serif

\newcommand{\ignore}[1]{}

\newcommand{\mat}[1]{\mathchoice{\displaystyle\mathbf#1}
{\textstyle\mathbf#1}{\scriptstyle\mathbf#1}
{\scriptscriptstyle\mathbf#1}}
\newcommand{\vek}[1]{\mathchoice{\displaystyle\boldsymbol#1}
{\textstyle\boldsymbol#1}{\scriptstyle\boldsymbol#1}
{\scriptscriptstyle\boldsymbol#1}}

\newcommand{\F}{\mathfrak} % math Fraktur
\newcommand{\C}{\mathcal}  % math Calligraphic
\newcommand{\mrm}{\mathrm}     % math roman upright

\newcommand{\di}{\mathop{}\!\mathrm{d}}

\newcommand{\ii}{\mathchoice{\displaystyle\mathrm i}
{\textstyle\mathrm i}{\scriptstyle\mathrm i}
{\scriptscriptstyle\mathrm i}}

\newcommand{\ip}[2]{\langle #1 , #2 \rangle}
\newcommand{\ipj}[1]{\langle #1 \rangle}

\newcommand{\bkt}[2]{\langle #1 | #2 \rangle}

\newcommand{\ns}[1]{| #1 |}
\newcommand{\nd}[1]{\| #1 \|}

\DeclareMathOperator{\tr}{tr}
\DeclareMathOperator{\diag}{diag}

}    %%% END IGNORE

\newcommand{\olsi}[1]{\,\overline{\!{#1}}} % overline short italic
\newcommand{\wht}[1]{\widehat{#1}}
	
\newcommand{\wtl}[1]{\widetilde{#1}}	
\newcommand{\wob}[1]{\olsi{#1}}	

\newcommand{\stoeq}{\overset{\text{m}}{=}}

\newcommand{\feq}[1]{Eq.(\ref{#1})} %[1]{#1} : [no of input parameters]{Parameter-1}
\newcommand{\feqs}[2]{Eqs.(\ref{#1}) and (\ref{#2})} %\ffigs{arg1}{arg2}
\newcommand{\feeqs}[2]{Eqs.(\ref{#1})--(\ref{#2})} %\ffigs{arg1}{arg2}
\newcommand{\feqss}[3]{Eqs.(\ref{#1}),(\ref{#2}) and (\ref{#3})} %\ffigs{arg1}{arg2}{arg3}
\newcommand{\fsec}[1]{Section~\ref{#1}}
\newcommand{\ffig}[1]{Fig.\ref{#1}} %\ffig{arg1}
\newcommand{\feT}[1]{Theorem~\ref{#1}}
\newcommand{\feP}[1]{Proposition~\ref{#1}}

\newcommand{\feD}[1]{Definition~\ref{#1}}
\newcommand{\feR}[1]{Remark~\ref{#1}}
\newcommand{\feC}[1]{Corollary~\ref{#1}}
\newcommand{\feX}[1]{Example~\ref{#1}}
\newcommand{\feXs}[2]{Examples~\ref{#1} and \ref{#2}} 
\newcommand{\feeXs}[2]{Examples~\ref{#1} -- \ref{#2}}

\newcommand{\Hf}[2]{\tensor[^#1]{#2}{}}
\DeclareMathOperator{\var}{var}
\DeclareMathOperator{\cov}{cov}
\newcommand{\Ex}{\D{E}}

\newcommand{\QCAlgrthm}{\E{N}}

\newcommand{\sptldom}{\C{G}}

\newcommand{\RR}{\D{R}}
\newcommand{\Rd}{\D{R}^d}

\newcommand{\CC}{\D{C}}

\newcommand{\Cn}{\D{C}^n}

\newcommand{\MMn}{\D{M}_n}
\newcommand{\MMin}[1]{\D{M}_{#1}}
\newcommand{\tpH}{{\ops{H}}}
\newcommand{\DMn}{\D{D}_n}

\newcommand{\smplspc}{\Omega}
\newcommand{\evt}{\omega}
\newcommand{\sigalg}{\F{F}}
\newcommand{\prob}{\D{P}}

\newcommand{\vA}{\vek{A}}
\newcommand{\vB}{\vek{B}}
\newcommand{\vC}{\vek{C}}
\newcommand{\vD}{\vek{D}}
\newcommand{\vE}{\vek{E}}

\newcommand{\vH}{\vek{H}}
\newcommand{\vI}{\vek{I}}
\newcommand{\vM}{\vek{M}}
\newcommand{\vP}{\vek{P}}
\newcommand{\vQ}{\vek{Q}}
\newcommand{\vR}{\vek{R}}
\newcommand{\vS}{\vek{S}}

\newcommand{\vU}{\vek{U}}
\newcommand{\vV}{\vek{V}}

\newcommand{\vZ}{\vek{Z}}
\newcommand{\vLbd}{\vek{\Lambda}}

\newcommand{\tH}{\tnb{H}}
\newcommand{\tI}{\tnb{I}}

\newcommand{\tQ}{\tnb{Q}}

\newcommand{\tU}{\tnb{U}}
\newcommand{\tV}{\tnb{V}}

\newcommand{\tX}{\tnb{X}}
\newcommand{\tY}{\tnb{Y}}
\newcommand{\tZ}{\tnb{Z}}

\newcommand{\tp}{\tnb{p}}

\newcommand{\va}{\vek{a}}
\newcommand{\vb}{\vek{b}}
\newcommand{\ve}{\vek{e}}
\newcommand{\vf}{\vek{f}}
\newcommand{\vg}{\vek{g}}

\newcommand{\vr}{\vek{r}}
\newcommand{\vu}{\vek{u}}
\newcommand{\vv}{\vek{v}}
\newcommand{\vw}{\vek{w}}
\newcommand{\vx}{\vek{x}}
\newcommand{\vy}{\vek{y}}
\newcommand{\vz}{\vek{z}}
\newcommand{\vrh}{\vek{\rho}}

\newcommand{\Alg}{\C{A}}
\newcommand{\Blg}{\C{B}}
\newcommand{\Clg}{\C{C}}
\newcommand{\Mlg}{\C{M}}
\newcommand{\Qlg}{\C{Q}}
\newcommand{\Hvk}{\C{H}}
\newcommand{\Kvk}{\C{K}}

\newcommand{\Lop}{\C{L}}
\newcommand{\LHA}{\Lop(\Hvk)}
\newcommand{\BHA}{\C{B}(\Hvk)}
\newcommand{\XA}[1]{\ops{X}(#1)}
\newcommand{\SA}[1]{\ops{S}(#1)}
\newcommand{\PA}[1]{\mrm{P}(#1)}
\newcommand{\UA}[1]{\mrm{U}(#1)}
\newcommand{\EA}[1]{\mrm{E}(#1)}
\newcommand{\BA}[1]{\mrm{B}(#1)}
\newcommand{\ZA}[1]{\mrm{Z}(#1)}

\newcommand{\DEN}[1]{\mrm{D}(#1)}
\newcommand{\rA}{\tns{A}}

\newcommand{\rD}{\tns{D}}

\newcommand{\rI}{\tns{I}}
\newcommand{\rM}{\tns{M}}
\newcommand{\rN}{\tns{N}}
\newcommand{\rP}{\tns{P}}

\newcommand{\rR}{\tns{R}}

\newcommand{\rU}{\tns{U}}

\newcommand{\rW}{\tns{W}}
\newcommand{\rX}{\tns{X}}
\newcommand{\rY}{\tns{Y}}
\newcommand{\rZ}{\tns{Z}}
\newcommand{\ra}{\tns{a}}
\newcommand{\rb}{\tns{b}}
\newcommand{\rc}{\tns{c}}
\newcommand{\re}{\tns{e}}
\newcommand{\rf}{\tns{f}}
\newcommand{\rg}{\tns{g}}
\newcommand{\rone}{\tns{1}}
\newcommand{\rnul}{\tns{0}}
\newcommand{\rp}{\tns{p}}

\newcommand{\rs}{\tns{s}}
\newcommand{\rt}{\tns{t}}
\newcommand{\rqq}{\tns{q}}
\newcommand{\rr}{\tns{r}}
\newcommand{\ru}{\tns{u}}
\newcommand{\rv}{\tns{v}}
\newcommand{\rx}{\tns{x}}
\newcommand{\ry}{\tns{y}}

\newcommand{\oL}{\Upsilon}

\newcommand{\obH}{\E{H}}
\newcommand{\svpi}{\tns{\vpi}}
\newcommand{\som}{\tns{\omega}}
\newcommand{\sal}{\tns{\alpha}}
\newcommand{\sbt}{\tns{\beta}}

\newcommand{\sphi}{\tns{\phi}}

\newcommand{\spsi}{\tns{\psi}}

\newcommand{\sta}{\tns{\tau}}
\providecommand{\D}[1]{\mathbb{#1}}

\providecommand{\NN}{\D{N}}

\newcommand{\citep}[1]{\cite{#1}}

\makeindex             % used for the subject index
\newcommand{\authorafm}{Antonio Falc\'o\thanks{Corresponding author.}}

\newcommand{\authorhgm}{Hermann G. Matthies}

\newcommand{\affilwire}{Institute of Scientific Computing, %\autheadcr
                        Technische Universit\"at Braunschweig \autheadcr
                        38092 Braunschweig, Germany, %\autheadcr
                        E-mail: \ttt{h.matthies@tu-braunschweig.de}}

\newcommand{\affilCEU}{Departamento de Matem\'aticas, F\'isica y Ciencias Tecnol\'ogicas \autheadcr
                        Universidad Cardenal Herrera-CEU, CEU Universities, San Bartolom\'e 55 \autheadcr
                        46155 Alfara del Patriarca (Valencia), Spain, %\autheadcr
                        E-mail: \ttt{afalco@uchceu.es}}

\newcommand{\thetitle}{%
  \texorpdfstring
  {Vistas of Algebraic Probability, \\ Quantum Computation and Information}
  {Vistas of Algebraic Probability, Quantum Computation and Information}
}

\newcommand{\thesubject}{46L53, 47L90, 46K10, 81R05, 81S25, 
                         46N50, 46L89, 81P68, 60A99}
\newcommand{\thekeywords}{algebraic probability, non-Kolmogorovean probability,
                           non-com\-muting random variables, quantum-like behaviour,
                          quantum channels and information, quantum computation}

\begin{document}

% ============================================================================
% connect to default LaTeX values

%\title*{\thetitle}
\title{\thetitle}
%\titlerunning{Probabilistic and Parametric Models}
% Use \titlerunning{Short Title} for an abbreviated version of
% your contribution title if the original one is too long

%\author{\theauthor}
\author[1]{\authorafm}
\author[2]{\authorhgm}

\renewcommand\Affilfont{\footnotesize}
\affil[1]{\affilCEU}
\affil[2]{\affilwire}

%\authorrunning{\theauthor}

%\institute{Hermann G. Matthies \at Institute of Scientific Computing, TU Braunschweig,
%           Brunswick, Germany, \email{wire@tu-bs.de}}

\date{}

\maketitle

\vspace{-3em}
%\epigraphhead[70]{%
\epigraph
    {\textit{\footnotesize Probability is the most important concept in modern science,
                        especially as nobody has the slightest notion what it means.}}
    {{\scriptsize Bertrand Russell, 1929}}
%}

%Probability is the most important concept in modern science, 
%especially as nobody has the slightest notion what it means. 
%---Bertrand Russell, 1929 Lecture 

% !TEX encoding = UTF-8 Unicode
% !TEX root = ../23_QC-algebra.tex
% RCSID:       $Id: abstract_QC-alg.tex,v 1.5 2025/03/30 20:55:23 hgm Exp $
% Author:      $Author: hgm $
% Contact:     wire@tu-bs.de
% ============================================================================
%% texfile{
%%  AUTHOR    = "$Author: hgm $",
%%  VERSION   = "$Revision: 1.5 $",
%%  DATE      = "$Date: 2025/03/30 20:55:23 $",
%%  FILENAME  = "$RCSfile: abstract_QC-alg.tex,v $"}
%
% =================================

\begin{abstract}
Kolmogorov's foundation of probability
takes measure spaces, $\sigma$-algebras, and probability measures as the foundation
of probability theory.  It is widely recognised that the classical Kolmogorovean 
view on probability is inadequate to deal with random phenomena involving quantum effects, 
or more widely with \emph{quantum-like} situations.  Thus one might look for
a wider, more encompassing view.  Such a formulation can be achieved by what may be termed an 
algebraic point of view.  This starts with algebras of random variables with
a distinguished linear functional, the state, which is interpreted as expectation.
Incidentally, this may be seen as a modern interpretation of the framework
used by the early practitioners of probability like the Bernoullis.
The algebraic view, in its modern form emanating from and extensively used in quantum physics, 
offers the potential to deal with both the classical and quantum-like behaviours,
but is not much appreciated in classical probability and uncertainty quantification,
although even here it may lead to new vistas and illuminate different aspects.
This origin gives a physics flavour to the language, but the subject is purely probability.
The key difference between classical and quantum-like probabilistic behaviour turns out 
to be commutativity, the lack of which leads to the effects encountered 
in quantum-like situations.
The advent of quantum computers %and quantum information channels 
is one example which may make it necessary even for classical practitioners of 
computational sciences to deal with quantum-like
behaviour. Here we endeavour to show the purely algebraic part of this approach.
By concentrating on finite dimensional algebras one can
avoid many difficult analytical questions, but at the same time still
show the main flavour and ideas of the algebraic point of view in classical probability, and
how this extends to quantum-like behaviour, and is actually applicable to quantum computing.

\tbf{Keywords: }\thekeywords

\tbf{MSC2020 Classification: }\thesubject
\end{abstract}

%  $Log: abstract_QC-alg.tex,v $
%  Revision 1.5  2025/03/30 20:55:23  hgm
%  little changes
%
%  Revision 1.4  2024/07/19 10:26:06  hgm
%  little changes
%
%  Revision 1.3  2024/03/28 15:40:59  hgm
%  litle changes
%
%  Revision 1.2  2023/12/27 07:49:43  hgm
%  first sketch
%
%  Revision 1.1  2023/12/07 16:34:03  hgm
%  from abstract_RV-alg.tex
%
%  Revision 1.0  2022/05/16 19:57:25  hgm
%  inital check in, from abstract_RV-alg.tex
%
%
%
%
%

%%% Local Variables: 
%%% mode: latex
%%% TeX-master: "../23_QC-algebra"
%%% End: 

%
\clearpage
\tableofcontents
\clearpage
\section{Introduction} \label{S:intro}
% !TEX root = ../23_QC-algebra.tex
% !TEX encoding = UTF-8 Unicode
% RCSID:       $Id: introduction_QC-alg.tex,v 1.20 2026/01/26 09:10:38 hgm Exp $
% Author:      $Author: hgm $
% Contact:     wire@tu-bs.de
% =================================

The classical probabilistic framework based on Kolmogorov's axioms 
\citep{Kolmogo1933, Kolmogo1950-en}
(see also e.g.\ \citep{Sullivan2015}) is typically 
used when performing uncertainty quantification (UQ).  It starts from 
sample spaces, $\sigma$-algebras of subsets of events and
probability measures, and can be used to define 
algebras of random variables (RVs) as measurable functions on these measure
spaces, and the expectation operator as integral of these RVs w.r.t.\ the
probability measure.  A different view is offered by an alternative route, 
where one uses
as fundamental concepts algebras of RVs, in physics lingo also called ``observables'', together 
with the expectation operator (e.g.\ \citep{segalKunze78}) as a positive linear
functional.  More to the point, the expectation is an example of a ``state''
(of the system under consideration), mathematically an element of the dual of
the algebra of RVs and what will take the r\^ole of probability distributions
in classical probability.  The necessity of this algebraic view arose from the
needs to describe quantum phenomena, or, to put it more generally, to describe
quantum-like behaviour (QLB), where the sequential order of observations matters.  
As it turns out, it is this property of non-commutativity or commutativity
in the algebra of RVs which describes the difference between QLB and the classical case. 
The algebraic view allows a unified picture of both classical and QLB.  And
this view does not distinguish between isomorphic objects resp.\ 
faithful representations, so that
much of the description and discussion can be done in an abstract framework,
which hence focuses on the essentials.  But it allows one also to switch to some
of the isomorphic concrete mathematical representations, if so desired.
This also allows one to infer properties in the abstract setting from well known
facts in one of the faithful representations.
Although a number of terms, like ``state'' or ``observable'', originate 
from quantum physics, it will hopefully become clear that algebraic 
probability is a purely mathematical theory, no physics is needed,
and when restricted to finite dimensional algebras it is actually just
part of (multi-)linear algebra.

In case the algebra of RVs is Abelian or commutative, one essentially
recovers equivalence with classical Kolomogorovean probability.  
Notably, the algebraic approach allows for non-commutative algebras of RVs, 
developed along with quantum theory, and is also important in order to deal 
with e.g.\ random matrices (e.g.\ \citep{MingoSpeicher2017}), random fields of tensors, and
quantum-like behaviour (QLB) (e.g.\ \citep{Baaquie2004, HoraObata07, YukalovSornette2009, 
Khrennikov2010, Bagarello2012, Busemeyer2012, BusemeyerBruza2012, HavenKhrennikov2013, 
Sornette2014, AshtianiAzgomi2015, YukalovSornette2016, Obata2017, Bagarello2019, 
Baaquie2020, FinEcmBsQntUnc2022, Khrennikov2023}), also outside of %the original
quantum theory.  
%What is important here is that it allows a description of
Here we also show how this nicely applies to
quantum computing (QC) (e.g.\ \citep{NielsenChuang2011}), and quantum channels and 
information (QI) (e.g.\ \citep{Werner2001, Barnett2009, Wilde2017}),
which incidentally is also an application of the theory in the simplest setting,
when the algebras of RVs may be taken as finite dimensional 
(e.g.\ \citep{BenyRicht15, Harlow2016}).  For the sake of simplicity and brevity 
we shall limit ourselves here to the finite dimensional setting and show the key concepts 
by concentrating on the algebraic aspects.  As will become clear very quickly, 
such probability algebras can be represented as algebras of linear maps on the algebra itself,
which can be equipped with the structure of a pre-Hilbert space.
Here the ``real'', or actually observable, or self-adjoint RVs correspond to self-adjoint maps.  
Thus, in the case of finite dimensional algebras, a faithful representation 
can be provided by a matrix algebra.  Although the focus here will be finite 
dimensional algebras, the presentation will still be kept
in the more abstract setting, as it brings out the algebraic content independently
of any representation.  The motivation for this kind of description is that it 
turned out that the specific realisation of the algebra does not matter, it is
the intrinsic algebraic properties which hold the probabilistic information,
whereas the concrete realisations or representations can be used as desired
for specific purposes.  In addition, this way offers the reader a smooth transition 
to approach the algebraic view also when dealing with continuous random variables. 

As will be shown later, the possible observations or samples of an observable or random 
variable are values in its spectrum, and thus the self-adjoint RVs have real valued samples.  
As the spectrum of self-adjoint linear maps on finite dimensional spaces is a discrete 
and finite subset of the real numbers, in order
to describe continuous RVs one has to deal with infinite dimensional algebras, so that linear
maps may have a continuous spectrum.  And to represent unbounded RVs, one has to deal 
with unbounded self-adjoint maps.  
Thus the concentration on finite dimensional algebras avoids the analytical difficulties
which otherwise appear when dealing with --- possibly unbounded --- linear operators 
on infinite dimensional spaces, while still occasionally offering a glimpse on
how this may be extended in the infinite dimensional case.
But we would like to point out that the algebraic view can naturally deal 
with systems with infinitely many degrees of freedom
(e.g.\ \citep{RedeiSummers2006p}, where the different types of von Neumann algebras are
investigated in this respect), stochastic processes, and random fields, or more 
generally with RVs with values in infinite dimensional spaces more directly; topics which can be
treated with classical probabilistic methods only in a somewhat circumlocutory fashion.

The treatment of QLB phenomena is not possible within the Kolomogorovean 
view --- which is equivalent to using only commutative algebras --- 
as it turns out that in the non-commutative case one can not define a global $\sigma$-algebra 
of events with consistently assigned probabilities, 
a notion which is at the basis of classical probability.  
This is due to the fact that for QLB phenomena not all observables or RVs are 
\emph{compatible}, i.e.\ they can not be realised or observed simultaneously --- in the 
same state of the system under consideration.   The possible probability assignments
depend on the ``context'', i.e.\ which other random variables (RVs) are observed
simultaneously.  Mathematically this is expressed by the
question of whether these RVs commute or not, and the Boolean algebra of events
is replaced by a weaker notion, the lattice of projections.  
As it turns out, there is no global ``sample space'', this is a 
``pointless'' probability theory.  This leads to consequences
which are ``strange'' or not possible from a classical point of view.  This is evidenced in
the theory in the violation of the Bell inequalities and by the connected Kochen-Specker 
theorem --- which we shall see is a purely probabilistic result --- 
two of the so-called ``no-go'' theorems of quantum theory.  
To wit, the 2022 Nobel prize in physics was awarded for 
``\dots, establishing the violation of Bell inequalities 
  and pioneering quantum information science \dots'',
the experimental confirmation of this fact.  But what looks strange in the
light of classical probability is exactly what is being harnessed in quantum information
and quantum computing.  The point of view taken here though is that the (non-commutative) 
algebraic view is a \emph{probability theory}, and not only a description of the physics of
quantum-physical phenomena. 

Looking back at the historical development of probability (e.g.\ \citep{Accardi00b}), 
its mathematical treatment began
by considering games of chance, and traces back to early work by Cardano, and further
seminal contributions in the 17th century by 
Fermat, Pascal, Huygens, and the Bernoullis in the early 18th century, culminating in 
Jacob Bernoulli's (1713) ``Ars Conjectandi'', for two modern appraisals 
of its significance see \citep{ShaferBern, SchneiderBern}.  
Starting with Huygens, Jacob Bernoulli and his nephew Daniel Bernoulli,
among others, clarified the meaning of probability and expectation. 
These early probabilists may be seen \citep{segalKunze78} as 
early proponents of the algebraic approach in their way of operating with RVs
algebraically like numbers according to certain intuitive rules, which were, 
however, never explicitly stated.  
Incidentally, what is mostly called probability today
was called ``chance'' then, and what was considered ``probable'' then, today would be called
a kind of Bayesian interpretation of probability \citep{ShaferBern}.  There is
even today a multitude of opinions of what probability actually ``is''
in the real world, not dissimilar to analogous discussion about the
interpretation of quantum theory (e.g.\ \citep{BBC}).  In both areas there is agreement
on how to use the formal machinery, but the interpretations of what this ``is'' differ.
The formalisation and analysis
of classical probability theory gained momentum with the works of Gauss and others,
and was brilliantly displayed with Laplace's analytical treatment of 
probability \citep{Laplace1812}.  
%One may also note that Laplace gave a first 
%general formulation of Bayes's theorem and hence of conditioning, a subject
%central to probability theory.

At the dawn of the new century in 1900 Hilbert announced a list of mathematical problems
which await a solution, and the 6th problem was to establish an axiomatic basis for 
probability and mechanics --- incidentally this was the same year when Planck announced 
his discovery of energy quanta, and which marked the start of quantum theory, which was 
to bring such changes to our understanding of probability, leading to quantum  
probability \citep{Accardi2018}.  For classical probability, such an axiomatic
basis, started by von Mises \citep{vonMises1928} in 1928,
was provided by Kolmogorov in 1933 with his so-called ``Grundbegriffe'' monograph
\citep{Kolmogo1933, Kolmogo1950-en}.  It put classical probability on a firm axiomatic
footing based on measure theory --- which emanated from the geometric problem of measuring
lengths, areas, volumes, etc., and had been brought to a mathematical maturation
in the preceding decades --- and defined RVs as derived objects, namely as measurable 
functions, and expectation as the integral w.r.t.\ a probability measure.

When Kolmogorov published his ``Grundbegriffe'' in 1933, he, in his own words,
expressed what many mathematicians were thinking about the formalisation of probability.
But the ``new'' probability in the form of quantum probability \citep{Accardi00b, Accardi2018} 
had already been formulated in its essential lines through contributions from Heisenberg
\citep{Heisenberg1925}, who in 1925 noted the non-commutativity of quantum observables, and this was 
immediately extended by his colleagues in Göttingen, Born and Jordan \citep{BornJordan1925},
and by all three together \citep{BornHsbgJord1926}.  While Heisenberg regarded the 
non-commutativity more as a nuisance \citep{Varad2011}, it was Dirac \citep{Dirac1925, Dirac1930} 
who may be credited with recognising its fundamental importance first.   Further contributions 
in this direction came from Hilbert \citep{HilbJvN-Nordh1928} together with von Neumann and 
Nordheim.  Von Neumann shortly afterwords worked out the Hilbert space formulation
\citep{Neumann1932, Neumann1955}, where the RVs are self-adjoint operators.  This was one
way to capture the non-commutative properties, stemming from the fact that it mattered in
which temporal order RVs were observed or measured.  

All this happened before the publication date of Kolmogorov's ``Grundbegriffe'', and von
Neumann's work indicated a possibility on how to build an axiomatic basis for both 
classical and quantum probability, but this seems to have evaded Kolmogorov at the time.
As Accardi rightly states \citep{Accardi00b, Accardi2018},  
there is a difference between describing the mathematical construction and deriving
it from a set of axioms.  And in the late 1920's and early 1930's the algebraic path 
seemed far too weak to use as an axiomatic foundation.  This was about to change
in the next decade.
 
It was Jordan \citep{Jordan1927} already in the late 20's, 
and later together with Wigner and von Neumann, 
who had started in the 1930's considering algebraic approaches \citep{JordJvN-Wig1934}
to quantum theory.  This was followed by von Neumann, who in the late 1930's and early 1940's
together with Murray started developing the algebraic method, all nicely summarised
in his collected works \citep{Neumann1961}.  This algebraic view of random variables,
in the form of algebras of observables with the expectation resp.\ state as
a linear functional, pioneered by von Neumann and Murray with their work on ``rings
of operators'' (in \citep{Neumann1961}), now provided a powerful framework for analysing 
quantum systems as well as classical ones.  
Subsequent developments, including the work of Gel'fand and Neumark (Naĭmark)
\citep{gelfandNaimark1943} and Segal \citep{segal1947}, led to the GNS-construction
(GNS for Gel'fand, Neumark, and Segal) and established the connection of abstract algebras
with the earlier Hilbert space approach used in quantum physics.

Although there have been several proposals on how to answer Hilbert's call for
axiomatisation in his 6th problem, starting from von Neumann himself (cf.\ again 
\citep{Neumann1961}), as well as e.g.\ \citep{segalPostQM1947, Mackey1963, 
Emch1972, Ludwig1985, Ludwig1987, Accardi2018, Ron2023}
to name but a few, we shall take a humbler path and approach the  algebraic view on RVs 
in an ``operational'' manner, by saying a bit informally what to do with RVs, not what 
they are.  This is then by design an abstract approach to establish rules for a mathematical
structure, and the usual ways of dealing both with classical (RVs are measurable functions,
i.e.\  a commutative or Abelian algebra) and QLB (RVs resp.\ observables are self-adjoint 
Hilbert space operators) appear as \emph{representations}.
This operational approach is very similar to what the early probabilists like the 
Bernoullis implicitly used, and actually most practically working statisticians do 
today, where the underlying measure spaces rarely appear, if at all, and the real work is 
with random variables and their distributions.

Our aim here is to give an appetiser of the view on the algebraic structure of 
probability theories, without delving too deep in algebraic theory,
but which may even in the Abelian case shed a different light
on the subject.  We hope to show that the study of different representations 
of probability may bring out different traits.  
This algebraic approach is also connected with a very geometric view, and it leads naturally
to the consideration of non-commutative algebras, which are a natural framework 
to consider QLB phenomena, or combinations of classical and QLB phenomena.  
We want to refrain from an attempt to 
re-create probability theory (even only for the Abelian case) from axioms,
and will develop the starting point more leisurely; 
for such axiomatic attempts see \citep{Jordan1927, HilbJvN-Nordh1928, Dirac1930, Neumann1932, 
Neumann1955, Neumann1961, JordJvN-Wig1934, segalPostQM1947, Mackey1963, 
Emch1972, Ludwig1985, Ludwig1987, Accardi1995, Accardi2018, Accardi2022}, 
mostly trying to develop quantum probability theory.
It is also not intended to take a puristic view and try to dispense with the 
notions of probability or $\sigma$-algebras,
but to embed this into a more general and deeper structure, and hence results
from measure and integration theory will be used where appropriate.
The algebraic theory has also been developed into a more detailed and deeper
analysis and understanding of what the ``value'' of a RV is, and what the formal
description of an observation, measurement, or sample of a RV is.

Our approach in this paper is twofold. Firstly, we aim to elucidate the algebraic 
view of probability in its simplest setting, emphasising its geometric and 
operational aspects.  Secondly, we touch on the implications of this framework by
applying it to quantum computing (QC) and quantum information (QI), highlighting its 
potential to accommodate quantum-like behaviour (QLB).  Our approach
tries to minimise the use of technicalities and generalities, albeit staying
mostly on an abstract level, trying to make statements simple and not the most general.  

No attempt has been made to be complete in any way, not with respect to 
the history of the subject, neither w.r.t.\ the selection of the associated topics,
nor in the selection of references, where obviously many
undoubtedly important contributions are not mentioned. 
Indeed, there are so many different aspects --- and open problems about them --- that 
a complete survey seems impossible, or at least extremely difficult, at the present time.
Examples of similar efforts are \citep{GhorbalSchuermann99, Mitchener2005,
Lehmann2009, BenyRicht15, Harlow2016}.
But we hope that this article may convince the reader that it is worthwhile to have different
ways of looking at the subject, and perhaps inspire her or him to consult 
the literature further, and possibly even pursue the concept in their own research.
We claim no originality, except for what has been left out from the presentation. 
The treatment in this paper is intended to be more descriptive than rigorous, 
focusing on key concepts and insights rather than technical details and proofs,
which are typically relegated to the references.  Although these ``vistas'' are
distinctly mathematical --- without diminishing the contribution originating 
and coming from physics ---
the aim is to avoid mathematical scaffolding, and make the exposition suitable 
for readers with diverse backgrounds in mathematics and computer science, the 
natural sciences, and engineering.

The plan of the paper is as follows. 
We begin by elucidating the algebraic view on random variables (RVs) in \fsec{S:formal}, 
highlighting its importance in dealing with non-commutative phenomena.   
This perspective allows for the representation of RVs as linear maps or 
operators, especially when a Hilbert space structure provided by the expectation
is added to the picture.  This leads to the description of properties of W*-algebras
and the structure of finite dimensional algebras.
In \fsec{S:op-represent}, after considering the important topic on how to formally
describe observations, and connecting this with the question of whether it is
possible to assign probabilities to events independent of observations --- which
is given a negative answer by the Bell-Kochen-Specker Theorem --- the focus shifts
to generalised observations or \emph{positive operator valued measures} (POVMs)
and the important topic of completely positive
maps and channels, which is the general description of information transmission.
A further example of the algebraic point of view is the connections with Krylov subspaces, 
orthogonal polynomials, and even the ladder operators from quantum mechanics (QM), 
including creation and annihilation operators and the (interacting) Fock space,
which again will be explored in its simplest setting of analysing just one
RV. % in \fsec{S:anal-one-rv}.
Quantum computing (QC)  and quantum information (QI) emerge as important subjects which
can be approached within this framework in its simplest purely algebraic setting in
\fsec{S:basic-QC}.  Here the important points from the previous development are collected
and used to define a \emph{quantum processing unit} (QPU) in an abstract way, and
to show how a program for such a QPU is formulated.  As an example,  the
Grover algorithm is described in this abstract setting in detail.
We conclude with \fsec{S:concl}.

%  $Log: introduction_QC-alg.tex,v $
%  Revision 1.20  2026/01/26 09:10:38  hgm
%  Completed plan of paper
%
%  Revision 1.19  2025/12/22 22:27:42  hgm
%  little changes
%
%  Revision 1.18  2025/11/24 18:25:15  hgm
%  tiny change
%
%  Revision 1.17  2025/07/08 21:45:26  hgm
%  little changes
%
%  Revision 1.16  2025/05/05 20:45:54  hgm
%  little changes
%
%  Revision 1.15  2025/04/17 00:10:11  hgm
%  little changes
%
%  Revision 1.14  2025/04/09 17:26:23  hgm
%  little changes
%
%  Revision 1.13  2025/04/03 21:21:59  hgm
%  commented text taken out
%
%  Revision 1.12  2025/03/30 20:55:44  hgm
%  little changes
%
%  Revision 1.11  2025/03/05 23:51:43  hgm
%  small changes
%
%  Revision 1.10  2025/01/16 20:27:49  hgm
%  small changes
%
%  Revision 1.9  2024/08/03 21:24:02  hgm
%  small changes
%
%  Revision 1.8  2024/07/19 10:29:27  hgm
%  little changes
%
%  Revision 1.7  2024/05/09 09:49:55  hgm
%  some changes
%
%  Revision 1.6  2024/04/20 08:04:49  hgm
%  little changes
%
%  Revision 1.5  2024/04/11 07:07:33  hgm
%  * almost finished *
%
%  Revision 1.4  2024/04/08 09:41:40  hgm
%  almost finished
%
%  Revision 1.3  2024/03/28 15:41:25  hgm
%  almost finished
%
%  Revision 1.2  2023/12/27 07:50:11  hgm
%  first sketch
%
%  Revision 1.1  2023/12/07 16:23:56  hgm
%  from introduction_RV-alg.tex
%
%  Revision 1.0  2022/05/16 20:10:52  hgm
%  inital check in, from introduction_QC-alg
%
%
%
%
%
%

%%% Local Variables: 
%%% mode: latex
%%% TeX-master: "../23_QC-algebra"
%%% End: 

%
\clearpage % opcional si quieres forzar que sea una página limpia
% ============================================================================
% Notation (1 page, matched to the paper's macros: \Alg, \Clg, \LHA, \Hvk, \EA,
% \SA, \DEN, \Ex, \tr, \ket/\bra, etc.)
% Requires: \usepackage{multicol,paralist} already in 23_QC-algebra.tex
% Suggested placement: after the Introduction (or at the start of Section 2).
% ============================================================================

\section*{Notation}
\addcontentsline{toc}{section}{Notation}

\begingroup
\small
\setlength{\parindent}{0pt}
\setlength{\columnsep}{18pt}

\begin{multicols}{2}
\begin{compactdesc}

\item[$\CC,\RR,\NN$]
Complex, real, natural numbers. Finite-dimensional Hilbert spaces are over $\CC$.

\item[$\Hvk$]
A (finite-dimensional, unless stated otherwise) complex Hilbert space.
The full operator algebra is $\LHA := \Lop(\Hvk)$ (bounded linear maps).

\item[$\Alg$]
A unital $*$-algebra (probability algebra). The unit and zero are denoted by
$\rone$ and $\rnul$. The involution is $\star$ (for matrices/operators this coincides
with the adjoint $\dagger$).

\item[$\Alg_{sa}$]
Self-adjoint (``real'') elements of $\Alg$: $\Alg_{sa}=\{\ra\in\Alg:\ra^\star=\ra\}$.
These are the observables.

\item[$\Clg \subseteq \Alg$]
A commutative (Abelian) $*$-subalgebra (a \emph{classical context}).
Its commutant is $\Clg'=\{\rb\in\Alg:\rb\ra=\ra\rb\ \forall\,\ra\in\Clg\}$.
A maximal Abelian $*$-subalgebra is a \textsc{MASA}.

\item[$\ra,\rb,\rx$]
Generic random variables (RVs) in $\Alg$; typically $\rx\in\Alg_{sa}$ for a real RV.
The spectrum of $\rx$ is denoted by $\sigma(\rx)$ (in a chosen representation).

\item[$\SA{\Alg}$]
States on $\Alg$: positive linear functionals $\svpi:\Alg\to\CC$ with $\svpi(\rone)=1$.
Expectation of $\ra$ in $\svpi$ is
\[
\Ex_{\svpi}(\ra) := \svpi(\ra)=\ip{\svpi}{\ra},
\]
where $\ip{\cdot}{\cdot}$ is the duality pairing on $\Alg^\star\times \Alg$.

\item[$\EA{\Alg}$]
Effects in $\Alg$: $\EA{\Alg}:=\{\re\in\Alg_{sa}:\rnul\le \re \le \rone\}$.
(Sharp events are typically projections; non-sharp events are general effects.)

\item[$\DEN{\Alg}_1$]
Densities (finite dimension): $\DEN{\Alg}_1:=\{\rho\in\Alg:\rho\succeq 0,\ \tr(\rho)=1\}$.
For $\Alg=\LHA\simeq M_d(\CC)$, states are represented by density matrices via
$\svpi_\rho(\ra)=\tr(\rho\,\ra)$.

\item[$\tr$]
Trace (finite dimension). The canonical faithful inner product is written
$\bkt{\ra}{\rb}_c$ (typically induced by $\tr$, e.g.\ $\bkt{\ra}{\rb}_c=\tr(\ra^\star\rb)$
in matrix realizations).

\item[$\Lop(U,V)$]
Linear maps between vector spaces $U,V$; write $\Lop(U)=\Lop(U,U)$.
Completely positive maps/channels appear as linear maps on operator algebras.

\item[$\Uppi_{\Hvk}$ (POVM), $\vP_{\Hvk}$ (PVM)]
A POVM on a finite outcome set $X$ is a map $\Uppi_{\Hvk}:2^X\to \EA{\LHA}$ with
$\Uppi_{\Hvk}(X)=I_{\Hvk}$ and countable additivity (finite here);
writing $\Uppi_{\Hvk}(\{i\})=E_i$, one has $E_i\succeq 0$ and $\sum_i E_i=I_{\Hvk}$.
A PVM $\vP_{\Hvk}$ is the projection-valued special case.

\item[$\E{F}:\Clg\to\LHA$]
The (linear) \emph{observation channel} associated with a POVM:
for $\vf=(f_1,\dots,f_n)\in \Clg\simeq \Cn$,
$\E{F}(\vf)=\sum_i f_i E_i$.

\item[Dirac notation]
$\ket{\psi}\in\Hvk$, $\bra{\psi}:=\ket{\psi}^\dagger$,
rank-one operators $\ket{\psi}\bra{\phi}\in\LHA$.

\item[$n$ qubits]
$\Hvk_1\simeq \CC^2$ with computational basis $\{\ket{0},\ket{1}\}$.
For $n$ qubits: $\Hvk_n:=(\CC^2)^{\otimes n}$ and $\Lop(\Hvk_n)\simeq M_{2^n}(\CC)$.

\end{compactdesc}
\end{multicols}

\vspace{-0.3em}
\noindent\emph{Standing convention.} Unless explicitly stated otherwise, we work in
finite dimensions, so that states admit density (trace) representations and all traces
(partial traces, where used) are well-defined.

\endgroup
\clearpage % opcional

\section{Algebraic Formulation of Probability} \label{S:formal}
% !TEX root = ../23_QC-algebra.tex
% !TEX encoding = UTF-8 Unicode
% RCSID:       $Id: rv-algebra_QC-alg.tex,v 1.34 2026/01/29 19:36:25 hgm Exp $
% Author:      $Author: hgm $
% Contact:     wire@tu-bs.de
% =================================

In the sequel we shall be mainly be concerned with numerical random variables (RVs),
as other RVs can be treated in terms of these.
Starting with the algebraic ideas on probability by focusing on random variables
and expectation / states, we first describe the basics of this algebraic language
and give some examples.  The first observation is that the algebra may be represented
by an algebra of linear maps --- the regular representation.   The expectation allows 
one to introduce an inner product and a norm based on it.  Now one has linear maps
on a Hilbert space --- the GNS-construction alluded to in \fsec{S:intro} --- 
and one may employ the spectral calculus of self-adjoint maps.  
This leads to further representations, as well as new norms and the $\Lp_p$-spaces.  
Additionally, correlation and independence, notions central
to probability, are given a geometrical interpretation.

Taking a closer look at the expectation entails examining the entire set of 
possible states.  
Observations or measurements naturally lead to non-commutative phenomena such as 
Heisenberg’s uncertainty relation and the Bell-Kochen–Specker theorem.  
All of these ideas converge on the notion of a duality between states and effects.

Observations or samples of a random variable (RV) will turn out to be elements of 
the dual space, specifically with values in the suitably defined spectrum of the RV.  
The algebraic structure also encompasses the idea of forming polynomials of RVs, so a 
key question is whether and how to extend polynomial operations to more general 
functional operations on RVs.  In a representation of probability algebras as linear 
maps on Hilbert spaces, real RVs --- viewed as the ``true'' observables --- correspond 
to self-adjoint operators, and the problem of computing functions of these RVs is 
then one of spectral calculus.

In infinite dimensional contexts, requiring the probability algebra to be a Banach or 
C*-algebra (see, for example, \citep{Sakai1971, Naimark1972, Arveson1976, Dix-C, segalKunze78,
Takesaki1, DorBel1986, BrattRob-1, Conway1990, Davidson1996, KadiRingr1-97, Blackadar2006}) 
enables holomorphic and continuous versions of the spectral calculus; for von Neumann or 
W*-algebras, one can also employ a Borel-measurable spectral calculus.  However, restricting
our attention to finite dimensional algebras means each RV can only attain finitely many 
(spectral) values.  Since any function on a finite set of real points can be interpolated 
by a polynomial, the polynomial functional calculus is fully adequate here, and more 
advanced constructions can remain in the background.  Basic information on algebras
in the context of finite dimensional linear algebra can be found in \citep{GreubLA1975}.  
A standard reference on finite dimensional associative algebras is \citep{DrozdKiri94}, 
and focusing on matrix algebras, one should mention \citep{HiaiPetz2014}.

We also look at how to combine probabilistic systems by looking at probability algebras 
in direct sums resp.\ direct products as well as tensor products. 
Finally, the general structure of such finite dimensional algebras is given
--- concretely as a faithful representation in a direct sum of matrix algebras.

%We additionally consider how to combine probabilistic systems by studying direct 
%and tensor products of probability algebras.  Concluding this section, we present 
%the general structure of such finite dimensional algebras, concretely realised via 
%faithful representations as direct sums of matrix algebras.

\subsection{Basics}  \label{SS:basics}
As mentioned in \fsec{S:intro}, the Bernoullis may be seen as early proponents
of an algebraic approach to probability.  They implicitly handled random variables
(RVs) like numbers; specifically, RVs could be added to each other, and they could
be multiplied by numbers and with each other following the usual rules of arithmetic.

\paragraph{Introduction to Probability Algebras:}
In modern mathematical language \citep{segal54-AJM, segalKunze78}, the Bernoullis' 
random variables (RVs) can be viewed as elements of a vector space \(\Alg\).  
Moreover, these elements can be multiplied by one another --- denoted \(\ra \cdot \rb\) 
for \(\ra, \rb \in \Alg\), but mostly, and also in the following except for these 
first few remarks, just denoted as a juxtaposition: $\ra \rb = \ra \cdot \rb$ --- 
and in the classical Bernoulli setting, this multiplication 
is commutative (\(\ra \cdot \rb = \rb \cdot \ra\)).  Constants, i.e.\ ordinary numbers, 
appear as a degenerate kind of random variable.  In contemporary terms, this means 
\(\Alg\) is not only a vector space, but also an associative, distributive, 
commutative (Abelian) algebra, whose unit element \(\rone\) represents the 
multiplicative identity: \(\rone \cdot \ra = \ra \cdot \rone = \ra\).  

The underlying idea is that if \(\ra, \rb \in \Alg\) are random variables, then 
expressions such as \(2\ra\), \(\ra + \rb\), and \(\ra \rb\) are themselves valid 
random variables.  If 
$\C{S}, \C{T} \subseteq \Alg$ are subsets of elements of $\Alg$, then
$\C{S} \C{T} := \C{S} \cdot \C{T} :=\{ \sum_j \rs_j \rt_j \mid \rs_j \in\C{S},\,\rt_j \in\C{T}\}
 \subseteq \Alg$ is a shorthand for the set of all sums of products where the first 
factor is in $\C{S}$ and the second one in $\C{T}$.
 
Historically, the Bernoullis worked with what in modern terms would be called 
\emph{expectation}, an \emph{averaging} procedure or \emph{mean}, 
that Laplace famously called ``mathematical hope.''  In more recent physics 
terminology, it is often referred to as a \emph{state}.  Mathematically, 
this expectation is a linear functional \(\svpi\) on \(\Alg\), i.e.\ 
an element of the dual space \(\Alg^\star\).  

Just as numbers themselves can be ``positive,'' random variables admit a partial 
order \(\ra \ge \rnul\), with \(\rone \ge \rnul\) and squares \(\rb^2 := \rb \rb\) 
being positive by definition.  If \(\ra\) is positive, \(\svpi(\ra)\) is 
non-negative, and therefore it is a \emph{positive} functional; 
moreover, \(\svpi\) takes the unit \(\rone \in \Alg\) to \(1 \in \RR\), 
making \(\svpi\) a \emph{normalised state}.
Note that one should really say that $\svpi$ is \emph{non-negative} or 
\emph{positive semi-definite}, but just as in the case of matrices and operators, 
or general elements $\ra \in \Alg$  of an abstract algebra $\Alg$, 
we are lazy and just say positive.  If one these really has to be positive and 
not just non-negative, we say \emph{positive definite}.

Random variables also have --- at least in the commutative setting --- a \emph{value},
that can be observed, measured, or sampled.  Mathematically, such an observation is 
again a linear functional \(\som \in \Alg^\star\), in fact it has reasonably to
satisfy all requirements from above to be a state.   One may regard the observational 
state \(\som\) as the ``state of the world'' at the time of observation.  
Crucially, \(\som\) also satisfies \(\som(\ra \rb) = \som(\ra) \som(\rb)\), meaning 
that the observed value of a product \(\ra \rb\) is the product of the observations of 
\(\ra\) and \(\rb\).  For now, no distinction is made between a random variable’s 
``true'' value and the uncertainty involved in measuring it; that refinement appears later.  

In modern language, these conditions mean that \(\som\) is a non-zero algebra homomorphism 
from \(\Alg\) into the base field, also called a \emph{character} or \emph{pure state}.  
Under reasonable conditions, such characters exist --- particularly in finite dimensional 
algebras.  As both \(\svpi\) and \(\som\) are elements of \(\Alg^\star\), one sometimes writes
\[
\svpi(\ra) = \ip{\svpi}{\ra} \quad\text{ and }\quad
\som(\ra) = \ip{\som}{\ra},
\]
and one may also write \(\ra(\som) := \ip{\som}{\ra}\) to emphasise classical sampling notation.

The Bernoullis used the real numbers as their underlying number field.  While imaginary 
and complex numbers were already known (with Cardano having used them in solving 
polynomial equations), it is possible to formalise Abelian probability algebras over 
the real field alone \citep{segal54-AJM}.  Yet, in keeping with the maxim of Painlevé 
and Hadamard --- that the shortest path between two truths in the real domain passes through 
the complex domain --- one often extends a real algebra \(\Alg_{\RR}\) to a complex 
one \(\Alg\).  Usually one sets
\[
         \Alg := \Alg_{\RR} \oplus \ii\, \Alg_{\RR}, % = \Alg_{\RR} \otimes \CC,
\]
with elements \(\ra + \ii\rb\), for \(\ra,\rb \in \Alg_{\RR}\) and imaginary 
unit $\ii \in \CC$, and the usual rules of complex arithmetic.  
This extension allows states and observations to be extended by complex linearity 
and also paves the way for non-commutative probability, as to deal with quantum-like behaviour 
(QLB) with non-commutative algebras, it is usually deemed necessary to consider complex algebras.
The real observables of primary 
interest (with real samples) are precisely the \emph{self-adjoint} elements of \(\Alg\), 
while the other elements can be seen as a convenient ``algebraic completion'' that 
makes the theory more elegant.

To accommodate complex conjugation \(\CC \ni z \mapsto z^*\), one needs an analogous 
operation in \(\Alg\).  Thus we require an \emph{adjunction} or \emph{involution} 
\(\star: \Alg \to \Alg\), \(\ra \mapsto \ra^\star\), satisfying
\begin{equation}  \label{eq:def-invo}
  (\ra^\star)^\star = \ra, 
  \quad 
  (\ra \rb)^\star = \rb^\star \ra^\star,
  \quad
  (z \ra + \rb)^\star = z^* \ra^\star + \rb^\star,
\end{equation}
for all \(\ra,\rb \in \Alg\) and \(z \in \CC\).  An algebra with this structure is
called an \emph{involutive algebra} or *-algebra.  In many familiar cases
(e.g.\ real matrix algebras), a natural involution exists (the transpose in the
real matrix case) --- which for the
moment will be denoted by \(\natural: \Alg_{\RR} \to \Alg_{\RR}\), 
\(\ra \mapsto \ra^\natural \) for \( \ra \in \Alg_{\RR} \) --- but one may also 
have trivial involutions like the identity.  If no non-trivial involution exists on 
\(\Alg_{\RR}\), one may define \(\ra^\natural = \ra\) for all \(\ra \in \Alg_{\RR}\).
Taking care of the natural involution, one then defines
\begin{equation}  \label{eq:def-invo-2}
    \bigl(\ra + \ii\,\rb\bigr)^\star := \ra^\natural - \ii\,\rb^\natural . 
\end{equation}
The second identity in \feq{eq:def-invo} is already formulated in such a way that
it can also be used for non-commutative algebras.  Elements which satisfy
$\ra = \ra^\star$ are called \emph{self-adjoint} or \emph{Hermitean}, 
whereas if they satisfy $\ra = -\ra^\star$, they are called \emph{skew-adjoint} or
\emph{skew-Hermitean}.  The constants $\rone, \rnul$ are self-adjoint, and every 
element of the  form $\ra^\star \ra$ or $\ra + \ra^\star$  is self-adjoint.   
For observations and states one requires *-linearity:
\(\som(\ra^\star) = \som(\ra)^*\) and \(\svpi(\ra^\star) = \svpi(\ra)^*\),  
ensuring real values on self-adjoint elements.

\begin{defi}[State, Expectation]     \label{def:state}
  Let \(\Alg\) be a unital associative *-algebra (not necessarily commutative).  
  A linear functional \(\svpi \in \Alg^\star\) is called a \emph{state} if it is  
%  \begin{itemize}
%    \item \textbf{Positive:} \(\svpi(\ra^\star \ra) \ge 0\) for all \(\ra \in \Alg\), and  
%    \item \textbf{Hermitean (self-adjoint):} \(\svpi(\ra^\star) = \svpi(\ra)^*\).  
%  \end{itemize}
  \begin{compactdesc}
    \item[Positive:] \(\svpi(\ra^\star \ra) \ge 0\) for all \(\ra \in \Alg\), and  
    \item[Hermitean (self-adjoint):] \(\svpi(\ra^\star) = \svpi(\ra)^*\).  
  \end{compactdesc}
  If additionally \(\svpi(\rone) = 1\),  \(\svpi\) is \emph{normalised}.  
  A state \(\svpi\) is \emph{faithful} when \(\svpi(\ra^\star \ra) = 0\)
   \emph{iff} \(\ra = \rnul\).  
  If \(\svpi(\ra \rb) = \svpi(\rb \ra)\) for all \(\ra,\rb \in \Alg\), then \(\svpi\) is 
  called a \emph{tracial} state or a \emph{trace}.
  
  The set of all normalised states is denoted by \(\SA{\Alg}\subset \Alg^\star\).  
  For \(\ra \in \Alg\), its \emph{expectation} under the state \(\svpi\) is denoted as 
  \[
    \Ex_{\svpi}(\ra) \;:=\; \ipj{\ra}_{\svpi} \;:=\; \svpi(\ra)  \;=\; \ip{\svpi}{\ra}.
  \]
\end{defi}

\begin{rem}  \label{rem:normalised-positive}
In finite dimensional algebras \(\Alg\) (and more generally \(\Ck^*\)-algebras), 
requiring \(\svpi(\rone) = 1\) already ensures \(\svpi\) is Hermitean and positive.  
\end{rem}

Now we are able to define:

\begin{defi}[Probability Algebra]\label{def:comm-prob-alg}
A \emph{probability algebra} is a complex, associative, distributive, and unital 
*-algebra \(\Alg\) (not necessarily commutative), equipped with a normalised state 
\(\svpi \in \SA{\Alg}\).  This will be denoted as the tuple \((\Alg,\svpi)\).
\end{defi}

One must distinguish between the elements of a probability algebra --- 
which are random variables (RVs) having expectations and thus constituting a probability 
law --- and the actual observables, which can be measured in different states.  As noted, 
only self-adjoint observables and random variables can truly be observed to yield real values.  
To formalise this distinction, we first specify what is meant by a *-homomorphism:

\begin{defi}[*-Homomorphism]\label{def:star_homomorphism}
Let \(\Blg\) and \(\Alg\) be two complex unital *-algebras. A \emph{linear} map 
$ \Phi: \Blg \;\to\; \Alg $ is called a \emph{*-homomorphism} if it is:
%\begin{enumerate}
%\item \textbf{(Linearity):} \(\phi(\ra + \rb) \;=\; \phi(\ra) \;+\; \phi(\rb)\) 
%      and \(\phi(z\,\ra) \;=\; z\,\phi(\ra)\) for all \(\ra,\rb \in \Blg\) and \(z \in \CC\).
%\item \textbf{(Multiplicativity):} \(\phi(\ra\,\rb) \;=\; \phi(\ra)\,\phi(\rb)\) 
%      for all \(\ra,\rb \in \Blg\).
%\item \textbf{(Involution-Preserving):} \(\phi(\ra^\star) \;=\; \phi(\ra)^\star\) 
%       for all \(\ra \in \Blg\).
%\end{enumerate}
\begin{compactdesc}
%\item[Linear:] \(\phi(\ra + \rb) \;=\; \phi(\ra) \;+\; \phi(\rb)\) 
%      and \(\phi(z\,\ra) \;=\; z\,\phi(\ra)\) for all \(\ra,\rb \in \Blg\) and \(z \in \CC\).
\item[Multiplicative:] \(\Phi(\ra\,\rb) \;=\; \Phi(\ra)\,\Phi(\rb)\) 
      for all \(\ra,\rb \in \Blg\).
\item[Involution-Preserving:] \(\Phi(\ra^\star) \;=\; \Phi(\ra)^\star\) 
       for all \(\ra \in \Blg\).
\end{compactdesc}
\end{defi}

With this in hand, one may define random variables abstractly:

\begin{defi}[Random Variable (RV)]  \label{def:rand-var}
A random variable  $\rR:\Blg \to \Alg$ is defined to be a *-homomorphism from an %complex, 
associative, distributive, and unital *-algebra $\Blg$ into a probability 
algebra $(\Alg, \svpi)$, cf.\ \feD{def:comm-prob-alg}. 
\end{defi}

The elements \(\rb \in \Blg\) --- at least the self-adjoint ones --- may thus be called
\emph{observables}.  This separates the observables \(\rb \in \Blg\) from the state 
\(\svpi\), which determines the probability law (see \feD{def:law_RV}).  
In particular, the state can change --- for example, when measurement or sampling 
(\feD{def:sample}) occurs, or when it is updated due to new information,
see \feX{ex:instr-cond} and \feX{ex:CEX-inst}.
%(cf.\ \feT{thm:Bayes}).  
Formally, such a change modifies the probability algebra 
from \((\Alg, \svpi)\) to \((\Alg, \svpi_{\sbt})\).  However, the observables 
\(\rb \in \Blg\) remain the same, even though the state and thus the probability law has changed.  

Often, the random variable is just the identity map \(\rR = \rI : \Alg \to \Alg\).  
In that situation, one may identify the observable with the RV itself and treat both 
as an element of \(\Alg\).  This identification is the most common scenario in the examples here.

\begin{defi}[Sample, Observation, Character]  \label{def:sample}
A \emph{sample} or \emph{observation} \(\som \in \SA{\Alg} \subset \Alg^\star\) 
is a \emph{character}, i.e.\ a non-zero *-homomorphism from \(\Alg\) into \(\CC\),
cf.\ \feD{def:star_homomorphism}.
Equivalently, it is a multiplicative state (and necessarily normalised).  
Denote the set of all characters by \(\XA{\Alg} \subset \SA{\Alg} \subset \Alg^\star\).  
In general, for a given \(\Blg\), it may happen that \(\XA{\Blg} = \emptyset\), 
i.e.\ there are no characters.
For an Abelian *-algebra \(\Alg\), the set of all characters \(\XA{\Alg}\) is also 
called the \emph{spectrum} of \(\Alg\).
\end{defi}

The notion of spectrum is pivotal in non-commutative probability.  We will return 
to it later, but for now we introduce the spectrum of a single random variable:

\begin{defi}[Spectrum]  \label{def:spec-one}
Let $\Alg$ be a unital algebra.
For \(\ra \in \Alg\), the \emph{spectrum} of \(\ra\) is defined as
\[
   \sigma(\ra) := \{ \lambda \in \CC \;\mid\; (\ra - \lambda \rone)\,
     \text{ is not invertible in } \,\Alg \}.
\]
(For the definition of the inverse \(\rb^{-1}\) of \(\rb \in \Alg\), such that 
\(\rb \rb^{-1} = \rb^{-1}\rb = \rone\), see \feD{def:alg-names}.)
The number $\vrho(\ra) = \sup \{ \ns{\lambda} \mid\, \lambda \in \sigma(\ra) \}$
is called the \emph{spectral radius}
\end{defi}

If \(\CC[\ra]\) denotes the algebra generated by \(\ra \in \Alg\) (all complex coefficient 
polynomials in $\ra$), we immediately get:

\begin{prop}[Spectrum and Sample]  \label{prop:spec-one}
If \(\som\) is a non-zero *-character (i.e.\ an observation or sample, cf.\ \feD{def:sample}), 
then no element of the form \(\ra - \som(\ra)\rone\) is invertible.  Hence, for any 
\(\som \in \XA{\CC[\ra]}\), one has \(\som(\ra) \in \sigma(\ra)\).
\end{prop}

\begin{proof}
If \(\rc \in \Alg\) is invertible, we have \(1 = \som(\rone) = \som(\rc \rc^{-1}) = \som(\rc)\,
\som(\rc^{-1})\).  Thus, \(\som(\rc)\neq 0\).  But for $\rc = \ra - \som(\ra)\rone$ one
has  \(\som(\rc)=\som(\ra - \som(\ra)\rone) = \som(\ra) - \som(\ra)\som(\rone) = 0\).  
Hence, \(\rc = \ra - \som(\ra)\rone\) cannot be invertible, implying \(\som(\ra)\in\sigma(\ra)\).
\end{proof}

What is seen here in \feP{prop:spec-one} is an important point, namely that the
values of observations like $\som(\ra)$ have to be elements of the spectrum $\sigma(\ra)$ 
of the observable $\ra \in \Alg$.

\begin{defi}[Commutator, Lie Algebra]   \label{def:commutat}
For an associative algebra \(\Alg\), the \emph{commutator} of two elements is
\[
   [\ra, \rb] := \ra\rb - \rb\ra, \quad \ra, \rb \in \Alg.
\]
This mapping \([\cdot,\cdot]: \Alg \times \Alg \to \Alg\) is antisymmetric and 
non-associative, but is linear in each argument, i.e.\ it is a product.  
The vector space \(\Alg\) with this product is a \emph{Lie algebra}.
If \([\ra,\rb] = \rnul\), then \(\ra\) and \(\rb\) commute.
\end{defi}

Having laid out the underlying motivation and definitions, we turn to examples
--- focusing first on commutative algebras.  As Paul Halmos advised, one should gather 
as many examples as possible; following David Hilbert, we start with the simplest ones.

\begin{xmpn}   \label{ex:complex-num}
Consider \(\CC\) as a complex algebra over itself, with involution given by complex 
conjugation.  Define a state by the identity functional \(\svpi_\CC(z)=z\).  
This yields a commutative probability algebra.  A sample or observation 
(i.e.\ a character) is simply the identity map \(\som_\CC(z) = z\).
\end{xmpn}

Allthough this algebra is ``trivial'' in the sense that its elements are not really random, 
it shows that the framework is not dealing with the empty set.

\begin{xmpn}[Hadamard Algebra]   \label{ex:complex-n-spc}
Consider \(\Cn\) with entry-wise complex conjugation as the involution.  
Define the multiplication \(\vy \odot \vz\) of two vectors \(\vy, \vz \in \Cn \) 
component-wise (this is called the \emph{Hadamard} product, although apparently
it goes back to \emph{Schur} \citep{Hackbusch_tensor}): 
\[
(\vy \odot \vz)_j := y_j z_j; \; j=1,\dots,n .
\]
The multiplicative identity is \(\ve=(1,1,\dots,1)\).  Note that this is also an
inner product space --- a finite dimensional Hilbert space --- with canonical inner product
$\bkt{\vy}{\vz}_c := \sum_j y^*_j z_j$.  Thus, for any \(\vz\in\Cn\), every 
linear functional on \(\Cn\) has the form \(\vz\mapsto \bkt{\vr}{\vz}_c\) for some 
\(\vr\in\Cn\).  One can define the \emph{canonical state} \(\sphi_{c}\in \SA{\Cn}\) by 
\(\sphi_{c}(\vz) = \frk{1}{n}\,\bkt{\ve}{\vz}_c\).  This is a commutative probability algebra corresponding to a uniform distribution on the index set 
\([n] := \{ j \in \D{N} \mid j \le n\} \); the vector $\frk{1}{n}\, \ve$ can be called a 
\emph{density} or a (uniform) discrete \emph{probability distribution} on $[n]$.

More generally, any discrete probability distribution or density \(\vrh=(\rho_i)\) on \([n]\) 
(\(\rho_i\ge 0\), \(\bkt{\vrh}{\ve}_c =1\)) defines a state 
\(\svpi_{\vrh}(\vz)=\bkt{\vrh}{\vz}_c\).  
The set of samples or observations (characters) is given by 
\(\som_i(\vz)=z_i=\bkt{\ve_i}{\vz}_c\), using the standard basis vectors \(\ve_i\).

An alternative viewpoint is to take a finite set \(\C{X}\) of cardinality \(n\).  
The \emph{algebra of functions} \(\C{F}(\C{X},\CC)\) under point-wise operations is isomorphic 
to \(\Cn\) via any enumeration \(\{x_1,\ldots,x_n\} = \C{X}\).  The function algebra
 \(\C{F}(\C{X},\CC)\) may also be seen as the \emph{algebra of continuous functions}
$\Ck(\C{X}, \CC)$ when $\C{X}$ carries the discrete topology.
\end{xmpn}

Clearly, \feX{ex:complex-num} is just the $n=1$ version of the above.

\begin{xmpn}[Polynomial *-Algebra]   \label{ex:polys}
Let \(\CC[\tns{X}]\) be the algebra of polynomials in one indeterminate \(\tns{X}\) with 
complex coefficients, where the involution on \(\tp(\tns{X})=\sum_{k=0}^n \alpha_k \tns{X}^k\) is 
\(\tp(\tns{X})^*=\sum_{k=0}^n \alpha_k^* \tns{X}^k\).  Define a state by 
\(\svpi(\tp(\tns{X}))=\alpha_0\), often written as \(\tp(0)\).  
This is a commutative probability algebra.
\end{xmpn}

\begin{xmpn}[Diagonal Matrix Algebra]   \label{ex:complex-diag}
Let \(\DMn(\CC)\subset \MMn(\CC)\) be the algebra of diagonal matrices over \(\CC\).  
The involution is entry-wise complex conjugation plus transpose \((\cdot)^\tpH\).
Recall that any linear functional on \(\MMn(\CC)\) has the form \(\vD\mapsto \tr(\vV\vD)\) 
for some \(\vV \in \MMn(\CC)\), due to the fact that $\bkt{\vV}{\vM}_F := \tr (\vV^\tpH \vM)$ 
defines the \emph{Frobenius} or \emph{Hilbert-Schmidt} inner product on $\MMn(\CC)$, 
making $\MMn(\CC)$ into a finite dimensional Hilbert space with subspace 
(in fact *-sub-algebra) $\DMn(\CC)$.
  
A \emph{canonical} or \emph{Frobenius} state \(\sphi_F\in \SA{\DMn(\CC)}\) 
is given by \(\sphi_F(\vD)=\frk{1}{n}\,\bkt{\vI}{\vD}_F\), a kind of uniform 
distribution with \emph{density matrix} $\frk{1}{n}\, \vI$.
This is a tracial state, cf.\ \feD{def:state}.
For a density matrix \(\vR=\diag(\vrh)\) with \(\vrh\in\RR^n\) a discrete 
probability distribution, define a state by \(\svpi_{\vR}(\vD)=\bkt{\vR}{\vD}_F\).
Observations (characters) take the form \(\som_i(\vD)=D_{ii}\).
This algebra is isomorphic to the component-wise or Hadamard algebra in \feX{ex:complex-n-spc}.
\end{xmpn}

\begin{xmpn}[Projection Algebra]   \label{ex:projections}
Let \(R = \{\vP_j\}_{j=1,\dots,m}\) be a set of commuting orthogonal 
projections in \(\MMn(\CC)\), including the identity \(\vI\).  
Thus \(\vP_j = \vP_j^\tpH = \vP_j^2\), and 
\([\vP_i,\vP_j] = \vek{0}\).  Using the same Frobenius 
state $\sphi_F$ as in \feX{ex:complex-diag}, then \(\CC[R]\) 
forms a finite dimensional commutative probability algebra.  Note that 
\[
\sphi_F(\vP_i) = \frk{1}{n}\, \bkt{\vI}{\vP_i}_F = \frk{1}{n}\,\tr \vP_i 
   = \frk{1}{n}\, \rank \vP_i .
\]
In case that $\vP_j - \vP_i$ is positive (semi-definite), denote this as 
$\vP_i \le \vP_j$.  This is a partial order on $\CC[R]$ (and on $\MMn(\CC)$).  
A non-zero projection $\vek{0} < \vP \in \CC[R]$ is called \emph{minimal}, 
if for any other projection $\vQ \in \CC[R]$ such that $\vQ \le \vP$, it holds
that either $\vQ = \vek{0}$ or $\vQ = \vP$.  

If both $\vP, \vQ  \in \CC[R]$ are
projections, then so is $\vR = \vP \vQ = \vQ \vP$, and one has $\vR \le \vP$ as 
well as $\vR \le \vQ$.  It is easily seen that $\vR_P = \vP - \vR = 
\vP (\vI - \vQ)$ and $\vR_Q = \vQ - \vR = \vQ (\vI - \vP)$ are also projections, 
that they are orthogonal: $\vR_P \vR_Q = \vek{0}$, and that they are orthogonal to
$\vR = \vP \vQ$: $\vR \vR_P = \vR \vR_Q = \vek{0}$.  Observe that the sum of these
$\vS := \vR_P + \vR_Q + \vR = \vP + \vQ - \vP \vQ$ is again a projection.  Using 
the notation from \feP{prop:boolean}, this may be written as
\[
   \vR = \vP \sqcap \vQ, \; \text{ and } \; \vS = \vP \sqcup \vQ = \vR_P + \vR_Q + \vR.
\]

  Note that in such a way one can find minimal projections 
  $\{ \vR_1, \dots, \vR_d \}=:B \subseteq  \CC[R]$, 
  mutually orthogonal to each other ($\vR_k \vR_j = \updelta_{kj} \vR_k$) ---
  where $\updelta_{ij}$ is the \emph{Kronecker}-$\updelta$ --- with 
  $d = \dim (\CC[R])$, which form a basis for $\CC[R]$, and 
  \emph{partition of unity}: $\sum_{k=1}^d \vR_k = \vI$.
\end{xmpn}

\begin{xmpn}[Commutative Matrix Algebra]   \label{ex:sa-mat}
If \(\vA = \vA^\tpH \in\MMn(\CC)_{sa}\) is self-adjoint, consider the commutative *-algebra 
\(\CC[\vA]\subset \MMn(\CC)\) generated by \(\vA\) and \(\vI\).  
With the Frobenius state from \feX{ex:complex-diag}, this is a commutative 
probability algebra.
\end{xmpn}

\begin{rem}  \label{rem:spec-dec-proj}
Any self-adjoint \(\vA\) can be diagonalised: \(\vLbd=\vU^*\vA\vU \in \DMn(\RR)\), 
for some unitary \(\vU \in \UA{\MMn(\CC)}\).  
Thus any function \(f(\vA)\) (particularly polynomials) 
shares the same diagonalising unitary.  Indeed, \(\vA\) has a spectral decomposition 
\(\vA=\sum_k\lambda_k\vP_k\) with \(\lambda_k\in\sigma(\vA)\subset\RR\) and with 
orthogonal projections $\vP_k$ onto the eigenspace of $\lambda_k$.  
The \(\vP_k\) are a \emph{partition of unity}
(cf.\ \feX{ex:projections}).  
Then \(f(\vA)=\sum_k f(\lambda_k)\vP_k\), and \(\CC[\vA]=\CC[\{\vP_k\}] =
\spn_\CC\{\vP_k\}\),
cf.\ \feX{ex:sa-mat}. 
\end{rem}

In his influential monograph \citep{Kolmogo1933, Kolmogo1950-en}, Kolmogorov begins 
with finite sample spaces and random variables taking finitely many values.  
The examples above reflect precisely this setting, and we remain mostly at that 
level here --- particularly when venturing into non-commutative territory ---
although we sometimes peek at infinite dimensional cases.

Commutative or Abelian algebras are essentially function algebras, and the
following \feX{ex:comm-fct-algebra} is an extension of the Hadamard algebra 
described at the end of \feX{ex:complex-n-spc}:

\begin{xmpn}[Function Algebra]   \label{ex:comm-fct-algebra}
For any non-empty set \(\C{X}\), 
consider \(\C{F}(\C{X},\CC)\), the vector space of complex-valued functions on \(\C{X}\).
Point-wise multiplication and complex conjugation form a commutative *-algebra 
structure with point-wise operations.  The algebraic unit is the 
function $\bbbone_{\C{X}} \equiv 1$.  If the cardinality of $\C{X}$ is finite,
this is the Hadamard algebra of \feX{ex:complex-n-spc}, otherwise
this is an infinite dimensional algebra.

States may be added to this in various ways, often by requiring possibly extra conditions 
like measurability or continuity of the functions, like in the more specialised
examples to follow.  
\end{xmpn}

The main importance of the next 
\feX{ex:simple-fcts} is to show that classical Kolmogorovean probability may be 
subsumed into the algebraic framework.

\begin{xmpn}[Algebra of Simple Functions]   \label{ex:simple-fcts}
Let \((\smplspc,\sigalg,\prob)\) be a classical probability space with sample 
space \(\smplspc\), \(\sigma\)-algebra \(\sigalg\), and probability measure 
\(\prob\).  For each \(\C{E}\in\sigalg\), let \(\bbbone_{\C{E}}\) be the indicator 
function.  Identify indicators that differ only on a nullset.  
Define \(\svpi_\prob(\bbbone_{\C{E}})=\prob(\C{E})\).  
Extend \(\svpi_\prob\) linearly to 
\(\Lp_{0 s}(\smplspc,\sigalg,\prob)=\spn_\CC\{\bbbone_{\C{E}}\mid \C{E}\in\sigalg\}\). 
As in \feX{ex:comm-fct-algebra}, with point-wise operations this is a commutative *-algebra, 
and \(\svpi_\prob\) is a state.  Thus \(\Lp_{0 s}(\smplspc,\sigalg,\prob)\) is a 
commutative probability *-algebra.
\end{xmpn}

This is essentially the infinite dimensional analogue of \feX{ex:projections}, 
where \(\bbbone_{\C{E}}\) acts like a projection.  If \(\smplspc\) has finite cardinality
and $\sigalg = 2^{\smplspc}$ is the power set, this is the special case \feX{ex:complex-n-spc}
of \feX{ex:comm-fct-algebra}.  Similarly, for any non-empty set \(\smplspc\), 
if \(\sigalg\) has finite cardinality, 
this reduces to \feX{ex:complex-n-spc}, as $\Lp_{0 s}(\smplspc,\sigalg,\prob)$
is finite dimensional.
Normally one would take completions of this algebra to obtain
mathematically more convenient structures, like in

\begin{xmpn}[Bounded RV Algebra]   \label{ex:Linfty-fcts}
Similarly, in the same measurable setting as \feX{ex:simple-fcts}, 
\(\Lp_\infty(\smplspc,\sigalg,\prob)\) --- the algebra of essentially bounded 
measurable complex functions --- has point-wise algebraic operations and involution by 
conjugation as in \feX{ex:comm-fct-algebra}, with state
\[
\svpi_\prob(f)=\int_{\smplspc}f(\evt)\,\prob(\di \evt).
\]
This is a commutative probability algebra, the completion of \feX{ex:simple-fcts} in the
$\Lp_\infty$-norm.
\end{xmpn}

Of course, \(\Lp_0(\smplspc,\sigalg,\prob)\) --- the *-algebra of all 
measurable complex functions, resp.\ the *-algebra of all classical random
variables (RVs) on $\smplspc$ --- is larger still, but here not every element has a 
finite integral.  A practical intermediate is:

\begin{xmpn}[All-Moments RV Algebra]   \label{ex:Linfty-minus}
Define 
\[
  \Lp_{\infty-}(\smplspc,\sigalg,\prob) 
  = \bigcap_{p\in\D{N}} \Lp_p(\smplspc,\sigalg,\prob),
\]
i.e.\ those random variables with finite moments of every order.  
Here $\Lp_p(\smplspc,\sigalg,\prob)$ is the completion of \feX{ex:simple-fcts} in 
the $\Lp_p$-norm, the \emph{Banach} space of RVs with finite moments up to order $p$.  
This is again a commutative *-algebra under point-wise operations and conjugation
as in \feX{ex:comm-fct-algebra}, with the same state \(\svpi_\prob\) as in
\feX{ex:Linfty-fcts}.  Hence \(\Lp_{\infty-}(\smplspc,\sigalg,\prob)\) is also a 
commutative probability algebra.  One thus has the chain of inclusions of *-algebras
\[
   \Lp_{0 s}(\smplspc,\sigalg,\prob)
   \;\subseteq\;
   \Lp_\infty(\smplspc,\sigalg,\prob)
   \;\subseteq\;
   \Lp_{\infty-}(\smplspc,\sigalg,\prob)
   \;\subseteq\;
   \Lp_0(\smplspc,\sigalg,\prob),
\]
with equality everywhere in the finite dimensional case.
\end{xmpn}

A further important setting arises when \(\sptldom\) is a compact topological space:

\begin{xmpn}[Uniform Algebra]   \label{ex:cont-fcts}
Let \(\sptldom\) be compact and \(\mu\) a finite Radon measure supported on 
\(\sptldom\).  Consider \(\Ck(\sptldom,\CC)\), the algebra of continuous complex 
functions on \(\sptldom\).  Under point-wise operations and conjugation as in 
\feX{ex:comm-fct-algebra}, and with the state 
\[
\svpi_\mu(\vphi)=\frac{1}{\mu(\sptldom)} \, \int_\sptldom\vphi(x)\,\mu(\di x),
\]
this is a commutative probability algebra.  A sample at \(x\in\sptldom\) is 
\(\som_x(\vphi)=\vphi(x)\).  This is a sub-algebra of \(\Lp_\infty(\sptldom,\F{B},\mu)\),
where $\F{B}$ is the Borel $\sigma$-algebra.  If \(\sptldom\) has finite cardinality, 
this reduces to \feX{ex:complex-n-spc}.
\end{xmpn}

Finally, we introduce a commutative sub-algebra that resides in any algebra:

\begin{defi}[Centre of an Algebra]   \label{def:centre}
The \emph{centre} of an algebra \(\Alg\) is 
\[
\ZA{\Alg} := \{\ra\in\Alg\mid[\ra,\rb]=\rnul\;\text{for all}\;\rb\in\Alg\}.
\]
In a unital algebra, we always have \(\CC[\rone]\subseteq \ZA{\Alg}\), where 
\(\CC[\rone]\) denotes the scalar multiples of the identity.
\end{defi}

\begin{xmpn}[Algebra Centre]   \label{ex:centre-alg}
For any probability algebra \(\Alg\), the centre \(\ZA{\Alg}\) is itself a 
commutative probability algebra.  If \(\ra\in\Alg\) is self-adjoint, the 
sub-algebra \(\CC[\ra]\subseteq\Alg\) (all complex polynomials in \(\ra\)) is also 
commutative (cf.\ \feR{rem:spec-dec-proj}).
\end{xmpn}

The smallest commutative sub-algebra is \(\CC[\rone]\), which is isomorphic to \(\CC\)
(\feX{ex:complex-num}); in this sense, \(\CC\) is naturally embedded in every unital algebra.

\paragraph{Non-Commutative Probability Algebra:}
We now extend the foregoing ideas to the non-commutative setting 
\citep{Dirac1930, Neumann1955, segal1947, segalPostQM1947, segal53-AoM, segal56-TAMS, 
segal56-AoM, segal65-BAMS, Naimark1972, Emch1972, segalKunze78, gudderHudson78, Conway1990, 
Meyer1995, Mitchener2005, Khrennikov2010, NielsenChuang2011, BenyRicht15, Accardi2018}.  
Recall from \feq{eq:def-invo} that the involution was defined in a way suitable for 
non-commutative algebras.  In this setting --- often associated with quantum-like behaviour (QLB)
--- the product of random variables (RVs) requires special attention.    Note that
sometimes we will call a commutative or Abelian algebra a ``classical algebra'', and a
non-commutative one a ``QLB-algebra'' (QLB --- Quantum Like Behaviour).

Following \citep{Accardi2018}, one can motivate the product of two RVs \(\ra\) and \(\rb\) 
by interpreting \(\rb\,\ra\) as a temporal ordering in which \(\ra\) is observed before 
\(\rb\).  When \(\rb\,\ra=\ra\,\rb\), the observables \(\ra\) and \(\rb\) are 
\emph{compatible}, implying that their order of observation does not matter and that 
they can effectively be measured simultaneously.  The involution corresponds to a time 
reversal (\feq{eq:def-invo}), so if \(\ra\) is self-adjoint (\(\ra=\ra^\star\)), 
the measurement has no further ``reverse'' aspect —-- such RVs represent the 
elementary observables of interest.

In this non-commutative setting, characters (i.e.\ non-zero *-homomorphisms into \(\CC\)) 
typically fail to exist on the entire algebra unless it is commutative.  As we will see 
(cf.\ the Heisenberg uncertainty relation, \feT{thm:uncert-rel}), non-commuting elements 
\(\ra,\rb\in\Alg\) (with \([\ra,\rb]\neq\rnul\)) are incompatible and cannot be jointly 
observed or measured.

%Most concepts and terminology from the commutative setting carry over, especially those 
%familiar from matrix algebras.
To speak about non-commutative algebras, some concepts and terminology are needed,
and most of them are familiar from matrix algebras.
Below is a concise summary of standard definitions and nomenclature:

\begin{defi}[Basic Definitions]  \label{def:alg-names}
Let \(\Alg\) be a complex associative unital *-algebra with involution \(\star\), 
and let \(\ra,\rb,\rc,\ru,\rv,\rr,\rp,\re,\rf \in \Alg\).  We use the following 
notation and terminology.

\begin{description}
\item[Unit and zero element.]  
The multiplicative unit is \(\rone\), and the zero element is \(\rnul\).

\item[Powers.] 
For \(\ra \in \Alg\) and \(n \in \D{N}_0\), define recursively
\[
\ra^0 := \rone, \qquad \ra^n := \ra \,\ra^{(n-1)} \quad \text{for } n>0.
\]
Clearly, \([\ra^n, \ra^m]=\rnul\) for any \(m,n \in \D{N}\).

\item[Algebra generated by elements.]  
\(\CC[\ra]\subset\Alg\) denotes the \emph{commutative} sub-algebra generated by 
all powers of \(\ra\).  If \(R=\{\ra_j : j\in J\}\) is a set of not necessarily 
commuting elements, 
\[
\CC\langle R \rangle \;=\;\CC\langle \ra_j : j\in J\rangle \;\subset\;\Alg
\]
is the (possibly non-commutative) sub-algebra of all polynomials in the \(\ra_j\), 
respecting the order of arguments.  If the \(\ra_j\) \emph{do} commute pairwise, 
the commutative sub-algebra they generate is written \(\CC[R]\subset\Alg\).

\item[Unital Sub-Algebra.]  Let $\Blg \subset \Alg$ be a sub-algebra, i.e.\ $\Blg$
is a subspace and for all $\rb, \rc \in \Blg$ it holds that $\rb \rc \in \Blg$.
If there is an element $\rone_{\Blg} \in \Blg$ which acts as an identity
on $\Blg$ ($\forall\, \rb \in \Blg:\, \rb \, \rone_{\Blg}=  \rone_{\Blg} \, \rb = \rb$),
then $\Blg$ is called a \emph{unital sub-algebra}.  Note that $\rone_{\Blg} \in \Blg$ 
can be different from $\rone \in \Alg$.

\item[Ideal.] A *-sub-algebra $\C{J} \subseteq \Alg$ is called a \emph{left ideal} 
\emph{iff} $\Alg \C{J} = \C{J}$, a \emph{right ideal} \emph{iff} $\C{J} \Alg = \C{J}$,
and a \emph{two-sided ideal} or just plain \emph{ideal} iff it is both a left
and right ideal.  An algebra $\Alg$ is called \emph{simple}, iff it has no non-trivial
ideals (ideals other than $\{ \rnul \}$ and $\Alg$) and $\Alg^2 := \Alg \Alg \ne \{ \rnul \}$.

\item[Commutant.]  
For a sub-algebra \(\Blg\subseteq\Alg\), its \emph{commutant} \(\Blg'\subseteq\Alg\) is
the sub-algebra
\[
\Blg' \;:=\;\{\ra \in \Alg : [\ra,\rb] = \rnul \ \text{ for all }\ \rb\in\Blg\} .
\]
In a unital algebra, \(\CC[\rone]\subseteq \Blg'\).  
Moreover, \(\ZA{\Alg}\subseteq \Blg'\) always, where \(\ZA{\Alg}\) is the centre of 
\(\Alg\), cf.\ \feD{def:centre}.  
Note that \(\ZA{\Blg}=\Blg\cap \Blg' \subset \Blg\), whereas the algebra 
\(\Blg'\) itself is generally \emph{not} a sub-algebra of \(\Blg\).
Observe that $\Blg \subseteq \Blg'' := (\Blg')'$.

\item[MASA --- Maximal Abelian Sub-Algebra.] The commutative resp.\ Abelian *-sub-algebra
$\Clg \subseteq \Alg$ is a \emph{maximal Abelian sub-algebra (MASA)} \emph{iff} it holds 
that for an Abelian *-sub-algebra $\Blg \subseteq \Alg$ with $\Clg \subseteq \Blg$, it 
follows that $\Blg = \Clg$.   Thus also $\Clg = \Clg'$.

\item[Mean and fluctuating part.]  
Given a state \(\svpi\colon \Alg\to\CC\), define the \emph{mean} of \(\ra\) by 
\[
  \wob{\ra} := \svpi(\ra)\,\rone, \quad \text{ and the \emph{fluctuating part} by } 
     \quad  \wtl{\ra} := \ra - \wob{\ra}.
\]
Hence \(\ra = \wob{\ra} + \wtl{\ra}\).  The algebra of scalar multiples of \(\rone\) 
is \(\wob{\Alg}:=\CC[\rone]\), and \(\wtl{\Alg}:=\ker\svpi\) is the vector space of all 
zero-mean elements.  Consequently, \(\Alg=\wob{\Alg}\oplus\wtl{\Alg}\).  
Viewing \(\CC\) embedded in \(\Alg\) via \(z\mapsto z\,\rone\), 
one can identify \(\wob{\Alg}\cong\CC\).

\item[Normal, self-adjoint, and skew-adjoint.] 
\(\ra\in \Alg\) is \emph{normal} iff \([\ra, \ra^\star]=\rnul\).  
It is \emph{self-adjoint} or \emph{Hermitean} iff \(\ra=\ra^\star\), 
and \(\rc\in \Alg\) is \emph{skew-adjoint}  or \emph{skew-Hermitean}
iff \(\rc=-\rc^\star\).  With $\ra$ self-adjoint, $\rc := \ii \ra$ is skew-adjoint.
Let \(\Alg_{sa}\subseteq\Alg\) denote the real vector space 
of self-adjoint elements.  Observe that the real and imaginary part
\[
\Re(\rb) \;:=\;\frac{1}{2}\bigl(\rb + \rb^\star\bigr),   % \tfrac12
\quad
\Im(\rb) \;:=\;\frac{1}{2 \ii}\bigl(\rb - \rb^\star\bigr)
\]
are self-adjoint, hence any element \(\rb\) decomposes as \(\rb=\Re(\rb)+\ii\,\Im(\rb)\).  
Thus \(\Alg=\Alg_{sa}\oplus\ii\,\Alg_{sa}=\spn_\CC(\Alg_{sa})\).  
%Normal elements commute with their own adjoint, and each self-adjoint element \(\ra\) is normal.

\item[Invertible elements.]
\(\ra\) is \emph{invertible} iff there is an element \(\rb \in \Alg\) with 
\(\ra\,\rb=\rb\,\ra=\rone\).  
In that case, \(\rb=\ra^{-1}\) is the \emph{inverse}.  
The multiplicative group of all invertible elements is denoted as \(\Alg^\times\).  
Clearly, the unit \(\rone\) is invertible (\(\rone^{-1}=\rone\)), whereas \(\rnul\) is not.

\item[Unitary elements.]  
\(\ru\) is \emph{unitary} iff \(\ru^\star=\ru^{-1}\).  In other words, 
\(\ru\ru^\star=\ru^\star\ru=\rone\).  All unitary elements are normal, 
and the multiplicative sub-group of all unitary elements is denoted by 
\( \UA{\Alg} \subset \Alg^\times\).

\item[Positive elements.]
A self-adjoint element \(\ra\) is \emph{positive}, written \(\ra\ge\rnul\), 
iff \(\ra\) can be expressed as \(\ra=\sum_j \rb_j^\star\,\rb_j\).  
The convex cone of positive elements is denoted as \( \Alg_+\subseteq\Alg_{sa}\).  
This defines a partial order: \(\ra\le\rc\) iff \(\rc-\ra\ge\rnul\).  

\item[Effects.]
A positive element \(\re\) is called an \emph{effect} iff \(0\le \re\le\rone\).  
Hence \(\re\) is self-adjoint and the set of all effects is denoted
\(\EA{\Alg}\subset\Alg_+\).

\item[Idempotents and projections.] 
\(\rv\in\Alg\) is an \emph{idempotent} iff \(\rv^2=\rv\).  If additionally \(\rv\) 
is self-adjoint (\(\rv=\rv^\star\)), it is called a \emph{projection}.  The set of 
projections is  \( \PA{\Alg}\subset\EA{\Alg}\subset\Alg_+\).  Two projections 
\(\rp_1,\rp_2\) are \emph{orthogonal} iff \(\rp_1\rp_2=\rnul\).  
For $\rp \in \PA{\Alg}$, \(\rp^\perp:=\rone-\rp\) is again a projection, 
and \(\rp\rp^\perp=\rnul\).

\item[Minimal projections.]
A projection \(\rp\) is \emph{minimal} in some subset $\C{P} \subseteq \Alg$ 
iff \(\rp>\rnul\) and there is no strictly 
smaller nonzero projection \(\rr \in \C{P}\) with \(\rnul<\rr<\rp\).  Similarly to
\feX{ex:projections}, this may be rephrased as saying that if there is
another projection $\rqq \in \C{P}$ with $\rqq \le \rp$, then either $\rqq = \rnul$
or $\rqq = \rp$.

\item[Isometries and Partial Isometries.]
$\rv \in \Alg$ is called an \emph{isometry}, iff \(\rv^\star \rv=\rone\) 
and \(\rv\,\rv^\star\) is a projection. 
\(\rf  \in \Alg\) is called a \emph{partial isometry}, iff 
\(\rf^\star \rf=\rp\) is a projection.  
In that case, \(\rf\,\rf^\star\) is also a projection.

\item[Bounded random variables.]
An element \(\ra\in\Alg\) is called \emph{bounded} if there exists a constant 
\(C_{\ra}>0\) such that
\[
\svpi\bigl(\rb^\star\,\ra^\star\,\ra\,\rb\bigr) \;\le\; C_{\ra}^2\,
   \svpi\bigl(\rb^\star\,\rb\bigr) \quad \text{ for all }\;\rb\in\Alg.
\]
The set of all bounded RVs is \(\BA{\Alg}\subseteq\Alg\), which is itself a probability 
sub-algebra.  One always has \(\CC[\rone]\subseteq\BA{\Alg}\).  In the finite dimensional case, 
\(\BA{\Alg}=\Alg\), so all elements are bounded.

\item[Internal automorphisms.]  
Let $\ru \in \UA{\Alg} \subset \Alg^\times \subset \Alg$ be a unitary,
then $\Psi: \Alg \to \Alg$, defined by $\ra \mapsto \ru^\star \ra \ru$, is a 
\emph{*-isomorphism}, an invertible *-homomorphism (\feD{def:star_homomorphism})
of the algebra onto itself, i.e.\  a \emph{*-automorphism}.  
As $\ru^\star \ra \ru$ is formed completely by expressions internal to the *-algebra, 
this is also called an \emph{internal} *-automorphism.

\end{description}
\end{defi}

An easy implication of the definitions is
\begin{prop}  \label{prop:normal-els}
For any $\ra \in \Alg$ one has $[\ra, \ra^\star] = 2 \ii \, [\Im(\ra),\Re(\ra)]$, and thus
$\ra$ is normal \emph{iff} $\Re(\ra)$ and $\Im(\ra)$ commute.  
Thus for a self-adjoint $\ra \in \Alg_{sa}$,
any $\rb \in \CC[\ra]$ in the commutative sub-algebra $\CC[\ra] \subseteq \Alg$ is normal.
\end{prop}

\begin{defi}[Structure constants]  \label{def:struct-const}
For an algebra $\Alg$, let $B = \{\rb_j\}_{(j \in J)} \subset \Alg$ be a basis.  
Then any product $\rb_i \rb_j \in \Alg$ of basis elements has an expansion 
\[
  \rb_i \rb_j = \sum_{k \in J} \tensor*[^B]{\gamma}{^k_{ij}}\; \rb_k .
\]
The $\tensor*[^B]{\gamma}{^k_{ij}} \in \CC$ are called the 
\emph{structure constants}  \citep{DrozdKiri94}
of the algebra $\Alg$ w.r.t.\ the basis $B$.
\end{defi}

For practical computations, knowledge of the structure constants in \feD{def:struct-const}
is important, as they determine the multiplicative structure of the algebra through
the bi-linearity of the product  \citep{DrozdKiri94}, and are sufficient to compute any product.

%\begin{rem}  \label{rem:struct-const}
%While the
%bi-linearity of the product of $\rx = \sum_i \xi_i \rb_i, \ry = \sum_j \eta_j \rb_j \in \Alg$
%is always given through bi-linearity of the product formula
%\[
%   \rx \ry = \left(\sum_i \xi_i \rb_i \right) \left(\sum_j \eta_j \rb_j \right) = 
%   \sum_{i,j,k} \xi_i \eta_j \, \tensor*[^B]{\gamma}{^k_{ij}} \rb_k =
%   \sum_k \zeta_k \rb_k ,\; \text{ with } \;
%   \zeta_k = \sum_{i,j} \xi_i \eta_j \, \tensor*[^B]{\gamma}{^k_{ij}},
%\]
%the structure constants cannot be chosen arbitrarily.  For an associative algebra
%\citep{DrozdKiri94} one requires for all $i, j, k, m \in J$:
%\begin{equation}   \label{eq:assoc-struc-const}
%  \sum_\ell \tensor*[^B]{\gamma}{^\ell_{i j}} \, \tensor*[^B]{\gamma}{^m_{\ell k}} =
%  \sum_\ell \tensor*[^B]{\gamma}{^\ell_{j k}} \, \tensor*[^B]{\gamma}{^m_{i \ell}} .
%\end{equation}
%  In case one has for all $i,j, k \in J$:
%$\tensor*[^B]{\gamma}{^k_{ij}} = \tensor*[^B]{\gamma}{^k_{ji}}$, the algebra is 
%commutative or Abelian.
%\end{rem}

\begin{rem}   \label{rem:bounded-ideal}
For the bounded elements of the algebra, with a bound $C_{\ra}$ as defined in 
\feD{def:alg-names}, set $\wht{C}_{\ra} =\inf \{ C_{\ra} \mid C_{\ra}\text{ is a bound}\}$
and $\C{I}_{\svpi} = \{ \ra \in \Alg \mid \wht{C}_{\ra} = 0 \}$
to be the \emph{ideal} (cf.\ \feD{def:alg-names}) of elements where the infimum over all 
possible bounds vanishes.  In case $\C{I}_{\svpi} = \{ \rnul \}$, the state $\svpi$ is called
\emph{non-degenerate}.   Otherwise, for $\C{I}_{\svpi} \ne \{ \rnul \}$, consider the
quotient *-algebra $\Alg / \C{I}_{\svpi}$, where the state corresponding to $\svpi$ then is
non-degenerate.

As later (cf.\ \feP{prop:alg-unif-n}) $\nd{\ra}_{\infty} = \wht{C}_{\ra}$ 
will be taken to define the operator
norm, the quotient construction makes sure that this is indeed a norm.  This removes
\emph{superfluous} elements from the algebra.
\end{rem}

The following \feP{prop:bounded-ideal} gives a few equivalent indicators for the
situation described in \feR{rem:bounded-ideal}, cf.\  \citep{A-S-math-05}.

\begin{prop}   \label{prop:bounded-ideal}
Let $(\Alg, \svpi)$ be a probability algebra. 
  Then the following statements are equivalent:

\begin{compactitem}

\item $\wht{C}_{\ra} = 0$ or $\ra \in \C{I}_{\svpi}$ (cf.\ \feR{rem:bounded-ideal}).

\item $\svpi(\rb \ra \rc) = 0$ for all $\rb, \rc \in \Alg$.

\item $\svpi(\rb^\star \ra \rb) = 0$ for all $\rb \in \Alg$.

\end{compactitem}

In case $\Alg$ is Abelian, or the state $\svpi \in \SA{\Alg}$ is \emph{tracial},
these statements are also equivalent with the following:
\begin{compactitem}

\item $\svpi(\ra^\star \ra) = 0$.

\item $\svpi(\ra \rb) = 0$ for all $\rb \in \Alg$.

\end{compactitem}
\end{prop}

In case $\C{I}_{\svpi} \ne \{ \rnul \}$, we shall henceforth tacitly assume that the 
factorisation described in \feR{rem:bounded-ideal} has been performed, 
and that in this case $\Alg$ has been replaced by $\Alg / \C{I}_{\svpi}$,
so that the statements in \feP{prop:bounded-ideal} apply only to $\ra = \rnul$.

The foregoing framework will make it possible that many standard 
constructions and concepts from commutative probability carry 
over naturally to the non-commutative realm, albeit 
with additional care due to the non-commuting product.

As will be seen later, certain probability algebras --- e.g.\ all finite 
dimensional ones --- are generated by their projections $\PA{\Alg}$,
cf.\ \feX{ex:projections}.  These are also important as they can be given 
an additional interpretation, namely with a projection representing an \emph{event}
--- see also \feX{ex:projections} and \feP{prop:boolean} ---
with a probability assigned to it; this assignment is also called the \emph{Born} rule:

\begin{defi}[Probability --- Born's rule]  \label{def:probability}
The \emph{probability}, in the state $\svpi \in \SA{\Alg}$, of observing an event 
described by a projection $\rp \in \PA{\Alg}$ is defined as
\begin{equation}   \label{eq:def-prob}
  \prob(\rp) := \svpi(\rp) = \Ex_{\svpi}(\rp).
\end{equation} 
\end{defi}

The following facts are directly implied by this and \feD{def:alg-names}:

\begin{prop}[Events]  \label{prop:proj-1}
  As for any positive element $\ra \ge \rnul$ one has $\svpi(\ra) = \Ex_{\svpi}(\ra) \ge 0$,
  this implies that for $\ra \le \rb$ 
\[
  \svpi(\ra) = \Ex_{\svpi}(\ra) \le \svpi(\rb) = \Ex_{\svpi}(\rb).
\]  
  Thus for a projection $\rp \in \PA{\Alg}$, one has $\rp^\perp \in \CC[\rp]$, and as 
  $\rnul \le \rp \le \rone$, this implies 
\[
   0 \le \prob(\rp) = \svpi(\rp) = \Ex_{\svpi}(\rp) \le 1 \;\text{ and }\;
   \prob(\rp) + \prob(\rp^\perp) = 1.
\]
For any character $\som$ on $\CC[\rp]$ one has $0 \le \som(\rp) = \som(\rp\rp)
=\som(\rp) \som(\rp) \le 1$, and hence
\[
   \text{ either }\, \som(\rp) = 0 \,\text{ or }\, \som(\rp) = 1, \; \text{ i.e. }\,
   \sigma(\rp) \subseteq \{ 0,1 \}. %\som(\rp) + \som(\rp^\perp) = 1.
\]
The case $\sigma(\rp) = \{0\}$ occurs for the constant $\rp \equiv \rnul$ (the impossible
event), and the case $\sigma(\rp) = \{1\}$ occurs for the constant $\rp \equiv \rone$ 
(the sure event); for any other (proper) projection
$\rnul < \rp < \rone$ one has $\sigma(\rp) = \{ 0,1 \}$.

Thus the observation $\som(\rp) = \ip{\som}{\rp} = \rp(\som) = 1$ can be interpreted 
in such way that the \emph{event} $\rp$ has been observed.  It is a \emph{sharp}
observable, there is no doubt that $\rp$ has been observed.
The probability of this happening was $\prob(\rp)$ before the observation, 
whereas the probability of 
observing the \emph{opposite event} $\rp^\perp$ was $\prob(\rp^\perp)= 1-\prob(\rp)$;
but $\rp^\perp$  with $\som(\rp^\perp) = \rp^\perp(\som) =  0$ 
has not been observed.
\end{prop}

The deliberations in \feX{ex:projections} may be continued in an abstract setting:

\begin{prop}[Boolean Algebra]  \label{prop:boolean}
  Let $R \subset \PA{\Alg}$, with $R := \{\rp_j\}, j \in J$ be a collection of 
  commuting projections including the identity $\rone$ --- 
  $\forall i,j\in J:\,[\rp_i, \rp_j] = \rnul$, cf. \feX{ex:projections} ---
  so that $\CC[R] \in \Alg$ is a commutative probability sub-algebra.
  Define for any projections $\rp_1, \rp_2 \in \PA{\CC[R]}$ as ``\emph{meet}'' or ``\emph{and}''
  or ``\emph{intersection}'' or ``\emph{infimum}'' 
  the symbol $\sqcap$; and as ``\emph{join}'' or ``\emph{or}'' 
  or ``\emph{union}'' or ``\emph{supremum}'' the symbol $\sqcup$; 
  and as ``\emph{complement}'' or ``\emph{negation}'' the symbol $\neg$, by setting:
\begin{align*}   
  \rp_1 \sqcap \rp_2 &:= \rp_1 \rp_2; \\
  \rp_1 \sqcup \rp_2 &:= \rp_1 + \rp_2 - \rp_1 \rp_2; \\
  \neg \rp_1 &:= \rp_1^\perp.
\end{align*}  
Then $\PA{\CC[R]}$ together with these operations is a \emph{Boolean algebra}, 
with greatest element $\rone$ and least element $\rnul$.
Together with the probability assignment from \feD{def:probability}, this is a
basic \emph{Kolmogorovean probability space}.
\end{prop}

The part to make this Boolean algebra into a $\sigma$-Boolean algebra when $J$ is an
infinite set, so that
the probability space can be defined on a $\sigma$-algebra, is a certain completeness
property, a subject not to be addressed here.  But for finite dimensional algebras,
when $J$ is necessarily of finite cardinality,
this is not needed, and the construction in \feP{prop:boolean} is a proper (finite,
but \emph{pointless})
Kolmogorovean probability space, with the projections being analogues of the elements 
of the algebra of sets.  Of course, it is well known that by Stone's theorem any such Boolean
algebra is isomorphic to a field of subsets of some set.

In continuation of \feR{rem:spec-dec-proj} and \feP{prop:boolean}, one may make the following

\begin{rem}[Matrix Observable, Projective Measurement]  \label{rem:spec-dec-proj-2}
Consider an observable $\vA \in \MMn(\CC)_{sa}$ ($\vA = \vA^\tpH$) with spectral decomposition 
$\vA = \sum_k \lambda_k \vP_k$ with $\lambda_k \in \sigma(\vA) \subset \RR$ 
%($\vLbd = \diag(\lambda_k)$) 
and thus commuting orthogonal projections $\vP_k$, which form a \emph{partition of unity}
$\sum_k \vP_k = \vI$, cf.\ \feR{rem:spec-dec-proj}.   The possible observations are the 
elements of the spectrum (cf.\  \feP{prop:spec-one}),  
sharp observations $\som_k(\vA) = \lambda_k$,
with the probability of observing $\lambda_k$ as $\prob(\vA = \lambda_k) := p_k =
\svpi(\vP_k) = \Ex_{\svpi}(\vP_k)$.  Observing $\som_k(\vA) = \lambda_k$
carries the same information as the observation $\som_k(\vP_k) = 1$, as necessarily
for $j \ne k$ one has $\som_k(\vP_j) = 0$, and we know that $\vA$ has the value $\lambda_k$.

Such \emph{sharp} observations of projections $\som(\vP)$ are also called a 
\emph{projective measurement},
and are the basic building blocks for all sharp observations.  In case  $\lambda_k$ is a simple
eigenvalue and hence the eigenspace is one-dimensional with unit eigenvector $\vv_k \in \Cn$,
the (minimal) projector is given by $\vP_k  = \vv_k \vv_k^\tpH$.  In this case the observation
can be written as $\som_k(\vA) = \tr (\vP_k \vA) = \vv_k^\tpH \vA \vv_k$ (cf.\ \feD{def:density-M}).
And in case of
a multiple eigenvalue with an eigenspace $V_k \subseteq \Cn$ of say dimension $d$, one can
find an ortho-normal basis $\{ \vv_{k_j} \}_{j=1\dots d}$ of $V_k$, so that the projector then
may be written as $\vP_k = \sum_j \vv_{k_j} \vv_{k_j}^\tpH$.  Thus the expected value can be written as 
$\Ex_{\svpi}(\vA) = \svpi(\vA) = \sum_k \lambda_k\, \prob(\vA = \lambda_k) = \sum_k \lambda_k p_k$.
\end{rem}

\begin{defi}[Projection Valued Measure (PVM)]   \label{def:PVM-M}
The assignment $\Uppi_M: 2^{\sigma(\vA)} \ni \{\lambda_k\} \mapsto \
\vP_{\lambda_k} := \vP_k \in \PA{\MMn(\CC)}$
is an example of a \emph{projection valued measure (PVM)} on the set of all
subsets of $\sigma(\vA) \subset \RR$.  
It satisfies  
\begin{align}   \label{eq:PVM-defi-1}
   \Uppi_M(\sigma(\vA)) =& \vI, \quad \text{ and for disjoint } \; 
         E_1, E_2 \subseteq \sigma(\vA):\\
   \Uppi_M(E_1 \uplus E_2) =& \Uppi_M(E_1) + \Uppi_M(E_2).   \label{eq:PVM-defi-2}
\end{align}
This implies $\Uppi_M(\emptyset) = \vek{0}$, and for not necessarily disjoint $E_1, E_2
\subset \sigma(\vA)$: 
\begin{align}   \label{eq:PVM-consq-1}
   \Uppi_M(E_1 \cup E_2) =& \Uppi_M(E_1) + \Uppi_M(E_2) - \Uppi_M(E_1 \cap E_2), \;\text{ and }\\
    \label{eq:PVM-consq-2}
   \Uppi_M(E_1 \cap E_2) =& \Uppi_M(E_1)\Uppi_M(E_2) = \Uppi_M(E_2)\Uppi_M(E_1).
\end{align}
From \feq{eq:PVM-consq-2} follows that the minimal projections $\vP_{\lambda_k} = \vP_k$ in 
the PVM are orthogonal to each other, and that they commute in general.  Thus
\[
  \spn \{ \Uppi_M(E) \mid E \subseteq  \sigma(\vA) \} = 
     \CC[\{ \Uppi_M(E) \mid E \subseteq  \sigma(\vA) \}] = \CC[\vA] \subset \MMn(\CC)
\]
is a commutative sub-algebra of $\MMn(\CC)$.

Together with the state $\svpi \in \SA{\MMn(\CC)}$ 
this leads to a discrete measure 
$2^{\sigma(\vA)} \ni \{\lambda_k\} \mapsto \Ex_{\svpi}(\vP_{\lambda_k}) = 
\prob(\vP_{\lambda_k}) = p_k \in \RR$.
\end{defi}

That in the above \feD{def:PVM-M} the PVM was chosen to be defined on the singletons
$\{\lambda_k\} \in 2^{\sigma(\vA)}$ was due to the development indicated in 
\feR{rem:spec-dec-proj-2}, via \feq{eq:PVM-defi-2} it can then be defined for
any subset $E \subset \sigma(\vA)$.  In fact, the $\sigma$-algebra $2^{\sigma(\vA)}$
is also a special case:

\begin{rem}[PVM on any $\sigma$-algebra]   \label{rem:PVM-more-X}
The more general definition of a PVM as in \feD{def:PVM-M} is that there is a
$\sigma$-algebra $\F{Y} \subseteq 2^Y$ of subsets of some set $Y$, so that
with subsets $E_1, E_2 \in \F{Y}$ the \feqs{eq:PVM-defi-1}{eq:PVM-defi-2} hold.

For a finite dimensional algebra as $\MMn(\CC)$ and any disjoint 
cover $\biguplus_k E_k = Y$, $(E_k \cap E_j = \emptyset, E_k \in \F{Y})$, 
there can obviously be only finitely many non-zero projections
in the sum in $\Uppi_M(Y) = \sum_k \Uppi_M(E_k) = \vI$.  So taking only minimal
projections (cf.\ \feD{def:alg-names} and \feR{rem:spec-dec-proj-2}) in the PVM for this
equation (corresponding to a maximal disjoint cover), 
one can find a finite set $X$ where the singletons $\{x\}, (x \in X)$, correspond
in a bijective fashion to the members of this maximal disjoint cover.
The $\sigma$-algebra $2^X$ --- generated by the singletons --- then is in 
one-to-one correspondence with $\F{Y}$, so that the minimal projections are
images of singletons $\{x\} \in 2^X$.  This is the setting which
will be assumed in the following.  In \feD{def:PVM-M} one has $X = \sigma(\vA) \subset \RR$.  

 The set $X$ 
is often called the \emph{outcome space}, meaning that when $\Uppi_M(\{x_k\}) = 
\vP_{x_k} = \vP_k$ has been observed, i.e.\  $\som(\vP_k) = 1$ for an observation
$\som$, the outcome of the observation is $x_k \in X$.
Note that for the mutually orthogonal projections ($\vP_k \vP_j = \updelta_{j k} \vP_k$)
in the PVM $\Uppi_M$, and considering \feP{prop:proj-1}, from $\som(\vP_k) = 1$ follows
that $\som(\vP_j) = 0$ for all $j \ne k$; where $\vP_j = \vP_{x_j} =  \Uppi_M(\{x_j\})$.
\end{rem}

\begin{defi}[Density Matrix, Vector State]   \label{def:density-M}
Recalling \feXs{ex:complex-diag}{ex:projections} and the definition of the Frobenius state 
$\sphi_{F}(\vA)= \frk{1}{n}\,\tr(\vA)$, one sees that any normalised state (cf.\ \feD{def:state})
$\sbt \in \SA{\MMn(\CC)}$ may be written as $\sbt(\vA) = \tr(\vR \vA)$.
The matrix $\vR$ has to be self-adjoint, positive, and of unit trace: $\sbt(\vI) = \tr (\vR \vI)
= \tr (\vR) = 1$.  Such a matrix is called a \emph{density} matrix.  
It thus has a spectral decomposition (cf.\ \feR{rem:spec-dec-proj}) 
$\vR = \sum_j \alpha_j \vr_j  \vr_j^\tpH$ with unit eigenvectors
$\vr_j$ and eigenvalues $\alpha_j \ge 0$, $\tr (\vR) = \sum_j \alpha_j = 1$.
The set of all such density matrices is denoted by
\begin{equation}   \label{eq:dens-mtx}
\DEN{\MMn(\CC)} := \{ \vR \in \MMn(\CC)_+ \mid   \tr \vR = 1 \} .
\end{equation}

Dealing with a matrix algebra like $\MMn(\CC)$, we may use a different notation for the
same state, depending on the situation; e.g.\ with $\vR \in \DEN{\MMn(\CC)}$ given, 
one has for all $\vA \in \MMn(\CC)$
\[
   \svpi_{\vR}(\vA) = \bkt{\vR}{\vA}_F = \tr \vR \vA =  n\, \sphi_F(\vR \vA).
\]

In case of a simple eigenvalue $\lambda_k$, the above observation $\som_k$ 
in \feR{rem:spec-dec-proj-2} is
$\som_k(\vA)  = \tr ((\vv_k \vv_k^\tpH) \vA) = \vv_k^\tpH \vA \vv_k$ with
density matrix $\vR_k = (\vv_k \vv_k^\tpH)$.  
In case of a multiple eigenvalue with eigenspace $V_k$, there is a unit vector $\vw_k \in V_k$ 
such that  $\som_k(\vA) = \tr ((\vw_k \vw_k^\tpH) \vA) = \vw_k^\tpH \vA \vw_k$.  
The associated density matrix for $\som_k$ is then either $\vR_k   = 
\vv_k \vv_k^\tpH$ or $\vR_k := \vw_k  \vw_k^\tpH$, a \emph{vector state}.
On an Abelian matrix algebra such as $\CC[\vA]$ from \feX{ex:sa-mat},
such a vector state is indeed a multiplicative
character if $\vv_k$ is an eigenvector of $\vA$.

For vector states like $\vr  \vr^\tpH$, one often calls the vector $\vr$ itself the
state.  Indeed, this is the origin of the name ``state'', as the vector $\vr$ can be regarded
as the state of some system modelled on $\Cn$.
\end{defi} 

\begin{prop}[Set of Density Matrices, Mixed State]   \label{prop:set-dens-M}
The set of density matrices $\DEN{\MMn(\CC)}$ (cf.\ \feD{def:density-M}) is closed and convex.
Geometrically speaking, it is the intersection of two closed convex sets, the positive cone
$\MMn(\CC)_+$ (cf.\ \feD{def:alg-names} and \feX{ex:projections}) and the hyperplane
$\C{T} = \{ \vQ \mid \tr \vQ = 1 \}$:
\begin{equation}   \label{eq:dens-mtx-cap}   
     \DEN{\MMn(\CC)} = \MMn(\CC)_+ \cap \C{T} .
\end{equation}
Denoting density matrices of rank $r$ by
$\DEN{\MMn(\CC)}_r = \{ \vR \in \DEN{\MMn(\CC)} \mid \rank \vR = r \}$, where $r=1,\dots,n$,
one has the \emph{disjoint union}
\begin{equation}   \label{eq:dens-mtx-rank}   
  \DEN{\MMn(\CC)} = \biguplus_{r=1}^n \DEN{\MMn(\CC)}_r .
\end{equation}
The extreme points of $\DEN{\MMn(\CC)}$ are the rank one matrices
\begin{equation}   \label{eq:dens-mtx-ext}  
     \mrm{ext}(\DEN{\MMn(\CC)}) = \DEN{\MMn(\CC)}_1 = \{  \vr  \vr^\tpH \mid \vr \in \Cn,
     \vr^\tpH \vr = 1 \}, 
\end{equation}
they are also called \emph{pure} or \emph{vector states}.  

According to the Krein-Milman theorem --- this also follows from the spectral 
decomposition in \feD{def:density-M} ---
a general density matrix $\vR \in \DEN{\MMn(\CC)}$ is the convex combination 
 of vector or pure states $\vr_j  \vr_j^\tpH \in \DEN{\MMn(\CC)}_1$, 
 it is often called a \emph{mixed state}.

There is a bijective correspondence between the set of states $\SA{\MMn(\CC)}$ and
the set of density matrices $\DEN{\MMn(\CC)}$, given by 
\[
 \SA{\MMn(\CC)} \ni \svpi_{\vR} \leftrightarrow \vR \in \DEN{\MMn(\CC)} \quad
 \Leftrightarrow \quad  \svpi_{\vR}(\vM) = \tr \vR \vM \quad \forall \vM \in \MMn(\CC).
\]
\end{prop}

\begin{prop}[Projection Lattice in $\MMn(\CC)$]  \label{prop:proj-matrix-measurement}
The projections $\PA{\MMn(\CC)} \subset \MMn(\CC)_{+}$ are a partially ordered set with 
the ordering from \feD{def:alg-names}, and may be given a \emph{lattice structure}. 
% For an orthogonal
%projection $\vP \in \PA{\MMn(\CC)}$, let $\C{P}_{\vP} = \im \vP = \vP(\Cn) \subseteq \Cn$ the 
%image or range space.  
Let $\vP_1, \vP_2 \in \PA{\MMn(\CC)}$ be two orthogonal projections.  The lattice structure
is established by
defining the \emph{infimum} as $\vP_1 \sqcap \vP_2 := \inf \{\vP_1, \vP_2\} := \vP_{\inf}$ the 
orthogonal projection $\vP_{\inf}: \Cn \to (\im \vP_1) \cap (\im \vP_2)$, and as \emph{supremum}
$\vP_1 \sqcup \vP_2 := \sup \{\vP_1, \vP_2\} := \vP_{\sup}$
the orthogonal projection $\vP_{\sup}: \Cn \to \spn \{ (\im \vP_1) \cup (\im \vP_2) \}
= \spn \{(\im \vP_1) + (\im \vP_2)\}$.  
 These definitions extend the ones which were given
in \feP{prop:boolean} for commuting subsets of $\PA{\MMn(\CC)}$.  

It is fairly obvious that the negation or complement $\vP^\perp = \vI - \vP$ for
a $\vP \in \PA{\MMn(\CC)}$ is the orthogonal projection onto the kernel 
$\ker \vP$ of $\vP$, and $\ker \vP \oplus \im \vP = \Cn$.  The greatest element in 
the lattice is again $\vI$ --- the identity --- and the least element is $\vek{0}$ 
--- the null matrix.  This lattice is denoted as $\opb{L}(\PA{\MMn(\CC)})$.

The \emph{atoms} in this lattice (for an atom $\vQ \in \opb{L}(\PA{\MMn(\CC)})$ there 
is no distinct other element $\vP \in \opb{L}(\PA{\MMn(\CC)})$ such that 
$\vek{0} < \vP < \vQ$) are the minimal 
projections of the form $\vQ = \vr \vr^\tpH$ with a unit vector $\vr \in \Cn$,
i.e.\ $\DEN{\MMn(\CC)}_1 \subset \PA{\MMn(\CC)}$ (\feP{prop:set-dens-M}).  
Every projection of rank $m \le n$ can be expressed as the sum of $m$ atomic projections
in $\DEN{\MMn(\CC)}_1$.

Due to Zorn's lemma, any commutative sub-algebra $\Clg$ (e.g.\ a PVM-generated Abelian
algebra $\Clg = \spn \{ \Uppi_M(E) \}$, cf.\  \feD{def:PVM-M}) can be embedded in or
extended to a MASA (maximal Abelian sub-algbra, cf.\ \feD{def:alg-names}).
A commutative sub-algebra $\Clg$ is a MASA \emph{iff} it is unitarily equivalent
to the commutative algebra of diagonal matrices  \(\DMn(\CC)\subset \MMn(\CC)\) from
\feX{ex:complex-diag}, i.e.\ $\exists\, \vU \in \UA{\MMn(\CC)}: \, \vU^\tpH \Clg \vU = 
\DMn(\CC)$, e.g.\ \citep{Landsman2017}.
\end{prop}

The advantage of an algebra is that all products and especially all powers are in the algebra.
As the state is defined on the whole probability algebra, this leads to the definition of moments:

\begin{defi}[Moments]  \label{def:moments}
The \emph{moments} of order $n \in \D{N}$ of $\ra$ are defined 
as the numbers $\rM_n(\ra) :=\svpi(\ra^n) = \Ex_{\svpi}(\ra^n)$.
For any finite set of RVs $\{\ra_j\}, j=1\dots m$, not necessarily distinct, and
which do not necessarily commute,
the \emph{mixed moments} of order $n_1,\dots,n_m \in \{ 1, \star \}$, $m \in \D{N}$, are 
defined as $\rM_{n_1 \dots n_m}(\ra_1,\dots,\ra_m) := \svpi(\ra_1^{n_1} \dots \ra_m^{n_m})$, 
where the ordering of the elements matters for non-commutative algebras.

If for two random variables $\ra_1, \ra_2$ it holds that for all $n$ the moments match,
$\rM_{n}(\ra_1) = \rM_{n}(\ra_2)$, denoting this as $\ra_1 \stoeq \ra_2$,
one says that they are \emph{moment equivalent}.
\end{defi}

Having defined the expectation of powers leading to moments, it is natural to extend
these definitions linearly to polynomials, leading to the definition of the
law or distribution of a RV: 

\begin{defi}[Law of a Random Variable]   \label{def:law_RV}
For a RV $\ra \in \Alg$ and a polynomial $\tp \in \CC[\tns{X}]$ in one unknown $\tns{X}$, 
one may compute $\svpi(\tp(\ra))$ --- e.g.\ with the help of the moments $\rM_n(\ra)$.  The 
\emph{law of $\ra$} is defined as the linear mapping $\tau_{\ra}: \CC[\tns{X}] \to \CC$:
$\tau_{\ra}: \tp \mapsto \tau_{\ra}(\tp) := \svpi(\tp(\ra)) = \Ex_{\svpi}(\tp(\ra))$.
\end{defi}

Just as for mixed moments in \feD{def:moments}, one can define the law of a collection
of non-commuting RVs by extending \feD{def:law_RV} to \emph{joint laws}, but this will
not be needed here any further.  The following fact follows obviously from linearity:

\begin{prop}   \label{prop:equal_mom-law}
  RVs with equal moments have equal laws, and they are moment equivalent.
\end{prop}

As RVs in finite dimensional algebras are always bounded (cf.\ \feD{def:alg-names}),
as will be seen shortly, it is convenient in the following to limit ourselves to bounded RVs.
From the kind of law defined in \feD{def:law_RV} on polynomials one comes very quickly to 
the classical notion of law as the distribution measure of a RV:

\begin{thm}[Distribution Law of a RV]   \label{prop:law-prob-m}
  For any self-adjoint bounded RV $\ra \in \Alg_{sa}$ there is a \emph{probability measure} 
  $\mu_{\ra}$ on $[-C_{\ra}, C_{\ra}] \subset \RR$, where $C_{\ra}$ 
  is a bounding constant of $\ra$ as in \feD{def:alg-names} (cf.\ \feP{prop:fin-dim-alg-gen}) 
  --- also called the law   or rather the \emph{distribution (law)} of the RV $\ra$ ---
  such that for all $\tp \in \CC[\tns{X}]$ one has
  \begin{equation}   \label{eq:distr_of_RV}
    \tau_{\ra}(\tp) = \int_{-C_{\ra}}^{C_{\ra}} \tp(x)\, \mu_{\ra}(\di x) .
  \end{equation}
  
  This allows one to extend the map $\tau_{\ra}$ from polynomials to the continuous functions
  on the interval $[-C_{\ra}, C_{\ra}]$,   $\wob{\tau}_{\ra}: \Ck([-C_{\ra}, C_{\ra}],\CC) \to \CC$ 
  via continuity, by defining for a continuous function $\vphi$:
  $\wob{\tau}_{\ra}(\vphi) = \int_{-C_{\ra}}^{C_{\ra}} \vphi(x)\, \mu_{\ra}(\di x) $. %;
%  and even further as in \feX{ex:cont-fcts} to $\Lp_\infty([-C_{\ra}, C_{\ra}],\F{B}, \mu_{\ra})$
%  and finally to $\Lp_1([-C_{\ra}, C_{\ra}],\F{B}, \mu_{\ra})$.
\end{thm}

The proof of this theorem (e.g.\ \citep{Mitchener2005}) follows from the 
Riesz-Markov-Kakutani representation theorem and the 
Stone-Weierstrass theorem.   Together with \feP{prop:equal_mom-law} one sees that 
this is the Hausdorff moment problem, which has the unique measure $\mu_{\ra}$ as solution.

As here the focus is on finite dimensional algebras, all RVs from such an algebra 
are not only automatically bounded, but, as seen in \feR{rem:spec-dec-proj-2} in connection
with the upcoming \feP{prop:fin-dim-alg-gen}, the values 
of such a RV are in a finite subset in $[-C_{\ra}, C_{\ra}]$; and on finite sets any 
distribution can be interpolated by a polynomial, and so no approximation is needed at all.

\begin{defi}[Convergence of RVs in Law or Distribution]  \label{def:conv_in_law}
  Let $\{\ra_n\}$, $n\in \D{N}$ be a sequence of RVs.  They \emph{converge in law}
  resp.\ \emph{converge in distribution} to a RV $\ra$ \emph{iff} $\tau_{\ra_n}(\tp)$ converges
  to $\tau_{\ra}(\tp)$ for all $\tp \in \CC[X]$.
\end{defi}

The following is implied directly by linearity:
\begin{prop}[Moments and Convergence in Distribution]  \label{prop:conv_in_law}
  The sequence $\{\ra_n\}$ converges in law resp.\ distribution to a RV $\ra$ \emph{iff} 
  for all $m \in \D{N}$ one has that $\rM_m(\ra_n)$ converges to $\rM_m(\ra)$ as $n\to \infty$.
\end{prop}

Now is the time for some non-commutative examples of probability algebras.  A very general
example is:

\begin{xmpn}[Commutant]   \label{ex:commutant}
  Given a probability sub-algebra $\Blg \subseteq \Alg$ of a probability algebra $\Alg$,
  its \emph{commutant} $\Blg^\prime$, cf.\ \feD{def:alg-names}, is a possibly non-commutative 
  probability sub-algebra, $\Blg^\prime \subseteq \Alg$.
\end{xmpn}

Observe that the elements in the commutant $\Blg^\prime$ commute only with all elements 
in $\Blg$, but not necessarily with each other.

\begin{xmpn}[Matrix Algebra]   \label{ex:complex-mat}
Let $\MMn(\CC)$ be the algebra of matrices with complex entries.  With $\vM
= (M_{ij}) \in \MMn(\CC)$, as involution take the Hermitean transpose $\vM^\tpH = 
(M_{ij})^\tpH := (M_{ij}^*)^\trpos = (M_{ji}^*)$,  cf.\ \feX{ex:complex-diag}.  And as 
already noted in \feX{ex:complex-diag}, any linear functional 
can be written as a trace: $\vM \mapsto \tr (\vV \vM)$ for some $\vV \in \MMn(\CC)$.
The induces the \emph{Frobenius} or \emph{Hilbert-Schmidt} inner product 
(and corresponding norm) $\bkt{\vB}{\vA}_F = \tr (\vB^\tpH \vA)$,
turning $\MMn(\CC)$ into a finite dimensional Hilbert space.

On $\MMn(\CC)$ one has a canonical \emph{Frobenius} state 
$\sphi_F(\vM) = \frk{1}{n}\,\tr \vM$ as in \feR{rem:spec-dec-proj-2}, 
where $\vV = \frk{1}{n}\, \vI$.  This is the \emph{unique tracial state}
on $\MMn(\CC)$, cf.\ \feD{def:state}.  It corresponds to the uniform distribution.
This non-commutative probability algebra has in general no characters when $n>1$, 
i.e.\ no samples or observations defined on
the whole algebra --- $\XA{\MMn(\CC)} = \emptyset$.

For a self-adjoint $\vA  \in \MMn(\CC)_{sa}$,
the sub-algebra $\CC[\vA] \subset \MMn(\CC)$ considered in \feR{rem:spec-dec-proj-2}
is a commutative sub-algebra of the full matrix algebra.  And on this commutative sub-algebra
there are characters, samples, or observations, i.e.\ $\XA{\CC[\vA]} \ne \emptyset$.  They
are pure states in the form of vector states like in \feD{def:density-M}, such that the
state vector is a normalised eigenvector of $\vA$.%  For $n=1$ this is \feX{ex:complex-num}.
\end{xmpn}

This is a typical example of a finite dimensional non-commutative probability algebra.
As will be seen later, any finite dimensional algebra is isomorphic to a direct sum 
of such algebras.  And as the algebraic view does not distinguish between isomorphic 
representations, sometimes arguments are carried out directly on this kind of 
faithful representation.

\begin{rem}[Matrix Schatten Norm]   \label{rem:Schatten-M}
When considering \feX{ex:complex-mat}, recall from linear algebra 
that each  matrix $\vB \in \MMn(\CC)$ has a \emph{singular value decomposition}
$\vB = \vU \vS \vV^\tpH$, where $\vU, \vV \in \UA{\MMn(\CC)}$ are unitaries
($\vU \vU^\tpH = \vU^\tpH \vU = \vI$) and $\vS = \diag(\vek{\vsigma})$ is the
diagonal self-adjoint matrix of non-negative \emph{singular values} 
$\vek{\vsigma} = (\vsigma_1,\dots,\vsigma_n)$, $\vsigma_j \in \RR_+$. Then for any $p\ge 1$
\begin{equation} \label{eq:Schatten-M}
     \nd{\vB}_{Sp} := \left(\sum_{j=1}^n \vsigma_j^p\right)^{\frk{1}{p}} = 
       \left(\tr \left[(\vB^\tpH \vB)^{\frk{p}{2}}\right]\right)^{\frk{1}{p}}
\end{equation}
is a unitarily invariant \emph{norm} on $\MMn(\CC)$, the \emph{Schatten $p$-norm}.
Extend this to $p = \infty$ by setting $\nd{\vB}_{S \infty} := \nd{\vB}_{\infty} = 
\max \{ \vsigma_j \mid 1 \le j \le n \}$, i.e.\ the \emph{operator norm} of 
$\vB \in \Lop(\Cn)$ when $\Cn$ is equipped with its canonical inner product 
(cf.\ \feX{ex:complex-n-spc}).  

Note that the Schatten-2-norm $\nd{\vB}_{S2} = \nd{\vB}_{F}$
is equal to the \emph{Frobenius} norm, and comes from the \emph{Hilbert-Schmidt} inner
product which makes $\MMn(\CC)$ into a \emph{Hilbert space}.  
The norm $\nd{\vB}_{S1}$ is also called the \emph{trace norm} or \emph{nuclear norm}.
As \emph{Banach spaces}, $(\MMn(\CC), \nd{\cdot}_{Sp})$ and $(\MMn(\CC), \nd{\cdot}_{Sq})$,
with $\frk{1}{p} + \frk{1}{q} = 1$ (extended to $p = 1, q = \infty$), are \emph{dual} to 
each other.  The operator- or Schatten-$\infty$-norm satisfies
$\nd{\vA \vB}_\infty \le \nd{\vA}_\infty \nd{\vB}_\infty$, and with it 
$(\MMn(\CC), \nd{\cdot}_\infty)$ is a \emph{Banach algebra} (cf.\ \feD{def:normed_algebras}).
\end{rem}

\begin{prop}[Density Matrix as Element of Matrix Algebra]  \label{prop:dens-in-alg-M}
If $\Clg \subset \MMn(\CC)$ is a unital *-sub-algebra, with states from $\MMn(\CC)$ 
restricted to $\Clg$ it becomes a probability algebra.  At the same time $\Clg$ is a closed
subspace of the finite dimensional Hilbert space $(\MMn(\CC),\bkt{\cdot}{\cdot}_F)$;
hence there is an orthogonal decomposition $\Clg \oplus \Clg^\perp = \MMn(\CC)$,
and $\Clg$ is a Hilbert space in its own right.  Thus any linear functional 
$\sbt \in \Clg^\star$ can be written as $\sbt(\vC) = \bkt{\vV}{\vC}_F$ for $\vC \in \Clg$
with some $\vV \in \Clg$.

Let $\vR \in \DEN{\MMn(\CC)}$ be a density matrix or mixed state $\vR$ as in 
\feD{def:density-M}, defining  a state  $\svpi_{\vR}(\vM) := \tr(\vR \vM)$ on $\MMn(\CC)$. 
Then there is a decomposition $\vR = \vR_C + \vR_\perp \in \Clg \oplus \Clg^\perp$, so that
$\bkt{\vR_\perp}{\vC}_F = 0$ for all $\vC \in \Clg$.
This means that for any state $\svpi_{\vR} \in \SA{\MMn(\CC)}$ resp.\ any density matrix
$\vR \in \DEN{\MMn(\CC)}$, there is a $\vR_C := \vR - \vR_\perp \in \Clg$, such that
$\svpi_{\vR}|_{\Clg}(\vC) = \bkt{\vR_C}{\vC}_F = \tr(\vR_C\, \vC)$ for all $\vC \in \Clg$.
Thus the density matrix can be always taken as an element of the sub-algebra, i.e.\
$\DEN{\Clg} \subset \Clg$.
\end{prop}

The next example is similar to \feX{ex:complex-mat}, 
and also typical of a possibly infinite dimensional
non-commutative probability algebra.

\begin{xmpn}[Bounded Operator Algebra]   \label{ex:complex-Hilbert}
Let $\Hvk$ be a complex Hilbert space with inner product $\bkt{\cdot}{\cdot}_{\Hvk}$, 
and $\xi \in \Hvk$ a unit vector.  For $A \in  \BHA$, a bounded linear map,
define a conjugation as the Hilbert space adjoint $A^\dagger$, i.e.\ 
$\bkt{\phi}{A \psi}_{\Hvk} = \bkt{A^\dagger \phi}{\psi}_{\Hvk}$.
With this involution, $\BHA$ becomes a unital *-algebra.
Recall that with the \emph{operator norm} $\nd{A}_\infty$, which satisfies
$\nd{AB}_\infty \le \nd{A}_\infty \nd{B}_\infty$ and $\nd{A^\dagger A}_\infty = \nd{A}_\infty^2$, 
the \emph{Banach space} $(\BHA, \nd{\cdot}_\infty)$ is a 
\emph{Banach algebra} (cf.\ \feD{def:normed_algebras}).

Let $\Alg_\xi \subseteq \BHA$ be a possibly non-commutative unital
complex sub-algebra closed under adjoints, and $\xi \in \Hvk$ a unit vector. 
%, such that for all $A \in \Alg_\psi$
%one has that $\bkt{\psi}{A\psi}_{\Hvk} = 0$ implies $A = 0$.  
Then, for $A \in \Alg_\xi$, one may define a %faithful 
state to be 
$\svpi_{\xi}(A) := \bkt{\xi}{A \xi}_{\Hvk}$.  This is a possibly non-commutative probability 
algebra, and the state $\svpi_{\xi}$ is an example of what is called a vector state,
i.e.\ a pure state (cf.\ \feR{rem:spec-dec-proj-2} and \feP{prop:states-prop}).  
The vector $\xi$ is usually called
the state(-vector) --- typically of some system whose observables are described by $\Alg_\xi$ 
--- which is the original use of the designation ``state''.
In case $\xi$ is a unit eigenvector of a self-adjoint $H \in \Alg_\xi \subseteq \BHA$,
$\som_{\xi}(A) := \bkt{\xi}{A \xi}_{\Hvk}$ is a character on the commutative 
sub-algebra $\CC[H] \subseteq \Alg_\xi \subseteq \BHA$.  Later, in the 
GNS-construction \feT{thm:GNS}, it will be seen that any state can be made to 
look like a vector state.  
\end{xmpn}

\begin{prop}[Trace]  \label{prop:trace}
In case the space $\Hvk$ is finite dimensional, it is possible to define a unique
kind of canonical faithful state.  Recall from linear algebra  (e.g.\ \citep{GreubLA1975})
that in the finite dimensional case on the algebra of linear maps $\LHA = 
\BHA$, there is a linear functional $\sta \in \LHA^*$, such that for 
$A, B \in \LHA$ one has $\sta(AB) = \sta(BA)$, unique up to a scalar factor.
It is automatically invariant under unitaries or internal automorphisms (cf.\ 
\feD{def:alg-names}, just like the usual trace on $\MMn(\CC)$), 
as for a unitary $U \in \UA{\LHA}$
one has $\sta(U^\dagger A U) = \sta(U U^\dagger A) = \sta(A)$.

If this functional $\sta$ is normalised for the identity $I \in\LHA$  to $\dim \Hvk$, 
then it is called the trace $\tr I := \sta(I) = \dim \Hvk$.  The inner product it 
induces on $\LHA$ is $\bkt{A}{B}_{HS} := \tr (A^\dagger B)$, the \emph{Hilbert-Schmidt} 
inner product.  With it one may define a faithful normalised state on $\LHA$, 
which we shall call the \emph{Frobenius} or \emph{canonical} state $\sphi_F$ in analogy to 
\feX{ex:complex-mat}, i.e.\ $\sphi_F(A) := \frk{1}{n}\, \tr A = \frk{1}{n}\, \bkt{I}{A}_{HS}$. 
It is the \emph{unique tracial state} (cf.\ \feD{def:state}), and represents a kind 
of uniform distribution, with density operator $\frk{1}{n}\, I$.
\end{prop}

\begin{rem}[Matrix Representation]   \label{rem:basic-matrix-rep}
Recall from linear algebra (e.g.\ \citep{GreubLA1975})
that in the finite dimensional Hilbert space case with $\dim \Hvk = n$, one 
may take an ortho-normal basis $B = \{ \psi_1, \dots, \psi_n \} \subset \Hvk$ and define a 
\emph{unitary} map $\Hf{B}{\tU}: \Hvk \to \Cn$ by $\Hf{B}{\tU}(\psi_j) = \ve_j$, where 
$\ve_j\in\Cn$ are the canonical unite vectors and $\Cn$ is equipped with the canonical 
inner product $\bkt{\cdot}{\cdot}_c$ as in \feX{ex:complex-n-spc}.  This in turn is 
well known to induce a *-isomorphism $\Hf{B}{\tH}: \BHA \to \MMn(\CC)$ between 
$\LHA = \BHA$ and $\Lop(\Cn) \cong \MMn(\CC)$ from  \feX{ex:complex-mat}, 
given by $\Hf{B}{\tH}(A) = \vM = (M_{ij}) = (\bkt{\psi_i}{A \psi_j}_{\Hvk})$.
\end{rem}

It is now fairly clear how to translate \feR{rem:Schatten-M} for \feX{ex:complex-mat} to 
the finite dimensional case of \feX{ex:complex-Hilbert}:

\begin{rem}[Operator Schatten Norms]   \label{rem:Schatten-H}
In case $\Hvk$ is a finite dimensional Hilbert space, for each operator
$B \in \BHA = \LHA$ and any $p\ge 1$ one may define
the unitarily invariant \emph{Schatten-$p$-norm}
\begin{equation} \label{eq:Schatten-H}
     \nd{B}_{Sp} := \left(\tr \left[(B^\dagger B)^{\frk{p}{2}}\right]\right)^{\frk{1}{p}} ,
\end{equation}
where $(B^\dagger B)^{\frk{p}{2}}$ has to be computed via functional calculus
(cf.\ \feP{prop:Banach-alg}).

Extend this to $p = \infty$ by setting $\nd{B}_{S \infty} := \nd{B}_{\infty}$, 
i.e.\ the \emph{operator norm} of \feX{ex:complex-Hilbert}.
Note that the Schatten-2-norm $\nd{B}_{S 2} = \nd{B}_{HS}$
is equal to the norm from the \emph{Hilbert-Schmidt} inner product,
which makes $\LHA$ into a \emph{Hilbert space}.  
The norm $\nd{B}_{S 1}$ is again called the \emph{trace norm} or \emph{nuclear norm}.
And as \emph{Banach spaces}, $(\LHA, \nd{\cdot}_{S p})$ and 
$(\LHA, \nd{\cdot}_{S q})$ with $\frk{1}{p} + \frk{1}{q} = 1$ are \emph{dual} 
to each other.  With the operator- or Schatten-$\infty$-norm, 
$(\BHA, \nd{\cdot}_\infty)$ is a \emph{Banach algebra} 
(cf.\ \feD{def:normed_algebras}), as Banach space (isomorphic to the) 
dual to the Banach space $(\BHA, \nd{\cdot}_{S 1})$.
\end{rem}

With the unitary $\Hf{B}{\tU}$ and the *-isomorphism $\Hf{B}{\tH}$ as defined in
the above \feR{rem:basic-matrix-rep}, it is now not difficult to translate the statements on 
$\MMn(\CC)$ from \feX{ex:complex-mat} given in \feR{rem:spec-dec-proj-2} about observables, 
PVMs in \feD{def:PVM-M}, densities in \feD{def:density-M}, the set of all densities in 
\feP{prop:set-dens-M}, and densities in the algebra in \feP{prop:dens-in-alg-M}, as well as
the lattice of projections in \feP{prop:proj-matrix-measurement} to this situation of 
the algebra of operators $\LHA = \BHA$ on a finite dimensional Hilbert space.
Without spelling everything out in detail, the main points are collected in:

\begin{coro}[Density Operators and Projection Lattice]  \label{coro:dens-in-alg-H}
Let $\Hvk$ be a finite dimensional Hilbert space with $\dim \Hvk = n$, 
and consider the unital *-algebra $\LHA = \BHA$.  
As $\LHA$ is a Hilbert space with the Hilbert-Schmidt inner product, any state 
$\svpi_R \in \SA{\LHA}$ can be represented by a 
\emph{density operator} or \emph{density matrix}
$R \in \DEN{\LHA} = \{ R \in \LHA_+\; \mid \; \tr R = 1 \}$ by 
$\svpi_R(A) = \bkt{R}{A}_{HS} = \tr (RA)$, which are in a
bijective correspondence with the set of states.

Recalling the rank of an operator $A \in\LHA$, which is given by 
$\rank A = \dim A(\Hvk) = \dim(\im A)$, and denoting 
the density operators of rank $r$ by $\DEN{\LHA}_r$, 
one has the \emph{disjoint union}
\[
  \DEN{\LHA} = \biguplus_{r=1}^n \DEN{\LHA}_r .
\]
The extreme points of $\DEN{\LHA}$ are the rank one operators
\[  
  \mrm{ext}(\DEN{\LHA}) = \DEN{\LHA}_1 = 
  \{  \bkt{x}{\cdot}_{\Hvk}\, x \mid x \in \Hvk, \bkt{x}{x}_{\Hvk} = 1 \}, 
\]
where 
$\bigl(\bkt{x}{\cdot}_{\Hvk}\, x\bigr): \; \Hvk \ni y \mapsto \bkt{x}{y}_{\Hvk}\, x \in \Hvk$.
They are also called \emph{pure} or \emph{vector states}, all other states are called
\emph{mixed states}, and according to the Krein-Milman theorem they are
a convex combination of vector or pure states.  Again, often the unit vector 
$x \in \Hvk$ is called the ``state'' or ``state vector'', 
this is actually the original usage of the term.

The pure states $\DEN{\LHA}_1 \subset \PA{\LHA}$ are orthogonal
projections of unit rank, and are the atoms in the projection lattice 
$\opb{L}(\PA{\LHA})$ of $\LHA$, cf.\ \feP{prop:proj-matrix-measurement}.  
Every projection of rank $m \le n$ can be expressed as the sum of $m$ projections
/ pure states in $\DEN{\LHA}_1$.

And any Abelian sub-*-algebra $\Clg \subseteq \LHA$ (e.g.\ PVM-generated) 
can be embedded in or extended to a \emph{maximal Abelian sub-algebra} (MASA).  
Such an Abelian sub-*-algebra $\Clg$ is a MASA \emph{iff} it is unitarily equivalent
to the algebra of diagonal matrices  \(\DMn(\CC)\subset \MMn(\CC)\) from
\feX{ex:complex-diag}, i.e.\ there is a unitary $\tV: \Hvk \to \Cn$ such
that  $\tV \, \Clg \, \tV^\dagger = \DMn(\CC)$.
\end{coro}

Before entering the subject of representations in the upcoming \fsec{SS:states}, a
first taste of which was provided by \feR{rem:basic-matrix-rep}, it is worthwhile to
look at some more examples of constructions of unital *-algebras using other mathematical
constructs like groups and tensor products.

\begin{xmpn}[Group Algebra]   \label{ex:group-algebra}
Let $G$ be a discrete group with group operation $g \boxtimes h \in G$ for $g,h \in G$.
Consider the free vector space $\CC G$ over $G$, i.e.\ the set of all formal finite linear 
combinations with complex coefficients with the group elements $g_j \in G$ as basis,
with generic elements like  $\phi = \sum_{j=1}^k z_j g_j$, with $z_j \in \CC$ and $g_j \in G$.
As involution $\natural$ on this vector space define for such $\phi$ the expression
$\phi^\natural = \sum_{j=1}^k z^*_j g^{-1}_j$, and as product for $\phi_1=\sum_{i=1}^k z_{1i} g_i, 
\phi_2 = \sum_{j=1}^\ell z_{2j} g_j \in \CC G$ set
\[
  \phi_1 \phi_2 := \sum_{i=1, j=1}^{k,\ell} z_{1i} z_{2j} (g_i \boxtimes g_j) .
\]
which makes $\CC G$ into the \emph{group algebra} of the group $G$.  Observe
that the product can also be written in the form of a \emph{convolution}.
Both the complex numbers $\CC$ and the group $G$ are naturally embedded in $\CC G$.
The unit of the algebra is the same as the group unit $e \in G$, and as state one may
take $\svpi_G(\phi) = \svpi_G(\sum_{j=1}^k z_j g_j) = \sum_{g_j = e} z_j$.  
Then this is a probability algebra.
\end{xmpn}

This is an algebra of bounded RVs.
For a finite group $G$, this is a finite dimensional algebra, and for a commutative
group $G$, this is a commutative algebra.

\begin{xmpn}[Tensor Matrix Algebra]   \label{ex:tensor-mat}
Let $(\Alg_c,\svpi_c)$ be any probability algebra with involution denoted by $\sharp$
--- the probability algebra for the components.  
Now consider $\MMn(\Alg_c)$, the set of matrices with entries from $\Alg_c$.  Define addition and
multiplication by complex scalars component-wise in the obvious fashion.  
Multiplication of two elements is done following the rules of matrix multiplication
for the algebra of matrices with complex entries:  Let $\tX = (\rX_{ik}), \tY = (\rY_{kj}) 
\in \MMn(\Alg_c)$ with $\rX_{ik}, \rY_{kj} \in \Alg_c$, $i,j,k = 1 \dots n$, then their product is 
\[
\tZ = \tX \tY = (\rZ_{i j}) = (\sum_k \rX_{ik} \rY_{kj}),
\]
where the ordering of the factors in the summands matters if $\Alg_c$ is not commutative.
Observe that this construction can also be denoted as $\Alg_c \otimes \MMn(\CC)$,
a tensor product algebra, as will be explained later in \feD{def:tensor-product}.
A general element in this view has the representation $\tZ = \sum_k \rZ_k \otimes \vZ_k$,
with $\rZ_k \in \Alg_c$ and $\vZ_k \in \MMn(\CC)$.

As involution $\flat$ on  $\MMn(\Alg_c)$ take for a $\tZ \in \MMn(\Alg_c)$
\[
\flat: \MMn(\Alg_c) \ni \tZ = (\rZ_{ij}) \mapsto \tZ^\flat := (\rZ^\sharp_{ji}) \in \MMn(\Alg_c), 
\]
this is very similar to what was done in \feX{ex:complex-diag} to define the
conjugation $\tpH$, but here using $\sharp$ instead of complex conjugation $*$.

Towards defining a state, observe that one may form a ``matrix of means'',
i.e.\ for  a $\tZ = (\rZ_{ij}) \in \MMn(\Alg_c)$ compute 
$\wob{\tZ}^{\svpi_c} := (\svpi_c(\rZ_{ij})) \in \MMn(\CC)$, a ``reduced state'' or 
``partial trace''.  In the representation $\tZ = \sum_k \rZ_k \otimes \vZ_k \in 
\Alg_c \otimes \MMn(\CC)$ from above, this partial trace is
$\wob{\tZ}^{\svpi_c} = \sum_k \svpi_c(\rZ_k) \vZ_k \in  \MMn(\CC)$.

As in \feX{ex:complex-mat}, let $\vV \in \MMn(\CC)$ be a self-adjoint positive definite
matrix with unit trace.
Then, for $\tZ \in \MMn(\Alg_c)$, one may define a state to be 
$\svpi_{(\vV, \svpi_c)}(\tZ) := \tr(\vV \wob{\tZ}^{\svpi_c}) = \svpi_{\vV}(\wob{\tZ}^{\svpi_c})$.  
This is a non-commutative probability algebra.  
It has in general no characters when $n>1$, i.e.\ no samples or observations.  
For $n=1$ this is just the component algebra $\Alg_c$, and for $\Alg_c = \CC$
this is \feX{ex:complex-mat}.
\end{xmpn}

This example is essentially a tensor product of some not specified algebra $\Alg_c$
with $\MMn(\CC)$.  Another frequent construction is to take one of the ``function algebras''
\feX{ex:complex-n-spc} or \feeXs{ex:comm-fct-algebra}{ex:cont-fcts}, denoted by 
$(\C{F}(\C{X}),\svpi_{\C{F}})$,
and take the tensor product with some unspecified probability algebra $(\Alg_v, \svpi_v)$
(cf. \  \feD{def:tensor-product}), this gives vectors resp.\ functions with values in $\Alg_v$.
In the examples alluded to above, the probability algebra $\Alg_v$ is $\CC$ 
(cf.\ \feX{ex:complex-num}).

\begin{xmpn}[Random Fields of Algebras]   \label{ex:func-tensor}
Here we take e.g.\ the commutative algebra from \feX{ex:cont-fcts}, $\Ck(\sptldom,\CC)$,
and define the new algebra $\Ck(\sptldom,\Alg_v)$ of continuous functions with values in
$\Alg_v$, with the algebraic operations performed point-wise.  If $\Alg_v$ is commutative,
this is an Abelian algebra, otherwise it is not.  The multiplicative unit is the function
with constant values $\rone_{\Alg_v}$, and the involution is also defined by using
the involution from $\Alg_v$ point-wise as well as complex conjugation.  
Just as in \feX{ex:tensor-mat} a matrix of
means or mean matrix was introduced, so one can define here a ``mean function''.
With a typical element $\vphi \in \Ck(\sptldom,\Alg_v)$ written as $\vphi: \sptldom \ni x 
\mapsto \vphi(x) = \ra(x) \in \Alg_v$, the ``mean function'' is again just applying a 
reduced state: $\wob{\vphi}^{\svpi_v}: \sptldom \ni x \mapsto \wob{\vphi}^{\svpi_v}(\ra(x)) = 
\svpi_v(\ra(x)) \in \CC$.
This means that in case $\sptldom \subset \Rd$ is some spatial domain, one may interpret
$\vphi \in \Ck(\sptldom,\Alg_v)$ as a \emph{random field} with values in $\Alg_v$, 
and $\wob{\vphi}^{\svpi_v} \in \Ck(\sptldom,\CC)$ as its \emph{mean field}.  
The total mean is then computed according
to $\svpi_{\sptldom}(\wob{\vphi}^{\svpi_v}) = \mu(\sptldom)^{-1} \int_{\sptldom}
\wob{\vphi}^{\svpi_v}(x)\, \mu(\di x) $, cf.\ \feX{ex:cont-fcts}.

Observe that this construction can also be denoted as $\Ck(\sptldom,\CC) \otimes \Alg_v$,
a tensor product algebra, as will be explained later in \feD{def:tensor-product}.
A general element then has the representation $\vphi = \sum_k \phi_k \otimes \ra_k$,
with $\phi_k \in \Ck(\sptldom,\CC)$ and $\ra_k \in \Alg_v$, and the partial trace resp.\
reduced state from above, producing the mean function, is given by
$\wob{\vphi}^{\svpi_v} = \sum_k \phi_k \otimes \svpi_v(\ra_k) =  \sum_k \svpi_v(\ra_k)\,\phi_k
\in \Ck(\sptldom,\CC)$.
\end{xmpn}

\paragraph{Dirac's Bra-Ket Notation:}  
Promoted by Dirac in 1939 \citep{Dirac1939},
this notation has caught on and is often used in matters related to quantum behaviour, 
so it is appropriate to mention it here shortly, as it will also be used later.
Let $\C{V}$ be a complex vector space, and $\C{V}^\star$ its dual space.  The duality pairing,
for a $\vv \in \C{V}$ and $\vx \in \C{V}^*$ has been denoted by $\ip{\vx}{\vv} := \vx(\vv)$,
a bilinear form, i.e.\ linear in both arguments.

Dirac proposed for a vector $\vv \in \C{V}$ to use a symbol called ``ket'', 
$\ket{\vv}$, to denote that vector $\ket{\vv} \in \C{V}$,
and even just $\ket{} \in \C{V}$ for a general vector, although this is rarely used.  One nice
feature is that inside the ket one can use an arbitrary label, like $\ket{007}, \ket{+},
\ket{01}, \ket{\%*\&}$, etc., instead of $\vv_{007}, \vv_+, \dots$. 
The surrounding ket makes it clear that this is an element of $\C{V}$.

The ket is the last part of the word ``bra-c-ket'', and the first part, the ``bra'' is written 
like $\bra{\vx}$, and it is a linear form $\vx \in \C{V}^\star$, i.e.\ 
$\C{V}^\star \ni \vx := \bra{\vx} := \bra{\vx}(\cdot) := \braket{\vx | \cdot}$. 
It can be combined with the ket to form a ``bra-ket'' $\braket{\vx | \vv}$, similarly to the
the duality pairing above, but with a difference, as the bra as well as the whole 
``bra-ket'' is anti-linear in the first argument:  for $\bra{\vz} = \bra{\vx + \ii \vy}$
with $\bra{\vx}$ and $\bra{\vy}$ from the underlying space $\C{V}$,
one has $\bra{\vz} = \bra{\vx} - \ii \bra{\vy}$, so that 
$\braket{\vz | \vv}$ is a sesqui-linear form, 
linear in the second and anti-linear in the first argument.
Observe that the vertical bar 
is usually only written once when a bra and a ket are combined to a bra-ket.  
%The assignment
%$\vx \mapsto \bra{\vx}=\braket{\vx |\cdot }$ is anti-linear, so that one may
%view this as an \emph{anti-linear} map $\C{V} \to  {\C{V}}^\star$
%into the dual space $\C{V}^\star$.

In case that $\Hvk$ is a Hilbert space with a sesqui-linear inner product 
$\braket{\ra | \rb}$ with $\ra, \rb \in \Hvk$, it is clear that
$\Hvk$ is isometrically isomorphic with its dual $\Hvk^*$ via the anti-linear 
Riesz-map $R: \Hvk \ni \ra \mapsto \bra{\ra} := R \ra \in {\Hvk}^*$  
to form the inner product $\braket{\ra | \rb} := \ip{R \ra}{\rb}$.  
Thus the bra is a linear map:
$R \ra = \bra{\ra} = \braket{\ra | \cdot} \in \Lop(\Hvk,\CC) = \Hvk^*$.  
From this, note that
another useful notation is the tensor product $\ket{\rb}\bra{\ra} =
\ket{\rb}\otimes\bra{\ra} \in \LHA$, 
which could be written as $\rb \otimes R \ra$, and which is a projection along 
$\ket{\ra}$ onto $\spn \{ \ket{\rb} \}$:
\[
   \ket{\rb}\bra{\ra}: \Hvk \ni \ket{\rx} \mapsto \braket{\ra | \rx} \, \ket{\rb} \in \Hvk .
\]
This notation is expanded generally for all kinds of tensor products.  
Let $\ra, \rb, \rc \in \Hvk$, and
consider the tensor product of $\Hvk$ with itself, where
an elementary tensor is often written as $\ra \otimes \rb \in \Hvk\otimes\Hvk = 
\Hvk^{\otimes 2}$.  In Dirac's notation, this can be just abbreviated to
$\ket{\ra} \ket{\rb} := \ket{\ra} \otimes \ket{\rb} = \ra \otimes \rb$, and even
$\ket{\ra \rb} := \ket{\ra}\ket{\rb}$.  Looking at e.g.\
$\ra \otimes \rb \otimes \rc \in \Hvk^{\otimes 3}$, this simply becomes
$\ket{\ra \rb \rc} := \ket{\ra} \ket{\rb} \ket{\rc} \in \Hvk^{\otimes 3}$.

Let $A \in \LHA$ be a linear map.  Then, for the inner product between
$\rc$ and $\ry = A \rx$, one may write:
\[
   \braket{\rc | \ry} = \braket{\rc \mid A \rx} = \bra{\rc} A \ket{\rx} = 
   \braket{A^\dagger \rc \mid \rx},
\]
where $A^\dagger \in \LHA$ is the adjoint of $A$.  For the special case from
above, where $A = \ket{\rb}\bra{\ra}$ (with $A^\dagger = \ket{\ra} \bra{\rb}$) 
one obtains with $\ry = A \rx = \braket{\ra | \rx} \ket{\rb}$ the following:
$
 \braket{\rc | \ry} = \bra{\rc} \; ( \ket{\rb}\bra{\ra} ) \; \ket{\rx} = 
 \braket{\rc | \rb} \braket{\ra | \rx},
$
showing the versatility and practicality of the notation.

In the finite dimensional complex Hilbert space $\Cn$ --- where one may denote an element
$\va \in \Cn$ also as $\ket{\va} \in \Cn$ ---  with the canonical inner product 
$\braket{\va | \vb}_c = \va^\tpH \vb$, the Riesz map can be written as $R: \va \mapsto \va^\tpH$.  
The tensor product $\ket{\vb} \bra{\va}$ is then equivalent to the expression 
$\vb \va^\tpH: \vx \mapsto (\va^\tpH \vx) \, \vb$.  The canonical basis $\ve_i$ may just
be written as $\ve_i = \ket{i}$.  A matrix $\vA \in \MMn(\CC)$ can be written as
$\vA = (A_{ij})_{i,j =1}^n = \sum_{i,j =1}^n A_{ij} \ket{i}\bra{j}$.

%\[
% \bra{\ell}\ket{22}; \bra{}; A \ket{v}; \bkt{a}{A b}; \braket{a|Ab}; \braket{A}; \set{a, b, c, d}
%\]

\subsection{Representations and Hilbert Space}  \label{SS:states}
The concept of representation, representing the algebra $\Alg$ as a sub-algebra of the
algebra of linear operators $\LHA$ on some vector space $\Hvk$,
turns out to be of great importance in algebraic probability.

\paragraph{General Representations:}
\begin{defi}[Representation]  \label{def:gen-repr}
Let $\Hvk$ be a vector space, and $\LHA$ the algebra of linear maps 
(operators) on $\Hvk$.
A representation of an algebra $\Alg$ is an algebra homomorphism $\oL$ of $\Alg$ into
$\LHA$, i.e.\ $\Alg \ni \ra \mapsto \oL(\ra) =: \oL_{\ra} \in \LHA$.
This means that $\oL: \Alg \to \LHA$ is linear, and
$\Alg \ni \ra \rb \mapsto \oL_{\ra \rb} = \oL_{\ra} \oL_{\rb} \in \LHA$.
The subspace $\im \oL = \oL(\Alg) \subseteq \LHA$ is a sub-algebra of $\LHA$.

For a unital algebra $\Alg$ one obtains that $\oL(\rone) = \oL_{\rone}$ has to act 
like the identity on $\im \oL$.
%, and one usually requires that 
%$\oL(\rone)  = I_{\Hvk} \in \LHA$, where $I_{\Hvk}$ 
%is the identity operator on $\Hvk$.

In case $\Alg$ is a *-algebra, and $\Hvk$ an inner product 
(pre-Hilbert) space (cf.\ \feX{ex:complex-Hilbert}),
then $\oL$ is a *-representation in case it is a *-homomorphism, 
i.e.\ if in addition for all $\ra\in\Alg$ one has 
$\oL(\ra^\star) = \oL_{\ra^\star} = \oL_{\ra}^\dagger = \oL(\ra)^\dagger$, 
where $\oL_{\ra}^\dagger \in \LHA$ is the adjoint of $\oL_{\ra}$.

If $\oL$ is injective, the representation is called \emph{faithful}.  In this case
the sub-algebra $\im \oL = \oL(\Alg) \subseteq \LHA$ is isomorphic to $\Alg$.
\end{defi}

We have already encountered observations or samples, i.e.\ non-zero *-homomorphisms 
or characters onto $\CC$ in \feD{def:sample}.   These are one-dimensional 
representations.  One of the most natural ones is the regular representation:

\begin{defi}[Regular Representation]  \label{def:reg-repr}
Let $\Alg$ be an algebra.
For each $\ra \in \Alg$, define a linear map $\oL_{\ra} \in \Lop(\Alg)$ 
%--- this denotes the linear, possibly unbounded, maps $\Alg \to \obH$ --- 
for all $\rb \in \Alg$ by $\oL_{\ra} \rb := \ra \rb$.
The homomorphism $\oL: \Alg \ni \ra \mapsto \oL(\ra) = \oL_{\ra} \in \Lop(\Alg)$ is 
called the \emph{(left) regular representation}.
%For unital algebras, one usually sets explicitly 
%$\oL(\rone) = \oL_{\rone} = I_{\Alg} \in \Lop(\Alg)$.
\end{defi}

As a concrete example, one might view \feX{ex:complex-diag} as the
regular representation of \feX{ex:complex-n-spc}.
The regular representation, \feD{def:reg-repr}, is faithful for unital algebras 
\citep{DrozdKiri94}:

\begin{thm}[Cayley]  \label{thm:Cayley-regrep}
Every unital algebra $\Alg$ admits a faithful representation, e.g.\ via the regular
representation $\oL$ from \feD{def:reg-repr}.  In other words, this means that $\Alg$ 
is isomorphic with the sub-algebra $\im \oL = \oL(\Alg) \subseteq \Lop(\Alg)$.
\end{thm}
\begin{proof}
If for $\ra, \rb \in \Alg$ it holds for the regular representation (\feD{def:reg-repr}) 
that $\oL_{\ra} = \oL_{\rb}$ in $\Lop(\Alg)$, 
then for a unital algebra $\oL_{\ra} \rone = \ra \rone = \ra = \oL_{\rb} \rone = \rb$,
and hence $\ra = \rb$; proving that the regular representation is faithful.
\end{proof}

As the interest here is mainly on *-representations, the focus will be in the sequel
on cases where the undelying vector space $\Hvk$ can be equipped with a Hilbert
space structure, cf.\ \feD{def:gen-repr}.  To exclude many almost pathological
situations, for the sake of simplicity one may also assume that the algebra
is a Banach *-algebra (cf.\ \feD{def:normed_algebras}).

\begin{prop}[Proper or Non-degenerate Representation]  \label{prop:reg-repr}
Let $\Alg$ be a Banach *-algebra, 
$\Hvk$ a Hilbert space, and $\oL: \Alg \to \BHA$ a *-representation.
Then the following are equivalent (cf.\ \citep{Takesaki1}):
\begin{compactitem}

\item $\spn \{ \oL_{\ra} \xi \mid \ra \in \Alg,\, \xi \in \Hvk\}$ is dense in $\Hvk$.

\item For any non-zero $\xi \in \Hvk$, there exists an $\ra \in \Alg$ such that
      $\oL_{\ra} \xi \ne 0$.

\end{compactitem}
If either one of these conditions is satisfied, the representation  
$\oL: \Alg \to \LHA$ is said to be \emph{proper} or \emph{non-degenerate}.
\end{prop}

Later in the GNS-construction to follow in \feT{thm:GNS}, 
the state will become a pure state, defined by a cyclic and separating vector,
so we define this for any representation:

\begin{defi}[Cyclic and Separating Vector]  \label{def:alg-Hilbert}
A representation $\oL: \Alg \to \BHA$ as in \feP{prop:reg-repr}, 
with $\Hvk$ a Hilbert space, is 
called a \emph{cyclic} representation (cf.\ \feX{ex:complex-Hilbert}) \emph{iff} 
there is a unit vector $\xi \in \Hvk$, also called a \emph{cyclic vector},
such that the \emph{orbit} $\{ \oL(\ra)\xi \mid \ra \in \Alg \} \subset \Hvk$ 
is \emph{dense} in $\Hvk$.

A unit vector $\xi \in \Hvk$ is called a \emph{separating} vector for the representation
$\oL: \Alg \to \BHA$ \emph{iff} for all $\ra \in \Alg$ it holds that 
$\oL(\ra) \xi = 0$ implies $\ra = \rnul$.
\end{defi}

Note that a cyclic representation is proper or non-degenerate, and that observations
(if they exist) are cyclic.

\begin{defi}[Irreducible Representation]  \label{def:irreduc-rep}
Given a representation $\oL: \Alg \to \BHA$ as in \feP{prop:reg-repr}, 
with $\Hvk$ a Hilbert space, a closed subspace $\Kvk \subset \Hvk$ is 
called an \emph{invariant} subspace of the representation $\oL$,
\emph{iff}  $\oL(\ra)(\Kvk) \subseteq \Kvk$ for all $\ra \in \Alg$.  In this
case the restriction $\oL(\ra)|_{\Kvk} \in \C{B}(\Kvk)$ yields a 
\emph{sub-representation} of $\oL$ on the Hilbert space $\Kvk$.  And it is easy
to see that the orthogonal complement $\Kvk^\perp \subset \Hvk$ is also an invariant
subspace of  the representation $\oL$, so that 
$\oL = \oL(\ra)|_{\Kvk} \oplus \oL(\ra)|_{\Kvk^\perp}$, i.e.\ the representation
$\oL$ is the \emph{direct sum} (cf.\ \feD{def:direct-sum-maps}) of two sub-representations.

If the only invariant subspaces of the  representation $\oL: \Alg \to \BHA$
are $\{0\}$ and $\Hvk$, the representation is called \emph{irreducible}; otherwise
it is called \emph{reducible}.
\end{defi}

It should be obvious that a cyclic representation (cf.\ \feD{def:alg-Hilbert})
is irreducible.  In fact one has \citep{Landsman2017}:

\begin{thm}[Schur’s Lemma]   \label{thm:irreduc-reps-cond}
For a representation $\oL: \Alg \to \BHA$ as in \feP{prop:reg-repr}
with $\Hvk$ a Hilbert space, \emph{irreducibility} (\feD{def:irreduc-rep}) is 
equivalent to each of the following statements:
\begin{compactitem}
\item  Any unit vector $\xi \in \Hvk$ is a \emph{cyclic vector} for $\oL$ 
      (cf.\ \feD{def:alg-Hilbert});
\item  $\oL(\Alg)^\prime = \CC[I]$ --- the commutant $\oL(\Alg)^\prime$ (\feD{def:alg-names})
      of the image $\oL(\Alg)$ are just the multiples of the identity, 
      this is called \emph{Schur's lemma}.
%\item  $(\oL(\Alg)^\prime)^\prime} = \BHA$ --- this statement for .
\end{compactitem}
\end{thm}

Irreducible representations are the basic building blocks 
of general representations:

\begin{prop}[Sum of Cyclic Representations]  \label{prop:sum-cyclic-repr}
Let $\Alg$ be a Banach *-algebra (cf.\ \feD{def:normed_algebras}), 
$\Hvk$ a Hilbert space, and assume that
$\oL: \Alg \to \BHA$ a proper, nondegenerate *-representation,
cf.\  \feP{prop:reg-repr}.  Then (cf.\ \citep{Takesaki1}) it is
the \emph{direct sum} (cf.\ \feD{def:irreduc-rep}) of \emph{cyclic representations},
cf.\ \feD{def:alg-Hilbert}.
\end{prop}

It is fairly obvious by now that any finite dimensional algebra $\Alg$ is isomorphic to a sub-algebra
of $\MMn(\CC)$ of \feX{ex:complex-mat}.  
%As that remark shows,

\begin{rem}[Matrix Algebra Representation]  \label{rem:fin-dim-mat-rep}
Often one starts out from the regular representation \feD{def:reg-repr}.
Then it all comes down to a choice of basis, and there is in general no canonical 
way of doing so.
A path which works in any algebra is
using the structure constants in \feD{def:struct-const}.
%while being careful to keep the linear vector isomorphism $\tV: \Alg \to \Cn$
%apart from the algebra-homomorphism $\tM: \Alg \to \MMn(\CC)$; in detail:  
But here we shall stay with the more restricted situation involving an
underlying Hilbert space structure, see also \feX{ex:complex-Hilbert} and the
resulting \feR{rem:basic-matrix-rep}.

\end{rem}

There are other ways of achieving a matrix representations \citep{Takesaki1}, but
they are connected to the structure theorems to be discussed later in 
\fsec{SS:funcs-RVs-norm-alg}.  Before refining the foregoing to the case of probability 
algebras, it is worthwhile to take a closer look at the states $\SA{\Alg} \subset \Alg^+$.

\paragraph{States of the Probability Algebra:}
In the previous representation results, \feT{thm:Cayley-regrep} and \feR{rem:fin-dim-mat-rep},
the representation was for valid for any algebra,
independent of the state $\svpi \in \SA{\Alg}$ which defines probabilities.  

To achieve a representation on a Hilbert space which takes care
of this, one has to consider the given state $\svpi$, to which we turn next.
Consider the dual space $\Alg^\star$ of $\Alg$.  Any $\sbt \in \Alg^\star$ defines a 
sesqui-linear form $b$ on $\Alg \times \Alg$ --- linear in the second argument and 
anti-linear in the first --- and a linear map $B: \Alg \to \Alg^\star$:
\begin{equation}  \label{eq:def-sesqi-gen}
  \forall \ra, \rb \in \Alg:\, b(\ra,\rb) := \sbt(\ra^\star \rb) = 
  \ip{\sbt}{\ra^\star \rb} =: \ip{B \ra^\star}{\rb},
\end{equation}
where one may recall that $\ip{\cdot}{\cdot}$ is the duality bracket on $\Alg^\star \times \Alg$.

For a such a map $B \in \Lop(\Alg, \Alg^\star)$, in reflexive situations where
$(\Alg^{\star})^\star \cong \Alg$ as for finite dimensional algebras, 
one may view the transpose or dual map $B^* \in \Lop((\Alg^\star)^\star, \Alg^\star)$
as a map $B^* \in \Lop(\Alg, \Alg^\star) \cong \Lop((\Alg^\star)^\star, \Alg^\star)$.
The map $B \in \Lop(\Alg, \Alg^\star)$ is called self-dual 
(or sometimes self-adjoint)
if  $B^* = B$.  Complementing \feD{def:state}, one is lead to

\begin{defi}  \label{def:pos-defnt}
In case $B$ is \emph{self-dual} (or self-adjoint), 
i.e.\ $\ip{B \ra^\star}{\rb} = \ip{B \rb}{\ra^\star} = \ip{B \rb^\star}{\ra}^*$,  
then $b$ and $\sbt$ are called self-adjoint (Hermitean), the self-adjoint functionals
are denoted by $\Alg^{sa} \subset \Alg^\star$.  This means that $\sbt(\ra)$ is real on 
self-adjoint $\ra$;  for $\sbt \in \Alg^{sa}$ one has $\sbt(\Alg_{sa}) \subseteq \RR$.

For $\sbt \in \Alg^\star$ one may define an involution $\natural: \sbt \mapsto \sbt^\natural$
such that for $\ra \in \Alg$ one has $\sbt^\natural(\ra) := \sbt(\ra^\star)^*$.
For self-adjoint $\sbt \in \Alg^{sa}$ one has $\sbt^\natural = \sbt$.

In case a self-dual $B$ is \emph{positive} ($\ip{B \ra^\star}{\ra} \ge 0$), 
the same is attached to $b$ and $\sbt \in \Alg^{sa}$, and in case
$B$ is \emph{positive definite} ($\ip{B \ra^\star}{\ra} > 0$ for $\ra \ne \rnul$),
so is $b$ and $\sbt$.

All such positive $\sbt \in \Alg^{sa} \subseteq \Alg^\star$ form a convex cone,
the \emph{dual cone} 
$\Alg^+ := \{ \sbt \in \Alg^{sa} \mid \ip{\sbt}{\ra} = \sbt(\ra) \ge 0 \quad 
\forall \ra \in \Alg_+ \}$,
dual to the positive cone $\Alg_+$, cf.\ \feD{def:alg-names}.

One may recall, cf.\ \feD{def:state}, that a self-adjoint and positive $\sbt \in \Alg^\star$ 
is called a \emph{state}, and if it is additionally positive
definite, $\sbt$ is also called \emph{faithful}, and  \emph{normalised} if $\sbt(\rone) = 1$.
\end{defi}

Note that the definition of the sesqui-linear form in \feq{eq:def-sesqi-gen}
is sometimes the other way around, i.e.\ linear
in the first argument and anti-linear in the second.

\begin{prop}[Geometry of the Set of States]  \label{prop:states-prop}
The self-adjoint functionals $\Alg^{sa} \subset \Alg^\star$ are a real vector space, and it 
holds that $\Alg^\star = \Alg^{sa} \oplus \ii \Alg^{sa}$ as a direct sum over $\RR$, as well as
$\Alg^\star = \spn_{\CC} \Alg^{sa}$, i.e.\ any functional is the complex linear combination
of two self-adjoint functionals, cf.\ \feD{def:alg-names}.  

The dual positive cone $\Alg^+ \subset \Alg^\star$ is a \emph{generating} cone, i.e.\
$\Alg^{sa} = \Alg^+ - \Alg^+$ over $\RR$ and $\Alg^\star = \Alg^+ - \Alg^+$ over $\CC$.
This means that any element $\sal \in \Alg^{sa}$ is the difference of two positive 
elements (corresponding to the \emph{Jordan decomposition} of measures):
$\sal = \sphi - \spsi$ with $\sphi, \spsi \in \Alg^+$, and hence any functional 
$\sbt \in \Alg^\star$ is the complex combination of four positive elements in $\Alg^+$,
cf.\ \feD{def:alg-names}.

The elements $\sbt \in \Alg^\star$ which satisfy $\sbt(\rone) = \ip{\sbt}{\rone} = 1$
form a hyperplane $\C{T} \subset \Alg^\star$.  Thus the set of normalised states
$\SA{\Alg} = \C{T} \cap \Alg^+$ is a convex set lying in the linear manifold $\C{T}$.  
Its extreme points are called
\emph{pure states}, all other states are called \emph{mixed states}.  
Vector states (cf.\ \feX{ex:complex-Hilbert}) are pure states,
as are characters $\som \in \XA{\Alg} \subset \SA{\Alg}$, if $\XA{\Alg}$ is non-empty.
\end{prop}

For more information on the geometry of the set of states, see \citep{AubrunSzarek2017}.
The next task is to define a trace functional on an abstract algebra $\Alg$:

\begin{prop}[Trace on $\Alg$]  \label{prop:fin-dim-alg-gen}
In case $\Alg$ is a finite dimensional algebra, one has that $\Alg = \BA{\Alg}$,
as from \feT{thm:Cayley-regrep} one sees that $\Alg$ is isomorphic to a sub-algebra
of $\Lop(\Alg) = \C{B}(\Alg)$, the linear maps of a finite dimensional space, as 
on finite dimensional spaces all linear maps are bounded and continuous \citep{Schaefer99}.

Now by the same argument as in \feP{prop:trace}, there is a unique tracial linear 
functional $\tr$ on $\Lop(\Alg) = \C{B}(\Alg)$, normalised to 
$\tr(I_{\Alg}) =  \dim \Alg$.  This one may use, via the faithful
regular representation $\oL: \Alg \to \Lop(\Alg)$ from \feD{def:reg-repr}, %.  To that end
to define the trace functional $\tr_{\Alg} \in \Alg^+$ on $\Alg$ by setting 
$\tr_{\Alg}(\ra) := \tr \oL_{\ra}$, or, more formally, $\tr_{\Alg} := \tr \circ \oL$.
In the sequel it is assumed that $\frk{1}{(\dim \Alg)}\,
\tr_{\Alg} \in \SA{\Alg}$ is non-degenerate (cf.\ \feR{rem:bounded-ideal}), 
resp.\ a faithful state (cf.\ \feP{prop:bounded-ideal}).

For a unital algebra one has that 
$\tr_{\Alg}(\rone) = \tr(\oL_{\rone}) = \tr(I) = \dim \Alg$.
and hence, according to \feR{rem:normalised-positive}, for a  *-algebra $\Alg$, 
$\sphi_c := \frk{1}{(\dim \Alg)}\, \tr_{\Alg} \in \SA{\Alg}$ is a faithful 
canonical state on $\Alg$.  
\end{prop}

%The results from this \feP{prop:fin-dim-alg-gen} will now be used to help define 
%a Hilbert space structure on $\Alg$, and with it a faithful Hilbert space representation
%even in the case where the original.

\paragraph{Associated Hilbert Space Representation:}
The \feR{rem:basic-matrix-rep} and \feR{rem:fin-dim-mat-rep} show how to map
a finite dimensional algebra onto a 
matrix algebra.  To achieve a *-representation for a unital *-algebra $\Alg$, one 
proceeds via a Hilbert space construction, and an associated representation,
the so-called GNS (Gel'fand - Neumark (Naĭmark)) - Segal) 
construction \citep{gelfandNaimark1943, segal1947, DorBel1986}.
%may proceed as in 
%\feP{prop:fin-dim-alg-gen}, and additionally choose a particularly simple basis.

\begin{prop}[Associated Hilbert Space]  \label{prop:state-Hilbert-spc}
Given a probability algebra $(\Alg, \svpi)$ with a faithful state $\svpi \in \SA{\Alg}$, 
define a positive definite sesqui-linear form
on $\Alg$ like in \feq{eq:def-sesqi-gen}, denoted as 
$\bkt{\ra}{\rb}_{2} := \svpi(\ra^\star \rb)$, an \emph{inner product}.

With this inner product, $\Alg$ becomes a pre-Hilbert space, and we denote its 
Hilbert space completion by $\Hvk_{\svpi} := \Lp_2(\Alg,\svpi)$, or by 
$\obH=\Lp_2(\Alg)$ for short if it is clear from the context which state
$\svpi \in \SA{\Alg}$ is meant.
This is the Hilbert space associated with the probability algebra $(\Alg,\svpi)$.  
The associated Hilbert norm is denoted by
$\nd{\ra}_2 := \sqrt{\bkt{\ra}{\ra}_{2}}$.

The direct sum $\Alg = \wob{\Alg} \oplus \wtl{\Alg}$ from \feD{def:alg-names} becomes
an \emph{orthogonal} direct sum, as with $\wob{\Alg} = \CC[\rone]$ for $\wtl{\ra} \in \wtl{\Alg}$ 
one has $\bkt{\rone}{\wtl{\ra}}_{2} := \svpi(\rone \wtl{\ra}) = \svpi(\wtl{\ra}) = 0$.
This orthogonal decomposition can be extended to $\obH = \wob{\obH} \oplus \wtl{\obH}$, with
$\wob{\obH} = \wob{\Alg} = \CC[\rone]$ and $\wtl{\obH}$ the completion of $\wtl{\Alg}$.

%Finite dimensional algebras $\Alg$ are already complete, so $\obH = \Alg$.
In case $\Alg$ is finite dimensional, it is already complete in all vector
Hausdorff topologies \citep{Schaefer99}, and thus one has $\obH = \Alg$ in this instance.
  Generally, $\Alg \subseteq \obH$ is a dense subspace, and  
the state $\svpi$ may be extended to $\obH$ by continuity.
\end{prop}

One such faithful (but un-normalised) state on a finite dimensional unital *-algebra $\Alg$ 
is the trace functional $\tr_{\Alg} \in \Alg^+$ introduced in \feP{prop:fin-dim-alg-gen}.
This may be used to introduce an alternative inner product on $\Alg$:

\begin{rem}[Canonical Associated Finite Dimensional Hilbert Space]  \label{rem:c-Hilbert-spc}
For a finite dimensional *-algebra $\Alg$,
according to \feP{prop:trace}, the trace on $\Lop(\Alg)$ induces
the \emph{Hilbert-Schmidt} inner product $\bkt{\cdot}{\cdot}_{HS}$ on $\Lop(\Alg)$.
%\[   
%  \bkt{\ra}{\rb}_c :=  \tr_{\Alg} (\ra^\star \rb) = \tr \oL(\ra^\star \rb) =
%  \bkt{\oL_{\ra}}{\oL_{\rb}}_{HS}   = \tr (\oL^\dagger_{\ra} \oL_{\rb}) = 
%     \tr (\oL_{\ra^\star} \oL_{\rb}),   
%\]
Via the trace $\tr_{\Alg} \in \Alg^+$ considered in \feP{prop:fin-dim-alg-gen}
for this finite dimensional case, 
a \emph{canonical} inner product on $\Alg$ may be introduced:
\[
\forall \ra, \rb \in \Alg: \; \bkt{\ra}{\rb}_c :=  \tr_{\Alg} (\ra^\star \rb) , 
\]
making the finite dimensional unital *-algebra $\Alg$ into a Hilbert space
$(\Alg,\bkt{\cdot}{\cdot}_c)$, which will be denoted by $\Hvk_c$.  
The regular representation \feD{def:reg-repr}
then satisfies following relation: for all $\ra, \rb \in \Hvk_c$,
\[
\bkt{\ra}{\rb}_c = \tr_{\Alg} (\ra^\star \rb) =  \tr (\oL_{\ra^\star} \oL_{\rb}) 
 = \tr (\oL_{\ra}^\dagger \oL_{\rb}) = \bkt{\oL_{\ra}}{\oL_{\rb}}_{HS} ,
\]
where the adjoint $\oL_{\ra}^\dagger$ is w.r.t.\ $\Lop(\Hvk_c)$.
This means that $\oL$ is a unitary map between 
the Hilbert space $\Hvk_c$ and the Hilbert sub-space
$\oL(\Alg)$ of $\Lop(\Alg)$ with the Hilbert-Schmidt inner
product $\bkt{\cdot}{\cdot}_{HS})$, cf.\ \feP{prop:trace}.

The connected canonical Frobenius state $\sphi_c$ defined 
in \feP{prop:fin-dim-alg-gen} on $\Alg$ is obviously a faithful state, 
so that all the statements from \feP{prop:state-Hilbert-spc} apply
to $(\Alg, \sphi_c)$.
\end{rem}

Not surprisingly, \feR{rem:Schatten-H} for \feX{ex:complex-Hilbert} and
\feR{rem:Schatten-M} for \feX{ex:complex-mat} have a version for
finite dimensional *-algebras:

\begin{rem}[Algebra Schatten Norms]   \label{rem:Schatten-A}
In case $\Alg$ is a finite dimensional *-algebra, following \feP{prop:fin-dim-alg-gen},
for each $\rb \in \Alg$ and any $p\ge 1$ one may define the \emph{Schatten-$p$-norm}
\begin{equation} \label{eq:Schatten-A}
     \nd{\rb}_{Sp} := 
     \left(\tr \left[(\oL_{\rb^\star} \oL_{\rb})^{\frk{p}{2}}\right]\right)^{\frk{1}{p}}
        =\left(\tr \left[(\oL_{\rb}^\dagger \oL_{\rb})^{\frk{p}{2}}\right]\right)^{\frk{1}{p}},
\end{equation}
and extend this to $p = \infty$ by setting $\nd{\rb}_{S \infty} := \nd{\oL_{\rb}}_{\infty}$, 
i.e.\ the \emph{operator norm} of $\oL_{\rb} \in \C{B}(\Alg) = \Lop(\Alg)$.
Note that the Schatten-2-norm $\nd{\rb}_{S2} = \nd{\rb}_{c}$
is equal to the norm from the canonical inner product from \feR{rem:c-Hilbert-spc}.  

The norm $\nd{\rb}_{S1}$ is again called the \emph{trace norm} or \emph{nuclear norm}.
And as \emph{Banach spaces}, $(\Alg, \nd{\cdot}_{Sp})$ and 
$(\Alg, \nd{\cdot}_{Sq})$ with $\frk{1}{p} + \frk{1}{q} = 1$ are \emph{dual} 
to each other.  With the operator- or Schatten-$\infty$-norm, 
$(\Alg, \nd{\cdot}_\infty)$ is a \emph{Banach algebra} (cf.\ \feD{def:normed_algebras}).
\end{rem}

Let $(\Alg,\sal)$ be a probability algebra.  In order to achieve a faithful 
Hilbert space representation, a faithful state is necessary.  To this end define
\begin{equation}  \label{eq:GNS-faithf-state}
\svpi = \begin{cases}
          \sal & \text{if $\sal$ is faithful}.\\
          \sphi_c & \text{from \feP{prop:fin-dim-alg-gen} otherwise}.
        \end{cases}
\end{equation}
This ensures that $\svpi \in \SA{\Alg}$ is a faithful state.  Using this
state $\svpi \in \SA{\Alg}$ from \feq{eq:GNS-faithf-state}, by now considering
the probability algebra $(\Alg,\svpi)$, one may now formulate a finite 
dimensional Hilbert space representation as the
simplified version of a  basic GNS (Gel'fand - Neumark (Naĭmark)) - Segal)
construction \citep{gelfandNaimark1943, segal1947, DorBel1986}:

\begin{thm}[Finite Dimensional GNS Construction]  \label{thm:GNS}
Let $(\Alg,\svpi)$ be a probability algebra where all elements are bounded $\Alg = \BA{\Alg}$
--- e.g.\ when $\Alg$ is finite dimensional --- with
a faithful state $\svpi \in \SA{\Alg}$, and according to \feP{prop:state-Hilbert-spc}
form the Hilbert space $\obH$ with inner product $\bkt{\cdot}{\cdot}_2$ through 
completion.  Combine this with the regular representation  \feD{def:reg-repr} to achieve
a faithful *-representation $\oL:\, \Alg \to \C{B}(\obH)$, cf.\ \feD{def:gen-repr},
with $\ra^* \mapsto \oL(\ra^\star) = \oL(\ra)^\dagger$, where $\oL(\ra)^\dagger$
is the Hilbert space adjoint to $\oL(\ra)$ w.r.t.\ $\bkt{\cdot}{\cdot}_2$.
In other words, $\Alg$ is *-isomorphic to $\oL(\Alg) \subseteq \Lop(\obH)$.  

In this associated Hilbert space framework, note that for $\rx \in \obH$:
\begin{equation}  \label{eq:exp-ipr}
    \Ex_2(\rx)  :=  \svpi(\rx) = \svpi(\rone^\star \rx)    =  \bkt{\rone}{\rx}_2,
\end{equation} 
with $\svpi$ continuously extended to $\obH$.
Hence, as $\oL(\ra) \rone = \oL_{\ra} \rone = \ra \rone = \ra$, one obtains
\begin{equation}  \label{eq:exp-means}
\Ex_{\oL}(\oL_{\ra}) :=  \bkt{\rone}{\oL_{\ra} \rone}_2 ,
\end{equation}
to define the corresponding \emph{vector state} $\Ex_{\oL}$ on $\C{B}(\obH)$.
This means that the state $\svpi \in \SA{\Alg}$ has been mapped onto a \emph{vector}
or \emph{pure state} for this representation in the associated Hilbert space algebra 
$\oL(\Alg) \subset \C{B}(\obH)$ with state vector $\xi = \rone \in \Alg \subseteq \obH$.
Note that the two random variables $\ra \in \Alg$ and $\oL_{\ra} \in \oL(\Alg)$
are \emph{moment equivalent}, $\ra \stoeq \oL_{\ra}$.

The vector $\xi = \rone \in \Alg \subseteq \obH$ in \feq{eq:exp-means} 
is a \emph{separating} vector for $\oL(\Alg)$, 
which according to \feD{def:alg-Hilbert} means that if $\oL(\ra) \rone = \rnul = \ra$, 
then  $\oL(\ra) = \vek{0}$, which is the case only for $\ra = \rnul$. 

The regular representation on $\obH$ just so constructed is also a \emph{cyclic} 
representation (cf.\ \feD{def:alg-Hilbert}),
as $\{ \oL(\ra) \rone = \ra \mid  \ra \in \Alg \} = \Alg$ %($\oL(\ra) \rone = \ra \rone = \ra$)
is by definition dense in $\obH$.
The vector $\xi = \rone \in \Alg \subseteq \obH$ is hence a \emph{cyclic} unit vector.
\end{thm}

One may recall that a vector state like \feq{eq:exp-means} is a pure state
of $\Lop(\obH)$, so this is an example of \emph{purification}, achieved through
 what is called the ``Church of the Larger Hilbert Space''
\citep{Smolin-Church} (cf.\ also \citep{Leifer-Church}).
It is accomplished through a representation in a ``larger'' Hilbert space, or rather
by embedding it as a sub-algebra $\oL(\Alg) \subseteq \Lop(\obH)$ of a ``larger'' algebra.
For such a state $\sbt(\ra) := \bkt{\xi}{\oL_{\ra}\, \xi}_2$, given as a vector state with
a normalised vector $\xi \in \obH$ as in \feq{eq:exp-means} in the above \feT{thm:GNS}, 
the vector $\xi$ itself is called a state or state vector
(of some system), and in fact this is the original use of the term ``state''.

In \feT{thm:GNS} a specific Hilbert space $\obH$ was constructed from the algebra $\Alg$
itself via the regular representation.  But the idea of a faithful pure representation 
is so useful, that one wants to use the term more generally:

\begin{defi}[GNS-Representation]  \label{def:GNS-rep}
Let there be given a probability algebra $(\Alg, \svpi)$ and  a 
\emph{faithful representation} $\Psi: \Alg \to \BHA$ of $(\Alg, \svpi)$ in 
$(\BHA, \som_\xi)$, where, as in \feX{ex:complex-Hilbert}, 
$\som_{\xi} \in \SA{\BHA}$ is the \emph{pure vector state}
$\som_\xi(A) = \bkt{\xi}{A \xi}_{\Hvk}$ for $A \in \BHA$.  
Moreover, the vector $\xi \in \Hvk$ is assumed to be a \emph{cyclic unit vector} 
for $\Psi(\Alg) \subseteq \BHA$ in the Hilbert space $\Hvk$. 
Then the tuple $(\Psi, \Hvk, \xi)$ resp.\  $(\Psi, \BHA, \som_\xi)$
is called a \emph{GNS-representation} of $(\Alg, \svpi)$.
\end{defi}

For the GNS-representations from \feT{thm:GNS}, there is also a matrix representation 
on $\MMn(\CC)$ to parallel \feR{rem:basic-matrix-rep}, and one does not have
to use the structure constants explicitly as in \feR{rem:fin-dim-mat-rep}.

\begin{coro}[GNS Matrix Representation]  \label{coro:GNS-matrix}
For a finite dimensional probability algebra  $(\Alg,\svpi)$ with a \emph{faithful} 
state $\svpi \in \SA{\Alg}$, equipped with the Hilbert space structure from
\feP{prop:state-Hilbert-spc}, and representation as in \feT{thm:GNS},  
the representation $\Hf{B}{\tH}: \LHA \to \MMn(\CC)$ from \feR{rem:basic-matrix-rep}
may now be used, given by $\Hf{B}{\tH}(\ra) = \vM = (M_{ij}) = (\bkt{\rb_i}{\ra \rb_j}_2)
 \in \MMn(\CC)$, where $B = \{ \rb_1, \dots, \rb_n \} \subset \obH$ is an ortho-normal 
basis of $\obH$ and $n = \dim_{\CC}\Alg$ the complex dimension of $\Alg$.
This defines the underlying \emph{unitary} map 
$\Hf{B}{\tU}: \Hvk \to \Cn$ by $\Hf{B}{\tU}(\rb_j) = \ve_j \in \Cn$.

In view of the fact that  $\Alg = \Alg_{sa} \oplus \ii\Alg_{sa}$ 
in \feD{def:alg-names}, it is clear that $\Alg_{sa}$ has real dimension also 
$\dim_{\RR}\Alg_{sa} = n$.
Thus one might choose a basis where the basis vectors are all self-adjoint, 
$\rb_j = \rb_j^\star$, starting with $\rb_1 = \rone  \in \RR[\rone]\subseteq \Alg_{sa}$, 
and completing this with $\{\rb_2,\dots,\rb_n\}$ of $\Alg_{sa}$ to an ortho-normal
self-adjoint real basis $B = \{\rb_1, \dots, \rb_n\} \subset \Alg_{sa} = \spn_{\RR} B$ 
of $\Alg_{sa}$.  This finally gives a complex but ortho-normal and self-adjoint basis of 
$\Alg= \spn_{\CC} B$ with $\svpi(\rb_1) = 1$ and $\svpi(\rb_j) = 0$ for $j>1$.  

%As now $\rb_i \rb_j = \rb_i^\star \rb_j^\star = (\rb_j \rb_i)^\star$, 
%the structure constants of this basis satisfy
%$(\tensor*[^B]{\gamma}{^k_{ij}}) = (\tensor*[^B]{\gamma}{^j_{ik}})^*$, and thus
%\[
%   \Hf{B}{\tH}(\rb_i)  = \tensor*[^i]{\vE}{} 
%   = (\tensor*[^i]{E}{_{kj}})
%   = (\tensor*[^B]{\gamma}{^k_{ij}}) = (\tensor*[^B]{\gamma}{^j_{ik}})^* 
%   = (\tensor*[^i]{E}{^*_{jk}}) = \tensor*[^i]{\vE}{^\tpH} = \Hf{B}{\tH}(\rb_i^\star).
%\]
%In this way the $\Hf{B}{\tH}(\rb_i)= \tensor*[^i]{\vE}{}$ are also self-adjoint in $\MMn(\CC)$,
%and $\Hf{B}{\tH}(\rone) = \tensor*[^1]{\vE}{} = (\tensor*[^1]{E}{_{kj}}) = \vI_n$, 
%i.e.\ $(\tensor*[^1]{E}{_{kj}}) = (\tensor*[^B]{\gamma}{^k_{1j}}) = (\updelta_{kj})$,
%where $\updelta_{kj}$ is the Kronecker-$\updelta$.  One may even go a step further and
%ensure that all basis vectors $\rb_j$ for $j>1$ are minimal projections 
%(cf.\ \feD{def:alg-names}), which may be found along the lines as in \feX{ex:projections}.

The matrix algebra representation from \feR{rem:basic-matrix-rep}
is thus a faithful *-representation,  $\Hf{B}{\tH}$ is
an injective *-algebra homomorphism (a *-monomorphism) of $\Alg$ onto 
$\Hf{B}{\tH}(\Alg) \subseteq \MMn(\CC)$, and $\Alg \cong \Hf{B}{\tH}(\Alg) \cong \oL(\Alg)$
are *-isomorphic.
%
%Note, as according to \feT{thm:GNS} $\rb_1 = \rone$ defines the pure
%vector state on $\oL(\Alg)$ corresponding to $\svpi \in \SA{\Alg}$, that with
%$\Hf{B}{\tU}(\rone) = \Hf{B}{\tU}(\rb_1) =  \ve_1 \in \Cn$, one has a pure vector state
%for $\Hf{B}{\tH}(\Alg) \subset \MMn(\CC) \cong \Lop(\Cn)$.
%In particular, for $\ra \in \Alg$ it holds that (denoting $\vM = \Hf{B}{\tH}(\ra)$)
%\[
%   \svpi(\ra) = \bkt{\rone}{\ra}_2 = \bkt{\rone}{\oL(\ra) \rone}_2 =
%         \bkt{\ve_1}{\Hf{B}{\tH}(\ra) \ve_1}_{\Cn} = \ve_1^\tpH \vM \ve_1 =  \vM_{11}.
%\]
\end{coro}

In \feT{thm:GNS} and up to here it was ensured through \feq{eq:GNS-faithf-state} 
that the state $\svpi \in \SA{\Alg}$ is faithful, even if the original state $\sal$ 
of the probability algebra $(\Alg,\sal)$ were not faithful. 
What \feT{thm:GNS} shows is that the GNS-construction then yields a faithful, cyclic
(cf.\ \feD{def:alg-Hilbert}), and hence irreducible (cf.\  \feD{def:irreduc-rep})
representation.

In case the original state $\sal$ of $(\Alg,\sal)$ is not faithful, 
%(i.e.\ the choice $\svpi = \sphi_c$ was taken in \feq{eq:GNS-faithf-state})
and one were to form the sesqui-linear form $\bkt{\ra}{\rb}_{\sal} = \sal(\ra^\star \rb)$,
this would not be an inner product, as it would not be positive definite
$\nd{\cdot}_2$ would not be a norm, and it could not be used to define a 
Hilbert space structure on $\Alg$ in \feP{prop:state-Hilbert-spc}.
For such a state $\sal$ which is not faithful
and the choice $\svpi = \sphi_c$ was taken in \feq{eq:GNS-faithf-state},
the inner product on $\obH = \Hvk_c$ is not connected to the probability 
structure defined by the state $\sal$.  One may stay with this representation
on $\Lop(\Hvk_c)$, and compute expectations using densities $\vR \in \DEN{\Lop(\Hvk_c}$
(cf.\ \feC{coro:dens-in-alg-H}). 
But one has non-zero elements $\oL_{\rb} \ne 0$ in the algebra 
$\oL(\Alg) \subseteq \Lop(\Hvk_c)$ which may \emph{vanish almost surely (a.s.)}, 
as $\tr(\vR \, \oL_{\rb}^\dagger \, \oL_{\rb}) = \sal(\rb^* \rb)$ may vanish.

On the other hand, if the state $\sal$ is not faithful and one deliberately takes
the choice $\svpi = \sal$ in \feq{eq:GNS-faithf-state}, this
means the that subspace --- actually it is a left ideal in the algebra $\Alg$ --- 
defined as $\C{J}_{\svpi} = \{ \ra \in \Alg \mid \svpi(\ra^\star \ra)  = 0\} \subset \Alg$
is non trivial, i.e.\ $\C{J}_{\svpi} \ne \{\rnul\}$.
In $\C{J}_{\svpi}$ are the elements which vanish a.s.\ as just above.
Note in connection with \feR{rem:bounded-ideal} and \feP{prop:bounded-ideal}, 
where the ideal  $\C{I}_{\svpi} = \{ \ra \in \Alg \mid 
    \sup_{\rx \ne \rnul} \{ \svpi((\ra \rx)^\star \ra \rx)\} = 0 \}$ 
of all elements where the least bounding constant vanished was assumed to be factored 
out, that $\ra \in \C{I}_{\svpi}$ implies $\ra \in \C{J}_{\svpi}$ by taking 
$\rx = \rone$; i.e.\ $\C{I}_{\svpi} \subseteq \C{J}_{\svpi}$.

In order to have a Hilbert space where these elements vanishing a.s. are
declared equivalent to zero, one may factor out $\C{J}_{\svpi}$
and consider the quotient space $\obH_{\svpi} := \cl (\Alg / \C{J}_{\svpi})$ instead,
the closure being in the induced quotient norm.
The sesqui-linear form induced on  $\obH_{\svpi}$ by $\svpi = \sal$ now is positive definite
and hence defines an inner product, and the elements in $\obH_{\svpi}$
are defined by equality a.s.  As $\C{J}_{\svpi}$ is a left ideal, the multiplication
from the left on $\Alg / \C{J}_{\svpi}$ is well defined, hence one may construct a
*-representation via the *-homomorphism $\oL_{\svpi}:\Alg \to \Lop(\Alg / \C{J}_{\svpi})$
induced by the left regular representation (\feD{def:reg-repr}) by following further as 
in \feT{thm:GNS}.  Observe that the *-representation $\oL_{\svpi}$ is not
injective and thus not faithful, but one may still find a *-isomorphic matrix
representation for $\oL_{\svpi}(\Alg)$ as in \feC{coro:GNS-matrix}.

In the infinite dimensional case there is no trace functional, so there is no
canonical state like $\sphi_c$ in order to achieve a faithful representation.
What helps here is a direct sum $\obH := \bigoplus_{\svpi \in \C{E}} \obH_{\svpi}$, where
$\C{E} = \mrm{ext}(\SA{\Alg})$ are the pure states of $\Alg$ --- the extreme points of the
set of states --- each of which can be shown \citep{Takesaki1, Landsman2017} to correspond 
to an irreducible GNS-representation on $\obH_{\svpi}$ (cf.\  \feD{def:irreduc-rep})
in order to achieve a faithful *-representation on $\obH$.

\paragraph{States and Densities:}
%
%What has been established for a finite dimensional probability algebra $\Alg$,
%by  \feP{prop:fin-dim-alg-gen} and \feT{thm:GNS}, as well as \feP{prop:FinDim-Hilbert-GNS},
%and \feP{prop:fin-dim-mat-rep} and  \feC{coro:GNS-matrix}, as well as 
%\feC{coro:fin-dim-alg-Hilbert}, are two faithful representations of $\Alg$.
%Once as a sub-algebra of $\LHA$, where $\Hvk = \Alg$ is the Hilbert space 
%equipped with the trace-based inner product $\bkt{\cdot}{\cdot}_c$.  The other time
%as a sub-algebra of $\Lop(\obH)$, where $\obH = \Alg$  is the Hilbert space 
%equipped with the inner product $\bkt{\cdot}{\cdot}_2$, based on the GNS-construction,
%which has the potential to be extended to infinite dimensional probability algebras.  
%Moreover, in a parallel fashion to these two Hilbert representations, 
%the finite dimensional probability algebra $\Alg$ has been represented in 
%$\MMn(\CC)$.

The representation based on the GNS-construction depended on the state $\svpi \in \SA{\Alg}$,
and produced a pure vector state, whereas the previous representation used the trace as 
a generic linear functional, which in this fashion is only possible in finite dimensions.  
In this latter case based on the trace, the considered state is represented by a 
\emph{density matrix} (cf.\ \feD{def:density-M}, as well as \feP{prop:set-dens-M}
and \feP{prop:dens-in-alg-M}) or a density operator (cf.\ \feC{coro:dens-in-alg-H}).
The idea of densities can be carried over to the abstract situation in the finite
dimensional case.

In \feP{prop:states-prop} the set of states was characterised as a subset of $\Alg^\star$,
the space of all linear functionals.  In finite dimensions one has even $\Alg = \Hvk$, 
and the Hilbert space structure allows one to find an equivalent to $\SA{\Alg}$ in the algebra
itself.

\begin{coro}[Density as Element of Algebra, Set of Densities and State Representation]
     \label{coro:dens-in-alg-A}
In case the probability algebra $\Alg$ is finite dimensional, it is equal as a vector space
$\Alg = \Hvk$ to the Hilbert constructed in \feR{rem:c-Hilbert-spc}.

Hence for any linear functional $\sal \in \Alg^\star$, there is an element $\ra_{\sal} \in \Alg$
such that $\sal(\rb) = \bkt{\ra_{\sal}}{\rb}_c = \tr_{\Alg} (\ra_{\sal}^\star \rb)$ for 
all $\rb \in \Alg$, using the trace based canonical inner product introduced on 
$\Alg$ in \feR{rem:c-Hilbert-spc}.

Define the density elements in $\Alg$ as 
$\DEN{\Alg} = \{ \rr \in \Alg_+\; \mid \; \tr_{\Alg} \rr = 1 \}$,
they may be called algebraic densities.  And any state $\svpi_{\rr} \in \SA{\Alg}$ 
can be written as $\svpi_{\rr}(\rb) = \tr_{\Alg} (\rr \rb)$ with $\rr \in \DEN{\Alg}$.
%Both expressions may be used here, depending on the circumstances.
The statements from \feC{coro:dens-in-alg-H} can now be translated very easily to this
abstract situation, so that there is no need spell them out in detail here.

The states $\SA{\Alg} \subset \Alg^+ \subset \Alg^\star$ are in one-to-one 
correspondence with the densities $\DEN{\Alg} \subset \Alg_+ \subset \Alg$
(cf.\ \feC{coro:dens-in-alg-A}):
\begin{align*}
    \SA{\Alg} \ni \svpi_{\rr} &\leftrightarrow \rr \in \DEN{\Alg} \\
    \forall \ra \in \Alg:\, \svpi_{\rr}(\ra) &= \tr_{\Alg}(\rr \ra) .
\end{align*}

The characterisation of the convex set of states $\SA{\Alg}$ in \feP{prop:states-prop}
can now easily be translated for the set of densities; it is a compact convex set, the
intersection $\DEN{\Alg} = \Alg_+ \cap \C{S}$ of the closed convex cone $\Alg_+$ and 
the hyperplane $\C{S} = \{ \rr \in \Alg_{sa} \mid \tr \rr = 1 \}$.
\end{coro}

As it is obvious how to reformulate \feC{coro:dens-in-alg-A} for concrete unital 
*-sub-algebras $\Blg \subset \MMn(\CC)$ or $\Clg \subset \LHA$, and has 
partly already been exposed (cf.\ \feD{def:density-M}, as well as \feP{prop:set-dens-M}
and \feP{prop:dens-in-alg-M}, or  \feC{coro:dens-in-alg-H}), there is no
need to repeat it here.
The statement in \feC{coro:dens-in-alg-A} can be viewed from another angle, in an
argument simplified to the present situation of finite dimensional algebras from
\citep{gudderHudson78}:

\begin{coro}[Radon-Nikodým for Functionals]  \label{coro:RN-alg}   
  Let $\Alg$ be a finite dimensional unital *-algebra, and let 
  $\sal \in \Alg^+ \subset \Alg^\star$ be self-adjoint and positive definite,
  i.e.\ a faithful unnormalised state.  Let $\sbt \in \Alg^\star$
  be an arbitrary functional.  Then there is a $B \in \Lop(\Alg)$,
  such that for any $\ra \in \Alg$ it holds that $\sbt(\ra) = \sal(B \ra)$.  
  $B$ is called the \emph{Radon-Nikodým} derivative of $\sbt$ w.r.t.\ $\sal$, 
  also denoted by $B = \di \sbt / \di \sal$.
   
  In case $\sbt$ is self-adjoint and positive (definite), so is $B$.  If in addition 
  $\sbt$ is normalised ($\sbt(\rone) = 1$, and thus $\sbt \in \SA{\Alg}$ is a state), 
  one has that $\sal(B \rone) = 1$, i.e.\ that $B$ is a density for $\sal$.  
  
  The trace $\tr_{\Alg} \in \Alg^+$ is such a self-adjoint and positive definite functional,
  so that for a state $\sbt \in \SA{\Alg}$ the Radon-Nikodým derivative 
  $R = \di \sbt / \di \tr_{\Alg} \in \DEN{\Alg}$ is a \emph{density}, such that
  for any $\ra \in \Alg$ one has $\sbt(\ra) = \tr_{\Alg}(R \ra)$.
    Note that one may write this in a symmetric form as
    $\sbt(\ra) = \tr_{\Alg}(R^{\frk{1}{2}} \ra R^{\frk{1}{2}})$.
\end{coro}

The last result concerning the density was
already stated in \feC{coro:dens-in-alg-A}.  So, in this case of the trace on a
finite dimensional algebra $\Alg$, the \feC{coro:RN-alg} is just another way of
looking at a well known fact.  Thus,
as in the finite dimensional case any state can be represented through densities, this
is often the preferred method of dealing with states, especially when working with concrete
representations such as sub-algebras of $\MMn(\CC)$ or $\LHA$.

\begin{rem}[Radon-Nikodým theorems]   \label{rem:RN-positive-cone}
This result from \feC{coro:RN-alg} can be extended to the case of infinite dimensional
algebras \citep{gudderHudson78},  but the functional $\sbt$ has to be absolutely 
continuous in a certain sense (not to be defined here) w.r.t.\ $\sal$.  This connects
\feC{coro:RN-alg} with the classical Radon-Nikodým theorem 
(e.g.\ \citep{Bobrowski2006/087}).  This one may see as a result translating
the relation between measures / functionals in the convex positive cone of the dual
space (cf.\ \feP{prop:states-prop}) into positive functions / operators resp.\
elements of the algebra, i.e.\ elements in the convex positive cone of the original space.
\end{rem}

\paragraph{Convergence:}
Starting from these Hilbert representations, the inner product 
$\bkt{\cdot}{\cdot}_2$ on $\obH := \Lp_2(\Alg,\svpi)$ --- which is the completion of $\Alg$
if the state $\svpi$ is faithful, and the completion of $\Alg / \C{J}_{\svpi}$
otherwise --- induces the norm $\nd{\cdot}_2$, with which
one may define a stronger notion of convergence for RVs than considered up to now:

\begin{defi}[Convergence in Mean Square]  \label{def:MS-convg}
Given a sequence $\{\ra_n\}$, $n \in \D{N}$ in $\obH$,
one says that it \emph{converges in mean square} --- or
in \emph{root mean square} (RMS) --- to $\rx \in \obH$ \emph{iff}
$\nd{\ra_n - \rx}_2 \to 0$.
\end{defi}

Later this will be extended to other powers than 2.  %One has immediately the

Before proceeding to to define the uniform norm on $\Alg$, 
now is the time for few definitions concerning normed algebras, e.g.\ cf.\
\citep{Naimark1972, segalKunze78, Arveson1976, Conway1990, Davidson1996, 
KadiRingr1-97, KadiRingr2-97, BrattRob-1, BrattRob-2, Schaefer99,
Takesaki1, Takesaki2, Takesaki3, Blackadar2006, Wolf2012, Landsman2017}:

\begin{defi}[Normed Algebras]   \label{def:normed_algebras}
Let $\Alg$ be a *-algebra with a norm defined on the vector space $\Alg$.
\begin{compactitem}
\item
In case the norm $\nd{\cdot}$ satisfies 
$\nd{\ra \rb} \le \nd{\ra} \nd{\rb}$ for all $\ra, \rb \in \Alg$,
then the algebra $\Alg$ is a \emph{normed algebra}.
This means that not only the vector space operations, but also the algebra
multiplication is continuous in the norm topology.

\item
If in addition $\Alg$ is a Banach space with the norm $\nd{\cdot}$,
it is called a \emph{Banach algebra}. 

\item
If a Banach algebra $\Alg$ satisfies \feq{eq:C*-alg-norm} in \feP{prop:alg-unif-n}, 
i.e.\ the so-called  \emph{$\Ck^*$-equality:}
$\nd{\ra^\star \ra} = \nd{\ra^\star}\nd{\ra} = \nd{\ra}^2$ for all $\ra \in \Alg$,
it is called a \emph{$\Ck^*$-algebra}.  

\item
If $\Alg$ satisfies \feq{eq:C*-alg-norm}, its norm-completion is a $\Ck^*$-algebra
denoted by $\Ck^*(\Alg)$.

\item
A unital $\Ck^*$-algebra, which, as a Banach space, is a dual space, 
i.e.\ there is a Banach space $\Blg$ such that $\Alg \cong \Blg^*$,
is called a \emph{$\Wp^*$-algebra} or a \emph{von Neumann} algebra,
\end{compactitem}
\end{defi}

Observe that for a finite dimensional normed *-algebra, 
one has $\Alg = \Ck^*(\Alg) = \BA{\Alg}$,
and $\Alg$ \emph{is a} $\Ck^*$-algebra as well.% as a $\Wp^*$-algebra.

The representation on a Hilbert space leads to

\begin{prop}[Uniform Norm on $\Lop(\obH)$]  \label{prop:alg-unif-n}
For the finite dimensional *-algebra $\Alg$, the faithful Hilbert *-representation
in  \feT{thm:GNS} shows that 
$\oL(\Alg) \subset \Lop(\obH) = \C{B}(\obH)$ is *-isomorphic to $\Alg$.  

In the finite dimensional case, all linear maps in $\Lop(\obH)$ on the space 
$\Hvk$ are bounded \citep{Schaefer99}, thus all $\ra \in \Alg$ are 
bounded, i.e.\ $\Alg = \BA{\Alg}$.  
%This is also implicated by the faithful 
%*-representations in $\MMn(\CC)$ (cf.\ \feP{prop:fin-dim-mat-rep} and 
%\feC{coro:GNS-matrix} or \feC{coro:fin-dim-alg-Hilbert}).
Hence one may take --- as in \feR{rem:bounded-ideal} --- the infimum 
$\wht{C}_{ra}$ of all possible bounding constants 
$C_{\ra}$ in \feD{def:alg-names}, and denote it by $\nd{\ra}_\infty$.  
This is the \emph{uniform} or operator norm on $\Lop(\obH)$, 
i.e.\ $\nd{\ra}_\infty := \nd{\oL_{\ra}}_{\Lop(\obH)}$, and it is 
thus a norm on $\Alg$.  Observe that 
$\nd{\ra}_\infty = \nd{\ra}_{S \infty}$, cf.\  \feR{rem:Schatten-A}.

It follows from these Hilbert *-representations that for 
$\ra, \rb \in \Alg$ one has $\nd{\ra \rb}_\infty  \le \nd{\ra}_\infty \nd{\rb}_\infty$
as this holds in $\Lop(\obH)$: 
$\nd{\oL_{\ra}\,\oL_{\rb}}_{\Lop(\obH)} \le \nd{\oL_{\ra}}_{\Lop(\obH)} 
\nd{\oL_{\rb}}_{\Lop(\obH)}$.  This makes 
$\Alg$ into a \emph{normed algebra}, cf.\ \feD{def:normed_algebras} below.  
Moreover, it satisfies also
\begin{equation}  \label{eq:C*-alg-norm}
\nd{\ra}_\infty^2 = \nd{\ra \ra^\star}_\infty = \nd{\ra}_\infty\, \nd{\ra^\star}_\infty ,
\end{equation}
as this is satisfied for $\Lop(\obH)$.  These results show that multiplication and
conjugation $\ra \mapsto \ra^\star$ are continuous in the uniform norm.
\end{prop}

With the metric defined by $\nd{\cdot}_\infty$ in \feP{prop:alg-unif-n}, 
one may now define an even stronger notion of convergence:

\begin{defi}[Uniform Convergence]  \label{def:infty-convg}
Given a sequence $\{\ra_n\}$, $n \in \D{N}$ in a bounded 
Banach probability algebra $\Alg = \BA{\Alg}$, 
one says that it \emph{converges almost uniformly} to $\rx \in \Ck^*(\Alg)$ iff
$\nd{\ra_n - \rx}_\infty \to 0$.
\end{defi}

From \feD{def:MS-convg} and \feD{def:infty-convg} one immediately  has
\begin{prop}[Uniform to RMS Convergence]  \label{prop:unif-RMS}
If a sequence $\{\ra_n\}$, $n \in \D{N}$ in a bounded 
Banach probability algebra $\Alg = \BA{\Alg}$
converges almost uniformly to $\rx \in \Ck^*(\Alg)$, 
then it converges also in mean square to $\rx \in \Ck^*(\Alg) \subseteq \obH$, 
cf.\  \feD{def:MS-convg}.   For the finite dimensional case, recall that
$\Alg = \BA{\Alg} = \Ck^*(\Alg) = \obH$ as sets.
\end{prop}

\paragraph{Correlation and Independence:}
With the inner product (now just denoted as $\bkt{\ra}{\rb}$)
and Hilbert space $\obH = \Lp_2(\Alg)$ from \feP{prop:state-Hilbert-spc}, 
some more quantities known from probability theory can be defined:

\begin{defi}[Correlation, Covariance]  \label{def:covariance}
For $\ra, \rb \in \Alg$, their inner product $\bkt{\ra}{\rb}$ is also called
their \emph{correlation}.  

Remembering the decomposition in \feD{def:alg-names} and \feT{thm:GNS} of RVs into their 
mean and the zero-mean fluctuating part, $\ra = \wob{\ra} + \wtl{\ra}$, and the associated
direct sum decomposition $\obH = \wob{\obH} \oplus \wtl{\obH}$, it is easy to
see that for all $\ra, \rb \in \obH$ one has $\bkt{\wob{\ra}}{\wtl{\rb}} = 0$,
and thus the direct sum decomposition is an \emph{orthogonal direct sum}.
Thus one has Pythagoras's theorem 
\[
\nd{\ra}_2^2 = \nd{\wob{\ra}}_2^2 + \nd{\wtl{\ra}}_2^2 = \Ex_{2}(\ra)^2 + \var_{\svpi}(\ra)
             = \svpi(\ra)^2 + \var_{\svpi}(\ra),
\]
where the quantity $\var_{\svpi}(\ra) := \nd{\wtl{\ra}}_2^2$ is called the \emph{variance}
of the RV $\ra$ (in the state $\svpi$).
For two RVs $\ra, \rb \in \obH$, the inner product of their zero-mean parts is 
their \emph{covariance}:
\begin{equation} \label{eq:covariance}
   \cov_{\svpi}(\ra, \rb) := \bkt{\wtl{\ra}}{\wtl{\rb}} = \bkt{\ra-\wob{\ra}}{\rb-\wob{\rb}} =
    \bkt{\ra-\Ex_{2}(\ra) \rone}{\rb-\Ex_{2}(\rb) \rone}.
\end{equation} 

Two RVs are \emph{uncorrelated} if $\cov_{\svpi}(\ra, \rb)= 0$; this means geometrically
that $\wtl{\ra}$ and $\wtl{\rb}$ are orthogonal, or that the two subspaces
$\spn_\CC \wtl{\ra}$ and $\spn_\CC \wtl{\rb}$ are orthogonal, also denoted as
$\bkt{\spn_\CC \wtl{\ra}}{\spn_\CC \wtl{\rb}} = \{ 0 \}$.
\end{defi}

Two probability sub-algebras $\Blg, \Clg \subset \Alg$ can not be orthogonal, as
both contain $\rone_{\Alg}$.  But the relevant idea that their zero-mean parts
are orthogonal is captured by (cf.\ \citep{HiaiPetz2014})

\begin{defi}[Complementarity]  \label{def:complementarity}
Let $\Blg, \Clg \subset \Alg$ be probability sub-algebras.  They are \emph{complementary}
or \emph{quasi-orthogonal} iff for $\rb \in \Blg$ and $\rc \in \Clg$ it holds that
$\cov_{\svpi}(\rb, \rc) = 0$. 
\end{defi}

Independence is another important notion from probability theory:

\begin{defi}[Independence]  \label{def:independ-tens}
Two RVs $\ra , \rb \in \Alg$ are \emph{independent} if they commute 
($[\ra, \rb] = \rnul$), and the probability sub-algebras they generate ($\CC[\ra]$
and $\CC[\rb]$) are complementary (cf.\ \feD{def:complementarity}).
%This means geometrically that the subspaces $\wtl{\CC[\ra]}$ and $\wtl{\CC[\rb]}$
%are orthogonal, $\bkt{\wtl{\CC[\ra]}}{\wtl{\CC[\rb]}} = 0$.

More generally, two probability sub-algebras $\Blg, \Clg \subset \Alg$ are independent
\emph{iff} they commute ($[\Blg, \Clg] = \{\rnul\}$) and are complementary or 
quasi-orthogonal (cf.\ \feD{def:complementarity}).

One may write this in such a way that for all $\rb \in \Blg, \rc \in \Clg$ one has
\begin{equation} \label{eq:def-indep}
\svpi(\rb \rc) = \Ex_{2}(\rb \rc) = \svpi(\rb)\svpi(\rc) = 
\Ex_{2}(\rb) \Ex_{2}(\rc).
\end{equation}
This kind of independence is sometimes also called \emph{tensor-independence} 
\citep{GhorbalSchuermann99, BenGhorbSchue02}.
\end{defi}

This means that for independent factors in \feq{eq:def-indep} the state acts 
like a multiplicative state resp.\ character.  
It should be obvious how to extend the above \feD{def:independ-tens}
to more than two RVs resp.\ sub-algebras.

From the \feD{def:independ-tens} of independence follows directly:
\begin{prop}  \label{prop:alg-gen-indep-subalg}
Assume that the sub-algebras $\Blg, \Clg \subset \Alg$ are independent.
Further let $\rb \in \Blg, \rc \in \Clg$.
The relation \feq{eq:def-indep} can also be expressed as the statement that 
\begin{equation}  \label{eq:def-indep-2}
   \svpi(\rb) = 0 \;\text{ and }\; \svpi(\rc) = 0 \quad \text{ implies }
   \quad \svpi(\rb \rc) = 0.
\end{equation}

For any three RVs $\ra, \rb \in \Blg$ and $\rc \in \Clg$ one has
\begin{equation}   \label{eq:thre-RVs-indep}
   \Ex_{2}(\ra \rc \rb) = \svpi(\ra \rc \rb) = \svpi(\rc) \svpi(\ra \rb) = 
   \Ex_{2}(\rc) \Ex_{2}(\ra \rb).
\end{equation}

Now assume additionally that the probability algebra $\Alg = \CC\ipj{\Blg, \Clg}$ 
is generated by the independent sub-algebras $\Blg, \Clg \subset \Alg$.  
This implies that the restrictions
$\svpi|_{\Blg}$ and $\svpi|_{\Clg}$ determine the state $\svpi$. 
\end{prop}

\paragraph{Sums and Products:}
When considering different system which interact in some way, one is lead
to consider the ways on how probability algebras can be combined to yield new
probability algebras, and how to interpret this probabilistically.

\begin{defi}[Direct Product]   \label{def:direct-product}
Given two probability algebras $\Blg$ and $\Clg$ with states $\svpi_b$ and $\svpi_c$
respectively, and involutions $\sharp$ and $\flat$, consider
their \emph{direct product} $\Alg = \Blg \times \Clg$.  It is a vector space of
elements denoted as $(\rb, \rc) \in \Alg$ with $\rb \in \Blg$ and $\rc \in \Clg$.

One now defines a (Hadamard) product as
\[
   (\rb_1, \rc_1) \odot (\rb_2, \rc_2) := (\rb_1\rb_2, \rc_1\rc_2)\;\text{ for }
     \rb_1, \rb_2 \in \Blg,\; \rc_1, \rc_2 \in \Clg ,
\]
with unit element $\rone_{\Alg} := (\rone_{\Blg}, \rone_{\Clg})$ and involution $\natural$
defined by $(\rb, \rc)^\natural := (\rb^\sharp , \rc^\flat)$.  

For a given $0 < \alpha < 1$, one may define the state 
$\svpi_\alpha(\rb, \rc) := \alpha \svpi_b(\rb) + (1 - \alpha) \svpi_c(\rc)$,
to obtain a new probability algebra $\Alg = \Blg \times \Clg$. 
$\Blg$ and $\Clg$ are naturally embedded in $\Alg$ as elements
of the form $(\rb, \rnul_{\Clg})$ resp.\ $(\rnul_{\Blg}, \rc)$.
\end{defi}

It is clear how to extend this to more than two factors.
The \feX{ex:complex-n-spc} may be seen as the $n$-fold direct product of \feX{ex:complex-num}.

By essentially only changing the notation, the same construction as in \feD{def:direct-product}
can be written for the \emph{direct sum} $\Alg_{\oplus} = \Blg \oplus \Clg$, as a finite
number of products and a finite number of summands produce isomorphic algebras.

\begin{defi}[Direct Sum]   \label{def:direct-sum}
Given two probability algebras $\Blg, \Clg$ with states $\svpi_b, \svpi_c$, 
and involutions $\sharp$ and $\flat$, consider
their \emph{direct sum} $\Alg = \Blg \oplus \Clg$.  It is a vector space of
elements denoted as $\rb \oplus \rc \in \Alg$ with $\rb \in \Blg$ and $\rc \in \Clg$.

One now defines a product --- written again as juxtaposition --- as
\[
   (\rb_1 \oplus \rc_1) (\rb_2 \oplus \rc_2) := (\rb_1\rb_2 \oplus \rc_1\rc_2)\;\text{ for }
     \rb_1, \rb_2 \in \Blg,\; \rc_1, \rc_2 \in \Clg ,
\]
with unit element $\rone_{\Alg} = (\rone_{\Blg} \oplus \rone_{\Clg})$ and involution $\natural$
defined by $(\rb \oplus \rc)^\natural := (\rb^\sharp \oplus \rc^\flat)$.  

For a given $0 < \alpha < 1$, one may define the state 
$\svpi_\alpha(\rb \oplus \rc) := \alpha \svpi_b(\rb) + (1 - \alpha) \svpi_c(\rc)$,
to obtain a new probability algebra $\Alg = \Blg \oplus \Clg$. 
$\Blg$ and $\Clg$ are naturally embedded as elements
of the form $(\rb \oplus \rnul_{\Clg})$ resp.\ $(\rnul_{\Blg} \oplus \rc)$.
\end{defi}

For a finite number of direct summands, it is obvious that the relation
$\Blg \oplus \Clg \ni \rb \oplus \rc \mapsto (\rb, \rc) \in \Blg \times \Clg$
provides the algebra *-isomorphism alluded to above.
One may note that in both cases (direct product and direct sum) 
the two embedded sub-algebras $\Blg, \Clg$ commute, and the 
sub-algebras $\Blg$ and $\Clg$ are sometimes called the \emph{sectors} of $\Alg$.
The interpretation of this construction is that of a \emph{mixture}:
with probability $\alpha$ one chooses
the algebra $\Blg$, and with probability $(1 - \alpha)$ the algebra $\Clg$.
%The \feX{ex:func-tensor} can be seen as an instance of this construction
%in \feD{def:direct-product}.

Considering maps on such direct products and direct sums, it is useful to have:

\begin{defi}[Direct Sum of Maps]   \label{def:direct-sum-maps}
If $B \in \Lop(\Blg)$ and $C \in \Lop(\Clg)$ are linear maps on $\Blg$ resp. $\Clg$, then
the direct sum $B \oplus C \in (\Lop(\Blg) \oplus \Lop(\Clg)) \subset \Lop(\Blg \times \Clg)$ 
is written as
\begin{align}   \label{eq:direct-prod-map}
   \text{on } \; \Blg \times \Clg: \; 
   B \oplus C: & \,(\rb, \rc) \mapsto \begin{bmatrix}  B & 0 \\ 0 & C \end{bmatrix} 
     \begin{bmatrix}  \rb \\ \rc \end{bmatrix} = ( B \rb,  C  \rc ) \; \text{ or }
     \\  \label{eq:direct-sum-map}   \text{on } \; \Blg \oplus \Clg: \;
   B \oplus C: & \,(\rb \oplus \rc) \mapsto  (B \rb \oplus C \rc) .   
\end{align} 
\end{defi}

To go beyond the idea of a mixture and express the idea that two combined components are 
interacting in one system, one uses the
\begin{defi}[Tensor Product of Algebras]   \label{def:tensor-product}
Given two unital *-algebras $\Blg, \Clg$ with 
involutions $\sharp$ and $\flat$, consider
their \emph{tensor product} $\Alg = \Blg \otimes \Clg$.  It is a vector space of
elements of the form $\sum_{j=1}^m z_j (\rb_j \otimes \rc_j) \in \Alg$ with 
$\rb_j \in \Blg$, $\rc_j \in \Clg$, and $z_j\in \CC$, a linear combination of elementary
tensors of the form $\rb \otimes \rc$.  
One now defines a bilinear product (linear in each entry) --- 
written again as juxtaposition --- on elementary tensors as
\[
   (\rb_1 \otimes \rc_1) (\rb_2 \otimes \rc_2) := (\rb_1\rb_2 \otimes \rc_1\rc_2)\;\text{ for }
     \rb_1, \rb_2 \in \Blg, \; \rc_1, \rc_2 \in \Clg ,
\]
and extends this by linearity.
The unit element is $\rone_{\Alg} := \rone_{\Blg} \otimes \rone_{\Clg}$, and the 
involution $\natural$ on $\Alg$ is
defined by $(\rb \otimes \rc)^\natural := \rb^\sharp \otimes \rc^\flat$, 
and again extended by linearity.  This defines a unital *-algebra structure on $\Alg$.  
\end{defi}

It is again clear how to extend this to more tensor factors.
The algebras $\Blg$ and $\Clg$ represent the component systems --- the elements of $\Blg$
resp.\ $\Clg$ are the RVs of the separate component systems, and those of
$\Alg = \Blg \otimes \Clg$ are the RVs of the combined compound system. 
The component algebras $\Blg$ and $\Clg$ are embedded into the compound algebra
$\Alg$ by identifying them as the generating and commuting sub-algebras 
$\Blg \otimes \CC[\rone_{\Clg}]$ and $\CC[\rone_{\Blg}] \otimes \Clg$ of $\Alg$,
with elements formed of elementary tensors of the form 
$\rb \otimes \rone_{\Clg}$ and $\rone_{\Blg} \otimes \rc$, respectively.

\begin{defi}[Tensor States --- correlated and entangled]   \label{def:tensor-states}
Assume given states $\svpi_b, \svpi_c$ on $\Blg, \Clg$, the components of
the compound system probability algebra $\Alg = \Blg \otimes \Clg$.

Given states $\svpi_b, \svpi_c$ on the components, one may define a
compound state $\svpi_a$ on $\Alg$ by
\begin{equation}  \label{eq:tens-state}
\svpi_a(\rb \otimes \rc) := \svpi_b \otimes \svpi_c(\rb \otimes \rc) =\svpi_b(\rb) \svpi_c(\rc),
\end{equation}
and extend this by linearity to all of $\Alg$.  This is called 
a \emph{product state} on $\Alg$.   If $\svpi_a = \svpi_b \otimes \svpi_c$,
the algebras $\Blg \otimes \CC[\rone_{\Clg}]$ and $\CC[\rone_{\Blg}] \otimes \Clg$  
are \emph{independent};
and the algebras $\Blg$ and $\Clg$ are termed \emph{uncorrelated} relative to $\svpi_a$.

In case 
\begin{equation}  \label{eq:separab-stat}
   \svpi_a = \sum_{j} \alpha_j\, \svpi_{b,j} \otimes \svpi_{c,j}; \quad 0 \le \alpha_j \le 1;
   \quad \text{ and } \sum_j \alpha_j = 1,
\end{equation}
 where $\svpi_{b,j}$ and $\svpi_{c,j}$ are states on $\Blg$ or $\Clg$ respectively
--- this is called a \emph{separable state}, a convex combination of product states --- 
the state $\svpi_a \in \SA{\Alg}$ is termed \emph{classically correlated}.
\emph{Otherwise} the state is called \emph{entangled}.

On the other hand,
let $\svpi_a \in \SA{\Alg}$ be any state on the compound probability algebra 
$\Alg = \Blg \otimes \Clg$.  As the algebras $\Blg, \Clg$ are naturally embedded in
$\Alg$ as $\rb \otimes \rone_{\Clg}$ for $\rb \in \Blg$, and as $\rone_{\Blg} \otimes \rc$
for $\rc \in \Clg$ one may, by setting
\begin{align} \label{eq:red-state-b}
\svpi_{b,r}(\rb) &:= \svpi_a(\rb \otimes \rone_{\Clg}),\; \text{ and} \\ 
  \label{eq:red-state-c}
\svpi_{c,r}(\rc) &:= \svpi_a(\rone_{\Blg} \otimes \rc);
\end{align}
 and extending by linearity,
define \emph{reduced} states on $\Blg$ and $\Clg$ respectively.  

These states $\svpi_{b,r} \in \SA{\Blg}$, $\svpi_{c,r} \in \SA{\Clg}$, on the components 
give rise to what is called a 
\emph{partial trace}, as linear maps  $\svpi_{\Blg} \in \Lop(\Alg,\Clg)$ 
resp.\ $\svpi_{\Clg} \in \Lop(\Alg,\Blg)$, 
by defining for an elementary tensor $\rb \otimes \rc \in \Alg = \Blg \otimes \Clg$
\begin{equation}  \label{eq:part-trace}
  \svpi_{\Blg}(\rb \otimes \rc) := \svpi_{b,r}(\rb) \rc \in \Clg, \quad  
  \svpi_{\Clg}(\rb \otimes \rc) := \svpi_{c,r}(\rc) \rb \in \Blg,
\end{equation}
and extending by linearity to all of $\Alg$.
\end{defi}

Again it is obvious how to extend this to more tensor factors.  This is one way how
probabilistic descriptions are extended in the algebraic picture.  The interpretation of
the uncorrelated case is: when the two components systems $\Blg$ and $\Clg$ are combined,
they do not interact.  Conversely, this means that interacting component systems 
\emph{can not} have a product state as in \feq{eq:tens-state} as a
global compound state.  In the classically correlated case \feq{eq:separab-stat}, 
it is important that
this is a \emph{convex combination of product states} \citep{Werner2001, Horodecki2001}.  
In fact, if one allows $\alpha_j \in \CC$, any linear form on 
$\Alg = \Blg \otimes \Clg$ could be written in that way.
The interpretation is that there is a classical probability
distribution --- given by the $\alpha_j$ --- which picks one of the uncorrelated
cases $\svpi_{b,j} \otimes \svpi_{c,j}$.    It is worthwhile to note that in case
one of the algebras $\Blg$ or $\Clg$ is Abelian or commutative, \emph{every} state
on the compound algebra $\Alg$ is separable or classically correlated.

It was already remarked in \feD{def:independ-tens} that that notion of independence is
closely linked to tensor products, and indeed the tensor product as given in
\feD{def:tensor-product} is the prime example of this kind of independence
\citep{BenGhorbSchue02}.  In fact, one has

\begin{prop}  \label{prop:indep-to-tensor}  
In case two (or more) sub-algebras $\Blg, \Clg \subset \Alg$
wich generate $\Alg = \CC\ipj{\Blg, \Clg}$ are tensor-independent 
as in \feD{def:independ-tens}, it is possible to find a representation of the 
algebra as a tensor product as in \feD{def:tensor-product}.  This means that
there is a *-isomorphism $\Phi: \, \Alg \to \C{D}$ with a probability algebra
$\C{D} = \C{E} \otimes \C{F}$, such that $\C{E}$ and $\C{F}$ are independent,
and additionally it holds that $\Phi(\Blg) = \C{E}$ and $\Phi(\Clg) = \C{F}$.
\end{prop}

The property which is most desired is given in
\feq{eq:def-indep}, or equivalently in \feqs{eq:def-indep-2}{eq:thre-RVs-indep}.
But this can also be achieved with another kind of product,
e.g.\ \citep{VoiculescuDykemaNica1992, HiaiPetz2000, Mitchener2005, 
MingoSpeicher2017, Speicher2017}, namely the \emph{free product}. 
It is defined by a universal property, similarly to the tensor product.
In this case a property similar to \feq{eq:def-indep} for free RVs
then defines \emph{free independence}.
Note that in \citep{GhorbalSchuermann99, BenGhorbSchue02} it is shown that
the tensor and the free product are the only two constructions to fulfill
certain requirements regarding the notion of independence.
The free product and connected developments are beyond the scope of this
paper, so we confine ourselves to these few words on this subject.

Now that the one-to-one correspondence between states and densities 
together with the trace has been firmly established in \feC{coro:dens-in-alg-A},  
it may be worthwhile to just touch on the tensor product \feD{def:tensor-product} 
and the topic of partial trace abstractly introduced in \feq{eq:part-trace} in 
\feD{def:tensor-states}.

\begin{rem}[Compound and Partial Trace:]  \label{rem:tensor-states-A}
Consider the tensor product of probability algebras $\Alg = \Blg \otimes \Clg$
as in \feD{def:tensor-states}, with densities $\rr_b \in \DEN{\Blg}$ and 
$\rr_c \in \DEN{\Clg}$.
The \emph{compound product} state corresponding to \feq{eq:tens-state} is for 
an elementary tensor $\rb \otimes \rc \in \Blg \otimes \Clg$ defined as
\begin{equation}  \label{eq:tens-dens-state-D}
   \tr_{\Alg} \bigl((\rr_b \otimes \rr_c)(\rb \otimes \rc)\bigr) :=
     \tr_{\Blg} (\rr_b \rb) \; \tr_{\Clg} (\rr_c \rc) ,
\end{equation}
and then extended by linearity.  This product state means that the corresponding density 
is $\rr_a = \rr_b \otimes \rr_c \in \DEN{\Alg}$, a \emph{product density}.

Echoing \feq{eq:separab-stat}, in case a density $\rr_a \in \DEN{\Alg}$ is given by
\begin{equation}  \label{eq:separab-stat-D}
   \rr_a = \sum_{j} \alpha_j\, \rr_{b,j} \otimes \rr_{c,j}; \quad 0 \le \alpha_j \le 1;
   \quad \text{ and } \sum_j \alpha_j = 1,
\end{equation}
with densities $\rr_{b,j}$ and $\rr_{c,j}$ on $\Blg$ or $\Clg$ respectively
(a convex combination of product densities),
this is a \emph{separable density} and $\Blg$ and $\Clg$ called 
\emph{classically correlated}. \emph{Otherwise} the density is called \emph{entangled}.

On the other hand, given a density $\rr_a \in \DEN{\Alg}$ on the compound probability algebra 
$\Alg = \Blg \otimes \Clg$, by analogy to \feqs{eq:red-state-b}{eq:red-state-c}
one defines \emph{reduced} densities or \emph{marginals} $\rr_{b,r} \in \DEN{\Blg}$ and 
$\rr_{c,r} \in \DEN{\Clg}$, by requiring for all $\rb \in \Blg$ and $\rc \in \Clg$
\begin{align} \label{eq:red-state-b-m}
 \tr_{\Blg} (\rr_{b,r}\, \rb) &= \tr_{\Alg} (\rr_a(\rb \otimes \rone_{\Clg})),\; \text{ and} \\ 
  \label{eq:red-state-b-n}
 \tr_{\Clg} (\rr_{c,r}\, \rc) &= \tr_{\Alg} (\rr_a(\rone_{\Blg} \otimes \rc)).
\end{align}

These reduced densities $\rr_{b,r}$ and $\rr_{c,r}$
on the components give rise to what is called (now truly) a
\emph{partial trace}, as linear maps  $\tr_{p,\Blg} \in \Lop(\Alg,\Clg)$ 
resp.\ $\tr_{p,\Clg} \in \Lop(\Alg,\Blg)$, by defining, similarly to 
\feq{eq:part-trace}, for elementary tensors:
\begin{align}  \label{eq:part-trace-M-m}
   \tr_{p,\Blg}:\, \Blg  \otimes \Clg \ni  \rb \otimes \rc  &\mapsto 
        \tr_{p,\Blg}(\rb \otimes \rc) := \tr_{\Blg}(\rr_{b,r}\rb)\, \rc \in \Clg, 
        \quad \text{ and} \\              \label{eq:part-trace-M-n}
   \tr_{p,\Clg}:\, \Blg \otimes \Clg \ni  \rb \otimes \rc  &\mapsto 
        \tr_{p,\Clg}(\rb \otimes \rc) := \tr_{\Clg}(\rr_{c,r}\rc)\, \rb \in \Blg  , 
\end{align}
and then extended by linearity.
\end{rem}

Again it is obvious how to reformulate \feR{rem:tensor-states-A} for concrete unital 
*-sub-algebras of $\MMn(\CC)$ or $\LHA$, so this
does not have to be spelt out here.

\subsection{Structure of Finite Dimensional Algebras}  \label{SS:funcs-RVs-norm-alg}
From \feP{prop:alg-unif-n} we have seen the $\nd{\cdot}_\infty$-norm introduced on $\Alg$
as the operator norm of a faithful *-representation.  As a finite dimensional $\Alg$
is complete in all Hausdorff vector space topologies \citep{Schaefer99}, when 
equipped with a norm it becomes a Banach space.
The combination of the norm with the multiplication leads to the various normed algebras
\citep{Naimark1972, segalKunze78, Arveson1976, Conway1990, Davidson1996, 
KadiRingr1-97, KadiRingr2-97, BrattRob-1, BrattRob-2, Schaefer99,
Takesaki1, Takesaki2, Takesaki3, Blackadar2006}, 
which were already defined in \feD{def:normed_algebras}.  Here we are
primarily interested in $\Wp^*$-algebras.

\paragraph{W*- or von Neumann Algebras:}
From  \feR{rem:Schatten-A} one may recall that in the finite dimensional case
the operator norm is equal to the Schatten-$\infty$-norm $\nd{\cdot}_{S\infty}$.

\begin{coro}   \label{coro:fin-dim-W}
As a Banach space, $(\Alg, \nd{\cdot}_{S\infty})$ is the \emph{dual} of
$(\Alg, \nd{\cdot}_{S1})$.  According to \feD{def:normed_algebras} it is thus
a $\Wp^*$-algebra.
\end{coro}

Often the name ``von Neumann algebra'' is reserved for sub-algebras of $\Lop(\obH)$,
whereas ``$\Wp^*$-algebra'' is used for the abstract concept.  Recall that $\C{B}(\obH)$
is a von Neumann or $\Wp^*$-algebra for any Hilbert space $\obH$.  And, due to the
faithful Hilbert *-representation $\oL$ in \feT{thm:GNS}
and \feP{prop:alg-unif-n}, a finite dimensional probability algebra $\Alg$ is norm 
*-isomorphic to a von Neumann sub-algebra $\oL(\Alg) \subseteq \Lop(\obH)$.

The normed algebras are important in general, as they come with an extensive 
functional calculus, especially in the form of Banach-, $\Ck^*$-, and $\Wp^*$-algebras.
Especially for self-adjoint elements of the algebra $\ra \in \Alg_{sa}$, 
one wants to be able to compute $f(\ra)$, where $f$ is a possibly complex-valued 
function defined on a subset of the real line,
like for example the exponential, the square root, the absolute value, or an
indicator function $\bbbone_{\C{E}}$ for $\C{E} \subset \RR$. 
This functional calculus allows one to proceed from polynomials on general 
and normed algebras to holomorphic functions on Banach algebras, on to 
continuous functions on $\Ck^*$-algebras, and finally to Borel-measurable 
functions on $\Wp^*$-algebras.

Luckily, as was remarked already several times
(e.g.\ \feR{rem:spec-dec-proj}), on finite dimensional algebras this all reduces
to a polynomial calculus, as the spectrum of self-adjoint elements --- being equivalent
to a linear map on a finite dimensional space --- consists only of a finite number of points,
and on a finite subset of $\RR$ any function can be written as a polynomial.
%\paragraph{Basic Properties of $\Wp^*$- resp.\ von Neumann-Algebrass}

Also, now is the point to collect few results which have been hidden in some assumptions
so far, namely  \feR{rem:bounded-ideal} and \feP{prop:bounded-ideal}, and to see
the significance in the assumption in \feP{prop:fin-dim-alg-gen}, 
and show their inter-dependencies, cf.\ \citep{AnnPrec-Yuan-1, Naimark1972}.  

\begin{prop}[Semi-Simple Algebra]  \label{prop:radical}
Let $\Alg$ be a finite dimensional unital *-algebra; then, given the trace 
$\tr_{\Alg} \in \Alg^\star$ as defined in \feP{prop:fin-dim-alg-gen}, for each
$\ra \in \Alg$ define the functional 
$\sal_{\ra}: \Alg \ni \rb \mapsto \sal_{\ra}(\rb) := % \frk{1}{(\dim \Alg)}
\, \tr_{\Alg} (\ra \rb) = \tr_{\Alg} (\oL_{\ra} \rb) \in \CC$.  The ideal 
\[
\text{\nmf rad}(\Alg) = \{ \ra \in \Alg \mid \sal_{\ra}\; \text{\nmf vanishes identically}\}
   \subseteq \Alg
\]
is called the (Jacobson) radical of the algebra.  It may also be defined as
\citep{Naimark1972}:
\[
\text{\nmf rad}(\Alg) = \{ \ra \in \Alg \mid (\rone + \rb \ra) \text{ \nmf{ is invertible
   for any }} \rb \in \Alg\}  \subseteq \Alg .
\]
  In case $\text{\nmf rad}(\Alg) = \{ \rnul \}$,
the algebra $\Alg$ is called \emph{semi-simple}.
 
The following statements are equivalent:
\begin{compactitem}

\item  $\Alg$ is semi-simple.

\item $\Alg$ has a faithful state.

\item $\Alg$ is a $\Wp^*$-algebra.

\item The trace $\tr_{\Alg} \in \Alg^\star$ is non-degenerate, cf.\ \feR{rem:bounded-ideal}.

\item The normalised trace functional on $\Alg$ is a faithful state, 
      cf.\ \feP{prop:bounded-ideal}.

\item $\Alg$ has a faithful *-representation on a Hilbert space, cf.\ \feT{thm:GNS}.

\end{compactitem}
\end{prop}

Hence, only in other words, up to now and also in the sequel, we have assumed that 
the algebra $\Alg$ is semi-simple, and \feP{prop:radical} notes a few equivalent 
properties.  As was already stated in \feR{rem:bounded-ideal}, in case 
$\text{\nmf rad}(\Alg) \ne \{ \rnul \}$, i.e.\ the algebra $\Alg$ is
not semi-simple, then the quotient unital *-algebra $\Alg / \text{\nmf rad}(\Alg)$
is semi-simple.  In the sequel, it will again be tacitly assumed that this
quotient is taken in case $\text{\nmf rad}(\Alg)$ is non-trivial.

Here we collect a number of results for such finite dimensional semi-simple 
$\Wp^*$-algebras:

\begin{prop}[General finite dimensional $\Wp^*$-Algebras]  \label{prop:Banach-alg}
Let $(\Alg, \svpi)$ be a finite dimensional $\Wp^*$-probability algebra 
with (uniform) norm $\nd{\cdot}_\infty$ and $\ra, \ra_1, \ra_2 \in \Alg_{sa}$ self-adjoint,
$\rb, \rx, \ry \in \Alg$ general elements, and $\rc \in \Alg$ normal (cf.\ \feD{def:alg-names}).  
Then
\begin{compactitem}
\item for a complex valued function $f$ defined on a domain $J \subseteq \CC$ ($f: J \to \CC$)
  which encompasses the spectrum $\sigma(\rc)\subseteq J$ of a normal element $\rc$, there is a 
  $\rg = f(\rc) \in \Alg$; 
\item for such functions there is a \emph{spectral mapping} 
  $\sigma(f(\rc)) = f(\sigma(\rc)) \subset \CC$; 
\item for such a normal element $\rc \in \Alg$ there is a unique injective *-homomorphism 
    $\Psi:\Ck(\sigma(\rc)) \to \Alg$ from the complex continuous functions on the spectrum 
    into the algebra $\Alg$, such that $\Psi(\bbbone_{\sigma(\rc)}) = \rone \in \Alg$
    and $\Psi(\mrm{id}_{\sigma(\rc)}) = \rc$, where $\mrm{id}_{\sigma(\rc)}: z \mapsto z$
    is the identity on $\sigma(\rc) \subset \CC$;
\item in particular it is possible to compute the square root of positive elements, 
    and the \emph{absolute value} or \emph{modulus} of an arbitrary element:
    $\ns{\rx} := (\rx^\star \rx)^{\frk{1}{2}} \in \Alg_+ \subset \Alg_{sa}$;
\item $\ra_1 \le \ra_2$ implies $\rb^\star \ra_1 \rb \le \rb^\star \ra_2 \rb$
    for any $\rb \in \Alg$,
    $\ns{\rx} \le \ns{\ry}$ implies $\nd{\rx}_\infty \le \nd{\ry}_\infty$, and
    $0 \le \ra_1 \le \ra_2$ implies $0 \le \ra_1^{\beta} \le \ra_2^{\beta}$
    for any $0 \le \beta \le 1$;
\item if $\ra_1$ is invertible and $0 < \ra_1 \le \ra_2$, then $\ra_2$ is invertible and
    $0 < \ra_2^{-1} \le \ra_1^{-1}$;
\item given the absolute value one can define the positive part 
   $\ra_+ := \frk{1}{2}(\ns{\ra} + \ra) \in \Alg_+$, and the negative part $\ra_- := 
   \frk{1}{2}(\ns{\ra} - \ra) \in \Alg_+$, such that $\ra = \ra_+ - \ra_-$, as well as  
   $\ra_+ \ra_- = \ra_- \ra_+ = \rnul$; this is  also called the
   \emph{Jordan-Hahn} decomposition; 
\item given the absolute value, one can find the \emph{polar decomposition}
    $\rb = \rg \ns{\rb}$, where $\rg \in \Alg$ is a partial isometry (cf.\ \feD{def:alg-names});
\item every $\ra \in \Alg_{sa}$ has a spectral decomposition: 
    \begin{equation}  \label{eq:gen-fdim-spec-dec}
        \ra = \sum_{\lambda_k \in \sigma(\ra)} \lambda_k \, \rp_{\lambda_k} ,
     \end{equation}
   with  $\lambda_k \in \sigma(\ra) \subset \RR$ and a complete set of mutually 
   orthogonal projections
   $\rp_{\lambda_k} \in \CC[\ra]$ ($\rp_{\lambda_k}\rp_{\lambda_j} = 
       \updelta_{kj}\rp_{\lambda_k}$),
   such that $\sum_{\lambda_k \in \sigma(\ra)} \rp_{\lambda_k} = \rone$ --- a 
   \emph{partition of unity}.   The assignment 
   \begin{equation}   \label{eq:PVM-defi-A}
      \Uppi_A: 2^{\sigma(\ra)} \ni \{\lambda_k\} \to \rp_{\lambda_k} \in \PA{\Alg}
   \end{equation} 
   is called an abstract \emph{projection valued measure} (PVM), cf.\ \feD{def:PVM-M},
   \feR{rem:PVM-more-X}, \feR{rem:spec-dec-PVM}, and \feR{rem:abstract-PVM}. 
   The algebra $\Alg$ contains all the projections $\rp_k \in \PA{\Alg}$ from 
   the spectral decomposition, in fact they are given explicitly by \emph{Lagrange 
   polynomials}, cf.\ \feR{rem:spec-dec-proj}:
   $\rp_{\lambda_k} = \prod_{j \ne k} (\ra - \lambda_j \rone)(\lambda_k - \lambda_j)^{-1}$.
   Just as in \feD{def:PVM-M}, the PVM generates the commutative sub-$\Wp^*$-algebra
   $\spn \{ \rp_{\lambda_k} \mid \lambda_k \in \sigma(\ra) \} = \CC[\{\rp_{\lambda_k} \}] 
   =\CC[\ra]$; 
\item as any element of the algebra $\Alg$ is the sum of two self-adjoint elements,
   it is clear from the above spectral decomposition that the algebra $\Alg$ 
   is generated by its projections $\PA{\Alg}$:
   $\Alg = \spn_{\CC}\, \PA{\Alg}$, and the projections $\PA{\Alg}$ can be
   equipped with a lattice structure as in \feP{prop:proj-matrix-measurement},
   which on commutative subsets coincides  with the one in \feP{prop:boolean}.
   This lattice is denoted by $\opb{L}(\PA{\Alg})$.
   As $\Alg$ is finite dimensional, from $\PA{\Alg}$ one can extract the
   minimal projections (cf.\ \feD{def:alg-names})  
   $M = \{ p \in \PA{\Alg} \mid p \text{ is minimal} \}$,
   such that $\Alg = \spn_{\CC}\, M$.  Any PVM-generated commutative sub-$\Wp^*$-algebra
   can be extended or embedded in a \emph{maximal Abelian sub-algebra} (MASA);  
\item \emph{Double Commutant Theorem, von Neumann:}  
  the image of the Hilbert *-represent\-ation $\oL(\Alg) \subset \Lop(\obH)$ is a 
  $\Wp^*$-sub-algebra of the $\Wp^*$-algebra $\Lop(\obH)$, it is its own 
  \emph{double commutant} (cf.\ \feD{def:alg-names}):
 \[  \oL(\Alg) = \Wp^*(\oL(\Alg)) = (\Wp^*(\oL(\Alg)))^{\prime \prime} = 
     (\oL(\Alg))^{\prime \prime}    \subset \Lop(\obH). \]
\end{compactitem}
\end{prop}

Given the faithful Hilbert space *-representation in \feT{thm:GNS}, most
of these statements are just well known facts from linear algebra, which can be transported 
to the $\Wp^*$-algebra $\Alg$ via the faithful representation; the only exception here may
be the double commutant theorem, cf.\ \citep{Neumann1961, Sakai1971, Takesaki1, Takesaki2, 
Takesaki3, KadiRingr1-97, KadiRingr2-97, Schaefer99, Blackadar2006}.

With the absolute value, one can define more norms, and associated metrics:

\begin{defi}[$\Lp_p$-spaces]  \label{def:Lp-space}
  Let $(\Alg, \svpi)$ be a finite dimensional probability algebra.  In case the state 
  $\svpi \in \SA{\Alg}$ is not faithful, $\Alg$ is replaced by $\Alg/\C{J}_{\svpi}$
  as in \fsec{SS:states}.
  For $1 \le p < \infty$ define the $\Lp_p$-norm of $\rb \in \Alg$  by
  \begin{equation}   \label{eq:Lp-norm-def}
    \nd{\rb}_p := \sqrt[p]{\svpi(\ns{\rb}^p)} = \Ex_{\svpi}(\ns{\rb}^p)^{\frk{1}{p}}
          = \svpi((\rb^\star \rb)^\frk{p}{2})^{\frk{1}{p}}.
  \end{equation}
  Observe the similarity with the Schatten-$p$-norms, cf.\ \feR{rem:Schatten-A}.
  Define the completion of $\Alg$ in those norms: $\Lp_p(\Alg,\svpi) = \cl_p(\Alg)$, 
  the $\Lp_p$-Banach spaces.   For $p=2$ this repeats  \feD{def:alg-Hilbert}.
  For $p=\infty$ one sets 
  $\Lp_\infty(\Alg,\svpi) = \cl_\infty(\Alg) = \Ck^*(\Alg) = \BA{\Alg} \equiv \Wp^*(\oL(\Alg))$,
  with $\nd{\ra}_\infty = \nd{\ra}_{\infty}$,  the operator norm of $\Lop(\Alg)$
  (cf.\ \feR{rem:Schatten-A}); this is an algebra.
  
  Convergence in those norms is called \emph{convergence in the $p$-th mean}.
  As in \feX{ex:Linfty-minus} one may form 
  \[
  \Lp_{\infty-}(\Alg,\svpi) = \bigcap_{p\in\D{N}} \Lp_p(\Alg,\svpi),
  \]
  which is again an algebra.
\end{defi}

Note that for integer values of $p$ the $\Lp_p$-norm can also be defined
by the last expression in \feq{eq:Lp-norm-def} without spectral calculus.

In general, one has for $1 \le p \le 2 \le q \le \infty$ the inclusions (natural 
continuous embeddings)
\[ \Lp_\infty(\Alg) \hookrightarrow \Lp_{\infty-}(\Alg) \hookrightarrow \Lp_q(\Alg) 
   \hookrightarrow \Lp_2(\Alg) 
   \hookrightarrow \Lp_p(\Alg) \hookrightarrow  \Lp_1(\Alg),  
\]
and $\Lp_q(\Alg)$ is the dual space to $\Lp_p(\Alg)$.  This little excursion was to
show how one may introduce these $\Lp_p$-spaces in the non-commutative case.
However, in finite dimensions the spaces $\Lp_p(\Alg)$ are all
the same as vector spaces, and equal to $\Alg$, but they each carry a different norm 
or metric.  Even so, as all vector space topologies on finite dimensional spaces are
equivalent, all the $p$-th mean convergences are equivalent in that case.

If the algebra is in addition Abelian or commutative, for example as an Abelian
sub-algebra $\Clg \subseteq \Alg$ of some possibly non-commutative probability algebra
$\Alg$, a bit more can be said:

\begin{prop}[Abelian $\Wp^*$-Algebras]   \label{prop:Banach-alg-comm}
In addition to the assumptions in \feP{prop:Banach-alg} assume that $\Blg \subseteq \Alg$ 
is a commutative or Abelian but not necessarily a $\Wp^*$-sub-algebra.  Then
\begin{compactitem} 
\item 
   the Abelian or commutative probability sub-algebra $\Blg \subseteq \Alg$ 
   can be embedded in a maximal Abelian sub-algebra, a MASA (cf.\ \feD{def:alg-names}), 
   and this is in fact its commutant $\Clg = \Blg'$, i.e.\
   $\Blg \subseteq \Blg'  = \Clg = \Blg^{\prime \prime} \subseteq \Alg$, 
   cf.\ \feP{prop:Banach-alg}.
   This MASA $\Clg \subseteq \Alg$ is a $\Wp^*$-sub-algebra, as it satisfies
   von Neumann's double commutant theorem in \feP{prop:Banach-alg};
\item
  every element $\rb \in \Blg$ is normal;
\item 
  the \emph{spectrum} $\Omega := \XA{\Blg} = \mrm{ext }\,  \SA{\Blg}$ is non-empty, 
  \emph{weak-*-compact}, and this
  are the \emph{extreme points} (pure states) of the weak-*-compact convex set of states
  $\SA{\Blg} \subset \Blg^*$.  Any element $\som \in \Omega$ is a *-algebra 
  homomorphism $\som: \Blg \to \CC$ (a sample);  
\item 
  (Krein-Milman) the state set $\SA{\Blg}$ is  the \emph{closure of the convex hull} 
  of the spectrum:   $\SA{\Blg} = \cl \co \XA{\Blg}$;
\item   
  for $\rb \in \Blg$ one has  $\lambda \in \sigma(\rb)$ iff there is a
  $\som \in \Omega: \som(\rb) = \ip{\som}{\rb} =  \rb(\som)  = \lambda$.  
  Or, in other words, the mapping (by abuse of notation 
  denoted again by $\rb$)   $\rb:\, \Omega \to \sigma(\rb) \subset \CC$ given by 
  $\rb: \som \mapsto \rb(\som)$ is continuous and surjective;
\item   
   (Gel'fand-Neumark) one has that $\Blg \cong \Ck(\Omega)$, i.e.\ the algebra is 
   isomorphic to the algebra of \emph{all continuous} complex-valued functions 
   on the weak-*-compact Hausdorff space $\Omega =\XA{\Blg}$. 
   These kind of algebras are mentioned in \feX{ex:cont-fcts}.  
   This isomorphism is called the \emph{Gel'fand transform};
\item 
   two Abelian $\Wp^*$-algebras $\Clg_1, \Clg_2$ are *-isomorphic \emph{iff} the spectra
   $\Omega_1 = \XA{\Clg_1}$ and $\Omega_2 = \XA{\Clg_2}$ are homeomorphic.
\item (Riesz-Markov-Kakutani) 
   as $\svpi \in \SA{\Blg} \subset \Blg^*$ is a positive continuous functional on 
   $\Blg \cong \Ck(\Omega)$, there is a \emph{Radon probability measure} 
   $\mu_{\svpi} \in \Ck(\Omega)^*$ such that for all $\ra \in \Blg$ one has 
   \[  
      \ip{\svpi}{\ra} = \svpi(\ra) = \int_{\Omega} \ra(\som) \, \mu_{\svpi}(\di \som) = 
     \int_{\Omega} \ip{\som}{\ra} \, \mu_{\svpi}(\di \som) . 
   \]
\item
   Conversly, any Radon probability measure $\nu \in \Ck(\Omega)^*$ induces in this way
   a state $\svpi_{\nu} \in \SA{\Blg}$.  This is another way of expressing that 
   $\SA{\Blg} = \cl \co \XA{\Blg}$, and it induces a correspondence between $\SA{\Blg}$ 
   and the Radon probability measures on $\Omega=\XA{\Blg}$;
\item (von Neumann) 
   there is an $\ra \in \Blg_{sa}$ such that any $\rc \in \Blg$ is a function 
   $f_{\rc}$ of $\ra$:  $\rc = f_{\rc}(\ra)$, and in  case $\Blg$ is finite dimensional, 
   this function can be expressed as a polynomial in $\ra \in \Blg_{sa}$.    
   This $\ra$ has a spectral decomposition 
   $\ra = \sum_{\lambda_j \in \sigma(\ra)} \lambda_j \, \rp_{\lambda_j}$, where
   the $\rp_{\lambda_j}$ are commuting, mutually orthogonal, minimal projections,
   i.e.\ a PVM;
\item there is a PVM $M = \{ \rp_j \}_{(j=1 \dots m)}$ of mutually orthogonal 
   minimal projections which are a partition of unity, such that $\Blg = \spn_{\CC} M$.  
   These projections can be taken as the ones from the 
   $\ra \in \Blg_{sa}$ just above, $\rp_j = \rp_{\lambda_j}$.
\end{compactitem}
\end{prop}

\paragraph{Structure of Finite Dimensional W*-Algebras:}
The idea here is to decompose any finite dimensional probability algebra into simpler
pieces, a purely algebraic result which does not use the state.  This may be seen as
a version of the Wedderburn–Artin theorem for rings.
These simpler pieces are the \emph{factors}, algebras which have simple resp.\ trivial centres:

\begin{defi}[Factor]   \label{def:factors}
A $\Wp^*$-algebra $\Alg$ is called a \emph{factor}, iff the centre  (cf.\ \feD{def:centre})
is trivial, i.e.\ it just consists of scalar multiples of the multiplicative unit: 
$\ZA{\Alg} = \CC[\rone]$.
\end{defi}

For any Hilbert space $\Hvk$, observe that $\BHA$ is a factor.  
Note that $\MMn(\CC)$ is a factor, indeed a factor of type $\mrm{I}_n$, e.g.\ cf.\
\citep{KadiRingr1-97, KadiRingr2-97, BrattRob-1, BrattRob-2,
Takesaki1, Takesaki2, Takesaki3, Blackadar2006}.
As will become clear shortly, any finite dimensional factor essentially looks like this.
To proceed, the minimal projections in the centre of the algebra will play a decisive role.  

\begin{thm}[Action of Minimal Projection]   \label{thm:action-minimal-p}
Let $\Alg$ be a finite dimensional unital $\Wp^*$-algebra.  Any ideal $\C{I} \subseteq \Alg$ 
in a finite dimensional $\Wp^*$-algebra has the form $\C{I} = \rqq \Alg \rqq = \Alg \rqq$ 
for some central projection $\rqq \in \ZA{\Alg}$, and $\rqq \in \Clg$ acts as a 
multiplicative unit $\rone_{\C{I}}$ on $\C{I}$.  

Let $\rp \in \ZA{\Alg}$ be a \emph{minimal} projection in the centre of the algebra.  
Then the ideal 
$\Clg := \rp \Alg \rp = \Alg \rp = \{ \rp \ra \rp \mid \ra \in \Alg\} \subseteq \Alg$
is a \emph{factor}, and its centre is $\ZA{\Clg} = \spn_{\CC} \{ \rp \} = \CC[\rone_{\Clg}]$,
cf.\ \citep{Takesaki1, Harlow2016}.
Such a factor $\Clg$ is *-isomorphic to a full matrix algebra $\D{M}_m(\CC)$ resp.\ to
$\Lop(\Kvk)$ with $\dim \Kvk = m$, cf.\ also \feT{thm:Murray-vN}.  The dimension 
$m = \dim \Kvk$ is the dimension of a MASA (maximal Abelian sub-algebra) 
$\Blg \subseteq \Clg$, cf.\ \citep{Takesaki1}, showing that the dimension of the 
factor $\Clg$ satisfies $\dim \Clg = m^2 = \dim \D{M}_m(\CC)$.
\end{thm}
%\begin{proof}
%It is obvious that $\Clg$ is a *-algebra where $\rp \in \Clg$ acts as a multiplicative unit, as
%in $\Clg$ one has $\rp = \rp \rone_{\Alg} \rp = \rone_{\Clg}$, and thus $\spn_{\CC} \{ \rp \} =
%\CC[\rone_{\Clg}] \subseteq \ZA{\Clg}$.  In case there is an $\rx \in \ZA{\Clg}$ with 
%$\rx \notin \spn_{\CC} \{ \rp \}$, then, without loss of generality,
% one may assume $\rx$ self-adjoint, as according 
%to \feD{def:alg-names} any element is the sum of two self adjoint elements, at least one of which
%is not in $\spn_{\CC} \{ \rp \}$.  According to \feP{prop:Banach-alg}, this $\rx$ has a 
%spectral decomposition $\rx = \sum_{j=1}^J \xi_j \rqq_j$ with non-zero projections 
%$\rqq_j \in \ZA{\Clg}$ (i.e.\ they commute with $\rp$) forming a 
%partition of the unit $\rone_{\Clg}$, such that $\sum_{j=1}^J \rqq_j = \rone_{\Clg} = \rp$.  
%At least one of the $\rqq_j$ has to be different
%than $\rp$, call this one $\rqq_1$.  As the sum has at least two non-zero terms (J > 1),
%this means that $\rnul < \sum_{j>1}^J \rqq_j = \rone_{\Clg} - \rqq_1 = \rp  - \rqq_1$, implying
%$\rnul < \rqq_1 < \rp$, contradicting the minimality of $\rp$, and finally proving that indeed
%$\ZA{\Clg} = \spn_{\CC} \{ \rp \} = \CC[\rone_{\Clg}]$.
%\end{proof}

Observe, as $\rp \in \ZA{\Clg}$ in \feT{thm:action-minimal-p} is minimal, according
to the first part of the theorem $\Clg$ has no non-trivial sub-ideals, and is thus
simple, cf.\  \feD{def:alg-names}.  Further, those *-isomorphisms in \feT{thm:action-minimal-p}
are representations on $\CC^m$ resp.\ $\Kvk$; and as a full matrix algebra $\D{M}_m(\CC)$
resp.\ the full operator algebra $\Lop(\Kvk)$ has no non-trivial proper invariant subspaces
(cf.\ \feD{def:irreduc-rep}), this representation of a factor is \emph{irreducible}.

\begin{thm}[Factor Decomposition]   \label{thm:factor-decomp-A}
Let $\Alg$ be a finite dimensional unital $\Wp^*$-algebra, with centre $\ZA{\Alg}$.  
The spectrum (cf.\ \feD{def:sample}) of this commutative sub-*-algebra, $\Omega=\XA{\ZA{\Alg}}$,
is a finite set, say  $\Omega = \{ \som_1, \dots, \som_K \}$.  For each $k \in 1,\dots,K$,
let $\rp_k \in \PA{\ZA{\Alg}}$ be the \emph{minimal central projection} such that 
$\rp_k(\som_j) = \updelta_{kj}$, cf.\ \citep{Takesaki1}.

One has $\sum_{k=1}^K \rp_k = \rone$, and with $\Clg_k = \Alg \rp_k = \rp_k \Alg \rp_k$
the algebra decomposes into factors --- in fact simple ideals --- 
as $\Alg = \bigoplus_{k=1}^K \Clg_k$, cf.\ \feT{thm:action-minimal-p}.

Each factor $\Clg_k$ is *-isomorphic to a full matrix algebra $\D{M}_{m_k}(\CC)$,
resp.\ to $\Lop(\Kvk_k)$ with $\dim \Kvk_k = m_k$, cf.\ \feT{thm:action-minimal-p}.
Hence one has the *-isomorphisms 
\begin{equation}   \label{eq:abst-algebra-factor-split}
  \Alg \cong \bigoplus_{k=1}^K \D{M}_{m_k}(\CC) \cong \bigoplus_{k=1}^K \Lop(\Kvk_k).
\end{equation}
In total these *-isomorphisms are bijective *-representations on $\bigoplus_{k=1}^K \CC^{m_k}$
resp.\ $\bigoplus_{k=1}^K \Kvk_{k}$, i.e.\ a direct sum of irreducible representations
(cf.\  \feD{def:irreduc-rep}).
Each $\CC^{m_k}$ resp.\ $\Kvk_{k}$ is an irreducible invariant subspace, and the
direct sum is an expression of intrinsic symmetries of the algebra.
\end{thm}

As we have previously seen in \feR{rem:fin-dim-mat-rep} that any finite dimensional 
algebra can be represented as a matrix sub-algebra, and that a finite dimensional
probability algebra is isomorphic to a von Neumann sub-algebra of $\Lop(\obH)$ (cf.\ 
\feT{thm:GNS}) for a finite dimensional Hilbert space $\obH$, resp.\ a matrix 
sub-algebra on the Hilbert space $\Cn$ (cf.\ \feC{coro:GNS-matrix}),
one has to say that the *-isomorphisms and representations in \feT{thm:action-minimal-p}
do not come via the GNS-construction, but via so-called \emph{matrix units}, 
cf.\ \citep{Takesaki1}.  

In \feT{thm:factor-decomp-A}  the algebra $\Alg$ satisfies
$\dim \Alg = n = \sum_{k=1}^K \dim \Clg_k = \sum_{k=1}^K m_k^2$
as $\dim \D{M}_{m_k} = m_k^2$, and the representation 
is on $\CC^m = \bigoplus_{k=1}^K \CC^{m_k}$ with $m = \sum_{k=1}^K m_k$.  It is in a
way the representation with the smallest possible $m$ for the given algebra $\Alg$.
On the other hand, in the GNS-construction of the representation in \feT{thm:GNS},
the \emph{regular representation} is on $\obH$ with $\dim \obH = \dim \Alg = n$, in the 
operator algebra of $\obH$ with $\dim \Lop(\obH) = n^2$.

Hence, comparing the GNS-construction in \feT{thm:GNS} with the factor representation 
in \feT{thm:factor-decomp-A}, for the underlying Hilbert spaces one has
$n = \dim \Alg = \sum_{k=1}^K m_k^2 \ge  m = \sum_{k=1}^K m_k$; and as 
in this relation equality holds only iff all
$m_k = 1$, this is another instance of \emph{dilation} resp.\ the ``Church of the
Larger Hilbert Space''.  For the algebras, the differences are even larger, as
$\dim \Lop(\obH) = n^2 > n = \dim \Alg$.
But one of the advantages of the GNS-construction is the existence of a cyclic
and separating vector, i.e.\ the ``purification''.

It is fairly obvious how to translate this to a von-Neumann-sub-algebra 
$\C{Q} \subseteq \LHA$ or a unital *-sub-algebra $\C{M} \subseteq \MMn(\CC)$,
and on these concrete algebras the \emph{multiplicity} of the representation enters,
and the isomorphism can be refined by \emph{unitary equivalence}.

\begin{thm}[Factor Classification]   \label{thm:factor-classifi}
Let $\Hvk$ be a $n$-dimensional Hilbert space, and let the von Neumann-sub-algebra 
$\C{F} \subseteq \LHA$  be a factor with centre $\CC[I_{\Hvk}]$.  
Then $\Hvk$ is \emph{unitarily equivalent} to a Hilbert tensor product of 
two finite dimensional Hilbert spaces $\Kvk \otimes \C{N} \equiv \Hvk$ ---
$\dim \Kvk = m, \dim \C{N} = \ell$ and $n = m\, \ell$ ---
such that the factor $\C{F}$ is \emph{unitarily equivalent} to 
$\Lop(\Kvk) \otimes \CC[I_{\C{N}}]$,
e.g.\ cf.\ \citep{BenyRicht15, Harlow2016}.  
Here $I_{\C{N}}$ is the identity on $\C{N}$.  
The factor $\C{F}$ is (as a *-algebra) *-isomorphic to $\Lop(\Kvk)$.
The commutant $\C{F}^\prime$ is unitarily equivalent to 
$\CC[I_{\Kvk}] \otimes \Lop(\C{N})$.
\end{thm}

A factor which has minimal projections (cf.\ \feD{def:alg-names}) is classed as a 
type $\mrm{I}$ factor.  There are also type $\mrm{II}$ and $\mrm{III}$ factors, which need
not concern us here.  It is clear that any factor of the type $\C{B}(\Kvk)$ with a 
Hilbert space $\Kvk$ is of type $\mrm{I}$, as the minimal projections are the
one-dimensional ones, namely $\ket{\xi}\bra{\xi} \in \PA{\C{B}(\Kvk)}$ with a unit
vector $\ket{\xi} \in \Kvk$.  More precisely, if $\dim \Kvk = \ell$, then
$\C{B}(\Kvk)$ is called a type $\mrm{I}_\ell$ factor, as is $\D{M}_{\ell}(\CC)$.
The \feT{thm:factor-classifi} holds in a wider context:

\begin{thm}[Murray - von Neumann]   \label{thm:Murray-vN}
  If the $\Wp^*$-algebra $\Alg$ is a factor of type $\mrm{I}$, it is *-isomorphic
  to $\C{B}(\Kvk)$ for some Hilbert space $\Kvk$, e.g.\ \citep{Kadison2004}.
\end{thm}

As all finite dimensional unital *-algebras are *-isomorphic to a sub-algebra of
some $\Lop(\Kvk)$ for a finite dimensional Hilbert space $\Kvk$ (cf.\ \feT{thm:GNS}
and \feT{thm:factor-decomp-A}), any finite dimensional factor is a type $\mrm{I}$ 
factor.  The factor $\C{F}$ in \feT{thm:factor-classifi} is a type $\mrm{I}_m$ factor,
as is the factor $\Clg$ in \feT{thm:action-minimal-p}.

Passing from the abstract Hilbert space case in \feT{thm:factor-classifi} to matrices
(cf.\ \feR{rem:fin-dim-mat-rep}) through a choice of an ortho-normal basis (cf.\ 
\feC{coro:GNS-matrix}), one arrives at matrix representations:

\begin{coro}[Matrix Algebra Factors]   \label{coro:matrix-factor}
Let the matrix *-sub-algebra $\C{M} \subset \MMn(\CC)$ be a factor with centre
$\CC[\vI_n]$, where $\vI_n$ is the $n \times n$ identity matrix.  Then there is a
unitary matrix $\vU \in \MMn(\CC)$ such that
\[
    \vU \C{M} \vU^\tpH = \D{M}_m(\CC) \otimes \CC[\vI_\ell],
\]
where $\vI_\ell \in \D{M}_\ell(\CC)$ is the $\ell \times \ell$ unit matrix, and $n = m\, \ell$;
e.g.\ cf.\ \citep{BenyRicht15}.  For the commutant $\C{M}^\prime  \subset \MMn(\CC)$ one has
$\vU \C{M}^\prime \vU^\tpH = \CC[\vI_m] \otimes \D{M}_\ell(\CC)$.
\end{coro}

%Now this result may be applied to a von-Neumann-sub-algebra $\C{Q}$:
%\begin{coro}   \label{coro:action-minimal-P-vN}
%For a unital von-Neumann-sub-algebra $\C{Q} \subseteq \LHA$ and a minimal projection 
%$P \in \PA{\LHA}$, consider the von-Neumann-sub-algebra $\C{F} := P\C{Q}P  
%\subseteq \Lop(\im P)$.  It is a factor with centre $\ZA{\C{F}} = \spn_{\CC} \{ P \}$.
%
%The minimal projection $P$ acts as as multiplicative unit in the factor $\C{F}$, 
%and as identity on the closed Hilbert subspace $\im P = P(\Hvk) \subseteq \Hvk$ 
%(i.e.\ $P_{| (\im P)} = I_{(\im P)}$).
%\end{coro}
%
%Translated to matrices, this means
%\begin{coro}   \label{coro:action-minimal-P-M}
%For a unital *-sub-algebra $\C{M} \subseteq \MMn(\CC)$ and a minimal projection 
%$\vP \in \PA{\MMn(\CC)}$, consider the *-sub-algebra $\C{P} := \vP\C{M}\vP  
%\subseteq \MMn(\CC)$.  It is a factor with centre $\ZA{\C{P}} = \spn_{\CC} \{ \vP \}$.
%
%The minimal projection $\vP$ acts as as multiplicative unit in the factor $\C{P}$, 
%and as identity on the closed Hilbert subspace $\im \vP = \vP(\Cn) \subseteq \Cn$ 
%(i.e.\ $\vP_{| (\im \vP)} = \vI_{(\im \vP)}$).
%\end{coro}
The foregoing results are now used for the projections in the centre of a sub-algebra
\citep{Takesaki1, Wolf2012, BenyRicht15, Harlow2016}, to parallel \feT{thm:factor-decomp-A}:

\begin{thm}[Hilbert Space Factor Decomposition]   \label{thm:factor-decomp-H}
Let $\C{Q} \subseteq  \LHA$ be a unital von Neumann-sub-algebra on a 
finite dimensional Hilbert space $\Hvk$.   Recall 
that the centre $\ZA{\C{Q}} \subseteq \C{Q}$ (cf.\ \feD{def:centre} and \feX{ex:centre-alg}) 
is a commutative von-Neumann-sub-algebra, spanned by a PVM $M = \{ P_k \}_{(k=1,\dots,K)}$ of 
mutually orthogonal \emph{minimal projections}, cf.\ \feP{prop:Banach-alg-comm}, 
cf.\  \feD{def:PVM-M} and  \feR{rem:spec-dec-PVM}.  
Setting $\Hvk_k := P_k(\Hvk) = \im P_k$, one
thus has $\Hvk = \bigoplus_{k=1}^K \Hvk_k$ as an orthogonal direct sum.

For each $k=1,\dots,K$, form the von-Neumann-sub-algebras $\C{F}_k := P_k\C{Q}P_k  
\subseteq \Lop(\Hvk_k)$, which according to \feT{thm:action-minimal-p} are all factors.
Then $\C{Q} = \bigoplus_{k=1}^K \C{F}_k$ (cf.\ \feD{def:direct-sum}), i.e.\  a direct sum
of factor algebras.  

From \feT{thm:factor-classifi} one has that each $\Hvk_k \equiv \Kvk_k \otimes \C{N}_k$ 
is \emph{unitarily equivalent} to a tensor product of two Hilbert spaces, so that
\begin{equation}   \label{eq:space-decomp-H}
   \Hvk= \bigoplus_{k=1}^K \Hvk_k \equiv \bigoplus_{k=1}^K \bigl(\Kvk_k \otimes \C{N}_k\bigr) .
\end{equation}
Further, when considering the factor $\C{F}_k \subseteq \Lop(\Hvk_k)$ as acting on $\Hvk_k$,
from \feT{thm:factor-classifi} follows that it is \emph{unitarily equivalent} to 
$\Lop(\Kvk_k) \otimes \CC[I_{\C{N}_k}]$.  Thus  
\begin{equation}  \label{eq:factor-split}
  \C{Q} =\, \bigoplus_{k=1}^K \C{F}_k \; \equiv \;
       \bigoplus_{k=1}^K \,\bigl(\Lop(\Kvk_k) \otimes I_{\C{N}_k}\bigr).
\end{equation}
\end{thm}
\begin{proof}
The space decomposition is seen quickly from
\[
  \Hvk = I_{\Hvk}(\Hvk) = \bigoplus_{k=1}^K P_k(\Hvk) = \bigoplus_{k=1}^K \im P_k = 
  \bigoplus_{k=1}^K \Hvk_k.
\]
The algebra decomposition follows from
\[
  \C{Q}  = I_{\Hvk} \C{Q} I_{\Hvk}  = \left( \sum_{k=1}^K P_k\right) \C{Q} 
    \left( \sum_{j=1}^K P_j\right) =     \sum_{k,j=1}^K P_k \C{Q} P_j 
    = \bigoplus_{k=1}^K P_k \C{Q} P_k ,
\]
as $P_k P_j = \updelta_{kj} P_k$, and one has for $ k \ne j$ that $P_k \C{Q} P_j =
\C{Q} P_k P_j = \{0\}$.
It thus reduces to $\bigoplus_{k=1}^K P_k \C{Q} P_k = \bigoplus_{k=1}^K \C{F}_k$.
\end{proof}

This is now easily translated into the picture and language of matrices, 
e.g. cf.\ \citep{BenyRicht15}:

\begin{coro}[Matrix Factor Decomposition]   \label{coro:factor-decomp-M}
Let $\C{M} \subseteq \MMn(\CC)$ be a unital *-sub-algebra.  Let $\C{Z} = \ZA{\C{M}}$ be 
its centre, and a PVM (cf.\ \feD{def:PVM-M}) of \emph{minimal projections}
 $M = \{ \vP_k \}_{k=1}^K \subset \C{Z}$, such that $\C{Z} = \spn M$.
According to \feT{thm:action-minimal-p}, each $\vP_k \C{M} \vP_k$ is a factor
and $\C{M} = \oplus_{k=1}^K \vP_k \C{M} \vP_k$, hence
there is a unitary matrix $\vU \in \MMn(\CC)$ such that 
\begin{equation}  \label{eq:algebra-factor-split}
   \vU \C{M} \vU^\tpH = \left( \oplus_{k=1}^K \C{M}_k \right) ,
\end{equation}
where each $\C{M}_k \subseteq \D{M}_{n_k}$ is a factor isomorphic to $\vP_k \C{M} \vP_k$, 
$n_k = \tr \vP_k$, and $n = \sum_{k=1}^K n_k$.  
From \feC{coro:matrix-factor} one may glean that one can choose $\vU$ such that each 
\begin{equation}  \label{eq:matrix-factor-strct}
\C{M}_k = \D{M}_{m_k}(\CC) \otimes \CC[\vI_{\ell_k}], \; \text{ with } \; n_k = m_k \ell_k.
\end{equation}

Equivalently, each $\vA \in \C{M}$ has, when transformed, according to 
\feq{eq:algebra-factor-split}, the block-diagonal form
\begin{equation}  \label{eq:matrix-factor-split}
   \vU \vA \vU^\tpH = \begin{bmatrix}  \vA_1 &  &  \\
          & \ddots &  \\
          & & \vA_K 
     \end{bmatrix} = \bigoplus_{k=1}^K \vA_k,
\end{equation}
where $\vA_k \in \C{M}_k \subseteq \D{M}_{n_k}(\CC)$, and with 
\feq{eq:matrix-factor-strct}  one has with certain $\vhat{A}_{m_k} \in \D{M}_{m_k}(\CC)$
\begin{equation}  \label{eq:matrix-factor-split-2}
  \vA_k = \vhat{A}_{m_k} \otimes \vI_{\ell_k} = \begin{bmatrix}  \vhat{A}_{m_k} &  &  \\
          & \ddots &  \\
          & & \vhat{A}_{m_k} 
     \end{bmatrix}.
\end{equation}
\end{coro}

\begin{rem}[Multiplicity]   \label{rem:multiplicity}
From \feT{thm:action-minimal-p} and \feT{thm:factor-decomp-A} the decomposition
is via \emph{*-isomorphisms}.  But observe, from the last two 
\feqs{eq:matrix-factor-split}{eq:matrix-factor-split-2} in \feC{coro:factor-decomp-M}, 
that for a random variable $\vA \in \C{M}_{sa}$, a value in the spectrum $\sigma(\vA)$ 
--- an eigenvalue and hence a possible value in a sample or observation ---
has to appear in one $\sigma(\vA_k)$ of the $\vA_k \in \C{M}_k \subseteq \D{M}_{n_k}$ in 
\feq{eq:matrix-factor-split}.  And the \feq{eq:matrix-factor-split-2} shows that this 
eigenvalue has at least multiplicity $\ell_k$. 

This multiplicity $\ell_k$ is established via \emph{unitary equivalence} based on the 
underlying Hilbert spaces $\Hvk_k$ resp.\ $\CC^{n_k}$, which is a more refined form of
*-isomorphism, and the multiplicity $\ell_k$ is a feature of the representation,
here the GNS-representation, cf.\ \feT{thm:GNS}, whereas the dimension $m_k$ 
is an \emph{intrinsic} property of the abstract algebra $\Alg$, independent of
any representation.
% More precisely, each $\C{M}_k$ in 
%\feqs{eq:algebra-factor-split}{eq:matrix-factor-strct} contains only RVs 
%$\vA_k \in \C{M}_k \subseteq \D{M}_{n_k}$ where each eigenvalue has 
%a multiplicity which is a multiple of $\ell_k$.
\end{rem}

%Now this can collected for an abstract algebra; let $\Alg$ be a finite dimensional 
%unital *-algebra with centre $\ZA{\Alg} \subseteq \Alg$.
%
%\begin{coro}[Algebra Factor Decomposition]   \label{coro:factor-decomp-A}
%Let $\Alg$ be a finite dimensional probability algebra, and $\C{Z} = \ZA{\Alg}$ its centre
%with \emph{minimal projection basis} $M = \{ \rp_k \}_{k=1}^K$.
%Thanks to \feT{thm:action-minimal-p} one knows that each $\Clg_k := \rp_k \Alg \rp_k$ is a
%factor, and $\Alg = \bigoplus_{k=1}^K \Clg_k$, cf.\ \feT{thm:factor-decomp-A}.  
%Additionally, from \feT{thm:factor-classifi}
%it follows that for each $k=1,\dots,K$ there are finite dimensional Hilbert spaces 
%$\Kvk_k, \C{N}_k$, such that $\Clg_k$ is also *-isomorphic to $\Lop(\Kvk_k)$, and in the 
%GNS-representation $\oL$ of $\Alg$ (cf.\ \feT{thm:GNS}), each $\oL(\Clg_k)$  is 
%\emph{unitarily equivalent} with $\Lop(\Kvk_k) \otimes \CC[I_{\C{N}_k}]$. 
%This means that one has the \emph{*-isomorphism}  \citep{Takesaki1, KadiRingr2-97} $\Alg \cong 
%\bigoplus_{k=1}^K \Lop(\Kvk_k)$, as well as the \emph{unitary equivalences}
%(\feC{coro:factor-decomp-M})
%\begin{equation}   \label{eq:abst-algebra-factor-split}
%  \oL(\Alg) \equiv \bigoplus_{k=1}^K \bigl( \Lop(\Kvk_k) \otimes \CC[I_{\C{N}_k}] \bigr)
%  \equiv  \bigoplus_{k=1}^K \bigl( \D{M}_{m_k}(\CC) \otimes \CC[\vI_{\ell_k}] \bigr).
%\end{equation}
%\end{coro}

These analogous decompositions, namely \feq{eq:factor-split} in \feT{thm:factor-decomp-H},
in \feC{coro:factor-decomp-M} the \feqs{eq:algebra-factor-split}{eq:matrix-factor-split},
and \feq{eq:abst-algebra-factor-split} in \feT{thm:factor-decomp-A}, 
all indicate, in similar ways, internal symmetries in
the algebra, for further information on this, e.g.\ cf.\ \citep{FaesslerStiefel1992} 
and the references therein.  This subject will not pursued further here except
for a few remarks which can be seen in an easy way by looking at 
\feq{eq:abst-algebra-factor-split} and \feeqs{eq:algebra-factor-split}{eq:matrix-factor-split-2}.  
From \feq{eq:abst-algebra-factor-split} it is seen that each $\CC^{m_k}$ in
$\CC^{m} = \bigoplus_{k=1}^{K} \CC^{m_k}$ is an irreducible invariant subspace,
with the decomposition representing the intrinsic symmetries of the algebra $\Alg$.
 
In \feeqs{eq:algebra-factor-split}{eq:matrix-factor-split-2}
in the transformed form one has $\Cn = \bigoplus_{k=1}^K \CC^{n_k}$, and from
\feq{eq:matrix-factor-split} it is obvious that each of the $\CC^{n_k}$ --- which is
the carrier for the representation of one of the factors $\C{M}_k$ --- is invariant under
all $\vA \in \C{M}$.  This is a blown-up version (by a factor $\ell_k$) of
the one in \feq{eq:abst-algebra-factor-split}, which results from the
representation resp.\ realisation as a sub-algebra of some ``larger''
full algebra $\MMn(\CC)$ on $\Cn$.   This is what causes the multiplicity mentioned
in \feR{rem:multiplicity}, in that \feqs{eq:matrix-factor-strct}{eq:matrix-factor-split-2} 
indicate that the representation of each factor $\C{M}_k$ on the corresponding 
$\CC^{n_k}$ may be further reducible to possibly $\ell_k$ repeated identical 
\emph{irreducible} representations or realisations of the factor $\D{M}_{m_k}(\CC)$, 
as $\CC^{n_k} = \bigoplus_{j=1}^{\ell_k} \CC^{m_k}$.

Note that up to now a possible
state on the algebra, which carries the probabilistic information, has not been used in
these structure results at all.  
%But the structure result will have a bearing on the
%probabilistic interpretation, a subject to be picked up after the definition of an
%observation or measurement has been generalised a bit.

\begin{rem}[Probabilistic Structure Interpretation]   \label{rem:interp-dir-sum}
Recalling the construction \feD{def:direct-product} of the direct product and the equivalent
one of the direct sum in \feD{def:direct-sum}, one sees that the above decompositions
are exactly this, direct sums of algebras, in fact factors.  Recalling the
interpretation of this construction, following the \feD{def:direct-sum} it was
stated that this is what one would call a mixture.  It is equivalent to a classical 
resp.\ commutative algebra sampling, and then deciding which of the summands to take
for the further action.

One might also recall, that in \feq{eq:abst-algebra-factor-split} in the above
\feT{thm:factor-decomp-A}, each summand $\Clg_k = \rp_k \Alg \rp_k$ is a factor, 
produced by one of the minimal projection in the centre of the algebra,
$\rp_k \in \ZA{\Alg}$, cf.\ \feT{thm:action-minimal-p}.  Now, if $\svpi \in \SA{\Alg}$
is the current state of the algebra, then each factor $\Clg_k$ in the direct sum is chosen 
with probability $\alpha_k = \svpi(\rp_k)$.  Obviously, as the minimal projections in the 
centre form a partition of unity, $\sum_k \rp_k = \rone$, one has  $\sum_k \alpha_k = 1$.
\end{rem}

\paragraph{Projective Measurements:}
The foregoing shows the importance of the projections when describing a probability algebra,
especially when realising that in the finite dimensional case they are always $\Wp^*$-algebras.
These are generated by their projections, and thus here is the appropriate point to
discuss the extended concept of measurement associated with projective measurements.

In \feD{def:PVM-M} the concept of a projection valued measure (PVM) was already
introduced on the special $\Wp^*$-algebra $\MMn(\CC)$, as well in an abstract
setting in \feP{prop:Banach-alg}.  These turned out to describe the
most basic measurements.  This development applies also one-to-one to the
sub-algebra representing $\Alg$ in \feC{coro:GNS-matrix}.  This has to 
be translated to the Hilbert space representation from the GNS construction in 
\feT{thm:GNS} with a finite dimensional Hilbert space $\obH$.  
It is then generalised to the abstract
setting of abstract $\Wp^*$-algebras in \feP{prop:Banach-alg}.

\begin{rem}[Projective Measurement and Projection Valued Measure (PVM) --- Hilbert 
Space Setting]  \label{rem:spec-dec-PVM}
Considering PVMs, analogous definitions to \feD{def:PVM-M} in the Hilbert space
setting can be phrased as follows: every self-adjoint 
$A \in \C{B}(\obH)_{sa}$ is known from spectral theory (e.g.\ \citep{segalKunze78})
to have a spectral decomposition analogous to \feq{eq:gen-fdim-spec-dec}:
\begin{equation}   \label{eq:spec-dec-H}
          A = \sum_{\lambda_k \in \sigma(A)} \lambda_k \, P_{\lambda_k} ,
\end{equation}
where the $P_{\lambda_k}$ is the orthogonal projection onto the eigenspace 
$\obH_{\lambda_k} \subseteq \obH$; it is given by a polynomial in $A$ 
(cf. \feP{prop:Banach-alg}).  These orthogonal, and also pairwise orthogonal, 
projections are a PVM, and generate, just as in \feD{def:PVM-M} and \feP{prop:Banach-alg}, 
a commutative sub-algebra which equals $\CC[A]$.  Each such commutative sub-algebra
can be extended to or embedded in a MASA, by breaking down the projections
in the PVM into \emph{minimal projections}.

As before, possible sharp observations are elements of the 
spectrum $\sigma(A)$, e.g.\  $\som_k(A) = \lambda_k$ with probability 
$p_k := \svpi(P_{\lambda_k})$ for some $\som_k \in \XA{\CC[A]}$,  
called a \emph{projective measurement} or \emph{sharp observation}.
Such a sharp observation is a character $\som_k \in \XA{\CC[A]}$,
 and thus has to be an extreme point of $\SA{\CC[\Alg]}$.  
In finite dimensions, these all have the form
$\som_k(A) =  \tr (\ket{\psi_k}\bra{\psi_k} A)$ of a vector state, 
where $\psi_k \in \obH$ is a unit eigenvector of $A$, i.e.\ $A \psi_k = \lambda_k \psi_k$.
As was mentioned before, the unit vector $\psi_k$ is often also called the state
or \emph{state vector}, and this is the original meaning of the concept, i.e.\
the system modelled on $\obH$ whose observables are described by $\Alg$ resp.\ $\oL(\Alg)$ 
was measured or is in the state $\psi_k$.

As already mentioned, such pairwise orthogonal projections as in \feq{eq:spec-dec-H}
are a partition of unity and form a PVM
\begin{equation}   \label{eq:PVM-defi-H}
   \Uppi_H: 2^X \ni \{ x \} \mapsto P_{x} \in \PA{\C{B}(\obH)},
\end{equation}
where $\sigma(A)$ has been replaced by a finite set $X$, cf.\ \feR{rem:PVM-more-X}.
If an observation $\som_x(P_x) = 1$ is made, one may say the outcome is $x \in X$.
\end{rem}

\begin{rem}[Abstract Projection Valued Measure (PVM)]    \label{rem:abstract-PVM}
Consider an observable $\ra \in \Alg_{sa}, \ra^\star = \ra$, with spectral decomposition as in
\feq{eq:gen-fdim-spec-dec}.  Possible observations are elements of the spectrum $\sigma(\ra)$,
e.g.\  $\som_k(\ra) = \lambda_k$ with probability $p_k := \svpi(\rp_{\lambda_k})$ for some 
$\som_k \in \XA{\CC[\ra]}$.  In this case it is known that for the corresponding projection
from the PVM $\sigma(\ra) \supset \{ \lambda_k \} \mapsto \rp_{\lambda_k} \in \PA{\Alg}$one has 
$\som_k(\rp_{\lambda_k}) = 1$ and for all other projections $\som_k(\rp_{\lambda_j}) = 0$
for $\lambda_j \ne \lambda_k$ (these are sharp observations).  
Hence knowing $\som_k(\rp_{\lambda_k}) = 1$ --- i.e.\ the
event encoded by $\rp_{\lambda_k}$ is true, or the system is in a state covered 
by $\rp_k$ --- is equivalent to knowing that $\ra$ has the value $\lambda_k$.  
Such sharp observations of a projection from a PVM
are again called \emph{projective measurements} resp.\  \emph{projective observations}.

A bit more general PVM is an assignment 
\begin{equation}   \label{eq:PVM-defi-AA}
  \Uppi_X: 2^X \ni \{ x \} \mapsto \Uppi_X(\{ x \}) = \rp_x \in \PA{\Alg}, 
\end{equation}
satisfying the requirements as laid out in \feD{def:PVM-M} and \feR{rem:PVM-more-X},
i.e.\ 
\begin{equation} \label{eq:PVM-X-part-unity}
    \sum_{x \in X} \, \rp_x = \rone_{\Alg}, \quad \text{and } \quad \forall x, y \in X:\;
        \rp_x \, \rp_y = \updelta_{x y} \, \rp_x .
\end{equation}
And, as already described in \feD{def:PVM-M}, \feP{prop:Banach-alg}, and 
\feR{rem:spec-dec-PVM}, the PVM generates a commutative sub-algebra
\[
   \spn \{\Uppi_X(\{x\}) \mid x\in X \} = \CC[\{\Uppi_X(\{x\}) \}] \subseteq \Alg .
\]
Each such commutative sub-algebra
can be extended to or embedded in a MASA, by breaking down the projections
in the PVM into \emph{minimal projections}.
\end{rem}

\begin{rem}  \label{rem:PVM-vs-pure_state}
  While in \feP{prop:proj-matrix-measurement} an atomic sharp observation was connected 
  with a vector state, i.e.\ a pure state,   now this has been generalised to 
  sharp observations connected to a PVM (cf.\ \feD{def:PVM-M}, \feq{rem:spec-dec-PVM} 
  in \feR{rem:spec-dec-PVM}, \feP{prop:Banach-alg}, and the 
  foregoing \feq{eq:PVM-defi-AA} in \feR{rem:abstract-PVM}), i.e.\ one
  may speak about an event represented by a projection $\vP_x \in \PA{\MMn(\CC)}$ resp.\
  $P_x \in \PA{\C{B}(\obH)}$ resp.\ $\rp_x \in \PA{\Alg}$ being true or false.
  Observing the sharp event $\rp_x$ then means that $x \in X$ has certainly occurred.
\end{rem}

As a final result connected with both the structure of algebras and PVMs, one may take a  
look at Gleason's theorem.  As one might recall from \feP{prop:Banach-alg}, a finite 
dimensional W*-algebra $\Alg$ is generated by its projections $\PA{\Alg}$.  The question
now arises, whether a state $\svpi \in \SA{\Alg}$ is determined by its behaviour
on the projections.  This is true in classical probability, i.e.\ for Abelian algebras.
The question in the general case is answered in the affirmative by Gleason's theorem, 
cf.\ e.g.\ \citep{Busch2003}:

\begin{thm}[Gleason]  \label{thm:Gleason}
 Let $\Hvk$ be a finite dimensional Hilbert space with $\dim \Hvk > 2$, 
 and consider the algebra $\LHA$  of all linear maps, and the lattice 
 $\opb{L}(\PA{\LHA})$ of all projections $\PA{\LHA}$
 (cf.\ \feP{prop:proj-matrix-measurement}).

A function $\gamma: \PA{\LHA} \to [0,1]$ which is such that for \emph{any}
given PVM
\[
\Uppi_X: 2^X \ni \{ x \} \mapsto P_x \in \PA{\LHA},
\]
defined on a finite set $X$, analogous to 
\feR{rem:spec-dec-PVM} and like in \feq{eq:PVM-X-part-unity}, it follows from 
\feq{eq:PVM-X-part-unity} that $\sum_{x \in X} \gamma(\Uppi_X(\{ x \})) = 1$ and that
for any subset $S \subseteq X$ it holds that 
\begin{equation}   \label{eq:Gleason}
   \gamma(\sum_{x \in S} \Uppi_X(\{ x \})) = \sum_{x \in S} \gamma(\Uppi_X(\{ x \}))  
\end{equation}
--- this means that it defines a probability measure on $X$ and on the PVM
$\Uppi_{X}$, namely $\prob(\{x\}) := \prob(P_x) = \gamma(P_x)$ like in 
\feP{prop:proj-1} on the Boolean algebra as in \feP{prop:boolean} --- 
can be \emph{extended} to a state $\gamma \in \SA{\LHA}$.

Conversely, a state $\svpi \in \SA{\LHA}$ gives rise to a 
probability measure on any PVM $\Uppi_X$ and on $X$, defined as
$\prob(\{x\}) := \prob(\Uppi_X(\{ x \})) := \svpi(\Uppi_X(\{ x \})) \in [0,1]$
on the Boolean algebra generated like in \feP{prop:boolean}.
\end{thm}

It should be clear that the theorem holds also for any finite dimensional probability
algebra which has no factor isomorphic to $\Lop(\Kvk)$ with $\dim \Kvk = 2$
(factor of type $\mrm{I}_2$) in their factor decomposition, 
cf.\ \feT{thm:factor-decomp-H} and \feT{thm:factor-decomp-A})

\paragraph{Structured Representation of States and PVMs:}
In the foregoing, the general structure of finite dimensional algebras was shown to
be a direct sum of factors, starting from the fundamental \feT{thm:factor-decomp-H} for
sub-algebras of $\LHA$, where $\Hvk$ is a finite dimensional Hilbert space, and
thus also for matrix sub-algebras of $\MMn(\CC)$ in \feC{coro:factor-decomp-M} with explicit
formulas in \feq{eq:matrix-factor-split}, as well as for general abstract algebras in
\feT{thm:factor-decomp-A}.  The basic results about faithful representations
and the GNS construction  \feT{thm:GNS} and
 \feC{coro:GNS-matrix} have been the basis for this.

The two concepts which have emerged so far  to describe the probabilistic content 
are on one hand PVMs as a general system of ``yes-no'' questions, as well as states to produce 
expectations for these questions.  It is important that states can be identified
via the canonical inner product given through the suitably normalised trace --- essentially
the unique tracial state (cf.\ \feD{def:state}) on a finite dimensional space ---
with elements in the algebra, the so-called density matrices resp.\ operators as
detailled in \feP{prop:dens-in-alg-M}, \feC{coro:dens-in-alg-H}, and \feC{coro:dens-in-alg-A}.
The general results on the geometry of states in \feP{prop:states-prop}  carry
over to the set of density matrices, cf.\ \citep{AubrunSzarek2017, Kremminger2022}.

For simplicity, the following discussion is carried out for the matrix representation
of a finite dimensional algebra $\Alg$ in $\MMn(\CC)$ as detailed in 
\feC{coro:factor-decomp-M} with explicit formulas in 
\feqs{eq:matrix-factor-split}{eq:matrix-factor-split-2}.  It is then a routine task
to translate this into the Hilbert space representation or formulate it for
the abstract algebra formulation.

\begin{coro}[Structure of Density Matrices]  \label{coro:struct-dens-mat}
Let $(\C{M} \subseteq \MMn(\CC), \svpi_{\vR})$ be a probability algebra.
A density matrix $\vR \in \DEN{\C{M}}$ of the sub-algebra $\C{M}$ 
representing the state $\svpi_{\vR} \in \SA{\C{M}}$
can be unitarily transformed to the block-diagonal form
\begin{equation}  \label{eq:dens-matrix-factor-split}
   \vU \vR \vU^\tpH = \begin{bmatrix}  \vR_1 &  &  \\
          & \ddots &  \\
          & & \vR_K 
     \end{bmatrix} = \bigoplus_{k=1}^K \vR_k,
\end{equation}
where each $\vR_k \in \D{M}_{n_k}(\CC)_+$.  Together with \feq{eq:matrix-factor-strct} from 
\feC{coro:factor-decomp-M}, one has for each of these the further block-diagonal form
with identical diagonal blocks $\vhat{R}_{m_k} \in \D{M}_{m_k}(\CC)_+$
\begin{equation}  \label{eq:dens-matrix-factor-split-2}
  \vR_k = \vhat{R}_{m_k} \otimes \frk{1}{\ell_k}\,\vI_{\ell_k} = \frac{1}{\ell_k}\,
    \begin{bmatrix}  \vhat{R}_{m_k} &  &  \\
          & \ddots &  \\
          & & \vhat{R}_{m_k} 
     \end{bmatrix},
\end{equation}
with $n = \sum_k n_k$, as well as for all $k:\; n_k = m_k \ell_k$.  Here the scaling
has been performed so that $\tr \vR_k = \tr \vhat{R}_{m_k}$.

According to \feC{coro:factor-decomp-M}, the form \feq{eq:dens-matrix-factor-split} comes
from a direct sum of algebras in \feq{eq:algebra-factor-split}, 
$\vU \C{M} \vU^\tpH = \oplus_{k=1}^K \C{M}_k$, with each $\C{M}_k \subseteq \D{M}_{n_k}(\CC)$
a factor unitarily equivalent to $\vP_k \C{M} \vP_k$, where $\vP_k \in \ZA{\C{M}}$ with 
$n_k = \tr \vP_k$ is one of the minimal projections in the centre of $\C{M}$.  
In \feR{rem:interp-dir-sum}, the probabilistic interpretation was that this corresponds
to a classical experiment choosing with probability $\alpha_k = \svpi_{\vR}(\vP_k)$
which of the summands to activate.  With the foregoing, we can conclude that
\[
   \alpha_k = \svpi_{\vR}(\vP_k) = \tr (\vR \vP_k) = \tr (\vP_k \vR \vP_k) 
      = \tr \vR_k = \tr \vhat{R}_{m_k} .
\]
\end{coro}

This algebraic structure, which holds for all densities $\vR \in \DEN{\C{M}}$,
can now be combined with the linear algebra structure shown in \feP{prop:dens-in-alg-M}.
The set of density matrices was shown to be the disjoint union of the densities of 
different rank, $\DEN{\C{M}} = \biguplus_{r=1}^n \DEN{\C{M}}_r$, where $\DEN{\C{M}}_r$ 
is the density matrices with rank $r$.

This combined structure can now be seen in the light of the geometric structure of 
$\DEN{\C{M}}$.  In \feP{prop:states-prop}
it was stated that the states $\svpi \in \SA{\Alg} \subset \Alg^+ \subset \Alg^{sa} 
\subset \Alg^\star$ are in the dual positive cone $\Alg^+$.  As states are self-adjoint, all
can be played out in the real space $\Alg^{sa}$.  So the normalisation condition
$\ip{\svpi}{\rone} = 1$ specifies a hyperplane $\C{T} \subset \Alg^{sa}$; and thus
$\SA{\Alg} = \Alg^+ \cap \C{T}$.  It should be noted that this hyperplane is orthogonal
to the subspace $\RR[\rone] \subset \Alg^{sa}$.  From \feP{prop:dens-in-alg-M} is is known that
for the matrix representation for any $\svpi \in \SA{\Alg}$ there is a unique density matrix 
$\vR_{\svpi} \in \MMn(\CC)_{sa}$.  The density matrices have to be positive and normalised, thus
$\DEN{\MMn(\CC)} = \MMn(\CC)_+ \cap \{ \vR \mid \tr \vR = 1 \}$ are again on the intersection of
positive cone and hyperplane.  Note again that the hyperplane $\{ \vR \mid \tr (\vR \vI) = 1 \}$
is perpendicular to the one dimensional subspace resp.\ line $\RR[\vI] \subset \MMn(\CC)_{sa}$.

\section{Developments from Algebraic Probability} \label{S:op-represent}
% !TEX root = ../23_QC-algebra.tex
% !TEX encoding = UTF-8 Unicode
% RCSID:       $Id: compl-repr_QC-alg.tex,v 1.26 2026/01/29 23:15:52 hgm Exp $
% Author:      $Author: hgm $
% Contact:     wire@tu-bs.de
% =================================

While up to now mainly the algebraic aspects of the subject were in the foreground,
only a minimum of analytic concepts had to be addressed.  By concentrating on
finite dimensional algebras, most of the analytical difficulties could be avoided.
It is by now clear that the possible observed valued of an algebraic observable
are elements of its spectrum.  Thus to deal with continuously distributed RVs,
the corresponding representation mappings have to be operators on an infinite
dimensional Hilbert space.  And  to deal with unbounded RVs, one has to address
unbounded operators in an infinite dimensional Hilbert space, with all the
difficulties this might involve.  The reader interested
in this area may consult the recent monograph \citep{Schmuedgen2020} and the
references therein.  

Here we continue the discussion of observations in finite dimensional 
probability algebras.  The subject of non-commutativity has surprising consequences,
like Heisenberg's uncertainty relation and the Bell-Kochen-Specker\footnote{Often it 
is just called the Kochen-Specker theorem.  But it turns out that 
{John Stewart Bell} actually proved the theorem in 1966, one year before {Simon Kochen} 
and {Ernst Specker}.} 
theorem.

From observations as samples or characters (cf.\ \feD{def:sample}) as well as their
generalisation, the projection valued measures (PVM, cf.\ \feR{rem:spec-dec-proj-2},
\feP{prop:Banach-alg}, and \feR{rem:abstract-PVM}),   we proceed by looking at 
\emph{valuations} in \fsec{SS:constraints}.  These operations can be seen as examples
of information transmission.  Even more generalised measurements or observations,
the \emph{Positive Operator Valued Measures} (POVM) 
are touched upon in \fsec{SS:POVM}.  This leads to a brief look on how to model
the transmission of information in general, a concept called 
\emph{channels}.  This is also a way to model input information from a classical system
--- a commutative probability algebra --- to a general, possibly non-commutative one,
modelling QLB-effects, and transmit this information to other such systems, and finally
read out or output (POVM measurement) in a classical system.  Another important
subject is the updating of probabilities and expected values due to information from 
observations; this can also be addressed in the picture of channels.

Finally, a subject of classical probability theory is touched upon, where the
methods of algebraic probability can be used, leading to new views on matters
like graphs, reaction networks, random walks, and Markov chains.   This is the
so-called quantum decomposition of classical commuting RVs,
which for analysis purposes are embedded in an algebra of non-commuting
operators on Fock spaces in \fsec{SS:Fock}.
%by limiting ourselves to bounded RVs, and
%discuss a bit further the representation of algebras and states, and connected with
%this the spectral calculus.  The whole may be seen as a special case of non-commutative
%integration, e.g.\ \citep{segal53-AoM, segal65-BAMS, Kostecki2014}.  It connects
%integration strongly with spectral theory, 
%e.g.\ \citep{gelfand64-vol4, segalKunze78, yosida-fa-1980, Kato1995, DautrayLions3}.
%Luckily, in finite dimensions this all just boils down to linear algebra.

\subsection{Observations and Valuations}  \label{SS:constraints}
In \feD{def:sample} observations of elements of a
commutative *-sub-algebra $\Clg \subseteq \Alg$ were defined, 
they turned out to be \emph{multiplicative}
states $\XA{\Clg}$, i.e.\ *-algebra homomorphisms.  As the image range $\CC$ is commutative,
it is clear that there are in general no non-zero one-dimensional *-algebra homomorphisms
on non-commutative probability algebras.  
Thus one might try to have fewer requirements \citep{segalPostQM1947}, and to define
a similar function at least on $\Alg_{sa}$, leading to the concept of valuation.

\begin{defi}[Observation and Value]   \label{def:obs-val}
A self-adjoint RV $\ra \in \Alg_{sa}$ has a definite value in a certain 
state $\sal \in \SA{\Alg}$, which could ideally be observed, 
if any one of the following equivalent conditions is satisfied.
\begin{compactenum}
\item $\var_{\sal}(\ra) = 0$ --- the state is dispersion free;
\item $\sal$ is pure, cf.\ \feP{prop:states-prop};
\item $\sal(\ra^2) = \sal(\ra)^2$.
\end{compactenum}
\end{defi}

That 1.) and 3.) are equivalent is obvious from $\var_{\sal}(\ra) = \sal((\ra - \sal(\ra)\rone)^2)
= \sal(\ra^2) - \sal(\ra)^2$, for the equivalence with 2.) see e.g.\  \citep{segalPostQM1947}.
From 3.) and $\sal(\rone) = 1$ --- which follows here from $\sal \in \SA{\Alg}$, but
which one would require for any reasonable value assignment --- one obtains with the
same kind of argument used in \feP{prop:spec-one} that $\sal(\ra) \in \sigma(\ra)$,
i.e.\ the ``values'' have to be in the spectrum of the RV.
Recall, cf.\ \feP{prop:states-prop}, that characters $\som \in \XA{\Alg}$ --- if
$\XA{\Alg}$ is non-empty --- satisfy obviously 3.), and thus also the others.

Recalling the action of characters on projections in \feP{prop:proj-1}, one sees that
for a projection $\rp \in \PA{\Alg} \subset \Alg_{sa}$ one has
$0 \le \sal(\rp) = \sal(\rp^2) = \sal(\rp)^2$, and thus $\sal(\rp)$ can only
have one of the two values $\sal(\rp) = 0$ or $\sal(\rp) = 1$, just like for characters.
And from \feP{prop:spec-one} it is clear that these two values are the only ones in
the spectrum,  $\sigma(\rp) \subseteq \{0, 1\}$, of a projection $\rp$.

A collection of self-adjoint RVs has jointly definite values in a certain state $\sal$
if the algebra they jointly generate has a state that satisfies \feD{def:obs-val}.
One may then obtain the following theorem \citep{segalPostQM1947, Holmes75}:

\begin{thm}  \label{thm:def-val-comm}
A collection of observables / RVs have jointly definite values in a certain state 
$\sal \in \SA{\Alg}$ --- depending on the collection and state --- 
\emph{iff} they are compatible, i.e.\ \emph{iff} they commute.  From \feD{def:obs-val} 
follows that in that case $\sal$ has to be pure.
On a commutative probability algebra, the pure states are characters.
\end{thm}

Another angle on this subject is obtained from the so called \emph{uncertainty relation},
cf.\ \citep{Sen2014}; originally due to Heisenberg, it was reformulated by Robertson,
and then extended by Schrödinger \citep{Schroedinger1930},
to include a second summand in the relation \feq{eq:uncert-rel} on the right hand side, 
which is shown here observing \feD{def:covariance}:

\begin{thm}[Heisenberg Uncertainty Relation]   \label{thm:uncert-rel}
Let $\Alg$ be a unital *-algebra.  For any state 
$\sal \in \SA{\Alg}$ and any two self-adjoint RVs $\ra, \rb \in \Alg_{sa}$ one has
\begin{equation}  \label{eq:uncert-rel}
    \var_{\sal}(\ra)\, \var_{\sal}(\rb) \ge  \ns{\frac{\sal([\ra,\rb])}{2}}^2 
%    +  \ns{\frac{\cov_{\sal}(\ra, \rb) + \cov_{\sal}(\rb, \ra))}{2}}^2. 
    +   \ns{\Re(\cov_{\sal}(\ra, \rb))}^2. 
\end{equation}
\end{thm}

%As $\var_{\sal}(\ra) = \nd{\wtl{\ra}}^2_{2,\sal}$, from the 
%Cauchy-Schwarz inequality one has 
%$\nd{\wtl{\ra}}^2_{2,\sal}\, \nd{\wtl{\rb}}^2_{2,\sal} \ge \ns{z}^2$,
%with $z = \bkt{\wtl{\ra}}{\wtl{\rb}}_{2,\sal} \in \CC$.  But one has 
%   $\ns{z}^2 = (\Re z)^2 + (\Im z)^2 \ge (\Im z)^2$.  
%Now 
%\begin{multline*}
%(\Im z)^2 = \frk{1}{4}\ns{z - z^*}^2 = \frac{1}{4}\ns{\bkt{\wtl{\ra}}{\wtl{\rb}}_{2,\sal} -
%\bkt{\wtl{\rb}}{\wtl{\ra}}_{2,\sal}}^2 = 
%  \frac{1}{4} \ns{\sal(\wtl{\ra}\wtl{\rb}) - \sal(\wtl{\rb}\wtl{\ra})}^2 =
%  \frac{1}{4} \ns{\sal([\ra,\rb])}^2 .
%\end{multline*}
\feT{thm:uncert-rel} is a consequence of the Cauchy-Schwarz inequality.
As may be gleaned from \feq{eq:uncert-rel}, two RVs can not have simultaneously 
a definite value in the same state unless $\sal([\ra,\rb])=0$, e.g.\ if they commute,
and the real part of their covariance vanishes.
Recall from \feD{def:covariance}, that if $\sal \in \XA{\Alg} \subset \SA{\Alg}$
is a character --- they exist generally only for commutative algebras --- its
covariance vanishes, as does its variance, as well as the value of any commutator,
i.e.\ $\sal([\ra,\rb])=0$ for all $\ra, \rb \in \Alg$.   Hence in this case 
both sides of \feq{eq:uncert-rel} vanish; this is the classical case.

The last two theorems, \feT{thm:def-val-comm} and \feT{thm:uncert-rel}, actually tell
us that states which provide a definite value to a RV, cf. \feD{def:obs-val},
have to be pure states (\feT{thm:def-val-comm}), and the RVs to which they apply
have to commute or be compatible.  So, to describe observations of a collection of
compatible RVs, one could always regard the commutative probability sub-algebra
$\Clg \subseteq \Alg$ which they generate.  
On such commutative algebras, characters exist, they are pure, and
they satisfy the uncertainty relation \feq{eq:uncert-rel} from \feT{thm:uncert-rel}
identically, such that both sides vanish.  Such a commutative or Abelian probability 
sub-algebra $\Clg$ is called a \emph{classical context}, as this describes situations where 
all RVs involved can be observed / measured or have a definite value simultaneously 
with the uncertainty relations being identically satisfied.

These Abelian sub-algebras can be given a partial order, given by set inclusion.
The poset of Abelian sub-algebras --- ordered by set inclusion --- 
of a probability algebra $\Alg$ are called its \emph{classical contexts}.
As for each such commutative or Abelian probability sub-algebra $\Clg$
it holds that $\Clg \subseteq \Clg'$, i.e.\ it is embedded in its 
commutant (cf.\ \feD{def:alg-names}), which, as it is
again Abelian, according to Zorn's lemma it follows that there are maximal
Abelian sub-algebras (MASAs, cf.\ \feD{def:alg-names}), and each chain of inclusions 
has a maximal element, namely such a MASA.  Thus any such classical context, as it is
an Abelian or commutative probability sub-algebra $\Clg$, can be embedded in a MASA, 
and this is in fact its commutant $\Clg \subseteq \Clg' \subseteq \Alg$, 
cf.\ \feD{def:alg-names} and \feP{prop:Banach-alg-comm}.

And obviously it can happen that some Abelian sub-algebra $\C{S} \subset \Alg$ is 
a subset of two (or more) different MASAs, e.g.\ $\C{S} \subset \Blg$
and $\C{S} \subset \Clg$, which do not commute, $[\Blg, \Clg] \ne \{\rnul \}$.
But the characters which give the observations on each Abelian sub-algebra 
($\XA{\Blg}$ and $\XA{\Clg}$) (classical context) may be different from each other.

The problem which can therefore arise --- as the characters which give the value or 
observations on each Abelian sub-algebra may be different from each other --- 
is that some observable / RV $\rx \in \C{S}$ gets different values depending on in 
which larger context (i.e.\ $\Blg$ or $\Clg$) the sub-algebra $\C{S}$ is considered.  
But on the other hand, one would expect that, if the RV $\rx$ ``has a value'', 
that this is the same value, independent of the context.
So more precisely, if say $\som_b \in \XA{\Blg}$ is one character, 
and $\som_c \in  \XA{\Clg}$ another one; the question then is whether one 
can avoid that $\som_b(\rx) \ne \som_c(\rx)$ for some $\rx \in \C{S}$.  

But, as characters  which could assign definite values to a RV generally do not 
exist on non-commutative algebras, one tries with less.  
To assign definite values, one needs a functional defined at least on
\emph{all} the ``real'' observables, $\Alg_{sa} \subset \Alg$,
satisfying the conditions 1.) or 3.) in \feD{def:obs-val} \citep{RajanVisser2019}: 

\begin{defi}[Valuation]  \label{def:valuation}
A \emph{valuation} $\tns{\upsilon}: \Alg_{sa} \to \RR$ is a functional satisfying for
compatible $\ra, \rb \in \Alg_{sa}$ (i.e.\ $[\ra, \rb] = \rnul$), and $z, w \in \CC$:
\begin{compactenum}
  \item $\tns{\upsilon}(z \ra + w \rb) = z\, \tns{\upsilon}(\ra) + w\, \tns{\upsilon}(\rb)$;
  \item $\tns{\upsilon}(\ra \rb) = \tns{\upsilon}(\ra) \tns{\upsilon}(\rb)$;
  \item $\tns{\upsilon}(\rone) = 1$.
\end{compactenum}
\end{defi}

Essentially, a valuation has to behave like a character, but only on commuting self-adjoint 
RVs, but on the other hand it has to be defined on \emph{all} of $\Alg_{sa}$ to be seen
as assigning a definite ``value'' .
From 1.) and 2.) in \feD{def:valuation} one can derive easily that $\tns{\upsilon}(\tp(\ra)) =
\tp(\tns{\upsilon}(\ra))$ for any polynomial $\tp \in \CC[X]$; hence for finite dimensional
algebras this is true for any function and not just polynomials, and under mild conditions
this can be established for almost all reasonable functions also for most
infinite dimensional algebras.
 
Recalling \feD{def:obs-val} and the discussion following it, it was established there
that the only values a projection $\rp \in \PA{\Alg}$ can have are in its spectrum
$\sigma(\rp) = \{ 0, 1 \}$.   And as any self-adjoint $\ra \in \Alg_{sa}$ is determined 
through the spectral decomposition (cf.\ \feP{prop:Banach-alg}) by its spectral projections, 
the same is also true for any such  self-adjoint $\ra$.   Therefore a valuation functional 
has to be dispersion free, and one should have $\tns{\upsilon}(\ra) \in \sigma(\ra)$.
Hence such a valuation could assign ``definite values'' for the all of 
$\Alg_{sa} \subset \Alg$.  But alas, this is in general \emph{not possible} 
\citep{RajanVisser2019, BudroniEtAl2023}, and that is the content of the next

\begin{thm}[Bell-Kochen-Specker]  \label{thm:BKS-thm}
   If $\Alg = \LHA$, the bounded linear operators on a Hilbert space $\Hvk$, 
   and $\dim \Hvk > 2$, \emph{no valuations exist} on $\Alg_{sa}$.
\end{thm}

This certainly also includes all algebras which contain a sub-algebra isomorphic to
$\LHA$ with $\dim \Hvk > 2$ (a factor of type $\mrm{I}_n$ with $n > 2$), here
it will not be possible to assign definite values to all RVs, the value 
\emph{depends on the context}. 
%An exception are commutative or Abelian algebras, as then all factors are one-dimensional, 
%i.e.\ in each factor one has $\Hvk \cong \CC$.

Although this theorem was derived in the realm of quantum mechanics, its formulation
and proof make only reference to a non-Abelian probability algebra, and it is elementary
linear algebra, although a bit combinatorially involved.  But it shows how radically
different commutative and non-commutative probability algebras are.

The proof of the theorem works by constructing a counter-example, i.e.\ for a
Hilbert space with $\dim \Hvk = 3$ to construct commutative sub-algebras of
$\LHA$ which show the impossibility of a valuation.  For $\dim \Hvk = 3$,
as already mentioned, the counter-example is quite involved.  But as an illustration
of the idea, for $\dim \Hvk = 4$ it is possible to give a much simpler counter-example.

Following a construction originally proposed in \citep{CabelloEtal1996}, take 
$\Hvk = \CC^4$ and define 18 vectors $\ket{j} \in \CC^4 \; (j=1,\dots,18)$, in 
Tables~\ref{table:KS-vect-1-9} and \ref{table:KS-vect-10-18}, and define the
orthogonal projections onto the one-dimensional subspaces which these vectors span:
\begin{equation}  \label{eq:18-proj}
   \vP_j := \frac{1}{\bkt{j}{j}} \ket{j}\bra{j} .
\end{equation}

\newcommand{\ph}[1]{\phantom{#1}}

\begin{table}[ht]
\centering 
\begin{tabular}{| c  c  c  c  c  c  c  c  c |} 
\hline\hline  
  \tiny{\ph{a}} & & & & & & & & \\ [-2.0ex]
  $\ket{\ph{1}1}$ & $\ket{\ph{1}2}$ & $\ket{\ph{1}3}$ & $\ket{\ph{1}4}$ 
   & $\ket{\ph{1}5}$ & $\ket{\ph{1}6}$ & $\ket{\ph{1}7}$
   & $\ket{\ph{1}8}$ & $\ket{\ph{1}9}$\\ [0.5ex] 
\hline 
  \tiny{\ph{a}} & & & & & & & & \\ [-2.0ex]
  $\begin{bmatrix} \ph{-}0 \\ \ph{-}0 \\ \ph{-}0 \\ \ph{-}1 \end{bmatrix}$  &  % #FFFF00 = | 1>
  $\begin{bmatrix} \ph{-}1 \\      -1 \\ \ph{-}1 \\      -1 \end{bmatrix}$  &  % #00FFEE = | 2>
  $\begin{bmatrix} \ph{-}0 \\ \ph{-}0 \\ \ph{-}1 \\ \ph{-}0 \end{bmatrix}$  &  % #AAFFAA = | 3>
  $\begin{bmatrix} \ph{-}1 \\      -1 \\      -1 \\ \ph{-}1 \end{bmatrix}$  &  % #FF55FF = | 4>
  $\begin{bmatrix} \ph{-}1 \\ \ph{-}1 \\      -1 \\ \ph{-}1 \end{bmatrix}$  &  % #99FFFF = | 5>
  $\begin{bmatrix} \ph{-}1 \\ \ph{-}1 \\ \ph{-}1 \\      -1 \end{bmatrix}$  &  % #AAAACC = | 6>
  $\begin{bmatrix} \ph{-}0 \\ \ph{-}1 \\ \ph{-}0 \\ \ph{-}0 \end{bmatrix}$  &  % #EE44EE = | 7>
  $\begin{bmatrix} \ph{-}1 \\ \ph{-}1 \\ \ph{-}1 \\ \ph{-}1 \end{bmatrix}$  &  % #EEDDAA = | 8>
  $\begin{bmatrix}      -1 \\ \ph{-}1 \\ \ph{-}1 \\ \ph{-}1 \end{bmatrix}$     % #EEDDFF = | 9>
                 \\[1.5ex] % [1ex] adds vertical space
\hline %inserts single line
\end{tabular}
\caption{Vectors $\ket{\ph{1}1}$ to $\ket{\ph{1}9}$} % title of Table
\label{table:KS-vect-1-9}
\end{table}

\begin{table}[ht]
\centering 
\begin{tabular}{| c  c  c  c  c  c  c  c  c |} 
\hline\hline  
  \tiny{\ph{a}} & & & & & & & & \\ [-2.0ex]
  $\ket{10}$ & $\ket{11}$ & $\ket{12}$ & $\ket{13}$ 
   & $\ket{14}$ & $\ket{15}$ & $\ket{16}$
   & $\ket{17}$ & $\ket{18}$\\ [0.5ex] 
\hline 
  \tiny{\ph{a}} & & & & & & & & \\ [-2.0ex]
  $\begin{bmatrix} \ph{-}1 \\ \ph{-}1 \\ \ph{-}0 \\ \ph{-}0 \end{bmatrix}$  &  % #FFFF88 = |10>
  $\begin{bmatrix} \ph{-}1 \\ \ph{-}0 \\ \ph{-}1 \\ \ph{-}0 \end{bmatrix}$  &  % #DD88DD = |11>
  $\begin{bmatrix} \ph{-}1 \\ \ph{-}0 \\      -1 \\ \ph{-}0 \end{bmatrix}$  &  % #FFDDDD = |12>
  $\begin{bmatrix} \ph{-}1 \\ \ph{-}0 \\ \ph{-}0 \\ \ph{-}1 \end{bmatrix}$  &  % #DDFFDD = |13>
  $\begin{bmatrix} \ph{-}1 \\ \ph{-}0 \\ \ph{-}0 \\      -1 \end{bmatrix}$  &  % #DD99BB = |14>
  $\begin{bmatrix} \ph{-}1 \\      -1 \\ \ph{-}0 \\ \ph{-}0 \end{bmatrix}$  &  % #FF5555 = |15>
  $\begin{bmatrix} \ph{-}0 \\ \ph{-}0 \\ \ph{-}1 \\ \ph{-}1 \end{bmatrix}$  &  % #55FF55 = |16>
  $\begin{bmatrix} \ph{-}0 \\ \ph{-}1 \\ \ph{-}0 \\      -1 \end{bmatrix}$  &  % #FF00FF = |17>
  $\begin{bmatrix} \ph{-}0 \\ \ph{-}1 \\      -1 \\ \ph{-}0 \end{bmatrix}$     % #5555FF = |18>
                 \\[1.5ex] % [1ex] adds vertical space
\hline %inserts single line
\end{tabular}
\caption{Vectors $\ket{10}$ to $\ket{18}$} % title of Table
\label{table:KS-vect-10-18}
\end{table}

These projections are now arranged into nine PVMs, $\Uppi_1$ to $\Uppi_9$, 
shown in Table~\ref{table:PVM-1-9}.
This is done in such a way that each PVM contains four atomic projections.  It is easy
to check that the vectors which generate the projections in each PVM are orthogonal,
so that the projections are themselves orthogonal to each other.  Each PVM generates,
as was already mentioned several times, cf.\ \feD{def:PVM-M}, \feP{prop:Banach-alg}, and
the discussion following \feR{rem:abstract-PVM}, a commutative sub-algebra of 
$\Lop(\CC^4) \cong \MMin{4}(\CC)$, i.e.\ a \emph{classical context}.

For a valuation it was already stated in the explanations following \feD{def:valuation}
that it can only assume values in the spectrum of a RV.  As for any orthogonal
projection $\vP$ one has $\sigma(\vP) = \{0,1\}$, and in a PVM $\sum_j \vP_j = \vI$, 
it is clear that in each sub-algebra
generated by one of the nine PVMs, three projections have to have a value of zero, and one has
to have a value of unity.  Thus, if one replaces in Table~\ref{table:PVM-1-9} the projections
by the value of the valuation, each column (classical context) must have three zeros
and one value of one.  Thus, there should be \emph{nine} ones in the Table~\ref{table:PVM-1-9},
one in each column, an \emph{odd} number.  

\begin{table}[ht]
\centering 
\begin{tabular}{| l  l  l  l  l  l  l  l  l |} 
\hline\hline  
  \tiny{\ph{a}} & & & & & & & & \\ [-2.0ex]
  $\Uppi_1$ & $\Uppi_2$ & $\Uppi_3$ & $\Uppi_4$ & $\Uppi_5$ & $\Uppi_6$ & $\Uppi_7$
   & $\Uppi_8$ & $\Uppi_9$\\ [0.5ex] 
\hline 
  \tiny{\ph{a}} & & & & & & & & \\ [-2.0ex]
  $\vP_{1}$ & $\vP_{1}$ & $\vP_{2}$ & $\vP_{2}$ & $\vP_{3}$ & 
      $\vP_{4}$ &  $\vP_{5}$ & $\vP_{5}$ & $\vP_{6}$\\
  $\vP_{3}$ & $\vP_{7}$ & $\vP_{4}$ & $\vP_{8}$ & $\vP_{7}$ & 
      $\vP_{8}$ &  $\vP_{6}$ & $\vP_{9}$ & $\vP_{9}$\\
  $\vP_{10}$ & $\vP_{11}$ & $\vP_{10}$ & $\vP_{12}$ & $\vP_{13}$ & 
      $\vP_{14}$ &  $\vP_{15}$ & $\vP_{11}$ & $\vP_{13}$\\
  $\vP_{15}$ & $\vP_{12}$ & $\vP_{16}$ & $\vP_{17}$ & $\vP_{14}$ & 
      $\vP_{18}$ &  $\vP_{16}$ & $\vP_{17}$ & $\vP_{18}$
                 \\[1.0ex] % [1ex] adds vertical space
\hline %inserts single line
\end{tabular}
\caption{PVMs 1 to 9 $\ph{\ket{1}}$\label{table:PVM-1-9}} % title of Table
\end{table}

On the other hand, one checks quickly that in Table~\ref{table:PVM-1-9} each projection 
appears \emph{twice}, obviously in different PVMs, i.e.\ classical contexts.  If a 
valuation function exists for $\D{M}_4(\CC)$, the value has to be independent of the 
fact in which classical context (PVM generated Abelian sub-algebra) the projection is 
located.  Thus, there should be an \emph{even} number of ones in  the 
Table~\ref{table:PVM-1-9}, and we have a \emph{contradiction} to the statement in 
the previous paragraph.  We may therefore deduce that no valuation function exists 
on $\D{M}_4(\CC)$, or any algebra containing a sub-algebra isomorphic to $\D{M}_4(\CC)$.

The interpretation is that situations like the one described above with 
$\som_b(\rx) \ne \som_c(\rx)$ can not be avoided if the algebra is sufficiently
large, and thus the RVs can not be given definite values.  This means that the value 
of a RV $\rx \in \Alg_{sa}$ depends on its \emph{context}, i.e.\ in which sub-algebra 
it happens to be embedded.  These facts, from a classical perspective disturbing, 
are on the other hand things to be exploited in quantum information and quantum computing.

\subsection{Generalised Observations and Channels}  \label{SS:POVM}
As a motivation for the following, recall that for a projection $\rqq \in \PA{\Alg}$ representing 
an event, in \feD{def:probability}  it was defined that $\D{P}(\rqq) := \Ex(\rqq)$ was the
probability of sharply observing that event ($\som(\rqq) = 1$).  Assume now that the event $\rqq$
can not be observed exactly or sharply, but that there is a small probability $\alpha$ that the
observation of the opposite event $\rqq^\perp = \rone - \rqq$ is mistaken for
an observation of $\rqq$.  The event of \emph{registering} the observation $\rqq$ is 
then $\re_1 := \alpha \rqq^\perp + (1 - \alpha) \rqq$,
while the opposite event is $\re_0 = \rone - \re_1 = \alpha \rqq + (1-\alpha) \rqq^\perp$.
For $0 < \alpha <1$ these are \emph{not sharp} events.

It is not difficult to see that $\rnul \le \re_1 \le \rone$, so $\re_1, \re_0 \in \EA{\Alg}$ 
are \emph{effects} (cf.\ \feD{def:alg-names}), and this is a simple example of a
\emph{Positive Operator Valued Measure} (POVM). % (on the subsets of $\{0,1\}$ in this case).

\paragraph{Positive Operator Valued Measures (POVMs):}
Generalising this little motivation, one defines a more general kind of measurement
by using effects to describe non-sharp events instead of projections for sharp events:

\begin{defi}[Positive Operator Valued Measure (POVM)]  \label{def:POVM-defi-A}
   Let $\Alg$ be a finite dimensional probability algebra, and let $(X, \F{X})$ be a 
  measurable space with base set $X$ and $\sigma$-algebra $\F{X}$.  
  An assignment $\Uppi_{\text{POVM}}: \F{X} \ni E \mapsto \Uppi_{\text{POVM}}(E) :=
  \re_E \in \EA{\Alg}$, such that
\begin{align}   \label{eq:POVM-defi-1}
   \Uppi_{\text{POVM}}(X) &= \rone, \quad \text{ and for disjoint } \;  E_1, E_2 \in \F{X}:\\
   \Uppi_{\text{POVM}}(E_1 \uplus E_2) &= \Uppi_{\text{POVM}}(E_1) + \Uppi_{\text{POVM}}(E_2),
         \label{eq:POVM-defi-2}
\end{align}
is called a positive operator valued measure (POVM).
This implies $\Uppi_{\text{POVM}}(\emptyset) = \rnul$.
\end{defi}

This is a generalisation of the sharp measurements which were considered before, the projection
valued measures (PVM), which can be seen as a special case of a POVM; one can say that
a PVM or a projection is a \emph{sharp} POVM resp.\ effect.  
Events are now described by the positive operators 
$\rnul \le \Uppi_{\text{POVM}}(E) = \re_E \le \rone$.
The \feD{def:probability} of the probability of \emph{observing an event} by a projection
is now generalised to the probability $0 \le p_E = \Ex(\re_E) \le 1$  of 
\emph{registering} the observation of the event $\re_E$ or outcome $E \in \F{X}$. 

For an observation $\som$, it was established in \feP{prop:proj-1} that 
the only observations or samples of a projection could be
zero or one, $\som(\rp) \in \sigma(\rp) \subseteq \{0,1\}$, which may be
interpreted as false and true.  With a POVM,
the observation can be any probability, $0 \le \som(\re_E) \le 1$.  One possible
interpretation is that this is the probability that the event $\re_E$ resp.\
the outcome $E \in \F{X}$ \emph{has been registered} as having occurred.

The POVM in turn defines a probability distribution on the $\sigma$-algebra $\F{X}$ by 
the assignment 
\begin{equation} \label{eq:prob-from-POVM}
   \prob_{\text{POVM}}: \F{X} \in  E \mapsto \prob_{\text{POVM}}(E) := 
      \prob_{\text{POVM}}(\re_E) := \Ex(\re_E) \in [0,1].
\end{equation}
This probability assignment generalises the one only for projections
previously established in \feD{def:probability}, and may be seen as an
extended Born rule.

As the projections are a special case of effects, $\PA{\Alg} \subset \EA{\Alg}$,
this makes the PVMs defined in \feD{def:PVM-M}, \feqs{eq:PVM-defi-A}{eq:PVM-defi-H}, and
\feR{rem:abstract-PVM} a special case of the abstractly defined POVM here.  It should
also be clear on how to formulate this concretely for the two algebras $\MMn(\CC)$
and $\BHA$ resp.\ $\LHA$ considered up to now in parallel to the
abstract setting.

\begin{rem}[POVM on any $\sigma$-algebra]   \label{rem:POVM-more-X}
For a PVM in a finite dimensional algebra $\Alg$ it was sufficient to consider finite
outcome sets $X$ with $\sigma$-algebra $2^{X}$, cf.\ \feR{rem:PVM-more-X}.  This is
different for a POVM, as the outcome space $X$ and the $\sigma$-algebra $\F{X}$ may be 
infinite.  Obviously, also something resembling $\sigma$-additivity has to be included
as well.  For the sake of simplicity though, and as it does not change anything in the
presentation here, it will simply be assumed in the following that $X$ is a finite
set and that the $\sigma$-algebra is $\F{X} = 2^{X}$ as before for PVMs.

Further note that the set of POVMs forms a convex cone in the vector space of
operator valued measures, and the normalisation requirement \feq{eq:POVM-defi-1}
defines a linear manifold in that vector space.  The normalised POVMs are thus
in the intersection of the positive cone and that linear manifold;
this is very similar to the geometry of states, as established in \feP{prop:states-prop}.

This set is clearly convex --- hence a convex combinations of normalised POVMs is
again a POVM --- and one can show that the \emph{extreme points} are precisely
those POVMs where the outcome space $X$ is finite \citep{ChiribellaEtal2010}. 
This leads to representation theorems analogous to the representation of density
matrices in terms of pure states.

Also in analogy to the situation with states and functionals in terms of the
Radon-Nikodým derivative (cf.\ \feR{rem:RN-positive-cone}), there are similar 
possibilities here \citep{ChiribellaEtal2010}, a topic which is outside the present scope.
\end{rem}

Gleason's \feT{thm:Gleason}, which showed that a function which assigns a probability
distribution for every PVM in $\PA{\LHA}$ can be extended to a state, has
an analogue for POVMs, cf.\ e.g.\ \citep{Busch2003}:

\begin{thm}[Gleason --- POVM]  \label{thm:Gleason-POVM}
 Let $\Hvk$ be a finite dimensional Hilbert space, 
 and consider the algebra $\LHA$
 of all linear maps, as well as the set of all effects $\EA{\LHA}$.
 
A function $\gamma: \EA{\LHA} \to [0,1]$, which is such
that for \emph{any} given POVM 
\[
\Uppi_{\text{POVM}}: 2^X \ni \{ x \} \mapsto E_x \in \EA{\LHA},
\]
defined on a finite set $X$, analogous to \feD{def:POVM-defi-A}
in \feqs{eq:POVM-defi-1}{eq:POVM-defi-2}, it
follows that $\sum_{x} \gamma(\Uppi_{\text{POVM}}(\{ x \})) = 1$ ---
which means that it satisfies the equivalent of \feq{eq:Gleason} and defines a 
probability measure on $X$ and on the POVM
$\Uppi_{\text{POVM}}$, namely $\prob(\{x\}) :=
\prob(E_x) = \gamma(E_x)$ like in \feq{eq:prob-from-POVM} --- 
can be \emph{extended} to a state $\gamma \in \SA{\LHA}$.

The converse, that any state induces a probability measure on a POVM, 
was already stated in \feq{eq:prob-from-POVM}.
\end{thm}

\begin{rem}   \label{rem:POVM-vs-PVM}
Generalising \feR{rem:PVM-vs-pure_state}, one may observe that we started in 
\feR{rem:spec-dec-PVM} by tying a sharp observation to an atomic projection,
then generalised this to sharp observations connected to a PVM, and now one
may speak about a \emph{non-sharp} event represented by an \emph{effect} $\re \in \EA{\Alg}$, 
having a probability $\Ex(\re)$ to be \emph{registered} as being true.
 
In case that $\som \in \SA{\Alg}$ is an actual observation or sample, 
i.e.\ a character $\XA{\Clg}$
of some commutative sub-probability algebra $\Clg \subseteq \Alg$ such that 
$\Uppi_{\text{POVM}}(\{x\}) \in \Clg$, then $\prob_{\som}(\re_x) := \som(\re_x)
= \ip{\som}{\re_x} = \re_x(\som) \in [0,1]$, not just in $\{0,1\}$ as for a PVM, 
can be seen as the probability that the event $\re_x$ (i.e.\ $x \in X$) 
\emph{has been registered as being observed}, 
whereas $\prob_{\text{POVM}}(\re_x) := \Ex(\re_x)$ may be 
interpreted as the probability that the event $\re_x$ (i.e.\ $x \in X$) 
\emph{will be observed}.
\end{rem}

\paragraph{Duality of Effects and States:}
As could be gleaned from the foregoing, in a probability algebra $\Alg$ we started 
from the duality between observables $\Alg_{sa} \subset \Alg$ and states $\SA{\Alg}
\subset \Alg^\star$, so that for an observable $\rx \in \Alg_{sa}$ and a state $\svpi
\in \SA{\Alg}$ the expectation of the observable in the state $\svpi$ was 
$\Ex_{\svpi}(\rx) = \svpi(\rx) = \ip{\svpi}{\rx}$.  Here the last expression is 
the duality pairing $\ip{\cdot}{\cdot}$ on $\Alg^\star \times \Alg$.
In \feX{ex:complex-Hilbert} the trace on $\LHA$ was defined for a finite
dimensional space $\Hvk$, and via the regular representation in $\Lop(\Alg)$
the canonical faithful inner product $\bkt{\cdot}{\cdot}_c$ on $\Alg^2$ was 
introduced in \feP{prop:fin-dim-alg-gen}.  Following the development for density
matrices \feP{prop:set-dens-M} and \feC{coro:dens-in-alg-H}, in \feC{coro:dens-in-alg-A}
density elements $\DEN{\Alg} \subset \Alg_+$ were introduced, so that for each
state $\svpi \in \SA{\Alg}$ there is a one-to-one correspondence
with an density element $\rr_{\svpi} \in \DEN{\Alg}$,
such that one may write with the canonical inner product
\begin{equation}   \label{eq:exp-duality-state-dens_mat-vs-obs}
  \Ex_{\svpi}(\rx) = \svpi(\rx) = \ip{\svpi}{\rx} = \bkt{\rr_{\svpi}}{\rx}_c \, \text{ on }
    \SA{\Alg} \times \Alg_{sa} \; \text{ resp. } \; \DEN{\Alg} \times \Alg_{sa}.
\end{equation}

When computing probabilities (cf.\ \feD{def:probability}), originally $\rx \in \PA{\Alg}$
had to be a projection.  Generalising this to PVMs, the \feq{eq:exp-duality-state-dens_mat-vs-obs}
then displays a duality between state resp.\ densities and PVMs to produce probabilities:
\begin{equation}   \label{eq:pr-duality-state-dens_mat-vs-PVM}
  \prob_{\svpi}(\rx) = \ip{\svpi}{\rx} = \bkt{\rr_{\svpi}}{\rx}_c \, \text{ on }
    \SA{\Alg} \times \PA{\Alg} \; \text{ resp. } \; \DEN{\Alg} \times \PA{\Alg}.
\end{equation}
Above, this was just further generalised to POVMs, where observations or samples
$\som \in \XA{\CC[\rx]}$ don't produce
sharp results like $1$ or $0$, but only a \emph{probability} that an event \emph{has been observed}.
So with the correspondence $\svpi \leftrightarrow \rr_{\svpi}$ between states and densities ---
as well as $\som \leftrightarrow \rr_{\som}$ --- one can observe the dualities
\begin{align}   \label{eq:pr-duality-state-dens_mat-vs-POVM}
  \prob_{\svpi}(\rx) &= \ip{\svpi}{\rx} = \svpi(\rx) = \bkt{\rr_{\svpi}}{\rx}_c \, \text{ on }
    \SA{\Alg} \times \EA{\Alg} \; \text{ resp. } \; \DEN{\Alg} \times \EA{\Alg}; \\
     \label{eq:obs-duality-state-dens_mat-vs-effect}
  \prob_{\som}(\rx) &= \ip{\som}{\rx} = \rx(\som) = \bkt{\rr_{\som}}{\rx}_c \, \text{ on }
    \XA{\CC[\rx]} \times \EA{\CC[\rx]} . %\; \text{ resp. } \\ 
%       \nonumber & \phantom{\prob_{\som}(\rx) = \ip{\som}{\rx}  = \rx(\som) 12345678} 
%            \DEN{\CC[\rx]}_1 \times \EA{\CC[\rx]} .
\end{align}

Just looking at effects seems to be sufficient to discuss probabilities prior to a sample
or observation in \feq{eq:pr-duality-state-dens_mat-vs-POVM}, as well as the probability
that something has been observed in \feq{eq:obs-duality-state-dens_mat-vs-effect}.
One might note that in the set of effects $\EA{\Alg}$, it is the partial order and
positivity which carries the main information (the structure of an ordered vector space), 
preserving the range and ordering of probabilities, 
a topic which will be taken up again when channels are shortly discussed.
This structure of the set of effects has been abstracted and generalised to 
\emph{effect algebras} --- e.g.\ cf.\ \citep{gudder1999, gudderEtal999, BarnumWilce2011}
and the references therein --- to describe a general kind of observable.
This in turn is the basis for theories of \emph{generalised probability}
(e.g.\ cf.\ \citep{JanottaHin2014, HolikEtal2021, Plavala2021}), which start from the
duality structure displayed in 
\feqs{eq:pr-duality-state-dens_mat-vs-POVM}{eq:obs-duality-state-dens_mat-vs-effect}.

This is that the effects $\rx$ are in a partially ordered set, usually a subset of an ordered
vector space $\C{E}$, and that they are in the ``interval'' $\rnul \le \rx \le \rone$, where
$\rnul \in \C{E}$ is the \emph{impossible event} and the origin of the vector space, and the 
``order unit'' $\rone \in \C{E}$ is the \emph{certain event}, and the above interval is a 
convex set, so that convex combinations of effects are again an effect.  This means in 
particular that effects are \emph{positive} $\rx \in \C{E}_+$.  Additions of effects are 
allowed as long as they stay below the order unit, i.e.\  the sum of $\rx_1$ and $\rx_2$ 
is an effect iff $\rx_1 + \rx_2 \le \rone$.  
The states $\svpi$ on the other hand live in the dual space $\C{E}^\star$ of
the effects, and the duality (a bilinear form) $p = \ip{\svpi}{\rx}$ is to be interpreted 
as the probability $0 \le p \le 1$ of the effect being observed in the state $\svpi \in \C{E}^\star$.  
Therefore the states have to be positive and normalised, i.e.\ $\ip{\svpi}{\rx} \ge 0$
(geometrically a cone in  $\C{E}^\star$) and  $\ip{\svpi}{\rone} = 1$ (geometrically a 
hyperplane in $\C{E}^\star$), so that the probabilities satisfy 
$0 \le p = \ip{\svpi}{\rx} \le 1$, and the set of states is convex, just as for effects.  
These few prescription are sufficient to be able to assign probabilities.  
Here, however, we shall not go further in this direction and stay within
the setting established so far.
%The states also form a convex set in the dual space, so that
%with $0 \le \alpha \le 1$ one has for any effect $\rx$
%\begin{multline*}
%   0 \le \ip{\alpha\svpi_1 + (1-\alpha) \svpi_2}{\rx} =
%     \alpha \ip{\svpi_1}{\rx} + (1-\alpha) \ip{\svpi_2}{\rx} \\ \le
%     \alpha \ip{\svpi_1}{\rone} + (1-\alpha) \ip{\svpi_2}{\rone} = p_1 + p_2
%    \le 1 .
%\end{multline*}
%Similarly for two effects $\rx_1, \rx_2$ and any state $\svpi$ it holds that
%\begin{multline*}
%   0 \le \ip{\svpi}{\alpha\rx_1  + (1-\alpha) \rx_2} =
%     \alpha \ip{\svpi}{\rx_1} + (1-\alpha) \ip{\svpi}{\rx_2} \\ \le
%     \alpha \ip{\svpi}{\rone} + (1-\alpha) \ip{\svpi}{\rone} = 1 .
%\end{multline*}

\paragraph{Positive Operator Valued Measures (POVMs) and Positive Maps:}
Although the idea of a POVM is a generalisation of the PVM, where the elements are projections
and not just effects, it turns out that these two concepts are very closely related.  This is
the content of the following ``Neumark(Naĭmark) Dilation Theorem'', another example of the 
``Church of the Larger Hilbert Space'', and it is usually formulated in the Hilbert space setting,
cf.\ e.g. \citep{Paulsen2003, Holevo2012}.  

\begin{thm}[Neumark(Naĭmark)]  \label{thm:Neumark-dilation}
Let $\Hvk$ be a finite dimensional Hilbert space, and
$\Uppi_{\Hvk}: 2^X \to \EA{\LHA}$ be a POVM as defined in \feD{def:POVM-defi-A}.
%with the cardinality of $\ns{X} = m$.

Then there exists another finite dimensional Hilbert space $\Kvk$, 
%with $\dim \Hvk \le \dim \Kvk \le m \, \dim \Hvk$,
an isometry $V \in \Lop(\Hvk, \Kvk)$ ($V^\dagger V = I_{\Hvk}$), and \emph{a PVM}
$\Uppsi_{\Kvk}:  2^X \to \PA{\Lop(\Kvk)}$ as defined for matrix algebras in
\feD{def:PVM-M}, formulated in \feP{prop:Banach-alg} (cf.\  \feq{eq:PVM-defi-A}) in a purely
algebraic setting, and spelt out for Hilbert spaces in \feR{rem:spec-dec-PVM}
(cf.\  \feq{eq:PVM-defi-H}), such that
\begin{equation}  \label{eq:Neumark-dilation}
  \forall E \in 2^X: \quad \Uppi_{\Hvk}(E) = V^\dagger \, \Uppsi_{\Kvk}(E) \, V .
\end{equation}
\end{thm}

What the \feq{eq:Neumark-dilation} says is that a POVM, when \emph{dilated} on a 
larger space resp.\ in a larger algebra, is really a PVM, or, in other words, 
that a POVM is the \emph{compression} of a PVM.  
Another way of saying this is that the POVM has been ``lifted'' to a PVM.
If the POVM value for $\{ x \} \subset X$ is the \emph{effect} 
$\Uppi_{\Hvk}(\{ x \}) = E_x \in \EA{\LHA}$, then there is a \emph{projection}
$\Uppsi_{\Kvk}(\{ x \}) = P_x \in \PA{\Lop(\Kvk)}$ such that $E_x = V^\dagger \, P_x \, V$.
One may also say that there is a system ($\Lop(\Kvk)$) with a sharp observable $P_x$,
but one sees only part of it, namely $V^\dagger(\Lop(\Kvk)) = \LHA$.

The form of the representation $E_x = V^\dagger \, P_x \, V$ is called the \emph{Kraus} form
\citep{Werner2001}, and the map $V$ --- which doesn't have to be necessarily an isometry --- is
called a Kraus map.  Observe that this can also be written as
\begin{equation}  \label{eq:Kraus-form}
  E_x = V^\dagger \, P_x \, V = (P_x \, V)^\dagger (P_x \, V) = K_x^\dagger K_x ,
\end{equation} 
with Kraus operators $K_x = P_x \, V \in \Lop(\Hvk, \Kvk)$, satisfying 
$\sum_{x \in X} K_x^\dagger K_x = \sum_{x \in X} E_x = I_{\Hvk}$. 

%A corollary to Neumark’s dilation theorem 
%provides a statistical interpretation of an 
%arbitrary resolution of the identity and 
%establishes that the generalised definition 
%of an observable as an effect is just an extension of the standard one as a projection.

%\begin{coro}[Kraus Quantum Randomisation]   \label{coro:Kraus-randomisation}
%\citep{Holevo2012} Let $E \in \EA{\LHA}$ be an effect.
%Then there exist a Hilbert space $\C{J}$, a unit vector $\psi \in \C{J}$, 
%and a projection (sharp observable) $P \in \PA{\Lop(\Hvk \otimes \C{J})}$, such that
%$E = \tr_{\C{J}} \bigl(I_{\Hvk} \otimes \ket{\psi}\bra{\psi} P \bigr)$. 
%\end{coro}
%

A POVM may be seen as a mechanism to transfer probabilistic information from a
general probability algebra to a probability distribution on a set $X$ as
in \feq{eq:prob-from-POVM}.  Such a transfer of information is called a
\emph{channel}.  To address channels more generally, we start with a 
motivating example connected to the preceding \feT{thm:Neumark-dilation}.

\begin{xmpn}[Observation Channel]  \label{ex:obs-channel}
Let $X$ be a finite set, and $n = \ns{X}$ its cardinality.  As $\CC^X \cong \Cn$, 
consider $\Clg = \Cn$, the commutative Hadamard algebra of \feX{ex:complex-n-spc}.
For a given POVM $\Uppi_{\Hvk}: 2^X \to \EA{\LHA}$  as in the above 
\feT{thm:Neumark-dilation} and $\vf \in \Cn$, $\Uppi_{\Hvk}(\{ i \}) = E_i$, define a map 
\[
\E{F}: \Clg \to \LHA \;\text{ by }\; 
     \vf = (f_1,\dots,f_n) \mapsto \sum_{i=1}^n f_i \, E_i.
\]
This is a linear map $\E{F}\in  \Lop(\Clg,\LHA)$, which maps 
\begin{compactitem}
 \item the identity $ \ve=(1,\dots,1) \mapsto \E{F}(\ve) = \sum_i E_i = I_{\Hvk}$ 
       onto the identity;
 \item adjoint elements $\vw^\star$ to adjoints $\E{F}(\vw^\star) = \sum_i w_i^* \,E_i=
      \bigl(\sum_i w_i\,E_i \bigr)^\dagger = \bigl(\E{F}(\vw)\bigr)^\dagger$,
      i.e.\ it is *-linear;
 \item positive $\vek{0} \le \vf \in \Clg_+$ to positive 
        $0 \le \E{F}(\vf) = \sum_i f_i\,E_i \in \LHA_+$;
 \item effects $\vek{0} \le \vg \le \ve$ to effects 
       $0 \le \E{F}(\vg) = \sum_i g_i\,E_i \le I_{\Hvk}$;
 \item and can be written as Kraus form: As $E_i \in \EA{\LHA} \subset \LHA_+$,
       it can by definition be written as $E_i = G_i^\dagger G_i$ with some 
       $G_i \in \LHA$ (e.g.\ $G_i = E_i^{\frk{1}{2}}$), and thus
       $\E{F}(\vf) = \sum_i f_i\,E_i = \sum_i G_i^\dagger \,f_i \, G_i$.
\end{compactitem}
This map transforms effects (possible measurements) in $\Clg$ to effects in $\LHA$,
where one can evaluate their probability with some state $\svpi \in \SA{\LHA}$: 
for $\vf \in \EA{\Clg}$ one has $\prob(\vf) = \svpi(\E{F}(\vf)) = \ip{\svpi}{\E{F}(\vf)}$.
In particular, choosing $\ve_i = (\updelta_{i j})$, the canonical unit vectors in $\Cn$, 
one obtains $p_i = \prob(\ve_i)$, a probability distribution on $X = \{1,\dots,n\}$.

Thus, defining the dual map $\E{F}^* \in \Lop(\LHA^*,\Clg^\star)$ as usual for
$\svpi_{\LHA} \in \LHA^*$ via
\[
  \ip{\E{F}^*(\svpi_{\LHA})}{\vg} = \ip{\svpi_{\LHA}}{\E{F}(\vg)} =
      \svpi_{\LHA}(\E{F}(\vg)) = p_{\vg},
\]
one has transferred the state $\svpi_{\LHA} \in \SA{\LHA}$ to a state
$\E{F}^*(\svpi_{\LHA}) =: \sal_{\Hvk} \in \SA{\Clg}$ which assigns the probability
$0 \le p_{\vg} = \sal_{\Hvk}(\vg) \le 1$ to the event / effect $\vg \in \EA{\Clg}$.
This is due to the fact that $\E{F}^*$ is also *-linear, positive, and state preserving:
\[ 
  1 = \ip{\svpi_{\LHA}}{I_{\Hvk}} =  \ip{\svpi_{\LHA}}{\E{F}(\ve)} =
 \ip{\E{F}^*(\svpi_{\LHA})}{\ve} = \ip{\sal_{\Hvk}}{\ve}.
\]

In case the state $\svpi_{\LHA}$ is represented by a density operator 
$R \in \DEN{\LHA}$ such that $\svpi_{\LHA}(A) = \tr R A$, one can define the
adjoint map $\E{F}^\dagger: \LHA \to \Clg$  by use of the Hilbert space structure 
on $\Clg = \Cn$ and $\LHA$, such that for $X \in \LHA$ and $\vx \in \Cn$:
\begin{multline*}    
  \bkt{X}{\E{F}(\vx)}_{c} =  \tr \bigl(X^\dagger \,\E{F}(\vx) \bigr) = 
    \sum_{i=1}^n \tr (X^\dagger E_i) x_i = \sum_{i=1}^n \tr (E^\dagger_i X)^* x_i \\
   =  \sum_{i=1}^n  \tr (G_i^\dagger G_i X )^* x_i 
   =  \sum_{i=1}^n  \tr ( G_i X G_i^\dagger)^* x_i = \bigl(\E{F}^\dagger(R)\bigr)^\tpH \vx
   = \bkt{\E{F}^\dagger(X)}{\vx}_{\Cn} .
\end{multline*}
Hence $\E{F}^\dagger(X) = [\tr ( G_1 X G_1^\dagger), \dots, \tr ( G_n X G_n^\dagger)]^\trpos
= [\tr (X E_1), \dots, \tr (X E_n)]^\trpos$.
It is easily seen that this maps density operators onto discrete probability distributions 
$\E{F}^\dagger(R) = \vrh = [\tr (R E_1), \dots, \tr (R E_n)]^\trpos = \sum_{i=1}^n 
\tr (R E_i) \ve_i \in \DEN{\Clg}$, and
$1= \tr R = \tr (R\, I) = \tr \bigl(R \,\E{F}(\ve) \bigr) = \bigl(\E{F}^\dagger(R)\bigr)^\tpH \ve
= \vrh^\tpH \ve$, i.e.\ it preserves the trace.
\end{xmpn}

To recap, the map $\E{F}: \Clg \to \LHA$, its dual $\E{F}^*$ and its 
adjoint $\E{F}^\dagger$, are *-linear, positive, and satisfy
\begin{compactitem}
 \item $\E{F}(\EA{\Clg}) \subseteq  \EA{\LHA}$, and $\E{F}(\ve) = I_{ \LHA}$
       ($\E{F}$ is unit preserving) (unital);
 \item $\E{F}^*(\SA{\LHA}) \subseteq  \SA{\Clg}$, i.e.\ $\E{F}^*$ is state preserving;
 \item $\E{F}^\dagger(\DEN{\LHA})\subseteq\DEN{\Clg}$, and $\E{F}^\dagger$ is 
        trace preserving (TP).
\end{compactitem}
Preserving the unit for $\E{F}$, and the state resp.\ trace preservation for
$\E{F}^*$ resp.\ $\E{F}^\dagger$, are two sides of the same coin.
These properties are apparently sufficient to allow assignments of probabilities
to the effects $\EA{\Clg}$ in $\Clg$, as just discussed above in the part on
the ``Duality of Effects and States''.  It is a transfer of probabilistic information,
as embodied in the state, to another system.

So, instead of performing observations on the system given by $\LHA$, one now 
observes on the system defined by $\Clg$.  The properties of *-linearity, positivity, 
and trace preservation were sufficient to achieve this.  
In \feD{def:star_homomorphism} in \fsec{S:formal} we
saw that *-homomorphisms could be used for this, cf.\ \feC{coro:CPUP-maps}.  
But the requirements here are weaker, the map does not have to be multiplicative, 
it only has to preserve positivity and the identity.  Incidentally, in the extreme
case where the POVM $\Uppi_{\Hvk}: 2^X \to \EA{\LHA}$ 
above is actually a PVM, the map $\E{F}$ \emph{does} become a \emph{*-homomorphism}.

Such a map as just described for the general $\E{F}$ is called a \emph{channel}. 
It transfers information from a system described by $\LHA$ to one 
modelled on the Hadamard algebra $\Clg = \Cn$, a classical commutative or Abelian
probability algebra.  The map $\E{F}$ is the channel in the \emph{Heisenberg} picture 
--- observe that it somehow points backwards --- and its dual $\E{F}^*$ and adjoint
$\E{F}^\dagger$, pointing forward in the direction of information transmittal, which
transfer states from $\LHA$ to $\Clg$, are the channel in the 
\emph{Schrödinger} picture.

Now to the formal definitions which are needed for such a channel: In the following, let
 $\Alg, \Blg, \Clg, \C{D}$ be finite dimensional unital *-algebras.

%\begin{defi}[Dual and Adjoint]   \label{def:dual-map}
%  Let   $\E{P} \in \Lop(\Alg,\Blg)$ be a *-linear map.  Its \emph{dual} map 
%  $\E{P}^*: \Blg^\star \to \Alg^\star$
%  is defined as usual for all $\ra \in \Alg, \rb \in \Blg$ and $\svpi \in \Blg^\star$ by
%\[ 
%   \ip{\svpi}{\E{P}(\ra)}_{(\Blg^\star \times \Blg)} = 
%       \ip{\E{P}^*(\svpi)}{\ra}_{(\Alg^\star \times \Alg)}.
%\]
%
%Its \emph{adjoint} map $\E{P}^\dagger \in \Lop(\Blg,\Alg)$ is defined, using the Hilbert
%space structure on $\Alg$ and $\Blg$ generated by the trace, by
%\[ 
%   \bkt{\rb}{\E{P}(\ra)}_{\Blg} =
%   \tr \bigl( \rb^\star \, \E{P}(\ra) \bigr) = 
%   \tr \bigl( (\E{P}^\dagger(\rb) )^\star \, \ra \bigr) = 
%   \bkt{\E{P}^\dagger(\rb)}{\ra}_{\Alg}.
%\]
%\end{defi}

\begin{defi}[Positive Map]   \label{def:positive-map}
  A *-linear map  $\E{P} \in \Lop(\Alg,\Blg)$ is \emph{positive}, \emph{iff} for all 
  positive $\ra \in \Alg_+$ one has that $\E{P}(\ra) \in \Blg_+$ is again positive,
  i.e.\ $\ra \ge \rnul_{\Alg} \; \Rightarrow \; \E{P}(\ra) \ge \rnul_{\Blg}$.
\end{defi}

\begin{defi}[Unit and Trace Preservation]   \label{def:trace-preserve}
   A *-linear map  $\E{P} \in \Lop(\Alg,\Blg)$ is \emph{unit preserving} (UP), or 
   \emph{unital}, or \emph{stochastic},  \emph{iff} $\E{P}(\rone_{\Alg}) = \rone_{\Blg}$.

  A *-linear map $\E{P}^\dagger \in \Lop(\Blg,\Alg)$ is \emph{trace preserving} (TP), 
  or \emph{doubly stochastic}, \emph{iff}   for all $\rb \in \Blg$ one has 
  $\tr_{\Blg} \rb = \tr_{\Alg} \bigl(\E{P}^\dagger(\rb)\bigr)$.
%  Sometimes this adjective is also attached to the underlying map $\E{P}$.
\end{defi}

From this one can show already in general some of the properties which were seen
in \feX{ex:obs-channel}.

\begin{prop}   \label{prop:pos-unit-trace}
Let $\E{P} \in \Lop(\Alg,\Blg)$ be a *-linear map.  
Then the following two statements are equivalent:
\begin{itemize}
\item The Heisenberg picture: $\E{P}$ is positive and unit preserving (unital);
%      \begin{compactitem}
%        \item $\E{P}$ is positive;
%        \item $\E{P}$ is unit preserving;
%      \end{compactitem}
      
\item The Schrödinger picture: $\E{P}^\dagger$ is positive and trace preserving.
%      \begin{compactitem}
%        \item $\E{P}^\dagger$ is positive;
%        \item $\E{P}^\dagger$ is trace preserving;
%      \end{compactitem}
\end{itemize}
Additionally, $\E{P}$ positive implies that it is monotone, 
i.e.\ if $\ra_1 \le \ra_2$ then $\E{P}(\ra_1) \le \E{P}(\ra_2)$;
e.g., cf.\ \citep{Holevo2012}.
\end{prop}

\paragraph{Completely Positive Maps:}
Given two *-linear positive maps $\E{P}: \Alg \to \Clg$ and $\E{Q}: \Blg \to \C{D}$, 
what one would like is to consider the
combined system $\E{P} \otimes \E{Q}: (\Alg \otimes \Blg) \to 
(\Clg \otimes \C{D})$, cf.\ \feD{def:tensor-product}.   It would be desirable for
the map $\E{P} \otimes \E{Q}$  to be positive if the component maps $\E{P}$ and $\E{Q}$ 
are.  For non-commutative algebras this is a non-trivial requirement, and, as counterexamples
show, this is unfortunately \emph{not always} the case;
e.g., cf.\ \citep{Davidson1996, KadiRingr2-97, Takesaki1, Werner2001, NielsenChuang2011, 
Holevo2012, Wolf2012, BenyRicht15, Wilde2017, Quillen2025}.
More than just positivity is required.
Luckily, it is sufficient to consider $\Blg = \C{D} = \MMn(\CC)$ and $\E{Q} = \vI_n$
(the identity in $\MMn(\CC)$) to check whether this holds in general.  
This is tensoring or ``interaction'' with a ``do-nothing'' ancilla system.

\begin{defi}[Completely Positive Maps]  \label{def:CP-maps}
A *-linear map $\E{P}: \Alg \to \Blg$ induces for every $n \in \D{N}$ *-linear 
maps $\E{P}_n :\Alg \otimes \MMn(\CC) \to \Blg \otimes \MMn(\CC)$ 
(cf.\ \feX{ex:tensor-mat}) by setting $\E{P}_n = \E{P} \otimes \vI_n$.

A positive *-linear map $\E{P}: \Alg \to \Blg$ is \emph{$n$-positive} iff 
$\E{P}_n$ is positive.  It is \emph{completely positive} (CP) iff it is 
$n$-positive for all $n \in \D{N}$.
\end{defi}

\emph{Completely Positive Trace Preserving} *-linear maps are often abbreviated 
as \emph{CPTP} maps; their adjoints, which are \emph{Completely Positive Unit Preserving},
are  often abbreviated as \emph{CPUP} maps.  They satisfy all the requirements for 
a successful information transmittal in a channel.
A moment's thought shows that (e.g.\ cf.\ \citep{Wolf2012})

\begin{coro}[Geometry of CP Maps and Compositions]   \label{coro:CP-composition}
The set of CP maps from a *-algebra $\Alg$ to another *-algebra $\Blg$, as a subset of
$\Lop(\Alg, \Blg)$, is a pointed convex cone.  Hence, convex combinations of CP 
maps are again CP maps.  The properties of \emph{unit preservation} (UP) and 
\emph{trace preservation} (TP) define linear manifolds in $\Lop(\Alg, \Blg)$, and hence 
are also preserved under convex combinations.

Let $\E{P}: \Alg \to \Blg$ and $\E{Q}: \Blg \to \Clg$ be two *-linear completely 
positive maps.  Then $\E{R} = \E{Q} \circ \E{P}$ is a  *-linear completely positive map.
If both $\E{P}$ and $\E{Q}$ are UP resp.\ TP, so is $\E{Q} \circ \E{P}$.
\end{coro}

Some basic results are given by

\begin{coro}[Some CP Maps]   \label{coro:CPUP-maps}
  A non-zero *-homomorphism (cf.\ \feD{def:star_homomorphism})  $\Phi: \Alg \to \Blg$ 
  is \emph{completely positive} and \emph{unit preserving}, cf.\ \citep{Davidson1996}.
  Thus every sample $\som \in \XA{\Alg}$, $\som: \Alg \to \CC$, is \emph{completely positive}
   and \emph{unit preserving}, cf.\ \feD{def:sample}.

  A *-linear positive map  $\E{P}: \Alg \to \Blg$, where $\Alg$ or $\Blg$ is Abelian or
  commutative, is \emph{completely positive}, cf.\ \citep{KadiRingr2-97, Wolf2012}.
  Thus every state $\svpi \in \SA{\Alg}$, $\svpi: \Alg \to \CC$, is \emph{completely positive}
  and \emph{unit preserving}, cf.\ \feD{def:state}.
\end{coro}

In the following, a few properties of CP-maps are collected \citep{Davidson1996, 
KadiRingr2-97, Takesaki1, Werner2001, NielsenChuang2011, 
Holevo2012, Wolf2012, BenyRicht15, Wilde2017, Quillen2025, Attal-notes}.

\begin{coro}[Cauchy-Schwarz for CPUP-maps]   \label{coro:Cauchy-Schwarz-CPUP}
  Let $\Alg$ be a finite dimensional unital *-algebra, and
  let $\E{T}^\dagger \in \Lop(\Alg)$ be a *-linear CPUP-map.  
  Then it holds that
\begin{equation}  \label{eq:Cauchy-Schwarz-CPUP}
  \forall \ra \in \Alg: \quad  \E{T}^\dagger(\ra^\star) \E{T}^\dagger(\ra) \le 
     \E{T}^\dagger(\ra^\star \ra) .
\end{equation}
e.g., cf.\ \citep{Wolf2012, Attal-notes}.  Observe that $\E{T}^\dagger$ is a *-homomorphism
\emph{iff} equality always holds in \feq{eq:Cauchy-Schwarz-CPUP}.

In case that there is a CP-map $\E{R} \in \Lop(\Alg)$ such that 
$\E{R} \circ \E{T} = \E{I}_{\Lop(\Alg)}$ (the identity on $ \Lop(\Alg)$),
then $\E{T}$ is an internal *-automorphism (cf.\ \feD{def:alg-names}).
\end{coro}

Note that these results show that the map $\E{F}$ in the above motivating \feX{ex:obs-channel}
is completely positive and unit preserving, and this is also shown by its Kraus form.

\begin{prop}[Kraus Form]    \label{prop:Kraus-props}
  Let $\Hvk$ and $\Kvk$ be finite dimensional Hilbert spaces, and let 
  $K \in \Lop(\Kvk,\Hvk)$ be a linear map.    Then the *-linear map 
  $\E{T}: \LHA \to \Lop(\Kvk)$, given in Kraus form by
\begin{equation}  \label{eq:Kraus-props}
   \E{T}: \LHA \ni A \mapsto \E{T}(A) = K^\dagger A K \in \Lop(\Kvk)
\end{equation}
is \emph{completely positive}, 
cf.\ \citep{Davidson1996, KadiRingr2-97, Takesaki1, Werner2001, Holevo2012, Wolf2012}.  
It is \emph{unit preserving} \emph{iff} $K^\dagger K = I_{\Kvk}$ ($K$ is an isometry).
It is \emph{trace preserving} \emph{iff} $K K^\dagger = I_{\Hvk}$.  
The map $K \in \Lop(\Kvk,\Hvk)$ is often called a \emph{Kraus map} or
\emph{Kraus operator}.

Note that when both $K^\dagger K = I_{\Kvk}$ and $K K^\dagger= I_{\Hvk}$
hold, then $K \in \Lop(\Kvk,\Hvk)$ is unitary, the Hilbert spaces
are isomorphic ($\Kvk \cong \Hvk)$,  and $\E{T}$ is a *-isomorphism of the
unital *-algebras $\LHA$ and $\Lop(\Kvk)$, cf.\ \citep{Attal-notes}.
%Conversely, any such *-isomorphism is of the form \feq{eq:Kraus-props} with
%a unitary $K \in \Lop(\Kvk,\Hvk)$.
\end{prop}

Together with \feC{coro:CP-composition}, one has that the convex combination
of unit preserving Kraus maps is completely positive and unit preserving.
This leads to

\begin{coro}[General Kraus Form]    \label{coro:gen-Kraus-form}
Let $\Hvk$ and $\Kvk$ be finite dimensional Hilbert spaces, and let
$X \supseteq \{x \} \mapsto K_x \in \Lop(\Kvk,\Hvk)$ be a collection of linear
maps ($X$ a finite set), such that 
\begin{equation}   \label{eq:gen-Kraus-factor}
  X \supseteq \{x \} \mapsto
  \Uppi_{K,X}(\{ x \}) =  F_x  = K_x^\dagger K_x \in \EA{\Lop(\Kvk)} 
\end{equation}
%where $F_x \in \EA{\Lop(\Kvk)}$
is a POVM (in particular $\sum_x F_x = I_{\Kvk}$).  Then
\begin{equation}   \label{eq:gen-Kraus-form}
   \E{T}_X:  \LHA \ni A \mapsto \E{T}_X(A) = 
      \sum_{x \in X}  K_x^\dagger A K_x \in \Lop(\Kvk)
\end{equation}
is a CPUP-map in \emph{Kraus form}.  Its adjoint, a CPTP-map, is given in Kraus form by
\begin{equation}   \label{eq:gen-Kraus-form-adj}
   \E{T}^\dagger_X:  \Lop(\Kvk) \ni B \mapsto \E{T}^\dagger_X(B) = 
      \sum_{x \in X}  K_x B K^\dagger_x \in \LHA .
\end{equation}
\end{coro}
Observe that the effects $F_x \in \EA{\Lop(\Kvk)}$ in a POVM 
(cf.\ \feq{eq:gen-Kraus-factor}) could be factored in 
many ways, given that the Kraus operators $K_x \in \Lop(\Kvk,\Hvk)$ are in no way unique.  
If $U_x \in \LHA$ are unitary, then the $Q_x = U_x K_x$ are also Kraus maps
with $F_x = Q_x^\dagger Q_x = K_x^\dagger U_x^\dagger U_x K_x = K_x^\dagger K_x$.
When $\Kvk = \Hvk$, one easy choice is $K_x = F_x^{\frk{1}{2}}$, as 
$F_x \in \EA{\Lop(\Kvk)} \subset \Lop(\Kvk)_+$.

When bases are chosen in the above Hilbert spaces to represent the algebras as 
sub-algebras of matrix algebras, as was discussed in \fsec{SS:states}, the criterion 
for complete positivity becomes checking the positivity of a matrix:

\begin{thm}[Choi]   \label{thm:Choi}
  Let $\tQ: \MMn(\CC) \to \D{M}_m(\CC)$ be a linear map, and let 
  $\vE_{ij} \in \MMn(\CC)$ be the matrix with all zero entries, except a $1$
  at position $(i,j)$ --- the canonical unit vectors in $\CC^{n \times n} \cong
  \MMn(\CC)$.  Then $\tQ$ is completely positive iff the \emph{Choi-matrix} 
\begin{equation}  \label{eq:Choi-matrix}
  \vC_{\tQ} = (\tI_{n \times n} \otimes \tQ)(\sum_{i,j = 1}^n \vE_{ij} \otimes \vE_{ij}) =
     \sum_{i,j = 1}^n \vE_{ij} \otimes \tQ(\vE_{ij}) \in \D{M}_{n m}(\CC)
\end{equation}
is positive, cf.\ \citep{Choi1975}.
\end{thm}

Now finally the result on completely positive maps and tensor products can be stated:

\begin{prop}[Tensor Products and CP Maps]   \label{prop:CP-tensor}
Let $\E{P}: \Alg \to \Clg$ and $\E{Q}: \Blg \to \C{D}$ be two *-linear completely positive maps.
Then the combined system map 
\[
  \E{P} \otimes \E{Q}: (\Alg \otimes \Blg) \to (\Clg \otimes \C{D})
\]
is *-linear and completely positive.  If both $\E{P}$ and $\E{Q}$ are unit preserving,
so is $\E{P} \otimes \E{Q}$.

In case $\Clg = \C{D}$ is an Abelian or commutative *-algebra, then the map 
$\E{R}: (\Alg \otimes \Blg) \to \Clg$ defined on elementary tensors 
$\ra \otimes \rb \in \Alg \otimes \Blg$ by 
\[
\E{R}(\ra \otimes \rb) := \E{P}(\ra) \E{Q}(\rb)
\]
is completely positive, and unit preserving in case both $\E{P}$ and $\E{Q}$ are;
e.g., cf.\ \citep{Takesaki1}.
\end{prop}

The completely positive maps are *-linear operators between algebras, and when
the algebras are represented as (sub-)algebras of operator algebras $\LHA$,
they are maps between these operator spaces, and are therefore sometimes referred to 
as \emph{super-operators}.  It is thus important to have methods which allow these
super-operators to be represented in terms of normal operators.  The Kraus-form
in \feP{prop:Kraus-props} may be seen as a first important result in this direction,
as it represents the super-operator $\E{T}$ in terms of ordinary linear Hilbert space
operators.  In the following, a few more such representation results will given.

A complete characterisation of completely positive maps is given by the following 
Stinespring Dilation

\begin{thm}[Stinespring]   \label{thm:Stinespring}
Let $\Alg$ be a finite dimensional unital *-algebra, $\Hvk$ a finite dimensional 
Hilbert space, and $\E{P}: \Alg \to \LHA$ a completely positive *-linear map.

Then there exists a finite dimensional Hilbert space $\Kvk$ and a unital *-homomorphism
$\Phi: \Alg \to \Lop(\Kvk)$ (cf.\ \feD{def:star_homomorphism}), 
as well as a linear map $V \in \Lop(\Hvk,\Kvk)$ from $\Hvk$ to $\Kvk$, such that 
\begin{equation}  \label{eq:Stinespring}
  \forall \ra \in \Alg: \quad \E{P}(\ra) = V^\dagger \, \Phi(\ra) \, V \; \in \LHA ;
\end{equation}
e.g., cf.\ \citep{Davidson1996, KadiRingr2-97, Takesaki1, Werner2001, 
NielsenChuang2011, Holevo2012, Wolf2012, BenyRicht15, Wilde2017, Quillen2025}.
If $V^\dagger V = I_{\Hvk}$ (i.e.\ $V$ is an isometry), then $\E{P}$ is unit preserving.

In the special case  $\Alg = \Lop(\C{J})$, where $\C{J}$ is a finite dimensional 
Hilbert space, and $\E{P}: \Lop(\C{J}) \to \LHA$ is CP and *-linear, 
there is a finite dimensional Hilbert space $\C{E}$, 
so that $\Kvk = \C{J}\otimes\C{E}$, and for the homomorphism $\Phi$ in
\feq{eq:Stinespring} one may take 
$\Phi: \, \Lop(\C{J}) \ni A \mapsto (A \otimes I_{\C{E}}) \in  \Lop(\Kvk) 
= \Lop(\C{J}\otimes\C{E})$.  Then
\begin{align}  \label{eq:Stinespring-H1}
    \forall A \in \Lop(\C{J}):\; \E{P}(A) &= V^\dagger (A \otimes I_{\C{E}}) V 
         \in \LHA,\\    \label{eq:Stinespring-H2}
    \text{and dually } \forall R \in \LHA: \; 
           \E{P}^\dagger(R) &= \tr_{\C{E}} (V R V^\dagger ) \in \Lop(\C{J}). 
\end{align}
The map $V \in \Lop(\Hvk,\Kvk)$ is often called a \emph{Stinespring map} or
\emph{Stinespring operator}.
\end{thm}

What the \feq{eq:Stinespring} says is that a *-linear CP map, when \emph{dilated} 
on an appropriate larger space resp.\ in a larger algebra, is really a 
*-homomorphism, or, in other words, that a *-linear CP map is the
\emph{compression} of a *-homomorphism.  Another way of saying this 
is that the CP map  has been ``lifted'' to a homomorphism.
This is yet another example of the ``Church of the Larger Hilbert Space'' theme.  
Note also the similarity to \feT{thm:Neumark-dilation}.  Another view of
\feq{eq:Stinespring} is that it says that any *-linear CP-map is a combination of
*-homomorphisms (\feC{coro:CPUP-maps}) and Kraus maps (\feP{prop:Kraus-props}).

Just as in the
discussion following \feP{prop:Kraus-props}, from \feC{coro:CP-composition}
it follows that the convex combination of such unit preserving Stinespring 
dilations is again CP *-linear and unit preserving. 
Here is a kind of converse of \feC{coro:gen-Kraus-form}: 

\begin{coro}[Kraus, Choi]   \label{coro:Choi-Stinespring}
  Let $\Hvk, \C{J}$ be finite dimensional Hilbert spaces, and 
  $\E{Q}: \Lop(\C{J}) \to \LHA$ a *-linear CP map.  Then there are 
  finitely many linear maps
  $V_j \in \Lop(\Hvk,\C{J})$, such that 
\begin{equation}  \label{eq:Choi-Stinespring}
  \forall A \in \Lop(\C{J}): \quad \E{Q}(A) = \sum_j V_j^\dagger \, A \, V_j \; 
  \in \LHA ;
\end{equation}
i.e.\ $\E{Q}$ can be brought into \emph{Kraus form}, cf.\ \citep{Choi1975,  Wolf2012}.  
Its adjoint may be gleaned from \feq{eq:gen-Kraus-form-adj}.

$\E{Q}$ is unit preserving \emph{iff} $\sum_j V_j^\dagger V_j = I_{\Hvk}$ 
(i.e.\ $\{j\} \mapsto V_j^\dagger V_j$ is a POVM), and $\E{Q}$ is trace preserving
\emph{iff} $\sum_j V_j V_j^\dagger = I_{\C{J}}$. 
\end{coro}

The importance of this representation in \feq{eq:Choi-Stinespring} is that the
map $\E{Q}: \Lop(\C{J}) \to \LHA$ between algebras of maps on Hilbert spaces
can be dealt with in the realm of maps on the Hilbert spaces themselves, just as in
\feqs{eq:Stinespring-H1}{eq:Stinespring-H2}.

The following \feT{thm:RN-single-CP} 
compares different CP-maps, cf.\ \citep{BelavkinStaszewski1986}:

\begin{thm}[Radon-Nikodým for CP-maps]  \label{thm:RN-single-CP}
As in \feT{thm:Stinespring}, let $\Alg$ be a finite dimensional unital *-algebra, 
$\Hvk$ a finite dimensional Hilbert space, and $\E{P}: \Alg \to \LHA$ 
a completely positive *-linear map.  From \feq{eq:Stinespring} one knows that
there exists a finite dimensional Hilbert space $\Kvk$ and a unital *-homomorphism
$\Phi: \Alg \to \Lop(\Kvk)$ (cf.\ \feD{def:star_homomorphism}), 
as well as a linear map $V \in \Lop(\Hvk,\Kvk)$ from $\Hvk$ to $\Kvk$, such that 
\[
  \forall \ra \in \Alg: \quad \E{P}(\ra) = V^\dagger \, \Phi(\ra) \, V \; \in \LHA.
\]

Let $\E{T}: \Alg \to \LHA$ be another CP *-linear map.  Then there is a self-adjoint
positive operator $R \in \Lop(\Kvk)_{+}$ in the commutant $\Phi(\Alg)'$ of
the sub-algebra $\Phi(\Alg) \subseteq \Lop(\Kvk)$, such that
\begin{equation}  \label{eq:RN-single-CP}
  \forall \ra \in \Alg: \quad \E{T}(\ra) = V^\dagger \, R\, \Phi(\ra) \, V 
   = (R^{\frk{1}{2}}\,V)^\dagger \, \Phi(\ra) \, (R^{\frk{1}{2}}\,V) \; \in \LHA.
\end{equation}
The  operator $R \in \Lop(\Kvk)_{+}$ may be seen as the \emph{Radon-Nikodým derivative} of
$\E{T}$ w.r.t.\ $\E{P}$, and the \emph{Stinespring operator} for $\E{T}$ is 
$R^{\frk{1}{2}}\,V \in \Lop(\Hvk,\Kvk)$. 
\end{thm}

\begin{rem}   \label{rem:RN-CP-connection}
Here it is worthwhile to point out some general structural similarities which 
arise when comparing states and functionals, as well as when comparing POVMs.  
Recalling the geometry of states
described in \feP{prop:states-prop} as living in the dual positive
cone $\Alg^+$, a similar pattern was observed for POVMs in \feR{rem:POVM-more-X}.  
In both cases, this leads to Radon-Nikodým type results --- \feC {coro:RN-alg} in the
case of functionals, and \feR{rem:POVM-more-X} in the case of POVMs --- when comparing
different elements of the positive cone.  It was already pointed out in 
\feR{rem:RN-positive-cone} that the Radon-Nikodým type statements translate this into
into positive operators resp.\ elements of the algebra, 
i.e.\ elements in the convex positive cone of the original space.

Here in \feT{thm:RN-single-CP} a similar pattern may be noted.  According to
\feC{coro:CP-composition}, CP maps form a positive cone, and the above theorem
tells us how the comparison between $\E{P}$ and $\E{T}$ is translated into a
positive element of the algebra $R \in \Lop(\Kvk)_+$.
\end{rem}

From \feT{thm:Stinespring} one can deduce \citep{Perez-GarciaEtAl2006, HiaiPetz2014} 
some results about the contractivity, as well as on the spectrum \citep{Wolf2012}
of positive and trace preserving maps.  These may be compared with similar
results in the usual Perron-Frobenius theory, which are important 
e.g.\ when considering Markov-chains.  First we need

\begin{defi}[Contractivity]   \label{def:contract}   
Let $\C{U}, \C{V}$ be vector spaces, equipped with the some norm, and
$\E{F} \in \Lop(\C{U}, \C{V})$ a linear map.  If it satisfies
\begin{equation}   \label{eq:contract-Sp}
 \forall \ra, \rb \in \C{U}:\; \nd{\E{F}(\ra) - \E{F}(\rb)}_{\C{V}} \le \nd{\ra - \rb}_{\C{U}}.
\end{equation}
then $\E{F}$ is \emph{non-expansive} or for short \emph{contractive}, to be precise
in the norm-combination $(\C{U},\nd{\cdot}_\C{U}), (\C{V},\nd{\cdot}_\C{V})$.
\end{defi}

\begin{prop}[Contractivity of CP maps]    \label{prop:contract}
Let $\Alg, \Blg$ be finite dimensional unital *-algebras, equipped with the
trace based canonical inner product, as well as with the 
Schatten-$p$-norms $\nd{\ra}_{Sp} = (\tr_{\Alg} \ns{\ra}^p)^{\frk{1}{p}} = 
(\tr_{\Alg} (\ra^\star \ra)^{\frk{p}{2}})^{\frk{1}{p}}$ for $\ra \in \Alg$, 
and similarly for $\Blg$, cf.\  \feR{rem:Schatten-A}.
For the sake of simplicity (some statements are true in more general settings), let  
$\E{T} \in \Lop(\Alg,\Blg)$ be a CPTP-map.  Or equivalently, let its adjoint 
$\E{T}^\dagger \in \Lop(\Blg,\Alg)$ be a CPUP-map, cf.\ \feP{prop:pos-unit-trace}.

Then $\E{T}$ is \emph{non-expansive} 
or \emph{contractive} in the Schatten-1- or  trace-norm 
(i.e.\ the Schatten-$1$-norm is to be used in \feq{eq:contract-Sp} on both sides), 
but \emph{not necessarily} in any of the other Schatten-$p$-norms for $p>1$.
In particular the trace $\tr_{\Alg}: \Alg \to \CC$ is Schatten-1-contractive.

Dually, its adjoint, the CPUP-map $\E{T}^\dagger \in \Lop(\Blg,\Alg)$ is 
\emph{contractive} in the Schatten-$\infty$-norm 
(i.e.\ the Schatten-$\infty$-norm is to be used in \feq{eq:contract-Sp} on both sides),
but \emph{not necessarily} in any of the other Schatten-$p$-norms for $p<\infty$. 
In case the map $\E{T}  \in \Lop(\Alg,\Blg)$ is in addition unital, then 
\feq{eq:contract-Sp} holds for all $p \ge 1$, i.e.\ $\E{T}$ and its adjoint $\E{T}^\dagger$
are contractive in \emph{all} Schatten-$p$-norms for $1 \le p \le \infty$.
\end{prop}

\begin{prop}[Perron-Frobenius Property]    \label{prop:Perron-Frobenius}
With the same notation and assumptions as above in \feP{prop:contract},
assume in the additionally that $\Blg = \Alg$, and further for simplicity that
$\E{T} \in \Lop(\Alg)$ is a *-linear CPTP-map.  Recall from linear algebra that 
$\E{T}$ and its adjoint $\E{T}^\dagger \in \Lop(\Alg)$ --- where $\Alg$ is equipped
with the trace-based canonical inner product (cf.\ \feP{prop:fin-dim-alg-gen}) ---
have the same spectrum, $\sigma(\E{T}) = \sigma(\E{T}^\dagger)$, and thus equal spectral 
radius  (cf.\ \feD{def:spec-one}), which in this case satisfies:
\begin{equation}   \label{eq:one-spec-rad}
     \vrho(\E{T})  = \vrho(\E{T}^\dagger) = 1,
\end{equation}
i.e.\ the eigenvalues are all in the unit circle in $\CC$, and positivity implies
that they are either real or come in complex conjugate pairs.  All eigenvalues
$\lambda \in \sigma(\E{T})$ on the unit circle --- the so called \emph{peripheral spectrum}
with $\ns{\lambda} = 1$ --- are \emph{semi-simple}, i.e.\ geometric and algebraic 
multiplicity agree. 

As $\E{T}$ is assumed positive and trace preserving, its adjoint $\E{T}^\dagger$ is 
positive and \emph{unital} (cf.\ \feP{prop:pos-unit-trace}), this implies that
$\E{T}^\dagger(\rone) = \rone$, and hence $\rone \in \Alg_+$
is a \emph{positive definite} eigenvector of $\E{T}^\dagger$ with eigenvalue 
$\lambda = 1 \in \sigma(\E{T}^\dagger) = \sigma(\E{T})$.

As $\E{T} \in \Lop(\Alg)$  is clearly continuous, and as
trace preserving means that for the closed convex set $\DEN{\Alg}$ it holds that 
$\E{T}(\DEN{\Alg}) \subseteq \DEN{\Alg}$, Brouwer's fixed point theorem implies that 
there is a fixed point (a density) $\rr_* \in \DEN{\Alg} \subset \Alg_+$ with 
$\E{T}(\rr_*) = \rr_*$.  This is hence a \emph{positive} eigenvector of 
$\E{T}$ for the eigenvalue $\lambda = 1 \in  \sigma(\E{T})$.

The following two statements are equivalent:
\begin{compactitem}
\item $\E{T}$ is \emph{irreducible}, i.e.\ if for any projection $\rp \in \PA{\Alg}$
      one has that if $\E{T}(\rp \Alg \rp) \subseteq \rp \Alg \rp$, then either $\rp = \rnul$
      or $\rp = \rone$.  %Note that this excludes $\E{T} = \oL_{\rone_{\Alg}}$, the identity. 
\item $\lambda = 1 \in \sigma(\E{T})$ is a \emph{simple} eigenvalue, 
      $\E{T}(\rr_*) = \rr_*$ is the unique fixed point in $\DEN{\Alg}$, and $\rr_*$ is
      a \emph{positive definite} eigenvector (i.e.\ $0 \notin \sigma(\rr_*)$,
      or $\tr_{\Alg} (\ra^\star \rr_* \ra) = 0$ only for $\ra = \rnul$).  
%\item There is a unique and positive definite $\rr_C \in \DEN{\Alg}$, 
%      such that for any $\rr  \in \DEN{\Alg}$ the \emph{Cesàro mean} converges to $\rr_C$:
%      \[
%         \lim_{n \to \infty} \frac{1}{n} \sum_{k=1}^n \E{T}^k(\rr) = \rr_C .
%      \]
\end{compactitem}

An irreducible CPTP-map $\E{T} \in \Lop(\Alg)$ implies that
\begin{compactitem}
%\item No other positive eigenvalue $0 < \lambda \le 1$ --- except $\lambda = 1$ ---
%      belongs to a positive eigenvector of $\E{T}$.
\item All eigenvalues $\lambda$ in the peripheral spectrum (i.e.\ $\ns{\lambda} = 1$) are
      \emph{simple}, and they are \emph{roots of unity}, i.e.\ of the form 
      $\lambda_k = \gamma^k$ for $k=0,\dots,m-1$, with $\gamma = \exp (2 \uppi\, \ii /m)$ 
      for some $m \in \D{N}$.
\item There is a unitary $\ru \in \UA{\Alg}$ such that 
      $\E{T}^\dagger(\ru^k) = \gamma^k \ru^k$ for $k=0,\dots,m-1$.
\end{compactitem}
\end{prop}

\paragraph{Channels}  %\label{SS:channels}
%
%Duality Effects, States, ,
After all this preparation, we are ready to discuss \emph{channels}, cf.\
\citep{Werner2001, NielsenChuang2011, Holevo2012, Wolf2012, BenyRicht15, Wilde2017, 
Quillen2025}.  Channels are the mathematical formalisation of a transfer of 
probabilistic information from one probability algebra to another one.

\begin{defi}[Channel]   \label{def:channel}
A channel from the probability algebra $\Alg$ to the probability algebra $\Blg$ is 
in the Heisenberg picture is a *-linear, completely positive, and unit preserving map 
(CPUP) $\E{P}: \Blg \to \Alg$, most importantly mapping effects (or POVMs and PVMs) 
of $\Blg$ to effects of $\Alg$.
Dually, the same channel in the Schrödinger picture, the adjoint, is a *-linear, 
completely positive, and trace preserving (CPTP) map $\E{P}^\dagger: \Alg \to \Blg$, essentially
mapping densities on $\Alg$ to densities on $\Blg$.
\end{defi}

Channels are a very versatile object, and can be used to model or describe many concepts
resp.\ operations in algebraic probability.  They are explored in connection
with quantum information and quantum computation.  We have already seen some
examples of CPTP maps --- which model channels according to the preceding \feD{def:channel} ---
in \feC{coro:CPUP-maps}.  We have also seen ways to construct CP maps in \feP{prop:Kraus-props}
in Kraus-form --- closely related to the structure theorems for CP maps, cf.\ 
\feT{thm:Stinespring} and \feC{coro:Choi-Stinespring} --- as well as through 
parallel compositions in \feP{prop:CP-tensor} and serial composition in \feC{coro:CP-composition}.

It is beyond the scope of the present work to discuss further topics connected with
the transmission of information, like coding, entropy, and channel capacities.  The 
interested reader is advised to consult the literature, e.g.\ 
\citep{Werner2001, NielsenChuang2011, Holevo2012, Wolf2012, BenyRicht15, 
Wilde2017, Quillen2025}, and the references therein.

\begin{xmpn}[Some Useful Channels]  \label{ex:channel-examples}
To show the versatility of the concept of channels, some \emph{examples} will be given,
all referring to *-linear CPTP- resp.\ CPUP-maps between finite dimensional algebras.

\begin{description}

\item[Positive Maps and Abelian Algebras:] One example already encountered were positive 
     *-linear maps involving an Abelian or classical algebra as domain or range, cf.\
     \feC{coro:CPUP-maps}; these are \emph{completely positive}. \\[0.3em]
     {\nmf \tbf{States:}}  as $\SA{\Alg} \subset \Alg^+$, they are positive unital maps from 
     an algebra $\Alg$ into $\CC$ (a commutative algebra, cf.\ \feC{coro:CPUP-maps}), 
     and are thus channels.  \\[0.3em]
     {\nmf \tbf{Classical Output:}} observables like $\E{F}$ from \feX{ex:obs-channel} are 
     another  such example; the output --- \emph{a classical output} --- is not just the 
     algebra  $\CC$ as for states, but any classical resp.\ Abelian or commutative algebra 
     $\Clg$ (cf.\ \feC{coro:CPUP-maps}), 
     and the map (in the Heisenberg picture) $\E{F}: \Clg \to \Alg$ can be realised 
     by a \emph{POVM}.  Recall that a \emph{PVM}, sometimes also called
     a \emph{von Neumann measurement}, is a \emph{sharp} (extreme) variant of a POVM, 
     where all the maps in the collection are mutually orthogonal projections.   \\[0.3em]
     {\nmf \tbf{Classical input:}} Similarly, a unital, positive, *-linear map $\E{I}$, 
     where the image is any classical (Abelian or commutative) algebra $\Clg$ is completely 
     positive,  cf.\ \feC{coro:CPUP-maps}.  Such a channel can be interpreted  in the 
     Heisenberg picture as a  \emph{classical input} to a general system $\Alg$, 
     i.e.\ $\E{I}: \Alg \to \Clg$.  
%     \paragraph{Classical to Classical:} Actually, any information processing by a channel $\E{C}$
%     involving classical information (Abelian algebras) at the ends --- input and output --- 
%     can be thought of as composed from a classical input $\Clg_I$ to a general 
%     system $\Alg$ through \emph{extension} $\C{E}: \Alg_1 \otimes \Alg_2 \to \Alg$ 
%     (see below), then any other unitary channel 
%     $\E{U}:\Blg \to \Alg_1 \otimes \Alg_2$, then a restriction $\E{R}:\C{R} \to \Blg$,
%     and finally a classical output $\E{F}$ to $\Clg_O$ like in \feX{ex:obs-channel}.
%     In the Heisenberg picture this means $\E{C}: \Clg_O \to \Clg_I$ is given by
%     $\E{C} = \E{I} \circ \E{E} \circ \E{U} \circ \E{R} \circ \E{F}$.

\item[*-Homomorphisms:]  Some other examples which were already encountered are the
     *-homo\-morph\-isms (\feD{def:star_homomorphism}), they are unit preserving.  
     The Stinespring dilation \feT{thm:Stinespring} shows that any channel can be 
     modelled in terms of  *-homomorphisms. If such a *-homomorphisms is an *-isomorphism, 
     it is an  \emph{invertible} channel, also possible to be used in the reverse direction
     as a channel.   \\[0.3em]
     {\nmf \tbf{Samples or Characters:}} as $\XA{\Clg} \subset \SA{\Clg}$, these are also
     states (\feD{def:sample}) and according to \feC{coro:CPUP-maps} they are channels, 
     but they are also \emph{*-homomorphisms} into the algebra $\CC$.    \\[0.3em]
     {\nmf \tbf{Automorphisms:}} A special case of *-homomorphism  
     are internal \emph{invertible} *-isomorphisms (channels), like
     *-automorphisms (unitary transformations or symmetries)
     of the form $\E{S}: \Alg \ni \ra \mapsto \ru \ra \ru^\star \in \Alg$, 
     where $\ru \in \UA{\Alg}$ is a unitary.  
     Such internal *-automorphisms typically indicate symmetries inherent in
     the system, where the symmetry group is represented on a subgroup of the unitaries, 
     a topic which will not be expanded here, e.g.\ cf.\ \citep{FaesslerStiefel1992}.
     They may also represent a \emph{quantum computation}, cf.\ \fsec{S:basic-QC}. \\[0.3em]
%     \tbf{Unitary Transformation:}
%     One *-automorphism which fits here and has the same mathematical form --- also embodying 
%     a symmetry --- is translation in time resp.\ \emph{unitary time-evolution}, e.g.\ as it 
%     occurs also during a \emph{quantum computation}, cf.\ \fsec{S:basic-QC}.\\[0.3em]
     {\nmf \tbf{Unitary Equivalences:}} Such transformations between Hilbert spaces 
     are a bit similar.  Let $\Hvk$ and $\Kvk$ be finite dimensional Hilbert spaces 
     with the same dimension, and a unitary $U: \Hvk \to \Kvk$.  Such unitaries 
     establish *-isomorphisms and hence \emph{invertible channels} between $\LHA$ 
     and  $\Lop(\Kvk)$:  for a $A \in \LHA$ one has the \emph{Kraus form} 
     (cf.\ \feP{prop:Kraus-props}) of $U A U^\dagger \in \Lop(\Kvk)$.  
     If $\Hvk = \Kvk$, this is a concrete form of an \emph{automorphism}.

\item[Expansion:] The \emph{expansion} is a \emph{tensoring} 
     of a probability-algebra
     $\Alg$ by a second one $\Blg$ with density $\rr_{\Blg} \in \DEN{\Blg}$, 
     cf.\ \feC{prop:CP-tensor}.  In the Heisenberg picture, 
     this is a CPUP-map $\E{E}: \Alg \otimes \Blg \to \Alg$
     defined on elementary tensors $\ra \otimes \rb \in \Alg \otimes \Blg$  by
     $\E{E}(\ra \otimes \rb) = \tr_{\Blg} (\rr_{\Blg} \rb)\; \ra$, and extended by linearity.
     The adjoint in the Schrödinger picture is a CPTP-map
     $\E{E}^\dagger: \Alg  \to \Alg \otimes \Blg$, for a state $\rr_{\Alg} \in \DEN{\Alg}$
     given simply by the tensorisation
     $\E{E}^\dagger(\rr_{\Alg}) = \rr_{\Alg} \otimes \rr_{\Blg} \in \DEN{\Alg \otimes \Blg}$.

\item[Restriction:] The \emph{restriction}, a CPUP-map $\E{R}: \Alg  \to \Alg \otimes \Blg$ 
     in the  Heisenberg picture, is simply $\E{R}(\ra) = \ra \otimes \rone_{\Blg}$.  
     For the adjoint in the Schrödinger picture, which is 
     a CPTP-map $\E{R}^\dagger: \Alg \otimes \Blg  \to \Alg$, this is just 
     a partial trace,  as already indicated in 
     \feqs{eq:red-state-b}{eq:red-state-c} in \feD{def:tensor-states}, and in
     \feqs{eq:red-state-b-m}{eq:red-state-b-n} in \feR{rem:tensor-states-A}:
     if $\rr_{\Alg,\Blg} \in \DEN{\Alg \otimes \Blg}$ is a state on the 
     tensor product $\Alg \otimes \Blg$, then the corresponding reduced state 
     on $\Alg$ is given by $\rr_{red,\Alg} = \E{R}^\dagger(\rr_{\Alg,\Blg}) = 
     \tr_{\Blg} \rr_{\Alg,\Blg} \in \DEN{\Alg}$.

\item[Combinations:] One possible way of combining CPUP-maps is \emph{convex combinations}
     of CPUP-maps, cf.\ \feC{coro:CP-composition}.   
     One may note that CPUP-maps have geometrically a
     very similar structure as states: the intersection of a cone with a linear sub-manifold.
     The same is true for their adjoints, their geometric structure is similar to that of 
     density matrices.
     
     Another very important way to combine channels is the \emph{concatenation}
     of CPUP-maps, cf.\ \feC{coro:CP-composition}; and, as already mentioned above,
     all channels can be built up like this, from a few basic building blocks, namely
     expansion and restriction and *-isomorphisms resp.\ unitary transformations.

%\item[Kraus Form:] Let $\Hvk, \Kvk$ be finite dimensional Hilbert spaces, and let 
%     $\E{Q}: \LHA \to \Lop(\Kvk)$ be a *-linear map in Kraus form --- 
%     cf.\ \feP{prop:Kraus-props},  \feC{coro:gen-Kraus-form}.   Such maps are completely
%     positive.  If $\E{Q}$ is unital or trace preserving, such maps in Kraus form are 
%     channels.  According to \feC{coro:Choi-Stinespring}, \emph{any} channel 
%     $\E{T}: \LHA \to \Lop(\Kvk)$ can be brought into this form.

\item[Conditional Expectation:]  
     This is a simplified discussion, as, depending on the exact requirements,
     conditional expectations in $\Ck^*$- or $\Wp^*$-algebras may not exist, or may 
     not be unique, e.g.\ see \citep{Umegaki1954, Tomiyama1957, Takesaki1972, 
     gudderHudson78, Gudder79, AccCecch82, Kadison2004, RedeiSummers2006p, 
     YukalovSornette2009, YukalovSornette2016}
     and the references therein.  Luckily, it is simpler for finite dimensional algebras
     \citep{gudderHudson78, Gudder79, AccCecch82, NielsenChuang2011, Wolf2012, HiaiPetz2014}.

     {\nmf Let $(\Alg, \svpi)$ be a finite dimensional probability 
     algebra with a faithful state $\svpi \in \SA{\Alg}$, 
     and $\Blg \subseteq \Alg$ a probability sub-algebra.}
%     , otherwise one has to pass to factors
%     $\Alg/ \C{J}_{\svpi}$ etc., as in \fsec{SS:states}.  
%     Also denote $\Ex(\cdot) = \svpi(\cdot)$.}
     
     \begin{defi}  \label{def:CEX}
     A positive unital map $\E{E} \in \Lop(\Alg, \Blg)$, which satisfies
     \begin{equation}  \label{eq:CEX}
       \forall \ra \in \Alg,\, \forall \rb \in \Blg:\, 
           \E{E}(\rb \ra) = \rb \, \E{E}(\ra), 
%         \forall \ra \in \Alg:\, \Ex(\ra) = \Ex(\E{E}(\ra))
     \end{equation}
     is called a $\svpi$-\emph{conditional expectation} \citep{HiaiPetz2014}, 
     conditioned on the sub-algebra $\Blg$.  It is sometimes also denoted
     $\E{E}_{\Blg}$, or even $\E{E}^{\Alg}_{\Blg}$, and often as $\Ex(\cdot | \Blg)$.
     \end{defi}
     
     {\nmf Some authors sometimes use slightly different, but finally equivalent conditions.
%     The following \feT{thm:dual-embed} is essentially a simplification of
%     \citep{Tomiyama1957, Takesaki1972, gudderHudson78, Gudder79, AccCecch82, 
%     Kadison2004, Wolf2012, HiaiPetz2014}
     What makes the matter simple in the case of finite dimensional algebras is that
     $\Blg$ is already complete in all vector space topologies.  Some of the properties
     of $\E{E}$  are collected in the following}
     
     \begin{thm}  \label{thm:dual-embed}
     In the case of finite dimensional algebras (and not only then), 
     the conditional expectation map $\E{E}$ \emph{exists} and is \emph{unique};
     it is actually the \emph{orthogonal projection} $P_{\Blg}$
     \citep{gudderHudson78, Gudder79, AccCecch82, HiaiPetz2014} of $\Alg$ onto 
     $\Blg$ in the associated Hilbert space inner product from $\svpi$, 
     cf.\ \feP{prop:state-Hilbert-spc}.  It can also be described as the 
     adjoint $\E{E} = P_{\Blg} = \E{J}^\dagger$ of the natural identity 
     embedding  $\E{J}: \Blg \hookrightarrow \Alg$.

%     \begin{thm}[Takesaki, Tomiyama]  \label{thm:dual-embed}
     Thus the map $\E{E}$ has operator norm less than unity (it is a contraction) 
     when the algebras are equipped with the Hilbert $\nd{\cdot}_2$-norms, and it is
     also a contraction (cf.\ \feP{prop:contract}) when the algebras are equipped with the 
     Schatten-$\infty$-norm (cf.\ \feR{rem:Schatten-A}).  Additionally it is
     \emph{completely positive} (CPUP), and hence a \emph{channel}.  

%     Equivalent with \feD{def:CEX} is that the linear map  $\E{E} \in \Lop(\Alg,\Blg)$ 
     It also satisfies    
     \begin{align}  \label{eq:CEX-2}
         \forall \ra \in \Alg,\, \forall \rb, \rc \in \Blg:\, 
           \Ex(\rb \ra \rc) &= \Ex(\rb \, \E{E}(\ra) \, \rc), \\
          \label{eq:CEX-1}
%       \forall \ra \in \Alg,\, \forall \rb, \rc \in \Blg:\, 
           \E{E}(\rb \ra \rc) &= \rb \, \E{E}(\ra) \, \rc .
     \end{align}
     
%     Additionally, for a *-linear and idempotent map $\E{E}: \Alg \to \Blg$, 
%     the following are equivalent:
%     \begin{compactitem}
%       \item $\E{E}$ is a conditional expectation;
%       \item $\E{E}$ is contractive in the Schatten-$\infty$-norm and completely positive;
%       \item $\E{E}$ is contractive in the Schatten-$\infty$-norm.
%     \end{compactitem}
     \end{thm} 
     
     {\nmf From this easily follows:  The conditional expectation $\E{E}(\ra)$
%     is also contractive in the $\nd{\cdot}_2$-norm, and, for $\ra \in \Alg$, it
     minimises the loss function $\Lambda_{\ra}(\rb) := \nd{\ra - \rb}_2^2$ over $\Blg$,
     and hence one has Pythagoras's theorem 
     \begin{align}  \label{eq:Pyth-condex}
        \nd{\ra}_2^2 &= \nd{\ra - \E{E}(\ra))}_2^2 + \nd{\E{E}(\ra)}_2^2,\; 
           \text{ and the Galerkin orthogonality}\\              \label{eq:weak-condex}
        0 &= \bkt{\rb}{\ra - \E{E}(\ra)}_2 \quad \forall \rb \in \Blg.
     \end{align}

     Two particular choices of sub-algebras deserve an extra note:
     \begin{compactitem}
     \item If $\Blg = \CC[\rone_{\Alg}]$, then
          $\Ex(\ra \mid \CC[\rone_{\Alg}]) = \Ex(\ra)\, \rone_{\Alg} = \wob{\ra}$.
     \item If $\Blg = \Alg$, then 
          $\Ex(\ra \mid \Alg) = \E{I}(\ra) = \ra$, where $\E{I}\in \Lop(\Alg)$
          is the identity.
     \end{compactitem}
     }

\item[Instrument:] This is a refinement of the \emph{observable} from \feX{ex:obs-channel}.
     There a \emph{CPUP}-map $\E{F}: \Clg \to \LHA$, where $\Clg \cong \Ck(X)$ is a 
     commutative algebra, was considered.  According to \feP{prop:Banach-alg-comm}, any
     commutative or Abelian $\Wp^*$-algebra $\Clg$ can be represented in such a way, where
     $X = \XA{\Clg}$ is the set of characters resp.\ samples.  This representation will
     be used here in the sequel.  For  $\Clg$ finite dimensional, $X = \XA{\Clg}$
     is a finite set, with $n = \ns{X} \in \D{N}$; hence 
     $\Clg \cong \Ck(X) \cong \CC^X \cong \Cn$ with unit 
     $\rone_{\Clg} = [1,\dots,1]^\trpos \in \CC^X \cong \Cn$, 
     and canonical unit vectors $\ve_x \in \CC^X$ with all components zero, 
     except unity at position $x \in X$
     --- this is the Hadamard algebra from \feX{ex:complex-n-spc}. 
      
     The observable from \feX{ex:obs-channel} transmits probabilistic 
     information from the algebra $\Alg = \LHA$ to the 
     classical commutative or Abelian algebra $\Clg \cong \Cn$.   
     To keep track of what happens to the observed system $\Alg$, 
     e.g.\ \citep{Werner2001, NielsenChuang2011, Wolf2012, Wilde2017}, one has
     to include the algebra $\Alg$ as part of the output, or even more generally, a 
     possibly different probability algebra $\Blg$.  In many cases it could be 
     identical to $\Alg$, or a sub-algebra  $\Blg \subseteq \Alg$, and represents 
     the observed system after the interaction with the instrument, which, 
     in the Heisenberg picture, is a *-linear CPUP-map
     \begin{equation}  \label{eq:instrument-H}
          \E{G}:\; \Clg \otimes \Blg \ni (\vf \otimes \rb)  \mapsto 
            \E{G}(\vf \otimes \rb) = \sum_{x \in X} f_x G_x(\rb) \in \Alg ,
     \end{equation}
     with $\vf = [\dots, f_x \dots]^\trpos \in \CC^X$, and then expanded by linearity.  
     Here each $G_x: \Blg \to \Alg$ is a *-linear completely positive map, 
     and preservation of the unit requires $\E{G}(\rone_{\Clg} \otimes \rone_{\Blg})
      = \rone_{\Alg}$.  This also results in 
     $\wob{\E{G}}(\rone_{\Blg}) :=  \sum_{x \in X} G_x(\rone_{\Blg}) = \rone_{\Alg}$,
     where the marginal map $\wob{\E{G}} := \sum_x G_x \in \Lop(\Blg, \Alg)$, which
     is CPUP, displays
     the changes to the system when the output on the system on $\Clg$ is ignored.
     Observe that these maps define the POVM for the observation: 
     $\rg_x := G_x(\rone_{\Blg}) \in \EA{\Alg}$.  In case the POVM is actually a PVM,
     the setup is called a \emph{von Neumann} instrument.
     
     The adjoint of $\E{G}$ in \feq{eq:instrument-H}, the channel in the Schrödinger 
     picture, is $\E{G}^\dagger: \Alg \to \Clg \otimes \Blg$, given by
     \begin{equation}  \label{eq:instrument-S}   
          \E{G}^\dagger(\rr_{\Alg}) = \sum_{x \in X} 
          \ve_x \otimes G^\dagger_{x}(\rr_{\Alg}),
     \end{equation}
     where $G^\dagger_{x}: \Alg \to \Blg$ is the adjoint of $G_x$, and $\ve_x \in \CC^X$
     is the canonical unit vector.
     
     Without the information from an observation, the interaction with
     the instrument leads to state changes, a new state on $\Clg \otimes \Blg$ is given
     by \feq{eq:instrument-S}
     \begin{equation}  \label{eq:instrument-S-sum}
       \rr_{\Clg \otimes \Blg} = \E{G}^\dagger(\rr_{\Alg}) = \sum_{x \in X} 
          \ve_x \otimes \rr_{\Blg,x} = \sum_{x \in X}  \ve_x \otimes G^\dagger_{x}(\rr_{\Alg}) .
      \end{equation}
     %where $\rr_{\Blg,x} := G^\dagger_{x}(\rr_{\Alg})$.  
     By taking partial traces
     one obtains the \emph{marginal} densities --- predicted without knowing the
     outcome of the observation.  On $\Clg$ it is the vector
     $\vrh_{\text{marg}} \in \CC^X$ with components $\rho_x = \tr_{\Blg} \rr_{\Blg,x} = 
     \tr_{\Alg} (\rr_{\Alg}\, \rg_x)$, and on $\Blg$ it is 
     $\rr_{\Blg,\text{marg}} = \wob{\E{G}}^\dagger(\rr_{\Alg}) := \sum_x \rr_{\Blg,x}$.
          
     On the other hand, assume that the outcome of the observation is $y \in X$,
     i.e.\ the effect $\rg_y$ has been observed.  This means that on $\Clg \cong \CC^X$
     the \emph{posterior} resp.\ \emph{observed} density is given by the vector 
     $\vrh_{\text{obs}} = \ve_y$, the density for the observation / character 
     $\som_y(\vf) = f_y$.  Moreover, this shows that in 
     \feqs{eq:instrument-S}{eq:instrument-S-sum} only the $x = y$ term is non-zero.
     
     The marginal density $\rr_{\Blg,\text{marg}}$ from above now is not consistent with 
     $\vrh_{\text{obs}}$, as it has to be proportional just to 
     $\rr_{\Blg,y} = G^\dagger_{y}(\rr_{\Alg})$; and trace preservation requires a unit trace:
     \begin{equation}   \label{eq:gen-post-den}
        \rr_{\Blg,\text{post}} = \frac{1}{\tr \rr_{\Blg,y}}\, \rr_{\Blg,y} .
     \end{equation}
     This \feq{eq:gen-post-den} is called the generalised \emph{Lüders'} or 
     \emph{Lüders-von Neumann} rule.
%     $\vrh_{\text{post}}$ and $\rr_{\Blg,\text{post}}$ are the marginal densities
%     of the posterior density 
%     \[
%        \rr_{\Clg \otimes \Blg,\text{post}} = \ve_y \otimes  \rr_{\Blg,\text{post}}.
%     \]
                 
\end{description}

     As can be shown \citep{Werner2001},
     any channel can be built as a \emph{concatenation} (cf.\ \feC{coro:CP-composition}) 
     from the *-homomorphisms resp.\ *-isomorphisms --- 
     unitary transformation being a special case --- and expansion 
     and restriction.  This is in some way also the content
     of the Stinespring dilation \feT{thm:Stinespring}.
\end{xmpn}

The idea of channels, i.e.\ noisy transmission of information, can, among other things, 
also be used to describe Markov chains \citep{Attal2010, Attal-notes}; in fact, any Markov 
chain can be described with a channel.

\paragraph{Instruments and Observations:} 
The last example in the list of possible channels, the
\emph{instrument}, deserves to be considered a bit more carefully.
There, in the Heisenberg picture, the marginal map which ``forgets'' about the
observation, was introduced from \feq{eq:instrument-H}% at $\rone_{\Clg} \otimes \rb$:
%this gives the map $\wob{\E{G}} \in \Lop(\Blg,\Alg)$ 
with $\wob{\E{G}}(\rb) = \E{G}(\rone_{\Clg} \otimes \rb) =  \sum_{x \in X} G_x(\rb)$,
where the $G_x \in \Lop(\Blg,\Alg)$ are completely positive.  It was already
pointed out that these maps define the POVM for the observation: 
$\rg_x := G_x(\rone_{\Blg}) \in \EA{\Alg}$.   The reverse question is to find
all the possible observations when given a CP map like $\E{T} = \sum_x \E{T}_x$
with all the $\E{T}_x$ completely positive.
As the Stinespring dilation \feT{thm:Stinespring} will be used, things will be 
formulated for operator algebras \citep{Werner2001}.  The result is in some  
sense analogous to the Radon-Nikodým version in \feC{coro:RN-alg}, and obviously
related to \feT{thm:RN-single-CP}:

\begin{thm}[Radon-Nikodým for sums of CP-maps]  \label{thm:RN-CP-maps}
  Let $\Hvk, \C{J}$ be finite dimensional Hilbert spaces, $X$ a finite set, and 
  $\E{T}_x:\; \Lop(\C{J})\to \LHA, x \in X$, *-linear completely positive maps,
  and set $\wob{\E{T}} = \sum_{x \in X} \E{T}_x$.  In the Stinespring form of
  \feq{eq:Stinespring-H1} in \feT{thm:Stinespring}, let $V \in \Lop(\Hvk, \C{J} \otimes \C{E})$
  be the Stinespring operator for $\wob{\E{T}}$, such that
  \[
     \wob{\E{T}}:\; \Lop(\C{J}) \ni A \mapsto
      \wob{\E{T}}(A) = V^\dagger (A \otimes I_{\C{E}}) V \in \LHA .
  \]
  Then there is a uniquely determined POVM, 
  $X \supset \{x\} \mapsto E_x \in \EA{\Lop(\C{E})}$,
  such that 
  \begin{equation}  \label{eq:RN-CP-maps}
     \forall x \in X:\;
      \E{T}_x(A) = V^\dagger (A \otimes E_x) V \in \LHA .
  \end{equation}
\end{thm}

An interesting conclusion may be drawn from \feq{eq:RN-CP-maps} when $\C{E}$ is 
one-dimensional, as then $\C{E} \cong \CC$ and thus $(\C{J} \otimes \C{E}) \cong \C{J}$.
The \feq{eq:RN-CP-maps} in this case reads $\E{T}_x(A) = V^\dagger (p_x A) V$, 
i.e.\ it is in Kraus form 
(cf.\ \feC{coro:gen-Kraus-form} and \feP{prop:Kraus-props})  with just one term.
The \feT{thm:RN-CP-maps} then tells us that $p_x \in [0,1]$ and $\sum_x p_x = 1$,
i.e.\ the only decomposition of $\E{T} = V^\dagger A V$ into CP maps $\E{T}_x$
is by positive multiples of $\E{T}$, and $\vrh_X = [\dots, p_x, \dots]^\trpos$ is a
probability distribution on $X$.  Moreover, in case $\E{T}$ is a channel and
thus unital, with its adjoint \feq{eq:Stinespring-H2} trace preserving, then
$V^\dagger V = I_{\Hvk}$ and $V$ is an isometry.  Especially in the case
$\C{J} = \Hvk$, the identity $\E{I}$ and symmetries are of this one term Kraus
form.  This leads to

\begin{coro}[No Perturbation --- No Information]  \label{coro:no-pert-no-info}
  With $\Clg \cong \CC^X$ and $\Hvk$ as before, let
  $\E{N}:\; \Clg \otimes \LHA \to  \LHA$ be an instrument such that
  $\E{N}(\rone_{\Clg} \otimes A) = \wob{\E{N}}(A) = U^\dagger A U$ with a unitary
  map $U \in \LHA$.  Then $\wob{\E{N}}^\dagger(R) = U R  U^\dagger$ is a unitary
  global state change.  Especially for $U = I_{\Hvk}$ it means that $\wob{\E{N}}$ and
  $\wob{\E{N}}^\dagger$ are the identity, and hence the state is not perturbed through the
  interaction with the instrument.  
  
  For a decomposition $\wob{\E{N}} = \sum_{x \in X} \E{N}_x$, the \feT{thm:RN-CP-maps} 
  in this case says that there
  is a \emph{unique} probability distribution  $\vrh_X = [\dots, p_x, \dots]^\trpos$ on $X$
  such that $\E{N}_x = p_x \wob{\E{N}}$.
  Thus, this probability $p_x$ is \emph{independent} of 
  the density operator $R \in \DEN{\LHA}$, and hence can not provide any 
  information on $R$.
\end{coro}

What this says, is that if one insists that the system represented by $\LHA$ 
is not perturbed, then there is one and only one probability distribution on $X$, 
uniquely dependent on the maps $\E{N}_x$, no matter what the state resp.\ density operator 
on $\LHA$; i.e.\ no information can be extracted by this kind of observation.

\paragraph{Updating and Conditional Expectation:}
In the following \feX{ex:instr-cond} the connection between the instrument channel
and the conditional expectation channel will be explored.  For that purpose we shall
look more closely at a special case of the instrument channel 
$\E{G}:\; \Clg \otimes \Blg \to \Alg$, namely for the sake of simplicity at the case 
where $\Blg = \Alg$, and the POVM associated with
the instrument, the observable $\{x\} \mapsto \rg_x = G_x(\rone_{\Blg}) \in \EA{\Alg}$
on $X$ for the probability algebra $\Alg$, is actually a PVM.  This
means that the $\{\rg_x\} \subset \PA{\Alg}$ are actually mutually orthogonal projections
and we have a von Neumann instrument, as well as that $\ns{X} \le \dim \Alg$.  

\begin{xmpn}[Update from an Instrument]  \label{ex:instr-cond}
To make this example more concrete, one could assume that a
PVM on $\Alg$ comes from an observation of some observable $\rc \in \Alg_{sa}$,
i.e.\ from its spectral decomposition $\rc = \sum_j \lambda_j \rp_j$, with spectral
projections $\rp_j \in \PA{\Alg}$, cf.\ \feP{prop:Banach-alg} and \feR{rem:abstract-PVM}.
Accordingly, we shall change the labelling of the PVM from $x \in X$ to $j =1,\dots,n$
with $n \le \dim \Alg$.  This PVM is combined with one on $\Clg \cong \Cn$, the canonical 
unit vectors $\ve_j$ as the standard PVM on $\Cn$.  This translates to starting with 
$\{j\} \mapsto \ve_j \otimes \rp_j  \in \PA{\Clg \otimes \Alg}$, a PVM on the tensor
product of the instrument.  We also want the CP-maps $G_x \in \Lop(\Alg)$ in 
\feq{eq:instrument-H} to evaluate as $G_j(\rone_{\Alg}) = \rg_j = \rp_j \in \PA{\Alg}$.  
They can be written in Kraus form as $G_j:\, \Alg \ni \rb \mapsto G_j(\rb) = \rp_j \rb \rp_j
\in \Alg$, which produces the correct result.  
Such maps are self-adjoint, so that $G^\dagger_j = G_j$ in \feq{eq:instrument-S}.
This also makes the marginal maps $\wob{\E{G}} = \sum_x G_x$ self-adjoint, i.e.\
$\wob{\E{G}} = \wob{\E{G}}^\dagger$, and both $\wob{\E{G}}$ and its adjoint are
both CPUP \emph{and} CPTP.

These marginal maps \emph{predict} the state / density before the observation,
or without considering the observation.  They are given at each position 
$j \in [n]$ by $\vrh_{\text{marg}} = [\dots, \rho_j, \dots]^\trpos$, with
\begin{multline} \label{eq:marg-post-prob} 
  \rho_j = \tr_{\Alg} \left[ (\ve_j^\tpH \otimes \rp_j^*) \E{G}^\dagger(\rr_{\Alg})\right]
    = \sum_{i = 1}^n \left[ (\ve_j^\tpH \ve_i) 
        \tr_{\Alg} (\rp_j^* \rp_{i} \rr_{\Alg} \rp_i) \right] \\
    = \tr_{\Alg} ( \rr_{\Alg} \rp_{j}) = \Ex_{\rr_{\Alg}}(\rp_j) = \prob_{\rr_{\Alg}}(\rp_j).
\end{multline}
%which does not take into account the information provided by the observation, 
%where the state was observed to be $\vrh_{\text{obs}}$.  Thus the  
The \emph{predicted} marginal density on $\Alg$ after the interaction with the observable 
without taking notice of the information provided by the observation is
in the Schrödinger picture
\begin{equation} \label{eq:marg-post-dens} 
  \rr_{\text{marg}} = \wob{\E{G}}^\dagger(\rr_{\Alg}) = 
    \sum_{j = 1}^n G^\dagger_{j}(\rr_{\Alg}) = \sum_{j = 1}^n \rp_{j}\, \rr_{\Alg}\, \rp_j .
\end{equation}
%is not consistent with the observed state $\vrh_{\text{obs}}$ on $\Clg$.
This flow of information for the prediction is depicted in \ffig{fig:CEX-1}.

\begin{figure}[H]
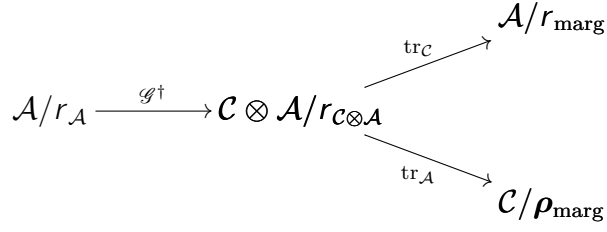
   
\[
\bfig
\morphism(0,300)/->/<800,0>[{\Alg/ \rr_{\Alg}}`{\Clg \otimes \Alg / \rr_{\Clg \otimes \Alg}};
{\E{G}^\dagger}] 
\Ctriangle(800,0)/<-``>/<800,300>[{\Alg/ \rr_{\text{marg}}}`
{\Clg \otimes \Alg / \rr_{\Clg \otimes \Alg}}`{\Clg/\vrh_{\text{marg}}};
{\tr_{\Clg}}``{\tr_{\Alg}}]
\efig
\]
\caption{Schrödinger picture / information flow due to interaction}
\label{fig:CEX-1}
\end{figure}

The Heisenberg picture dual to \feq{eq:marg-post-dens}, the change in the observable
$\ra \in \Alg_{sa}$, just due to the interaction with the instrument,
without taking notice of or evaluating the information provided by the observation, 
follows from \feq{eq:instrument-H} to be 
\begin{equation}   \label{eq:marg-post-obs}
\ra_{\text{marg}} = \wob{\E{G}}(\ra) = \sum_{j = 1}^n G_{j}(\ra)
    = \sum_{j = 1}^n \rp_{j}\, \ra\, \rp_j .
\end{equation}
Observe that both $\wob{\E{G}}$ and its adjoint $\wob{\E{G}}^\dagger$ fulfil the
assumptions in \feT{thm:RN-CP-maps}, which means that each $G_j$ resp.\ $G_J^\dagger$
could be described in the Stinespring form with a Radon-Nikodým effect.

Assume now that a certain $k \in [n] = \{1,\dots,n \}$ is the outcome
of the observation, i.e.\ the projection $\rp_k = G_k(\rone_{\Alg}) \in \PA{\Alg}$
has been observed.  The observation / sample on $\Clg = \Cn$ is thus given by the 
\emph{pure state} / rank-one \emph{observed} density  
$\vrh_{\text{obs}} := \ve_k \in \DEN{\Cn}_1$, the density representation of a 
character $\som_k \in \XA{\Clg}$ with $\som_k(\vf) = f_k$.

The information from this is that now the predicted density $\vrh_{\text{marg}} \in \Clg$ in 
\feq{eq:marg-post-prob} is not consistent with the observation and has thus 
to be changed to the \emph{observed} density $\vrh_{\text{obs}}$.
This means that also the predicted marginal density $\rr_{\text{marg}} \in \Alg$ in 
\feq{eq:marg-post-dens} is not consistent with the observation, and has to be adjusted
to incorporate the information from the observation.

Taking into account of this information that $k \in [n]$ was observed 
means that in \feqs{eq:marg-post-dens}{eq:marg-post-obs} only the term $j = k$ is
non-zero.  For the density this means that the \emph{posterior density} 
$\rr_{\text{post}}$ on $\Alg$ has to be proportional to
$G^\dagger_{k}(\rr_{\Alg}) = \rp_{k}\, \rr_{\Alg}\, \rp_k$.  Trace preservation requires 
that the \emph{posterior} density conditioned on the information of the observation thus 
is given by the so-called \emph{Lüders} or \emph{von Neumann - Lüders} 
rule \feq{eq:gen-post-den}
\begin{equation} \label{eq:Lueders} 
  \rr_{\text{post}} = 
      \frac{1}{\tr_{\Alg} G^\dagger_{k}(\rr_{\Alg})} \, G^\dagger_{k}(\rr_{\Alg}) 
     = \frac{1}{\tr_{\Alg} (\rr_{\Alg} \rp_k)} \, \rp_{k}\, \rr_{\Alg}\, \rp_k =
      \frac{1}{\prob_{\rr_{\Alg}}(\rp_k)}  \,  \rp_{k}\, \rr_{\Alg}\, \rp_k ,
\end{equation}
by observing that 
$\tr_{\Alg} (\rp_k \rr_{\Alg} \rp_k) = 
    \tr_{\Alg} (\rr_{\Alg} \rp_k)= \prob_{\rr_{\Alg}}(\rp_k)$.

One sees directly in \feq{eq:Lueders} that the \emph{prior} density from 
\feq{eq:marg-post-dens}, $\rr_{\Alg} \in \DEN{\Alg}$, is changed (by projection) 
due to the observation as little as possible --- just so much as to be consistent 
with the new information contained in the observation --- to arrive at the 
\emph{posterior} density, which connects it with \feT{thm:dual-embed}.  
As in the Heisenberg picture \feq{eq:marg-post-obs} also only the term $j = k$ is 
non-zero, an observable $\ra$ is changed to 
\begin{equation} \label{eq:Lueders-obs} 
  \Alg \ni \ra \mapsto G_k(\ra) = \rp_{k}\, \ra \, \rp_k  = \ra_{\text{post}} \in \Alg.
\end{equation}
\end{xmpn}

What may be noted already here is that the update to take account of the information
that there was either an interaction with the instrument, or both an interaction and
an observation, hast two different forms of being expressed: one is the Heisenberg picture
\feqs{eq:marg-post-obs}{eq:Lueders-obs} where observables, PVMs, and effects are changed
while the state is kept, and the other is the Schrödinger picture 
\feqs{eq:marg-post-dens}{eq:Lueders} where the state resp.\ the density is changed.

Both approaches give the same probabilities, both ways are the same 
update, e.g.\ from \feqs{eq:marg-post-dens}{eq:marg-post-obs} one obtains for all 
$\rr_{\Alg} \in \DEN{\Alg}$ and $\ra \in \Alg_{sa}$
\begin{multline}  \label{eq:same-prob-update}
   \tr (\rr_{\text{marg}}\, \ra) = \tr (\wob{\E{G}}^\dagger(\rr_{\Alg})\, \ra)
   = \sum_{j=1}^n \tr  ( G_j^\dagger(\rr_{\Alg})\, \ra ) = \sum_{j=1}^n \tr 
   \rp_j \, \rr_{\Alg}\, \rp_j \, \ra \\
   = \sum_{j=1}^n \tr \rr_{\Alg}\, \rp_j \, \ra \, \rp_j = \sum_{j=1}^n \tr  (
      \rr_{\Alg}\, G_j(\ra) )   = \tr (\rr_{\Alg} \, \wob{\E{G}}(\ra)) =
      \tr (\rr_{\Alg} \, \ra_{\text{marg}}) .
\end{multline}

A bit disturbing is the fact that the change in the observable \feq{eq:Lueders-obs}
as a map $G_k: \Alg \to \Alg$ is completely positive, but \emph{not unit preserving}
($G_k(\rone_{\Alg}) = \rp_k$), and thus does not fit under the channel paradigm to cover
the update taking into account the information from the observation.

One way to rectify this is with the concept of conditional expectation discussed above in 
the example list of possible channels, by redefining $G_k: \Alg \to \Blg$ with a yet
to be defined sub-algebra $\Blg \subseteq \Alg$.  The \emph{conditioning fact}   
is here that an observation of the PVM ($\{j\} \mapsto \rp_j$) is to take place, and this
interaction with the instrument defined by the PVM will lead to new information,
and this new information implies an update on the observables, i.e.\ \feq{eq:Lueders-obs}.
In the next \feX{ex:CEX-inst}, this all goes  into the definition of the sub-algebra $\Blg$, 
which is needed to apply \feD{def:CEX}.

Conditional expectation in probability algebras is a lively subject, e.g.\ see
\citep{Umegaki1954, Tomiyama1957, Takesaki1972, gudderHudson78, Gudder79, AccCecch82, 
Kadison2004, RedeiSummers2006p, YukalovSornette2009, YukalovSornette2016} 
and the references therein.
One can make the requirements very strict, with the effect that a conditional
expectation only exists under certain circumstances for infinite dimensional
algebras, cf.\ \citep{RedeiSummers2006p}; luckily for finite dimensional algebras
they do exist and there is even quite explicit descriptions \citep{Wolf2012}.
On the question of whether Lüders' rule \feqs{eq:Lueders}{eq:Bayes-CEXed} is 
the right way to formulate conditional expectation \feq{eq:Bayes-CEX-post}, 
see also \citep{CassinelliZanghi1983, CassinelliZanghi1984, Ozawa1984, Bobo2013, 
PerezLuis2022} and the references therein.

\begin{xmpn}[Update as Conditional Expectation]  \label{ex:CEX-inst}
The conditional expectation $\E{E}_{\Blg}(\ra)$ in \feq{eq:CEX} of some 
observable $\ra \in \Alg_{sa}$ is going to be a function 
of that observable and the PVM $\{j\} \to \rp_j \in \PA{\Alg}$ which it 
is being conditioned on.
This sub-*-algebra can thus be taken, in case $\rp_j$ is observed, as the one
generated by expressions of the kind $\rp_j \, \rb \, \rp_j \in 
(\rp_j \, \Alg \, \rp_j) =: \Blg_j$ for any $\rb \in \Alg$.  This leaves all 
possibilities for such a function for any $\rb \in \Alg$, and at the same times 
insures that self-adjoint elements $\ra \in \Alg_{sa}$
are mapped onto self-adjoint elements $\rp_j \, \ra \, \rp_j \in \Blg_{j,sa}$.

Observe that for $j \ne m$ the sub-algebras $\Blg_j = \rp_j \, \Alg \, \rp_j$ and 
$\Blg_m = \rp_m \, \Alg \, \rp_m$ are orthogonal.  Hence 
\[
   \C{N} = \bigoplus_{j = 1}^n \Blg_j   \subset \Alg
\]
would be a candidate sub-algebra, but it is too big.  One only wants the 
sub-algebra of $\Blg \subset \C{N}$ generated by expressions of the form 
$\sum_{j = 1}^n \rp_{j}\, \rb \, \rp_j $ in \feq{eq:marg-post-obs}, where each 
summand has the same $\rb \in \Alg$ in the middle.  Thus one may set
\begin{equation}  \label{eq:Bayes-sub-alg}
   \Blg := \CC \left\langle \left\lbrace \sum_{j = 1}^n \rp_{j}\, \rb \, \rp_j \mid 
     \rb \in \Alg \right\rbrace  \right\rangle \subset \C{N} \subset  \Alg .
\end{equation}

Hence it is clear that $\E{J}: \Blg \to \Alg$, where $\E{J}(\sum_{j = 1}^n 
\rp_{j}\, \rr\, \rp_j) = \sum_{j = 1}^n \rp_{j}\, \rr\, \rp_j \in \Alg$  
is the identity embedding of $\Blg$ into $\Alg$.  From \feT{thm:dual-embed} one 
may then see that its adjoint $\E{E}_{\Blg} := \E{J}^\dagger$, the orthogonal
projection onto $\Blg$, is the sought after conditional expectation. 

Note that these expressions are identical to $\wob{\E{G}} = \wob{\E{G}}^\dagger$
as used in \feX{ex:instr-cond} above; they will not be used here as they may lead
to confusion.  The reason is that the point of view in \feX{ex:instr-cond} was that
there is a flow of information from $\Alg$ to $\Clg \otimes \Blg \subset \Clg \otimes \Alg$
with a trace preserving map, whereas here we use the \emph{unital} conditional expectation
$\E{E}_{\Blg}:\, \Alg \to \Blg$, and thus the flow of information is in the direction
of the trace preserving $\E{J} = \E{E}_{\Blg}^\dagger:\, \Blg \to \Alg$, as information
from the observation is to be transferred to the whole algebra $\Alg$.

For an observable $\ra \in \Alg_{sa}$, the expression of the \emph{conditional expectation}
\begin{equation}  \label{eq:Bayes-CEX}
  \Ex(\ra | \Blg) := \E{E}_{\Blg}(\ra) = \E{J}^\dagger(\ra) = \ra_{\text{marg}} =
   \sum_{j = 1}^n \rp_{j}\, \ra \, \rp_j    \in \Blg
\end{equation}
contains the information after the interaction with the instrument, and prior to the
information from the observation.  It contains all the possibilities which might result 
from an observation, i.e.\ all the possibilities how any $\ra$ may be updated once some 
$\rp_k$ has been observed.  Its adjoint in the Schrödinger picture is the already
mentioned embedding $\E{J} = \E{E}^\dagger_{\Blg} = \Ex^\dagger(\cdot|\Blg):\, \Blg \to \Alg$.
The whole situation is shown in \ffig{fig:CEX-2}.
%\begin{equation}  \label{eq:Bayes-CEX-adj}
%  \Blg \ni \rr_{\Blg} = \sum_{j = 1}^n \rp_{j}\, \rr\, \rp_j \mapsto
%  \E{E}^\dagger_{\Blg}(\rr_{\Blg})  = \E{J}(\rr_{\Blg}) = 
%  \sum_{j = 1}^n \rp_{j}\, \rr\, \rp_j \in \Alg .
%\end{equation}

\begin{figure}[H]   
\[
\bfig
\morphism(0,0)/<-/<600,0>[{\Alg}`{\Blg};{\Ex^\dagger(\cdot|\Blg)}] 
\morphism(600,0)/<-/<600,0>[{\Blg}`{\Clg \otimes \Blg};{\tr_{\Clg}}] 
\morphism(1200,0)/<-/<600,0>[{\Clg \otimes \Blg}`{\Clg};{\tr_{\Blg}^\dagger}] 
\efig
\]
\caption{Conditional expectation induced information flow due to interaction}
\label{fig:CEX-2}
\end{figure}

After the observation of some $\rp_k$ --- this means a pure state / character 
$\som_k$ on $\Clg$ --- due to the new information, just the term $j=k$
is left in \feq{eq:Bayes-sub-alg}, and this becomes the projection (cf.\ \feq{eq:Lueders-obs})
\begin{equation}  \label{eq:Bayes-CEX-post}
  \Alg \ni \ra \mapsto \Ex(\ra | \Blg_k) := 
       \rp_k \, \ra\, \rp_k = \ra_{\text{post}} \in \Blg_k, 
\end{equation}
and in contrast to the conditional expectation in \feq{eq:Bayes-CEX}, this could be
called the \emph{post conditional expectation}, as it is after the conditioning information
has been taken into account.  But it usually also called conditional expectation,
sometimes leading to confusion on whether \feq{eq:Bayes-CEX} or \feq{eq:Bayes-CEX-post}
is meant.  Note that, as $\Ex(\rone_{\Alg} | \Blg_k) = \rp_k = \rone_{\Blg_k}$, the map
in \feq{eq:Bayes-CEX-post} is CPUP.

The new state on the algebra --- after the new information due to the 
observation $\rp_k$ --- can then finally be found through the adjoint of 
\feq{eq:Bayes-CEX-post},  it is the embedding
\begin{equation}  \label{eq:Bayes-CEXed}
  \Blg_k \ni \rr_{\text{post}} = 
  \frac{1}{\prob_{\rr_{\Alg}}(\rp_k)}  \,  \rp_{k}\, \rr_{\Alg}\, \rp_k 
  \mapsto  \Ex^\dagger(\rr_{\text{post}} | \Blg_k) = 
     \frac{1}{\prob_{\rr_{\Alg}}(\rp_k)}  \,  \rp_{k}\, \rr_{\Alg}\, \rp_k \in \Alg,
\end{equation}
confirming Lüders' rule \feq{eq:Lueders}.  This \emph{posterior density} 
then defines a new state on $\Alg$
which is consistent with the new information from the observation $\rp_k$.
Thus Lüders' rule \feqs{eq:Lueders}{eq:Lueders-obs} can be seen in 
the light of conditional expectation and the updating following an observation.
In classical commutative probability this is connected with Bayes's theorem and 
is often called \emph{Bayesian updating}.  One may keep the same term also
in non-commutative probability, although Bayes's theorem does not hold generally
in this instance  \citep{CassinelliZanghi1984, Accardi00b, Bobo2013, 
PerezLuis2022, Accardi2018, Accardi2022}, and it is not the basis for the update.

The expectation computed with this new density on $\Alg$, 
with  \feqs{eq:Lueders}{eq:Lueders-obs}, is
\begin{multline}  \label{eq:Bayes-CEXpected}
   \Alg \ni \ra \mapsto \Ex(\ra | \rp_k) := \tr_{\Alg} [ \rr_{\text{post}} \ra ] =
   \frac{1}{\prob_{\rr_{\Alg}}(\rp_k)} \, \tr_{\Alg} [ \rp_{k}\rr_{\Alg}\rp_k\ra ] \\ = 
 \frac{1}{\prob_{\rr_{\Alg}}(\rp_k)}\,\tr_{\Alg}[ \rp_{k}\rr_{\Alg}\rp_k \rp_k\ra\rp_k ]
   = \tr_{\Blg} [ \rr_{\text{post}} \ra_{\text{post}} ] \in \CC ,
\end{multline}
and this could be called the \emph{conditioned expectation}, as it is the expectation
\emph{after} the new information of the conditioning event has been taken into account.
The functional $\Ex(\cdot | \rp_k) \in \SA{\Alg}$ is the \emph{new state} to reflect this.
Note that in this new state $\rp_k$ is a \emph{sure} event, as  $\Ex(\rp_k | \rp_k) =
\frac{1}{\prob_{\rr_{\Alg}}(\rp_k)} \, \tr_{\Alg} [ \rp_{k}\rr_{\Alg}\rp_k \rp_k ] = 
\frac{1}{\prob_{\rr_{\Alg}}(\rp_k)} \, \tr_{\Alg} [ \rr_{\Alg}\rp_k ]= 1$.

This last case from \feq{eq:Bayes-CEXpected} is shown in \ffig{fig:CEX-3}.

\begin{figure}[H]   
\[
\bfig
\morphism(0,00)|a|/<-/<600,0>[{\Alg}`{\Blg_k};{\Ex^\dagger(\cdot|\Blg_k)}] 
\morphism(600,00)|a|/<-/<600,0>[{\Blg_k}`{\Clg \otimes \Blg_k};{\tr_{\Clg}^{\phantom{\dagger}}}] 
\morphism(1200,00)|a|/<-/<600,0>[{\Clg \otimes \Blg_k}`{\Clg};{\tr_{\Blg_k}^\dagger}] 
\morphism(1800,00)|a|/<-/<600,0>[{\Clg}`{\CC};{\som_k^\dagger}] 
%\morphism(0,100)|a|/{@{>>}@/^1.5em/}/<2400,0>[{\Alg}`{\CC};{\Ex(\cdot|\rp_k)}] 
\efig
\]
\caption{Conditional expectation induced information flow due to observation}
\label{fig:CEX-3}
\end{figure}

One may also observe that according to  \feC{coro:RN-alg}, the conditioned expectation 
state $\Ex(\cdot | \rp_k) \in \SA{\Alg}$ has a \emph{Radon-Nikodým} derivative
$C := \di \Ex(\cdot | \rp_k) / \di \Ex(\cdot) \in \Lop(\Alg)$ w.r.t.\ the \emph{prior}
state $\Ex(\cdot)$, so that for all $\ra \in \Alg$ one has $\Ex(\ra | \rp_k) = \Ex(C \ra)$.
An explicit form of $C$ may be found in \citep{Wolf2012}.
\end{xmpn}

This \feX{ex:CEX-inst} shows how the interaction with the instrument --- via the
conditional expectation --- and the observation --- via the conditioned expectation ---
can both be brought under the channel paradigm.  One may also note how the flow of 
information is from the instrument interaction (in \ffig{fig:CEX-2}) resp.\ the 
observation $\som_k$ to the original algebra $\Alg$ of observables in \ffig{fig:CEX-3}.  

One might add that in both instances, i.e.\ the one in \ffig{fig:CEX-2} and the one in
\ffig{fig:CEX-3}, the last mapping into the original algebra $\Alg$ is the identity
embedding and thus a purely formal
one.  This is due to the fact that according to \feqs{eq:Bayes-CEX}{eq:Bayes-CEX-post}
all the elements $\ra \in \Alg$ are updated and mapped into a sub-algebra of $\Alg$.
This is either just due to the instrument interaction (\ffig{fig:CEX-2}) in 
\feq{eq:Bayes-CEX} a mapping into the sub-algebra $\Blg$, or in case of an additional
observation (\ffig{fig:CEX-3}) in \feq{eq:Bayes-CEX-post} into the sub-algebra $\Blg_k$.
And as \feqss{eq:marg-post-dens}{eq:Lueders}{eq:Bayes-CEXed} show, the corresponding
densities $\rr_{\text{marg}} \in \Blg$ and $\rr_{\text{post}} \in \Blg_k$ are also
in the corresponding sub-algebra to provide probabilities according to 
\feqs{eq:same-prob-update}{eq:Bayes-CEXpected}.  Hence, any further operation
might just as well proceed from one of the probability sub-algebras, either $\Blg$ or 
$\Blg_k$, depending on whether the information from the observation has been taken 
into account or not, and the original probability algebra $\Alg \supset \Blg \supset \Blg_k$
is too big and not needed any more.

\paragraph{Conditional Probability:}
Like in  \feD{def:probability}, \emph{conditional probability} will be based on
conditional expectation.  It was already mentioned that this concept is still
being debated, e.g.\ see
\citep{Umegaki1954, gudderHudson78, Gudder79, AccCecch82, Kadison2004, RedeiSummers2006p,
YukalovSornette2009, YukalovSornette2016} and the references therein.

The question is whether Lüders' rule \feqs{eq:Lueders}{eq:Bayes-CEXed} is 
the right way to formulate conditional expectation \feq{eq:Bayes-CEX-post},
and conditional probability based on it, e.g.\ cf.\ 
\citep{CassinelliZanghi1983, CassinelliZanghi1984, Ozawa1984, Bobo2013, PerezLuis2022}.
Like many extensions into a wider realm, often it is not possible to keep all properties of
some construction when extending it.
It is clear that in order to take account of new information in the form of what is in many
instances called \emph{data assimilation}, such a concept is needed, cf.\ e.g.\ 
\citep{FreemannGiannakisMintzEtAl2022}, but, on the other hand, Bayes's theorem does not
hold in the non-commutative case \citep{CassinelliZanghi1984, Bobo2013, 
PerezLuis2022, Accardi2018, Accardi2022}.   This may be seen as the defining difference
between commutative and non-commutative probability theory \citep{Accardi00b}.

Using the analogue to \feD{def:probability}, one can define what
might be called \emph{conditional probability}
by employing the new state $\Ex(\cdot | \rp_k) \in \SA{\Alg}$ from \feq{eq:Bayes-CEXpected}.
For $\rqq \in \PA{\Alg}$ one can just define a probability as before in \feD{def:probability}: 
\begin{equation}   \label{eq:cond-prob}
   \prob(\rqq | \rp_k) := \Ex(\rqq | \rp_k).
\end{equation}
And in case $\rqq$ and $\rp_k$ commute, it behaves like a classical Kolmogorovean
conditional probability, as 
\[
  [\rqq, \rp_k] = \rnul \quad \Rightarrow \quad
  \prob(\rqq | \rp_k) = \frac{1}{\prob_{\rr_{\Alg}}(\rp_k)}\,
  \tr_{\Alg}[ \rp_{k}\rr_{\Alg} \rp_k \rqq \rp_k ] = 
  \frac{\prob_{\rr_{\Alg}}(\rqq \rp_k)}{\prob_{\rr_{\Alg}}(\rp_k)},
\] 
and $\prob_{\rr_{\Alg}}(\rqq \rp_k)$ may be
seen as the \emph{joint probability} of $\rqq, \rp_k \in \PA{\Alg}$.
But in case they do not commute and are not compatible, there is no joint probability, 
and some properties are lost; 
this is discussed extensively in e.g.\  \citep{Bobo2013} and the references therein.

In this context, it is worthwhile to note that the classical Kolmogorovean
conditional probability has a unique property.  To paraphrase 
\citep{CassinelliZanghi1983, Bobo2013},
given a classical probability measure $\prob$ on a $\sigma$-algebra $\sigalg$, and two 
events $A, B \in \sigalg$, the classical conditional probability, defined as usual by the 
ratio $\prob(A | B) := \prob(A \cap B)/\prob(B)$ and using the joint probability
$\prob(A \cap B)$, is characterised by the fact that it is the \emph{only} 
probability measure defined on all of $\sigalg$, such that for $A \subseteq B$
the conditioning on $B$ just involves a renormalisation of the initial probability 
measure; i.e.\ $\prob(A | B) = \prob(A \cap B)/\prob(B) = \prob(A)/\prob(B)$.

Now assume that $\Alg = \LHA$, where $\Hvk$ is a finite dimensional Hilbert space
with $\dim \Hvk > 2$ and density operator $R \in \DEN{\Alg}$, and let 
$\opb{L}(\PA{\Alg})$ be the lattice of all orthogonal 
projections, cf.\ \feP{prop:proj-matrix-measurement}.  Further, let $P \in \PA{\Alg}$ 
be an orthogonal projection, and consider the sub-lattice 
$\opb{L}(\le P) := \{ Q \in \PA{\Alg}, Q\le P \}\subset \opb{L}(\PA{\Alg})$ of
all projections less or equal to $P$.  Note that, as for $Q \in \opb{L}(\le P)$ one
has $Q \le P$, it follows that $QP = PQ = Q$, and thus all $Q \in \opb{L}(\le P)$
commute with $P$.  Then it holds that \citep{CassinelliZanghi1983, Bobo2013}:

\begin{thm}[Existence and Uniqueness of Lüders Conditional Probability] \label{thm:Lueders}
In the setting just sketched, assume that $\prob_R(P) = \Ex_R(P) = \tr (RP) \ne 0$ 
(cf.\ \feD{def:probability}).
In an analogous fashion to the classical definition and to the definition in
\feq{eq:cond-prob}, define for any $Q \in \opb{L}(\le P)$ the function 
$\Pr(Q) := \prob_R(Q) / \prob_R(P) = \prob_R(QP) / \prob_R(P)$; obviously this 
is just a renormalisation of the initial probability measure $\prob_R(Q)$.  

Then there is a \emph{unique} 
extension $\prob_R(\cdot | P)$ of $\Pr(\cdot)$ to all of $\PA{\Alg}$,
%\begin{compactitem}
%  \item $\Pr(\cdot)$ is a probability on $\opb{L}(\le P)$;
%  \item there is an extension $\prob_R(\cdot | P)$ of $\Pr(\cdot)$ to all of $\Alg$;
%  \item the extension is \emph{unique}, 
        it is derived from the state given by the conditioned expectation 
        (cf.\ \feq{eq:Bayes-CEXpected})  $\Ex(\cdot | P) \in \SA{\Alg}$ on $\Alg$ 
        --- it thus satisfies \feq{eq:Gleason} --- and  
        whose density operator $R_P$ is given by Lüders' rule (cf.\ \feq{eq:Lueders}) 
        $R_P := P R P / \prob_R(P)$, such that for all $Q \in \PA{\Alg}$
        \begin{equation}   \label{eq:cond-prob-1}
           \prob_R(Q | P) = \Ex_R(Q | P) =\tr (R_P Q) = \frac{\tr (P R P Q)}{\prob_R(P)}.
        \end{equation}
%        In case the density $R_{\psi} = \ket{\psi}\bra{\psi} \in \PA{\Alg}$ comes from a 
%        pure state given by the unit vector $\psi \in \Hvk$, this may be written as
%        \[
%           \prob_{\psi}(Q | P) := \prob_{R_{\psi}}(Q | P) = 
%              \frac{\bkt{P\psi}{Q P\psi}}{\bkt{P\psi}{P \psi}}.
%        \]
%\end{compactitem}
\end{thm}

It should be clear that the theorem holds also for any finite dimensional probability
algebra which has no factor isomorphic to $\Lop(\Kvk)$ with $\dim \Kvk = 2$
(factor of type $\mrm{I}_2$) in their factor decomposition, 
cf.\ \feT{thm:factor-decomp-A} and \feT{thm:factor-decomp-H}. 

This \feT{thm:Lueders} is often taken as reason enough to define conditional probabilities 
like in \feq{eq:cond-prob}, based on conditional expectation like in \feq{eq:Bayes-CEXpected}.
But, as noted already, some additivity properties are lost due to non-commutativity
and Bayes's theorem does not hold --- one reason being that it is not possible to
formulate \emph{joint} probabilities in every case \citep{Accardi00b, Accardi2018, Accardi2022}
--- so that sometimes an extreme sounding position is 
argued \citep{Bobo2013}, in that there actually is \emph{no} fully viable definition 
of conditional probability in the non-commutative case.  We will not enter further 
into this discussion here.

\paragraph{Channel-State Duality:}
The next result, which establishes a one-to-one relationship between channels and 
ordinary linear maps on Hilbert spaces themselves, is similar to the ones mentioned
previously; in fact it is closely related to \feT{thm:Choi}, \feT{thm:Stinespring},
and \feC{coro:Choi-Stinespring}.  Here one of the simpler versions is presented,
for variations on this theme, and different versions
of it, cf.\ \citep{FrembsCavalcanti2024}.

\begin{thm}[Jamiołkowski-Choi]  \label{thm:Choi-Jamio}
Let $\Hvk$ (with $\dim \Hvk = n$ and orthonormal basis $\{ \ket{v_j} \}_{j=1}^n$) 
and $\Kvk$ (with $\dim \Kvk = m$) be two finite dimensional Hilbert spaces, 
and let $\E{Q}: \LHA \to \Lop(\Kvk)$ be a linear map (a super-operator).  
Denote by
\[
\ket{W} = \frac{1}{\sqrt{n}}\, \sum_{j=1}^n \ket{v_j} \otimes \ket{v_j}
  \in \Hvk \otimes \Hvk
\]
a (maximally entangled) unit vector in $\Hvk^{\otimes 2}$, and by
$\tnb{P} = \ket{W} \bra{W}$ the orthogonal projector onto it's subspace.  Then
\begin{equation}  \label{eq:Ch-JA-AtoH}
  \Lop(\LHA, \Lop(\Kvk)) \ni
    \E{Q} \mapsto \tnb{Q} = (\E{Q} \otimes I_{\Hvk}) \tnb{P} \in \Lop(\Kvk\otimes\Hvk)
\end{equation}
is a linear isomorphism $\E{Q} \mapsto \tnb{Q}$ between those spaces of mappings indicated.  

The inverse mapping $\tnb{Q} \mapsto \E{Q}$ is given by 
mapping a $\tnb{Q} \in \Lop(\Kvk\otimes\Hvk)$ onto  
\begin{equation}  \label{eq:Ch-JA-HtoA}
  \E{Q}:\; \LHA \ni A \mapsto
    \E{Q}(A) = n\,\tr_{\Hvk}(\tnb{Q}(I_{\Kvk}\otimes A^\dagger)) \in \Lop(\Kvk) .
\end{equation}
\end{thm}

Observe that the isomorphism of the 
Jamiołkowski-Choi-\feT{thm:Choi-Jamio} applies
not just to channels, but to any $\E{Q} \in \Lop(\LHA, \Lop(\Kvk))$.
The importance of this isomorphism is that it allows to read off properties of the
map $\E{Q} \in \Lop(\LHA, \Lop(\Kvk))$ by looking at the ``normal''
Hilbert space operator $\tnb{Q} \in \Lop(\Kvk\otimes\Hvk)$ of the bi-partite system
$\Kvk\otimes\Hvk$.  These correspondences
of properties are collected in Table~\ref{table:Prop-Ident}.

\begin{table}[ht]
\centering 
\begin{tabular}{| c | c | c |} 
\hline\hline  
  \tiny{\phantom{a}} & & \\ [-1.5ex]
  Operator $\tnb{Q}$ & Map $\E{Q}$ & Property \\ [0.5ex] 
\hline 
  \tiny{\phantom{a}} & & \\ [-1.8ex]
$\tnb{Q} = \tnb{Q}^\dagger$ & $\E{Q}(A^\dagger) = \E{Q}(A)^\dagger $ 
                 & Hermitean, *-linearity  \\
$\tnb{Q} \ge 0 $ & CP & (complete) positivity \\ 
$\tr_{\Hvk}(\tnb{Q}) = \frk{1}{m}\, I_{\Kvk}$ & $\E{Q}(I_{\Hvk}) = I_{\Kvk} $
                 & unital, unit preserving (UP) \\
$\tr_{\Kvk}(\tnb{Q}) = \frk{1}{n}\, I_{\Hvk}$ & $\E{Q}^\dagger(I_{\Kvk}) = I_{\Hvk} $
                 & trace preserving (TP) \\
%$\tr(\tnb{Q}) = 1$ & $\tr(\E{Q}(I_{\Hvk}))=m,\; \tr(\E{Q}^\dagger(I_{\Kvk}))=n$ & normalisation \\
[1ex] % [1ex] adds vertical space
\hline %inserts single line
\end{tabular}
\caption{Correspondence of Properties} % title of Table
\label{table:Prop-Ident} % is used to refer this table in the text
\end{table}

One may observe that for a map to be a channel 
$\E{Q} \in \Lop(\LHA, \Lop(\Kvk))$, 
it has to possess all the properties in the middle column of 
Table~\ref{table:Prop-Ident}.  This means that the corresponding Hilbert space 
operator $\tnb{Q} \in \Lop(\Kvk\otimes\Hvk)$ has all the properties in the
first column of Table~\ref{table:Prop-Ident}.

\begin{coro}[Channel-State Duality]  \label{coro:Choi-Jamio}
Assume now that $\E{Q} \in \Lop(\LHA, \Lop(\Kvk))$ is a completely positive
unit preserving map (CPUP), i.e.\ a \emph{channel}, checking all points in the second 
column of Table~\ref{table:Prop-Ident}.
This means that the corresponding Hilbert space operator 
$\tnb{Q} \in \Lop(\Kvk\otimes\Hvk)$ satisfies all points in the first column of
Table~\ref{table:Prop-Ident}.

First, note that the fourth and fifth row of Table~\ref{table:Prop-Ident} show that
the partial traces $\tr_{\Hvk}(\tnb{Q})$ and $tr_{\Kvk}(\tnb{Q})$ are density
operators on $\Lop(\Kvk)$ and $\LHA$, respectively.  Moreover,
\[
  \tr(\tnb{Q}) = \tr_{\Kvk}(\tr_{\Hvk}(\tnb{Q})) = \tr_{\Hvk}(\tr_{\Kvk}(\tnb{Q})) = 1.
\]

Together with the other properties in the first column of Table~\ref{table:Prop-Ident},
it becomes clear that the corresponding Hilbert space operator 
$\tnb{Q} \in \Lop(\Kvk\otimes\Hvk)$ is a \emph{density operator} on the
Hilbert space $\Kvk\otimes\Hvk$.  Thus the \feqs{eq:Ch-JA-AtoH}{eq:Ch-JA-HtoA} 
implement a \emph{duality} of channels between systems $\Lop(\Kvk)$ and
$\LHA$, and states on the combined system $\Lop(\Kvk\otimes\Hvk)$, 
represented by density operators, the so called \emph{channel-state duality}.
\end{coro}

\subsection{Interacting Fock Space and Quantum Decomposition}  \label{SS:Fock}
In this sub-section the idea is to embed an algebra of commuting random variables,
i.e.\ a classical algebra, into a larger, non-commutative algebra, which can help
in analysing those random variables.  For the sake of simplicity, here we only look
at one random variable, a case which still transmits the flavour of the subject.
Connections will come up between the Fock space, orthogonal polynomials, Krylov
subspaces, and the Lanczos method.  These methods have led to new views on diverse
classical subjects, like graphs, reaction networks, random walks, and Markov chains.
For a broader survey, cf.\ \citep{Meyer1995}.

\newcommand{\annh}[1]{\tensor{#1}{^-}}
\newcommand{\crea}[1]{\tensor{#1}{^+}}
\newcommand{\pres}[1]{\tensor{#1}{^o}}

\paragraph{Fock Space and Ladder Operators:}
The following idea arose in quantum physics to describe the so-called second 
quantisation, so to introduce and motivate it, some physical terms will be used.  
Given a Hilbert space $\C{J}$  to describe
the states resp.\ modes of some kind of elementary particle, 
the Fock space is defined as $\C{F}(\C{J}) := \bigoplus_{k \ge 0} \Kvk_k$,
with the \emph{sectors} $\Kvk_k := \C{J}^{\otimes k}$,
e.g.\ cf.\ \citep{Attal-notes}.
This is the \emph{full} or \emph{Boltzmann\-ian}
\emph{Fock space} of \emph{distinguishable} particles resp.\ modes.
Each tensor power resp.\ sector in this \emph{non-interacting} Fock space is equipped with
the natural inner product from the tensor power, and similarly then the direct sum.  
The one-dimensional space $\C{J}^{\otimes 0} \cong \CC$ is often modelled as 
$\spn_{\CC} \Psi_0$, where the unit vector $\Psi_0$ is assumed to be orthogonal to
$\C{J}$, and is called the \emph{vacuum vector}.
For elementary particles like Fermions or Bosons, one takes the 
anti-symmetric tensor product for the Fermionic resp.\ the symmetric tensor product  
for the Bosonic Fock space in order
to accommodate the \emph{Pauli exclusion principle} resp.\ the
\emph{indistinguishability} of individual Bosons.  
The symmetric tensor product resp.\ Bosonic Fock space leads 
to Gaussian measures \citep{Janson1997}, cf.\ also \feT{thm:q-Fock-moments}.

As state one can take the vector or pure state $\svpi_{\Psi_0}(A) := \bkt{\Psi_0}{A \Psi_0}$
(also called the vacuum state) for $A \in \Lop(\C{F}(\C{J}))$, giving a probability algebra 
$(\Lop(\C{F}(\C{J})),\svpi_{\Psi_0})$ with GNS-representation 
$(\C{F}(\C{J}),\Psi_0)$.
An important set of operators in that probability algebra are the \emph{ladder operators}, 
i.e.\ the \emph{creation operator} $\crea{A}$ and its adjoint the \emph{annihilation operator} 
$\annh{A} = (\crea{A})^\dagger$, and the \emph{number operator} $N$.  
The latter is the easiest to describe and the same in all variants of the Fock space: 
each sector $\Kvk_n$  resp.\ tensor power $\C{J}^{\otimes n}$ is an eigenspace, 
and for $\psi_n \in \Kvk_n$ one defines $N \psi_n := n \psi_n$.  
In this spectral resolution, other \emph{diagonal} operators
are easily described:  Let $\{ \alpha_n \}_{n=0}^\infty$ be a sequence of real numbers,
then the \emph{operator function} $\Hf{{\vek{\alpha}}}{N}$ of the number operator $N$ is again
defined on each sector $\Kvk_n$ by $\Hf{{\vek{\alpha}}}{N} \psi_n := \alpha_n \psi_n$.
Similarly, via such a functional calculus, one defines on each sector $\Kvk_n$ the
diagonal operator $\; \Hf{{\vek{\alpha}}}{(N+I)} \psi_n := \alpha_{n+1} \psi_n$.

The \emph{creation and annihilation} operators are \emph{ladder} resp.\ \emph{shift} 
operators.  The creation operator for a new mode $\chi \in \C{J}$ is defined on 
each sector $\Kvk_n = \C{J}^{\otimes k}$ for elementary tensors 
$\psi_n = \bigotimes_{k=0}^n \vphi_k$ (with $\vphi_k \in \C{J}$ and 
$\psi_0 = \vphi_0 = \Psi_0$) by
%\[
$  \crea{A}(\chi): \, \Kvk_n \ni \psi_n \mapsto 
     \psi_n  \otimes \chi \in \Kvk_{n+1}$.
%\]  
Similarly, its adjoint, the annihilation operator, is given by $\annh{A}(\chi) \Psi_0 = 0$ 
for $n=0$, and for  $n>0$ by 
%\[
$  \annh{A}(\chi): \,  \Kvk_n \ni  \bigotimes_{k=0}^n\vphi_k \mapsto
     \bkt{\chi}{\vphi_n}\; \bigotimes_{k=0}^{n-1} \vphi_k \in \Kvk_{n-1}$.
%\]

These ladder operators are defined for some mode or state $\chi \in \C{J}$ for each 
sector $\Kvk_n$ in an analogous way, without regard to the number of modes, 
using the natural mode-number independent inner
product on the sector $\Kvk_n = \C{J}^{\otimes n}$; 
explaining the \emph{non-interacting} character.
%  Often one considers just the
%sub-*-algebra generated by the operators $\{N, \crea{A}(\chi), \annh{A}(\chi)\}$
%(for all $\chi \in \C{J}$) of the probability algebra $(\Lop(\C{F}(\C{J})),\svpi_{\Psi_0})$. 

In case $\dim \C{J} = 1$, i.e.\ $\C{J} = \spn \{ \Psi_1 \}$, all tensor powers resp.\ sectors
$\C{J}^{\otimes n} = \spn \{ \Psi_1^{\otimes n} \}$ are also one-dimensional.  Hence one may, 
for the sake of simplicity, just take as a unitarily equivalent substitute for $\C{F}(\C{J})$
a separable Hilbert space $\Hvk$, the closure of $\spn \{ \Psi_0, \Psi_1, \dots \}$, 
where the the $\{\Psi_n\}$ are an orthogonal sequence, and set $\Kvk_n := \spn \{ \Psi_n \}$,
cf.\ e.g.\  \citep{AccardiBozejko1998, HoraObata07, Obata2017}; this avoids unnecessarily 
dealing with tensor products.

\paragraph{One-mode Interacting Fock Space:}
As stated above, here only the special case of an one-mode interacting Fock space
built from one-dimensional sectors is considered, 
but which still is useful when analysing just one random variable,
cf.\ e.g.\ \citep{AccardiBozejko1998, AccKuoStan04, AccKuoStan05, AccardiNahni2003,
 HoraObata07, Obata2017}, and which shows how the inner product and definition of ladder
operators on each $n$-particle resp.\ -mode sector is changed such that the 
interaction is taken care of.

\begin{defi}[Jacobi sequence]   \label{def:Jacobi-seq}
Let $\{ \lambda_n \}_{n=0}^\infty$ be a sequence of non-negative real numbers with
$\lambda_0 = 1$, and the property that if for some $m_0 \ge 0$ one
has $\lambda_{m_0 + 1} = 0$, this implies $\lambda_n = 0$ for all $n > m_0$.
This is called a \emph{Jacobi sequence}.
If such a $m_0 \in \D{N}$ exists, the sequence is called of \emph{finite type}, otherwise
of \emph{infinite type} ($m_0 = \infty$).
\end{defi}

A first Jacobi sequence will be used to define the interacting inner products
on each sector of the Fock sum.

\begin{defi}[One-mode interacting Fock space]   \label{def:interact-Fock}
As indicated above, a concrete one-mode interacting Fock space is defined 
in the simplest manner by taking a complex infinite dimensional separable Hilbert 
space $\Hvk$ with inner product $\bkt{\cdot}{\cdot}_{\Hvk}$, and a complete orthogonal system
$\{ \Psi_0, \Psi_1, \dots \}$, and letting for $n \in \D{N}_0$ the corresponding summand 
resp.\ sector be $\Kvk_n := \spn \{ \Psi_n \}$.  It is often convenient to extend this by 
setting $\Psi_{-1} = 0$ and $\Kvk_{-1} = \spn \{ \Psi_{-1} \} = \{ 0 \}$.

Let $\{ \lambda_n \}$ be a Jacobi sequence as in \feD{def:Jacobi-seq}.
  The inner product
on the sector $\Kvk_n$ is defined for $n \ge 0$ and $\xi_n, \eta_n \in \Kvk_n$ as
\begin{equation}   \label{eq:interact-innprod}
   \bkt{\xi_n}{\eta_n}_n := \lambda_{n} \bkt{\xi_n}{\eta_n}_{\Hvk} .
\end{equation}
It will be assumed that the $\Psi_n$ in the orthogonal system are 
normalised by $\bkt{\Psi_n}{\Psi_n}_n = 1$.
Let $m_0$ be the extended natural number from \feD{def:Jacobi-seq} which indicates
for $\{ \lambda_n \}$ a Jacobi sequence of finite or infinite type.  Then the
algebraic direct sum
\begin{equation}   \label{eq:interact-FockSpace}
   \Gamma :=  \bigoplus_{n=0}^{m_0} \Kvk_n ,
\end{equation}
a pre-Hilbert space
equipped with the inner products \feq{eq:interact-innprod} on each sector,
yields the direct sum inner product $\bkt{\cdot}{\cdot}_{\Gamma}$, and
defines the basic \emph{one-mode interacting full Fock space}.
\end{defi}

\begin{defi}[Number and Ladder Operators]   \label{def:ladder}
Let $\{ \omega_n \}_{n=0}^\infty$ be a new Jacobi sequence, dependent on 
the Jacobi sequence $\{\lambda_n \}$ from \feD{def:interact-Fock} by defining
$\omega_0 = 1$, and 
for $n>0$: $\omega_n = \lambda_{n} / \lambda_{n-1}$, with the understanding
that in the situation $ 0 / 0$ one sets $\omega_n = 0$.
%\note{\\ Or is it $\omega_n = \lambda_{n+1} / \lambda_{n-1}$ ?}

In the algebra $\Lop(\Gamma)$, the \emph{number operator} $N$ and the 
\emph{diagonal operators} $\Hf{{\vek{\alpha}}}{N}$ for a sequence of real numbers 
$\vek{\alpha} = \{ \alpha_n \}$ are defined as above on each sector $\Kvk_n$
--- each sector $\Kvk_n$ is an eigenspace --- by
\begin{align}   \label{eq:number-def}
   N: \Kvk_n \ni \Psi_n  &\mapsto  n\; \Psi_{n} \in \Kvk_{n}; 
   \\  \label{eq:alpha_N-def}
   \Hf{{\vek{\alpha}}}{N}: \Kvk_n \ni \Psi_n  &\mapsto 
       \alpha_{n}  \; \Psi_{n} \in \Kvk_{n};
      \\  \label{eq:alpha_N1-def} \text{note also }\;
   \Hf{{\vek{\alpha}}}{(N+I)}: \Kvk_n \ni \Psi_n  &\mapsto 
       \alpha_{n+1}  \; \Psi_{n} \in \Kvk_{n}.   
\end{align}

The ladder operators, the \emph{creation and annihilation operators} $\crea{B}, \annh{B} 
\in \Lop(\Gamma)$, now taking into account the interaction of the number of modes, 
are defined on each sector by
\begin{align}   \label{eq:crea-def}
   \crea{B}: \Kvk_n \ni \Psi_n  &\mapsto \sqrt{\omega_{n+1}}\; \Psi_{n+1} \in \Kvk_{n+1}; 
   \\  \label{eq:annh-def}
   \annh{B}: \Kvk_n \ni \Psi_n  &\mapsto \sqrt{\omega_{n}}  \; \Psi_{n-1} \in \Kvk_{n-1}.   
\end{align}
The operators $N$, $\Hf{{\vek{\alpha}}}{N}$, and $\Hf{{\vek{\alpha}}}{(N+I)}$ are self-adjoint, 
whereas the ladder operators $\crea{B}$ and $\annh{B}$ are adjoints of each other.
\end{defi}

Note that one often starts with a Jacobi sequence $\{ \omega_n \}$ like in \feD{def:ladder}
and defines the Jacobi sequence $\{ \lambda_n \}$ for \feD{def:interact-Fock}
by $\lambda_0 = 1$ and $\lambda_{n} := \lambda_{n-1} \omega_n$ for $n  > 0$; thus giving
$\lambda_{n} = \prod_{k=0}^n \omega_k$.

\begin{defi}[One-mode interacting Fock Probability Algebra]   \label{def:interact-Fock-pr}
Taking the basic one-mode interacting full Fock space from \feD{def:interact-Fock},
and the ladder operators from \feD{def:ladder}, one calls
the quadruple $\Gamma_{\{ \lambda_n \}} := (\Gamma, \Psi_0, \annh{B}, \crea{B})$
the interacting Fock system associated to the Jacobi sequence $\{ \omega_n \}$.

$(\Lop(\Gamma), \svpi_{\Psi_0})$ is called the associated interacting Fock probability 
algebra with GNS-repre\-sentation $(\wob{\Gamma}, \Psi_0)$.  Often just the algebra
generated by $\annh{B}, \crea{B}$, and $N$ is considered.
\end{defi}

One easy consequence is:
\begin{prop}   \label{prop:comm-rel-basic}
Let $\Gamma_{\{ \lambda_n \}}$ be an interacting Fock system.  
Define the associated Jacobi sequence $\vek{\omega} = \{\omega_n\}$ 
according to \feD{def:ladder}, and denote by $\wtl{\vek{\omega}}$ the
same sequence, except for $\wtl{\omega}_0 = 0$.   Then
\begin{equation}   \label{eq:comm-rel}
   \crea{B}\annh{B}  = \Hf{{\wtl{\vek{\omega}}}}{N}, \quad \text{ and } \; 
   \annh{B}\crea{B}  = \Hf{{\vek{\omega}}}{(N+I)} .
\end{equation}
Additionally, it holds that for all $n \ge 0$:
\begin{equation}   \label{eq:power-crea-rel}
   (\crea{B})^n \,\Psi_0 = \sqrt{\omega_n \cdots \omega_0}\; \Psi_n 
       = \sqrt{\lambda_n}\; \Psi_n\, , \; \text{ so that } \; 
       \nd{(\crea{B})^n \,\Psi_0}_{\Gamma} = \sqrt{\lambda_n} \, .
\end{equation}
\end{prop}   

Note that with the sequence $\vek{\gamma} = \{\gamma_n \}_{n=0}^\infty$, 
given by  $\gamma_0 := \omega_1$ and for $n>0$ by $\gamma_n := \omega_{n+1} - \omega_n$,
one can describe the interacting \emph{commutation relations}
\begin{equation}  \label{eq:comm-rels-gamma}
  [\annh{B}, \crea{B}] = \annh{B}\, \crea{B} - \crea{B} \,  \annh{B} =
     \Hf{{\vek{\gamma}}}{N} .
\end{equation}

One way to change the probability distribution and define a Jacobi sequence 
is by ``deforming'' the
commutation relations \citep{BozejkoKuemSpeich1997}:

\begin{defi}[$q$-deformed Commutation Relations]   \label{def:q-comm-rel}
For $-1 \le q \le 1$, define the associated Jacobi sequence according to 
\feD{def:ladder} of the \emph{$q$-Fock probability algebra} by
$\{ \omega_0 =1, \omega_n = [n]^q \}$ (where for $n>0$ the expression
$[n]^q := \sum_{k=0}^{n-1} q^k$ are the \emph{$q$-numbers of Gauss}), 
and $\{ \lambda_{n}  = \prod_{k=0}^{n} \omega_k \}$ for $n \ge 0$,
yielding the \emph{$q$-deformed commutation relation}
\[
   \annh{B}\, \crea{B} - q\, \crea{B} \,  \annh{B} = I .
\]

Special cases are
\begin{description}

\item[$q= \phantom{-}1$ --- The Bosonic One-Mode Fock Probability Algebra] with the
      \emph{canonical commutation relations} (CCR): 
      $\annh{B}\, \crea{B} - \crea{B} \,  \annh{B} =
      [\annh{B}, \crea{B}] = I$.  The Jacobi sequence is explicitly
      $\{\omega_n\} = \{1,1,2,3,4,5,\dots\}$ (for $n> 0$: $\omega_n = n$), i.e.\
      it is of \emph{infinite type}.
      
\item[$q= \phantom{-}0$ --- The Free One-Mode Fock Probability Algebra]  defined by the 
      \emph{free} commutation relations: $\annh{B}\, \crea{B} = I$.  
      The Jacobi sequence is explicitly $\omega_n \equiv \lambda_n \equiv 1$ for
      all $n \ge 0$, i.e.\ it is of \emph{infinite type}.
      
\item[$q= -1$ --- The Fermionic One-Mode Fock Probability Algebra] 
      generated by the \emph{canonical
      anti-com\-muta\-tion relations} (CAR) $\annh{B}\, \crea{B} + \crea{B} \,  \annh{B} = I$,
      yielding the associated Fock probability space as $\Gamma_{\{\lambda_n \}} \cong \CC^2$
      --- a \emph{Qubit}.  The Jacobi sequences are explicitly $\{ \omega_0 = \omega_1=1, 
      \omega_2 = \omega_3 = \dots = 0\}$ and $\{ \lambda_0 = \lambda_1= 1, 
      \lambda_2= \lambda_3 = \dots = 0\}$, i.e.\  they are of \emph{finite type}.
      With the identification $\Gamma_{\{\lambda_n \}} \cong \CC^2$, one obtains
      \[
        \Psi_0 = \begin{bmatrix} 0 \\ 1 \end{bmatrix}, \; 
        \Psi_1 = \begin{bmatrix} 1 \\ 0 \end{bmatrix}, \quad  
        \crea{B} = \begin{bmatrix} 0 & 1 \\ 0 & 0 \end{bmatrix}, \quad
        \annh{B} = \begin{bmatrix} 0 & 0 \\ 1 & 0 \end{bmatrix}.
      \]
\end{description}
\end{defi}   

Before entering into a short discussion on orthogonal polynomials and various polynomial 
algebras, it is already possible now to indicate some of the surprising results
\citep{HoraObata07, Obata2017}.

\begin{thm}[$q$-Fock Moments]   \label{thm:q-Fock-moments}
For the one-mode interacting $q$-Fock probability algebra of \feD{def:q-comm-rel},
denote the Fock space by $\Gamma(q)$ and the ladder operators as functions of $q$: 
$\crea{B}(q)$ and $\annh{B}(q)$.   Now consider the RV 
$Z(q) := \crea{B}(q) + \annh{B}(q) \in \Lop(\Gamma(q))$,
and the sub-*-probability algebra $\CC[Z(q)]$ it generates.

In the special cases of \feD{def:q-comm-rel}, the RV $Z(q) \in \CC[Z(q)]$ is moment
equivalent to some well known distributions:
\begin{compactdesc}

\item[The Bosonic $q=1$ case:] The \emph{standard normal Gaussian} random variable $\gamma$
     and $Z(1)$ are moment equivalent: 
     \[
       \frac{1}{\sqrt{2 \uppi}} \; \exp (-x^2/2) \;\sim\; \gamma \quad\stoeq\quad Z(1).
     \]
     
\item[The Free $q= 0$ case:]  The random variable $-2 \le \upsilon \le 2$ with 
     distribution according to \emph{Wigner's semi-circle law} and $Z(0)$ are 
     moment equivalent:
     \[
       \frac{1}{2 \uppi} \; \sqrt{4-x^2} \;\sim\; \upsilon \quad\stoeq\quad Z(0).
     \]

\item[The Fermionic $q= -1$ case:] The \emph{Bernoulli (-1,+1)} random variable $\beta$
     and $Z(-1)$ are moment equivalent: 
     \[
       \frac{1}{2} \; (\updelta_{-1}(x) + \updelta_{+1}(x)) \;\sim\;
           \beta \quad\stoeq\quad Z(-1).
     \]

\end{compactdesc}
\end{thm}

\paragraph{Orthogonal Polynomials:}  
This is a classical subject, and some well known
facts are first shortly recalled:  Let $\nu$ be a Borel probability measure on the real
line $\RR$ having moments of any order.  Using this, define the usual complex
$\Lp_2(\RR,\nu)$-space with corresponding inner product and norm:
\[
     \Ex_{\nu}(f) = \int_{\RR} f(x) \, \nu(\di x), \quad 
     \bkt{f}{g}_{\Lp_2} = \Ex_{\nu}(f^*\, g), \; \text{ and } \;
      \nd{f}_{\Lp_2} = \sqrt{\Ex_{\nu}(f^*\, f)}.
\]
Consider the monomial functions $\bbbone_{\RR}, x, x^2, \dots, x^n \dots \in \Lp_2(\RR,\vu)$,
and perform the \emph{Gram-Schmidt} orthogonalisation process starting with the
constant function $\bbbone_{\RR}$, to obtain 
$\bkt{\cdot}{\cdot}_{\Lp_2}$-orthogonal polynomial functions $P_n(x) \in \Lp_2(\RR,\vu)$. 
They are inductively defined as $P_0(x) = \bbbone_{\RR}$, and for $n>0$ by
\begin{equation}   \label{eq:ortho-poly}
   P_n(x) = x^n - \sum_{j=0}^{n-1} \,
          \frac{\bkt{P_j}{x^n}_{\Lp_2}}{\bkt{P_j}{P_j}_{\Lp_2}}\, P_j(x) .
\end{equation}
It is convenient to set $P_{-1} \equiv 0$.
Similarly to, and following \feD{def:Jacobi-seq}, we define

\begin{defi}[Jacobi-Szegö Coefficients or Parameters]   \label{def:Jacobi-coeff}
Let $\vek{\omega} = \{ \omega_n \}_{n=0}^\infty$ and $\vek{\alpha} = 
\{ \alpha_n \}_{n=0}^\infty$ be a pair of real sequences, such that 
$\vek{\omega} =\{ \omega_n \}$ is a \emph{Jacobi sequence} according 
to \feD{def:Jacobi-seq}, and $\alpha_0 = 1$.
If $\{ \omega_n \}$ is of \emph{finite type} with some $m_0 \in \D{N}_0$, it is 
assumed that $\alpha_{n} = 0$ for all $n > m_0 + 1$.  Such a pair of sequences 
are called \emph{Jacobi-Szegö coefficients} 
or \emph{Jacobi-Szegö parameters} of \emph{finite type}.  Otherwise, the 
\emph{Jacobi-Szegö coefficients} resp.\ \emph{Jacobi-Szegö parameters} are called
of \emph{infinite type}.
\end{defi}

Some well known facts on orthogonal polynomials will be needed in the sequel:
\begin{thm}[Orthogonal Polynomials]   \label{thm:ortho-poly}
Given a probability measure $\nu$ with all moments as above, and the corresponding polynomials
as in \feq{eq:ortho-poly}, define $m_0$ by $1 + m_0 = \ns{\supp \nu}$ 
(the extended number $\ns{\supp \nu}$ is the cardinality of the support of the measure); 
if $m_0 \in \D{N}$, the measure $\nu$ is called of \emph{finite type}, 
otherwise of \emph{infinite type}.  Then

\begin{enumerate}

\item All the polynomial functions $\bbbone_{\RR}, x, x^2, \dots, x^n, \dots$ 
      are linearly independent for a measure of infinite type, 
      whereas for a measure of finite type, the  polynomial functions 
      $\bbbone_{\RR}, x, \dots, x^{m_0}$ are independent, and the Gram-Schmidt process in
     \feq{eq:ortho-poly} stops at $m_0$.
     
\item The polynomial functions $P_n(x) = x^n + \dots$ are monic polynomials of degree 
     $n \in \D{N}_0$.   They are orthogonal, i.e.\ for $0 \le n, k \le m_0$:
      $\bkt{P_n}{P_k}_{\Lp_2} = \updelta_{nk} \prod_{j=1}^n \omega_j = \updelta_{nk} \lambda_n$,
      cf.\ \feD{def:ladder}.
          
\item There exist \emph{Jacobi-Szegö coefficients} $\vek{\omega} = \{ \omega_n\},
     \vek{\alpha} = \{\alpha_n \}$  (cf.\ \feD{def:Jacobi-coeff}) of 
     corresponding type (finite or infinte), such that for $0 \le n \le m_0$
     \begin{align}  \label{eq:3-term-poly}
        x\, P_{n}(x) = P_{n+1}(x) + \alpha_{n+1} \, P_{n}(x) + \omega_{n}\, P_{n-1}(x) .
     \end{align} 
     The Jacobi-Szegö coefficients are completely defined by the Gram-Schmidt process 
     \feq{eq:ortho-poly}, and orthogonality relation in the preceding item.

\item Moreover, $\alpha_1= \Ex_{\nu}(x)$ and $\omega_1 = \Ex_{\nu}((x-\alpha_1)^2) =
      \var_{\nu}(x)$.  In case $\nu$ is symmetric (i.e.\ $\nu(-\di x) = \nu(\di x)$), all 
      \emph{odd} moments vanish, i.e.\ $\Ex_{\nu}(x^{2k - 1}) = 0$ for $k \in \D{N}$.  
      If all odd moments vanish, then $\alpha_n = 0$ for $n>0$.

\end{enumerate}
\end{thm}

Denote by $\E{P} := \mrm{P}(\RR,\nu) \subset \Lp_2(\RR,\nu)$ the pre-Hilbert space 
of all polynomial functions, still equipped with the same inner product as $\Lp_2(\RR,\nu)$.
Clearly, the polynomial functions $\{ P_n \}$ from above are an orthogonal basis of $\E{P}$.  
Introducing normalised polynomial functions $\{ Q_n \}$, defined by 
$Q_n = P_n / \sqrt{\lambda_n}$, the three-term recurrence \feq{eq:3-term-poly} 
reads (still $Q_0 = \bbbone_{\RR}, Q_{-1} \equiv 0$): for all $n$ with $0 \le n \le m_0$:
\begin{equation}  \label{eq:n-3-term-poly}  
        x\, Q_{n}(x) =  \sqrt{\omega_{n+1}}\, Q_{n+1}(x) + \alpha_{n+1} \, Q_{n}(x) 
             + \sqrt{\omega_{n}}\, Q_{n-1}(x) .
\end{equation}

\begin{coro}[$q$-Fock Polynomials]   \label{coro:q-Fock-poly}
In the special cases of the one-mode $q$-Fock probability algebra considered in
\feT{thm:q-Fock-moments}, the orthogonal polynomials $\{ P_n \}$ are well known:

\begin{compactdesc}

\item[For the Bosonic $q=1$ case:]  these are the ``probabilist's'' \emph{Hermite polynomials}.

\item[For the free $q=0$ case:]  the \emph{modified Chebyshev polynomials of the 
      second kind}.

\item[For the Fermionic $q=-1$ case:]  these are just the constant and linear function, as
      the space is only two-dimnsional..

\end{compactdesc}
\end{coro}

\paragraph{Unitary Equivalences:}
Recall the polynomial algebra $\CC[X]$ from \feX{ex:polys}.  Using the
\feD{def:reg-repr}, note that $\CC[X]$ is represented in $\Lop(\CC[X])$ 
as usual via the regular representation
\begin{align}  \label{eq:poly-repr1}
  \oL: \; \CC[X] \ni \tnb{p} &\mapsto \oL_{\tnb{p}} \in \Lop(\CC[X]), \; \text{ where}\\
  \label{eq:poly-repr2}
  \oL_{\tnb{p}}: \; \CC[X] \ni \tnb{q} &\mapsto   \tnb{p}\tnb{q} \in  \CC[X].
\end{align}

\begin{prop}[Probability Algebra --- Polynomials]   \label{prop:c-alg-polys}
Let $(\Alg, \svpi)$ be a probability algebra, and $\ra \in \Alg_{sa} \cap \BA{\Alg}$
a bounded observable.  For brevity, set $\E{R} := \CC[X]$ and $\E{S} := \CC[\ra]$.
Define an injective map $V:  \E{R} \to \E{S}$ by mapping 
$1 \equiv X^0 \in \E{R}$ to $\rone \in \E{S}$, 
  and further for $n > 0$ the corresponding powers
\begin{equation}  \label{eq:def-V-polyX}
  V: \; \E{R} \ni X^n \mapsto \ra^n \in \E{S},
\end{equation}
i.e.\ mapping the formal monomial to the corresponding power of $\ra$.
The \emph{law} $\tau_{\ra}$ of $\ra$ (cf.\ \feD{def:law_RV}) may be used to define a state
$\Ex_{\tau_{\ra}}(\tnb{p}) := \tau_{\ra}(\tnb{p}) = \svpi(\tnb{p}(\ra))$ on $\E{R} = \CC[X]$.
Consider $\E{S} = \CC[\ra] \subseteq \Alg$ as a pre-Hilbert space with the inner product 
from $\svpi \in \SA{\Alg}$, as well as $\E{R}= \CC[X]$ with the inner product from
the above state derived from $\tau_{\ra}$.
Then the map $V$ from \feq{eq:def-V-polyX} may be extended to a unitary map
between the Hilbert space completions:
$V: \; \wob{\E{R}} = \cl_{\tau_{\ra}} \CC[X] \to \cl_{\svpi} \CC[\ra] = \wob{\E{S}}$.

According to  \feT{thm:GNS}, the state corresponding to this on $\Lop(\wob{\E{S}})$
is $\Ex_{\oL,\svpi}(\oL_{\rx}) = \bkt{\rone}{\oL_{\rx} \rone}_2 =: 
\Ex_{\svpi}(\rx) = \svpi(\rx)$.  For the representation \feqs{eq:poly-repr1}{eq:poly-repr2},
on $\Lop(\wob{\E{R}})$, the corresponding state is
$\Ex_{\oL, \tau_{\ra}}(\oL_{\tnb{p}}) = \bkt{X^0}{\oL_{\tnb{p}} X^0}_{\tau_{\ra}}
= \Ex_{\tau_{\ra}}(\tnb{p})$.

The probability algebra $(\CC[X], \tau_{\ra})$ has a GNS representation 
(cf.\ \feD{def:GNS-rep}) on $(\wob{\E{R}},X^0)$, and the probability algebra
$(\CC[\ra], \svpi)$ has a GNS representation on $(\wob{\E{S}}, \rone)$.
\end{prop}

Now, take the \emph{distribution law} $\mu_{\ra}$ of the RV $\ra \in \Alg_{sa}$, which
according to \feP{prop:law-prob-m} is a probability measure on $\RR$ with all finite moments,
and set in the construction described in \feT{thm:ortho-poly} $\nu = \mu_{\ra}$.

In a completely analogous manner to $\CC[X]$ in \feqs{eq:poly-repr1}{eq:poly-repr2},
also the polynomial function space $\E{P} = \mrm{P}(\RR,\nu)$ may be represented in 
$\Lop(\E{P})$ by the regular representation map 
$Y:\, \E{P} \to \Lop(\E{P})$.  Note that a state can be
defined on $\im Y \subset \Lop(\E{P})$ for $f \in \E{P}$ by 
$\Ex_{Y,\nu}(Y_f) := \bkt{\bbbone_{\RR}}{Y_f \, \bbbone_{\RR}}_{\Lp_2} = \Ex_{\nu}(f)$.
The probability algebra $\E{P} = \mrm{P}(\RR,\nu) = \mrm{P}(\RR,\mu_{\ra})$ thus
has a GNS representation on $(\wob{\E{P}},\bbbone_{\RR})$.  In the finite dimensional
case when $m_0 \in \D{N}_0$ and the measure $\nu$ is of finite type,
$\E{P} = \wob{\E{P}} = \Lp_2(\RR,\nu)$.

\begin{prop}[Polynomial Representation]   \label{prop:equiv-gen_poly-poly}
  Define an injective map $R:  \E{P} \to \E{R} = \CC[X]$ by mapping 
  $\bbbone_{\RR} \in \E{P}$ to $1 \equiv X^0 \in \CC[X]$, 
  and further for $0 < n \le m_0$ the corresponding powers
\begin{equation}  \label{eq:def-R-poly2}
  R: \; \E{P} \ni x^n \mapsto X^n \in \E{R} = \CC[X],
\end{equation}
i.e.\ mapping the polynomial function to the corresponding formal polynomial.

%Set $\E{R} = \im R \subseteq \CC[X] $, and let $\tnb{f} \in \E{P}$ be a
%formal polynomial.  Define a state on $\E{R}$ via 
%$\Ex_{R}(\tnb{f}) := \Ex_{\nu}(R^{-1}(\tnb{f}))$, and from this inner product
%and norm on $\E{R}$, making it into a pre-Hilbert space.  Then t
The map $R$ in
\feq{eq:def-R-poly2} can be extended to a unitary map on the Hilbert space
completions, i.e.\ $R:\; \wob{\E{P}} = \cl_{\nu} \E{P} \to \wob{\E{R}} = \cl_{\tau_{\ra}} \E{R}$.
%For the regular representation of $\CC[X]$ in \feqs{eq:poly-repr1}{eq:poly-repr2} one
%obtains a pure or vector state by $\Ex_{\oL, R}(\oL_{\tnb{p}}) := 
%\bkt{X^0}{\oL_{\tnb{p}} X^0} = \Ex_{R}(\tnb{p})$.

For a polynomial function $f \in \E{P} = \mrm{P}(\RR,\nu)$,
let $R(f) = \tnb{f} \in \E{R}$ be a formal polynomial.  Then 
$M = Y \circ R^{-1}: \E{R}\to \Lop(\E{P})$ is a representation
of the sub-algebra $\E{R} \subseteq \CC[X]$ in $\Lop(\E{P})$.
Especially $M_X := M(X)$ is the linear map equivalent of multiplying a polynomial function
$f \in \E{P}$ by the polynomial function $x \in \E{P}$:
\begin{equation}  \label{eq:mult-x-operator}
   M_X: \; \E{P} \ni f(x) \mapsto x f(x) \in \E{P}.
\end{equation}
\end{prop}

The next topic is to consider the normalised three-term recurrence relation
\feq{eq:n-3-term-poly} in a different light, by establishing some unitary
equivalences.

\begin{prop}[Interacting Fock Space --- Polynomial Functions]   \label{prop:IFS-poly_fcts}
Consider a probability measure $\nu$ on $\RR$ with all moments
finite, and the induced Jacobi-Szegö coefficients $\vek{\omega} = \{ \omega_n\},
\vek{\alpha} = \{ \alpha_n \}$ 
as in \feT{thm:ortho-poly}, as well as the orthonormal 
polynomials in \feq{eq:n-3-term-poly}.  For the pre-Hilbert one-mode 
interacting Fock space $\Gamma$ in \feq{eq:interact-FockSpace} in
\feD{def:interact-Fock}, take the Jacobi sequence $\{ \lambda_n = \prod_{k=0}^n \omega_k \}$
defined with the Jacobi-Szegö coefficients of the measure $\nu$.  Define a map 
\begin{equation}  \label{eq:IFS-polyf}
  U:\; \Gamma \to \E{P} \; \text{ via } \;
    U:\; \Psi_n \mapsto Q_n; \; 0 \le n \le m_0 .
\end{equation}
This map preserves orthogonality and lengths, and hence can be extended to a unitary map
on the Hilbert space completions, i.e.\ 
$U:\; \wob{\Gamma} = \cl_{\Psi_0} \Gamma \to \wob{\E{P}}$.
\end{prop}

Note that in the three cases of \feT{thm:q-Fock-moments} the probability measures
are symmetric, so that all odd moments vanish and $\alpha_n = 0$ for $n>0$.
Otherwise, next to creation and annihilation operators from \feqs{eq:crea-def}{eq:annh-def}
in \feD{def:ladder}, one additional operator is needed on the one mode interacting 
Fock space $\Gamma$, namely the \emph{preservation operator}
\begin{equation}  \label{eq:pres-operator}
    \pres{B} :=  \Hf{{\vek{\alpha}}}{(N+I)} \in \Lop(\Gamma) ,
\end{equation}
a function of the number operator $N \in \Lop(\Gamma)$, 
already introduced in \feq{eq:alpha_N1-def} in \feD{def:ladder}.
With this one can set 
\begin{align}   \label{eq:crea-P}
\crea{C} &:= U \crea{B} U^\dagger \in \Lop(\E{P}) , \\   \label{eq:annh-P}
\annh{C} &:= U \annh{B} U^\dagger \in \Lop(\E{P}) , \\   \label{eq:pres-P}
\pres{C} &:= U \pres{B} U^\dagger \in \Lop(\E{P}) ,
\end{align}
to define creation, annihilation, and preservation operators (CAP-operators) 
matching $\crea{B}, \annh{B}, \pres{B}$ on $\Lop(\Gamma)$.  From here,
in an analogous way, define in $\Lop(\E{R})$ the corresponding CAP-operators
\begin{equation}  \label{eq:oper-A-crea}
 \crea{A} := R\, \crea{C}\, R^\dagger, \;\annh{A} := R\, \annh{C}\, R^\dagger,\; 
 \pres{A} := R\, \pres{C}\, R^\dagger \in \Lop(\E{R}), 
\end{equation}
as well as the CAP-operators in $\Lop(\E{S})$
\begin{equation}  \label{eq:oper-D-crea}
 \crea{D} = V\, \crea{A}\, V^\dagger, \; \annh{D} = V\, \annh{A}\, V^\dagger,\; 
 \pres{D} = V\, \pres{A}\, V^\dagger \in \Lop(\E{S}).
\end{equation}

\begin{prop}[Operator Three Term Recurrence]   \label{prop:oper-3-term}
Rewriting \feq{eq:n-3-term-poly} with the help of \feq{eq:mult-x-operator}:
\begin{equation}  \label{eq:oper-3-term-poly}
        M_X\, Q_{n}(x) =  \crea{C}\, Q_{n}(x) + \pres{C} \, Q_{n}(x) 
             + \annh{C}\, Q_{n}(x) \in \E{P},
\end{equation}
one sees that 
\begin{equation}  \label{eq:oper-3-term-poly-C}
Y_x = M_X = \crea{C} + \pres{C} +  \annh{C}.  
\end{equation}
This is called the \emph{quantum decomposition} of the RV $x \in \E{P}$.  In a similar manner,
\begin{align}  \label{eq:oper-3-term-poly-A}
        \oL_X &=  \crea{A}\, + \pres{A} \,  + \annh{A} \in \Lop(\E{R}), \\
          \label{eq:oper-3-term-poly-D}
        \oL_{\ra} &=  \crea{D}\, + \pres{D} \,  + \annh{D} \in \Lop(\E{S}),
\end{align}
are the \emph{quantum decompositions} of the RVs $X \in \CC[X] = \E{R}$ and 
$\ra \in \CC[\ra] = \E{S}$ respectively, e.g.\ 
cf.\ \citep{AccardiBozejko1998, AccardiNahni2003, AccKuoStan04, AccKuoStan05}.
\end{prop}

\begin{prop}[Equivalent Random Variables]   \label{prop:equiv-RVs-QD}
The following moment equivalences and equalities between RVs have been established:
\begin{multline}  \label{eq:mom-equiv-RVs-QD}
  \ra \in \Alg \; \stoeq  x \in \E{P} \; \stoeq \; X \in \E{R} \; \stoeq  \;
  Z := \crea{B} + \pres{B}  + \annh{B} \in \Lop(\Gamma) \\
  \stoeq  \; Y_x = M_X \in \Lop(\E{P}) \; \stoeq  \; \oL_X \in \Lop(\E{R})
  \; \stoeq  \; \oL_{\ra} \in \Lop(\E{S}).
\end{multline}
\end{prop}

The ladder and CAP-operator approach and interacting Fock spaces have been used by 
several authors to investigate orthogonal polynomials, also in many and even in an
infinite number of variables,
e.g.\ cf.\ \citep{AccardiBozejko1998, AccKuoStan04, AccKuoStan05, AccardiNahni2003,
 HoraObata07, Robin2013, Obata2017, AccardiEtal2017, FilipukEtal2018, DhahriEtal2020,
 AccardiLu2020, AccardiLu2021}.
The techniques alluded to in the present section, which use the quantum decomposition
in \feP{prop:oper-3-term} and the embedding of a classical RV in a larger non-commutative
algebra generated by the creation, annihilation, and number operators, resp.\ the
CAP-operatotrs, have been
employed in the spectral analysis of (growing) graphs  \citep{HoraObata07, Obata2017},
in stochastic mechanics \citep{BaezBiamonte2019}, and in a variety of other classical
topics \citep{Bagarello2012}.  For a historical appraisal of the mathematical connections,
see \citep{AccardiLu2020}, and for further developments and generalisations,
see \citep{AccardiLu2021}.

\paragraph{Finite Jacobi Matrices and Favard's Lemma:}
Consider the three-term recurrence \feq{eq:n-3-term-poly} on $\E{P}$, and take the
basis $\{ Q_k \}_{k=0}^{n}$ in the $n+1$-dimensional subspace $\E{P}_n \subseteq \E{P}$,
truncated after $Q_n$. Then $\vek{e}_0 = [1, 0, \dots, 0]^\trpos \in \RR^{n+1}$ 
corresponds to the 
starting vector $Q_0 = \bbbone_{\RR}$, and the other canonical basis vectors $\vek{e}_n$
correspond to $Q_n$.  Let $\vek{M}_x$ be the matrix representation
of the multiplication operator $M_X$ \feq{eq:mult-x-operator} in that basis.
Then $\E{P}_n = \E{K}(\vek{M}_x, \vek{e}_0, n)$ is the \emph{Krylov} subspace generated by
the symmetric operator $M_X$ with starting vector $\vek{e}_0 \equiv Q_0 = \bbbone_{\RR}$ 
after $n$ steps.
Approximating through the truncation --- effectively forcing $Q_{n+1} = 0$ ---
one obtains on $\RR^{n+1} \equiv \E{P}_n$ the well known three-diagonal 
\emph{Jacobi matrix} $\vek{M}_x$, composed from the Jacobi-Szegö coefficients, 
representing and approximating $M_X$ in that basis:
\begin{equation}  \label{eq:Lanczos}
  M_X \approx \vek{M}_x =
  \vek{T}_{J,n} = \begin{bmatrix}  \alpha_1 & \sqrt{\omega_1} &  & &  \\
       \sqrt{\omega_1} & \alpha_2 & \sqrt{\omega_2} &  &  \\
        & \ddots & \ddots & \ddots  &  \\
        & &  \sqrt{\omega_{n-1}} & \alpha_{n} & \sqrt{\omega_{n}} \\
        & & &  \sqrt{\omega_{n}} & \alpha_{n+1}
  \end{bmatrix} .
\end{equation}
This is also in essence the result of the well known Lanczos algorithm ---  which need 
not be reproduced here --- for approximating the spectrum of a symmetric operator 
(like $M_X$), e.g.\ cf.\ \citep{Fischer}.  Further, in order to compute the
spectrum of a tridiagonal matrix, there are very effective numerical algorithms,
e.g.\ cf.\ \citep{Fischer}.  One may observe that
the Lanczos algorithm, which produces the Krylov subspace and the representation
in \feq{eq:Lanczos}, effectively computes the Jacobi-Szegö coefficients 
$(\vek{\omega} = \{ \omega_n \}, \vek{\alpha} = \{ \alpha_n \})$.  
The relation between a probability measure with all finite moments
and the Jacobi-Szegö coefficients, in one and many dimensions, is the content of what 
is called Favard's Lemma (only a simple version is stated here), e.g.\ cf.\
\citep{Dhahri2-2014, AccardiEtal2017, Obata2017, DhahriEtal2020} and the references therein:

\begin{thm}(Favard)   \label{thm:Favard}
To any probability measure $\nu$ on the real line with finite moments of any order,
there are associated Jacobi-Szegö coefficients 
$(\vek{\omega} = \{ \omega_n \}, \vek{\alpha} = \{ \alpha_n \})$,
the mapping being surjective.

On the other hand, given two such Jacobi-Szegö parameter sequences --- or in other words, 
given a sequence of polynomials satisfying a three-term recurrence relation like 
\feq{eq:3-term-poly}, where the coefficients are Jacobi-Szegö parameters as in
\feD{def:Jacobi-coeff} --- there is a state on the algebra of polynomials,
given by a probability measure $\nu$ on $\RR$ with all finite moments, such
that the polynomials from the three-term recurrence are 
orthogonal in the inner product induced by the state.
\end{thm}

\section{Modelling Quantum Processor Units (QPUs)} \label{S:basic-QC}
%\input{\thetext/quant-comp_QC-alg}
% !TEX root = ../23_QC-algebra.tex
% !TEX encoding = UTF-8 Unicode
% RCSID:       $Id: quant-comp-a_QC-alg.tex,v 1.2 2026/01/15 21:04:52 hgm Exp $
% Author:      $Author: hgm $
% Contact:     wire@tu-bs.de
% =================================

%\note{This text is an alternate version.}

\subsection{Preliminaries}\label{SSS:plan-QC}
Quantum computing is a new and rapidly developing field that uses the 
principles of quantum mechanics to perform computations \citep{NielsenChuang2011}.
The idea is that certain types of problems can be solved more efficiently 
using quantum computers than with classical computers.
The actual devices --- the quantum computers --- employed are composed of components
which are such that the algebra of physical observables is non-commutative.  
These components are in their basic form quantum mechanical systems with a finite
number of discrete quatum states; usually they have just two states and are
described by \emph{qubits}, or quantum bits, which are the basic units of 
information in quantum computing.  A quantum computer is then composed of
a number $n \ge 1$ of interacting qubits which are operated on by so-called
\emph{quantum gates}, which constitutes the actual \emph{quantum algorithm}.

Qubits are similar to the traditional bits used in classical computers, 
but they can exist in a superposition of the two states, meaning that they
can represent multiple values at the same time. This kind of inherent parallelism
allows quantum computers to perform certain types of calculations much faster than
classical computers, such as factoring large numbers and searching through
large databases.
Quantum computing has many potential applications, including but not 
limited to cryptography, optimisation, simulation, and machine learning.

\ignore{     %%% BEGIN IGNORE

}    %%% END IGNORE

The current state-of-the-art is defined by
Noisy Intermediate-Scale Quantum (NISQ) devices.
A NISQ (Noisy Intermediate Scale Quantum) computer is a type of 
quantum computer that is still under development and lacks some of the 
capabilities of hopefully more advanced quantum computers. It refers to the fact 
that these early quantum computers are prone to errors, noise and other 
issues that can affect their performance and accuracy.  These are \emph{noisy qubits}
which are subject to decoherence, which means they lose their entangled quantum state
properties over time due to interactions with the environment, such as thermal
fluctuations, electromagnetic noise, or other sources of error, as well as only
\emph{limited quantum error correction}, still \emph{slow quantum gates}, a
still quite \emph{limited number of qubits}, and a \emph{lack of quantum algorithms}.

A Quantum Processor Unit (QPU) is a hardware component in a quantum 
computer that performs the actual computations. It consists of one or more
quantum processors, which are responsible for executing quantum 
instructions and performing calculations based on quantum gates,
which are the basic building of a quantum circuit (also called quantum code).
The QPU is \emph{the} essential component of a quantum computer, as it is 
responsible for performing the computations that make quantum computing 
possible. It is important to note that QPUs are still under development 
and there is ongoing research to improve their performance and capabilities.
But the specific type of QPU is not essential for the quantum algorithms,
as the algorithms reference purely the abstract algebraic properties of the
quantum gates.  In this way the representation of quantum algorithms is
independent of the actual hardware implementation of the QPU, similarily as
the representation of classical algorithms is independent of the hardware
implementation of classical computer and CPU (central processing unit).

\subsection{From Algebraic Probability to Quantum Computation}   \label{SS:Ap2QC}
  In the following Dirac's \emph{bra-ket} notation explained
at the end of \fsec{SS:basics} will be used exetnsively.
From the above very sketchy description of a QPU, one may deduce that according to the
rules of quantum mechanics one qubit will be
described with the bounded operator algebra (cf.\ \feX{ex:complex-Hilbert})
of a complex 2-dimensional Hilbert space $\Hvk$ ($\dim \Hvk = 2$) --- or more
to the point that the observables are the self-adjoint elements $\LHA_{sa}$
of the full finite dimensional operator algebra $\LHA$ --- where
one identifies a basis of two orthogonal unit vectors $B = \{\ket{0}, \ket{1} \}$,
$\spn B = \Hvk$, which are the \emph{states} where the qubit \emph{is set}
($\ket{1}$) resp.\ \emph{not set}  ($\ket{0}$).
Obviously, as described in \feX{ex:complex-Hilbert}
and \feC{coro:dens-in-alg-H}, these unit vectors resp.\ more precisely
the corresponding density
operators $\ket{0}\bra{0}$ and  $\ket{1}\bra{1}$ define \emph{pure states}
(cf.\ \feP{prop:set-dens-M}) in $\DEN{\LHA}_1$.

It is customary to choose a computational basis identification of $\Hvk$ and
$\CC^2$ by setting
\begin{equation}   \label{eq:qubit-basis}
\ket{0} := \begin{bmatrix} 1 \\ 0  \end{bmatrix}, \quad
\ket{1} := \begin{bmatrix} 0 \\ 1  \end{bmatrix}.
\end{equation}
As described in \feR{rem:basic-matrix-rep}, this gives a *-isomorphic
matrix representation of the one-qubit algebra $\LHA$ on $\MMin{2}(\CC)$,
where the latter is decribed in \feX{ex:complex-mat} with its Hilbert-Schmidt inner
product based on the trace and the associated Frobenius state, with more details
in \feeXs{ex:complex-diag}{ex:sa-mat}.  The density matrices and
vector states for this matrix algebra are described in \feD{def:density-M}, and
the set of density matices --- which represents all possible states ---
in \feP{prop:set-dens-M}.

In a QPU, $n > 1$ qubits interact in a \emph{quantum register} when a quantum
algorithm is executed.  According to \feD{def:tensor-product} the appropriate algebra
for this situation is the $n$-fold \emph{tensor product} of the one-qubit algebra:
$\Qlg := (\MMin{2}(\CC))^{\otimes n} = \MMin{2^n}(\CC)$, the matrix algebra on the
Hilbert space $\CC^N$, where $N = 2^n$.  The elements $\rA \in \Qlg$ are often
represented by tensor products of one-qubit matrices $\rA = \bigotimes_{k=1}^n \vA_k$,
with $\vA_k \in \MMin{2}(\CC)$.

A quantum algorithm
involves manipulating the qubits in a way that allows 
for the solution of certain types of problems and enables the use of 
quantum computing for a wide range of applications. The implementation of 
this  process will depend on the specific problem being solved 
and may involve manipulating the qubits using classical electrical signals
or other techniques.  From this and the previous remarks regarding NISQ computers,
one may deduce that the whole process involves input from the classical world to
a quantum device (see the Classical Input Channel of \feX{ex:channel-examples}
in \fsec{SS:POVM}), the computation (a general channel taking account of all
the error sources alluded to above, i.e.\ a CPUP map, cf.\ \feD{def:channel}),
and an output to the classical world
(see  the Classical Output Channel of \feX{ex:channel-examples} and in more detail
in \feX{ex:obs-channel}) typically realised as a positive operator valued
measure (POVM) (cf.\ \feD{def:POVM-defi-A}).

But to describe a quantum algorithm, one does not have to deal with a \emph{real}
QPU, but one may consider an \emph{ideal} one; here the computation channel can
be considered as an internal *-automorphism (cf.\ \feX{ex:channel-examples}) realised
by unitary transformations, a reversible channel according
to the rules of quantum mechanics;
see also in this connection the Stinespring dilation \feT{thm:Stinespring}.
Similarly, instead of a general POVM one only need consider the idealised version
of a projection valued measure (PVM) (cf.\ \feD{def:PVM-M}
and the part on projective measurements at the end of \fsec{SS:funcs-RVs-norm-alg});
observe also the Neumark dilation \feT{thm:Neumark-dilation}.

%\todo{Should there be something how quantum gates\\
 %     generate (are universal in) $\UA{\MMin{2^n}(\CC)}$ ?}

Two complementary universality statements are relevant here: (i) exact universality of
two-level embedded $SU(2)$ (resp.\ $U(2)$) subgroups for $SU(2^n)$ (resp.\ $U(2^n)$) via
QR/Givens-type factorizations, and (ii) discretization/compilation from a finite dense
$SU(2)$ gate set lifted through the same embeddings (e.g.\ via Solovay--Kitaev).
We refer to \citep{FalcoFalcoPomaresMatthies2026_ElemGatesEmbeddings} for the geometric
embedding-based formulation and the associated universality and discretization results.

The density matrices $\DEN{\Qlg}$, which are a one-to-one representation of
the states $\SA{\Qlg}$, were described in \feD{def:density-M}.
From \feP{prop:set-dens-M}, recall \feq{eq:dens-mtx-rank}, which shows that
the set of density matrices
is the disjoint union of of densities with ranks of $r = 1, \dots, N$:
\[ \DEN{\Qlg} = \biguplus_{r=1}^N \DEN{\Qlg}_r , \]
with the extreme points of $\DEN{\Qlg}$, also called the \emph{pure states},
being the rank-one densities (cf.\ \feq{eq:dens-mtx-ext})
\[
   \mrm{ext}(\DEN{\Qlg} = \DEN{\Qlg}_1 =
   \{  \ket{\Psi}  \bra{\Psi} \; : \; \ket{\Psi} \in \CC^N,
       \;  \bkt{\Psi}{\Psi} = 1 \} .
\]

In the light of the basis choice \feq{eq:qubit-basis} and
Dirac's bra-ket notation, an orthonormal basis
for $\CC^N = \CC^{2^n} = (\CC^{2})^{\otimes n}$ can be generated by the
tensor products of the basis vectors of $\CC^{2}$.  These are
\begin{multline}  \label{eq:n-qubit-basis}
  \ket{00\dots 00} = \ket{0}\ket{0}\dots\ket{0}\ket{0} =
     \ket{0}\otimes\ket{0}\otimes\dots
     \otimes\ket{0}\otimes\ket{0}; \;
     \ket{00\dots 01} = \ket{0}\ket{0}\dots\ket{0}\ket{1};\\
 \ket{00\dots 010} = \ket{00\dots 0}\ket{1}\ket{0}; \quad \dots \quad ;
  \ket{11\dots 10} = \ket{11 \dots 1}\ket{0}; \; \ket{11\dots 11}.
\end{multline}

It is fairly obvious that a basis vector $\ket{x_0x_1\ldots x_{n-1}}$
in \feq{eq:n-qubit-basis}, where $x_i \in \{0,1\}$ for all
$0 \le i \le n-1$, is the binary representation $(k)_2$ of
an integer $0 \le k \le N-1$, where $(k)_2 = x_0x_1\ldots x_{n-1}$.
This allows us to associate each integer labeled ket $\ket{k}$ with the
canonical vector $\ket{(k)_2} \in \CC^{2^n}$ from \feq{eq:n-qubit-basis}, as
it is not difficult to see that the set $\{\ket{k}:k\in \D{Z}_{2^n}\}$ is
the canonical basis of $\CC^{2^n}$, literally by interpreting the tensor
product as a \emph{Kronecker product}.

As already mentioned, for the output channel here we want to
use a PVM (cf.\ \feD{def:PVM-M}),
here composed of the projections associated with the unit
vectors in \feq{eq:n-qubit-basis}:
\begin{equation}   \label{eq;QPU-PVM}
   \D{Z}_N = \D{Z}_{2^n} \ni k \mapsto \rP_k := \ket{k}\bra{k} \in \PA{\Qlg} .
\end{equation}
It is not difficult to see that the $\{\rP_k\}$ are commuting
(in fact orthogonal) minimal projections which form a projection
algebra (cf.\ \feX{ex:projections}), and even a $N$-dimensional
\emph{maximal Abelian self-adjoint algebra} (MASA, cf.\ \feD{def:alg-names})
$\Mlg := \spn_{\CC} \{\rP_k\} = \CC[\{\rP_k\}] \subset \Qlg$.
This commutative sub-algebra is the \emph{classical context} for the
output / measurement / observation, cf.\ \fsec{SS:constraints}.

The projections can be considered as \emph{events}
which signify the bit pattern $(k)_2 = x_0x_1\ldots x_{n-1}$, and
(cf.\ \feP{prop:proj-1}) with \emph{probabilities} of being observed
in a state $\svpi_{\vrh}$ with density
matrix $\vrh \in \DEN{\Qlg}$ given according to the
Born rule (cf.\ \feD{def:probability})
\begin{equation}   \label{eq:QPU-prob}
\prob(\rP_k) = \svpi_{\vrh}(\rP_k) = \tr \vrh \rP_k = \bra{k} \vrh \ket{k} .
\end{equation}
They actually form a
\emph{Boolean algebra} according to \feP{prop:boolean} (the classical output).

%Observation \feD{def:sample}  
%[Hadamard Algebra]   \feX{ex:complex-n-spc}
%[Diagonal Matrix Algebra]   \feX{ex:complex-diag}
%[Projection Algebra]   \feX{ex:projections}
%[Commutative Matrix Algebra]   \feX{ex:sa-mat}
%SpecDec of self-adj Matrix     \feR{rem:spec-dec-proj}
%MASA \feD{def:alg-names}
%[Projection Lattice in $\MMn(\CC)$]  \feP{prop:proj-matrix-measurement}
%[Probability --- Born's rule]  \feD{def:probability}
%[Events]  \feP{prop:proj-1}
%[Boolean Algebra]  \feP{prop:boolean}
%[Matrix Observable, Projective Measurement]  \feR{rem:spec-dec-proj-2}
%[Projection Valued Measure (PVM)]   \feD{def:PVM-M}
%[Projection Lattice in $\MMn(\CC)$]  \feP{prop:proj-matrix-measurement}

\begin{rem}[QPU --- Interacting Fock Space]  \label{rem:JacobiSeq-QC}
Now we connect this with the interacting Fock space from \fsec{SS:Fock}.  Define
a \emph{Jacobi sequence} of finite type (cf.\ \feD{def:Jacobi-seq}):
\begin{equation}   \label{eq:Jacobi-QC}
  \forall k \text{ with } 0 \le k \le N-1 :\,  \lambda_k = 1;\quad \text{and }
  \forall j \ge 0:\, \lambda_{N + j} = 0 .
\end{equation}
With this Jacobi sequence \feq{eq:Jacobi-QC}, construct the one-mode interacting
Fock space (cf.\ \feD{def:interact-Fock}) with the \emph{sectors}
$\Kvk_k = \spn \{\ket{k}\}$ for $0 \le k \le N-1 = 2^n - 1$.
This set-up then defines the self-adjoint number operator $\rN \in \Qlg$
(cf.\ \feq{eq:number-def} in \feD{def:ladder}) with action
$\rN \ket{k} = k \ket{k} \in \Kvk_k$ for $0 \le k \le N-1 = 2^n - 1$.
It thus has the spectral decomposition (cf.\ \feR{rem:spec-dec-proj})
\begin{equation}   \label{eq:numberOP-QC}
\rN = \sum_{k=0}^{N-1} k \, \ket{k}\bra{k}
\end{equation}
and $(\rN + \rI)$ generates (cf.\ \feX{ex:sa-mat}) the MASA
$\Mlg = \CC[\rN] = \CC[\{\rP_k\}] \subset \Qlg$.

The ladder operators, i.e.\ the creation and annihilation operator
$\crea{B}, \annh{B}$, defined by \feqs{eq:crea-def}{eq:annh-def} in
\feD{def:ladder}, can be seen as increasing resp.\ decreasing the
binary representation of the quantum register.
\end{rem}

\begin{defi}[$n$-qubits digital quantum computer]  \label{def:n-bits-QPU}
An $n$-qubits digital quantum computer is a pair 
$(\rA,\svpi_{\rho_0})$ where $\rA \in \Mlg_{sa} \subset \Qlg$ is an
non-singular observable that represents the
digital quantum computer and is given by
\begin{equation}   \label{eq:specDec-QC}
\rA = \sum_{k=0}^{2^n-1} \lambda_{k} \, \ket{k} \bra{k},
\end{equation}
with distinct non-zero eigenvalues $\lambda_k \neq \lambda_s$ for $k\neq s$,
Such an observable generates (cf.\ \feX{ex:sa-mat}) the MASA
$\Mlg = \CC[\rA] \subset \Qlg$.

The event $\{\rA = \lambda_k\}$ corresponds to the observable
\begin{equation}   \label{eq:event-k-QC}
\rP_{\{\rA = \lambda_k\}} = \ket{k} \bra{k} = \rP_k,
\end{equation}
for each $k \in \D{Z}_{2^n}$.

Note that the number operator $(\rN+\rI)  \in \Mlg_{sa} \subset \Qlg$
derived from \feq{eq:numberOP-QC} in \feR{rem:JacobiSeq-QC} with eigenvalues
$\lambda_k = k+1$ for $k \in \D{Z}_{2^n}$ is such an observable.
\end{defi}

Observations $\som \in \XA{\Mlg}$ (cf.\ \feD{def:sample}) of the MASA $\Mlg$
(cf.\ \feR{rem:spec-dec-proj-2}) can also be represented with the
minimal one-dimensional projections $\{\rP_k\} \subset \PA{\Mlg}$.
This is achieved by realising that they are also
density matrices $\{\rP_k\} \subset \DEN{\Mlg}$ of pure states, and in
fact multiplicative states, i.e.\ characters.  When the event
$\rP_k = \ket{k} \bra{k}$ is observed, for a $\rA \in \Mlg$
with a spectral decomposition $\rA = \sum_{j=0}^N \mu_{j} \, \ket{j} \bra{j}$
as in \feq{eq:specDec-QC}, they have the form
\begin{equation}   \label{eq:obs-M-QC}
  \som_{k}(\rA) = \tr(\rP_k \rA) = \bra{k} \rA \ket{k} =
%  \tr\left(\ket{k} \bra{k} (\sum_{j=0}^N \mu_{j} \, \ket{j} \bra{j})\right)
  \sum_{j=0}^N \mu_{j} \, \ns{\bkt{k}{j}}^2 = \mu_{k},
\end{equation}
as $\bkt{k}{j} = \updelta_{k,j}$.  This naturally also gives
$\som_k(\rP_j) = \updelta_{k,j}$.

\begin{prop}   \label{prop:rank-one-probability}
  Let be $\ket{u} \in \CC^N$ be a unit vector and
  $\svpi_{\vrh} \in \SA{\Qlg}$ with density matrix $\vrh \in \DEN{\Qlg}$
  be a quantum state. Then, considering the rank-one
  observable $\rA = \ket{u} \bra{u}$, 
\begin{equation}   \label{eq:genBorn-QC}
 \svpi_{\vrh}(\rA) =   
 \svpi_{\vrh}(\ket{u} \bra{u}) = \tr \vrh \rA =
 \tr(\vrh \ket{u} \bra{u}) = \bra{u} \vrh \ket{u}.
\end{equation}
Furthermore, in case $\vrh \in \DEN{\Qlg}_1$ is a pure state,
where $\vrh = \ket{\Psi} \bra{\Psi}$ for some unit vector $\ket{\Psi} \in \CC^N$,
then (cf.\ \feq{eq:QPU-prob})
\begin{equation}   \label{eq:uBorn-QC}
  \svpi_{\ket{\Psi} \bra{\Psi}}(\ket{u} \bra{u}) = (\bkt{\Psi}{u})^2 = \cos^2 \theta_{\Psi,u},
\end{equation}
where $\theta_{\Psi,u}$ is the angle between the vectors $\ket{\Psi}$ and $\ket{u}$.

This means that the law (cf.\ \feD{def:law_RV}) of the quantum random
variable $\rA = \ket{u}\bra{u} \in \Qlg_{sa}$ in this pure state
results in the distribution (cf.\ \feP{prop:law-prob-m}) given by \feq{eq:uBorn-QC}
\begin{equation}   \label{eq:P1-QC}
  \prob_{\ket{\Psi}\bra{\Psi}}\left(\ket{u}\bra{u} = 1\right) =
      (\bkt{\Psi}{u})^2 = \cos^2 \theta_{\Psi,u},
\end{equation}
and hence
\begin{equation}   \label{eq:P2-QC}
    \prob_{\ket{\Psi}\bra{\Psi}}\left(\ket{u}\bra{u} = 0\right) = \sin^2 \theta_{\Psi,u}.
\end{equation}
\end{prop}

The following is an easy exercise in linear algebra:

\begin{prop} \label{prop:rank-one-probability1}
Let $\svpi_{\vrh} \in \SA{\Qlg}$ be a quantum state represented by the density
matrix $\vrh \in \DEN{\Qlg}_r$ for some $r=1,2,\ldots,N.$
Then, for any unitary matrix
$\rU \in \UA{\Qlg}$, it holds that (cf.\ \feX{ex:channel-examples})
$\rU \vrh \rU^\tpH \in \DEN{\Qlg}_r$ is a new rank-$r$ density matrix
transformed by the channel / internal automorphism given by $\rU$.

Similarly, in case $\rA \in \Qlg_{sa}$ is an observable,
$U\rA U^\tpH$ is still an observable.
Furthermore, the probability law of  $U\rA U^\tpH$ under the state
$\svpi_{\vrh}$ is the same as the probability law of
$\rA$ under state $\svpi_{U\rho U^\tpH}$.
\end{prop}

The above \feP{prop:rank-one-probability1}
allows one to introduce the next definition.
First, consider the map $\QCAlgrthm$ to express one step of a QPU: 
\begin{equation}   \label{eq:Qalgrthm-QC}
  \QCAlgrthm : \UA{\Qlg} \times \DEN{\Qlg}  \longrightarrow \DEN{\Qlg}, \quad 
   \QCAlgrthm(U,\vrh):= \rU \vrh \rU^\tpH.
\end{equation}
From \feP{prop:rank-one-probability1}, $\QCAlgrthm$ is well defined.
For each fixed $\rU \in \UA{\Qlg}$, the map $\QCAlgrthm_{\rU} :=
\QCAlgrthm(\rU,\cdot)$ is called the quantum channel associated with
the unitary matrix / internal automorphism defined by
$\rU$, cf.\ \feX{ex:channel-examples}. 

\begin{defi}[Quantum Code]  \label{def:QCode-QC}
  For a fixed quantum state $\svpi_{\vrh_0} \in \SA{\Qlg}$ represented by
  a density matrix $\vrh_0 \in \DEN{\Qlg}$, a quantum code
  or quantum algorithm is
  a finite sequence $\rU_0, \rU_1, \ldots, \rU_{\ell} \in  \UA{\Qlg}$, where 
  $\rU_0=\rI$ is the identity matrix, and $\ell$ is the length of the
  quantum code.  The quantum code is used to perform quantum computations
  on the quantum state density matrix $\vrh_0$ using the channel
  defined in \feq{eq:Qalgrthm-QC} as follows:
\begin{equation}   \label{eq:QCode-QC}
  \vrh_{\ell} = \QCAlgrthm_{\rU_{\ell}}(\QCAlgrthm_{\rU_{\ell-1}}(\cdots
  \QCAlgrthm_{\rU_1}(\QCAlgrthm_{\rU_0}(\vrh_0))\cdots)).
\end{equation}
\end{defi}

A $n$-qubits digital quantum computer,
given the quantum code $\rU_0, \rU_1, \ldots, \rU_{\ell} \in  \UA{\Qlg}$,
acts as follows:
\begin{enumerate}
\item The quantum computer is initialized in the state $\svpi_{\vrh_0}$.

\item The quantum processing unit $\QCAlgrthm$ is applied to the quantum
  state $\svpi_{\vrh_0}$ represented by the density matrix
  $\vrh_0 \in \DEN{\Qlg}$ by using the quantum code
  $\rU_0, \rU_1, \ldots, \rU_{\ell} \in  \UA{\Qlg}$, and transforms it 
  as in \feq{eq:QCode-QC} in \feD{def:QCode-QC} into a new quantum state
  $\svpi_{\vrh_\ell}$ represented by the density matrix
  $\vrh_\ell \in \DEN{\Qlg}$.
%  \begin{equation}   \label{eq:QCode2-QC}
%     \vrh_{\ell} = \QCAlgrthm_{\rU_{\ell}}(\QCAlgrthm_{\rU_{\ell-1}}(\cdots
%       \QCAlgrthm_{\rU_1}(\QCAlgrthm_{\rU_0}(\vrh_0))\cdots)).
%  \end{equation}

\item The measurement of the observable $\rA \in \Mlg_{sa}$ from
  \feD{def:n-bits-QPU} --- resp.\ on the PVM defined by its spectral
  decomposition --- is performed in
  the quantum state $\svpi_{\vrh_{\ell}}$.  Thus, the outcome of the quantum
  code given by the $n$-qubits digital quantum computer results in
  the law of the quantum random variable $\rA$ under the state $\svpi_{\vrh_{\ell}}$:
  \begin{equation}   \label{eq:finalSt-QC}
    \prob_{\vrh_{\ell}}(\rA = \lambda_j) = \svpi_{\vrh_{\ell}}(\rP_j)
    = \bra{j} \vrh_{\ell} \ket{j} = \bra{j} \rU_{\ell} \cdots \rU_1 \rU_0 \vrh_0
    \rU_0^\tpH \rU_1^\tpH \cdots \rU_{\ell}^\tpH \ket{j},
  \end{equation}
  for each $j \in \D{Z}_{2^n}$.  For the observation itself, if the event $\rP_k$
  is observed, from \feq{eq:obs-M-QC} one has that $\som_k(\rA) = \lambda_k$.
\end{enumerate}

\begin{rem}  \label{rem:statePrep-QC}
  %A commercial digital
  Often a quantum computer uses a pure initial state
  $\svpi_{\vrh_0} \in \SA{\Qlg}$ given by the density matrix
  $\vrh_0 = \ket{0}\bra{0} \in \DEN{\Qlg}_1$.

  The law (cf.\ \feD{def:law_RV}) of the quantum random
  variable $\rA  \in \Mlg_{sa}$ in this pure state is characterised by
  the distribution (cf.\ \feP{prop:law-prob-m}) given in accordance with
  \feq{eq:uBorn-QC} (cf.\ also \feqs{eq:P1-QC}{eq:P2-QC} in
  \feP{prop:rank-one-probability}) by
  \begin{equation}   \label{eq:initialSt-QC}
    \prob_{\vrh_0}(\rA = \lambda_k) = \prob(\rP_k) = (\bkt{0}{k})^2 = \updelta_{0,k},
    \quad k \in \D{Z}_{2^n}.
  \end{equation}

  Hence, often the first steps of a quantum algorithm have the purpose
  of \emph{state preparation}, in order to achieve an appropriate starting
  state for the quantum algorithm.
\end{rem}

Next, we  illustrate a search problem that can be solved by using a
digital quantum computer.

%\subsubsection{Grover's algorithm}
\subsection{Grover's Algorithm}  \label{SS:Grover}
Grover's algorithm is a quantum algorithm that finds with high
probability the unique input to a black box function that produces
a particular output value, using just $O(\sqrt{N})$ evaluations of
the function, where $N$ is the size of the function's domain.
Grover's algorithm is an example of a quantum algorithm that provides
a quadratic speedup over the best classical algorithm.

Consider the set $\{0,1\}^n$ of all binary strings of length $n$.
Let $f:\{0,1\}^n \to \{0,1\}$ be a function that is constant on all
inputs except for one, which is called the \emph{marked element}.
The goal of Grover's algorithm is to find a quantum code, namely 
$\rU_0, \rU_1, \dots, \rU_{\ell}$, which provides an unitary matrix
$\rU_f := \rU_0 \rU_1 \cdots \rU_{\ell}$ that acts on the quantum state
$\svpi_{\vrh_0} \in \DEN{\Qlg}_1$ such that $\svpi_{\rU_f \vrh_0 \rU_f^\tpH}$ is
a quantum state that has a high probability of measuring the marked element.

Quantum computers start with a pure state $\svpi_{\vrh_0} \in \DEN{\Qlg}_1$
given by $\vrh_0 = \ket{0}\bra{0}$.
%, where $0 \in \mathbb{Z}_{2^n}.$ Recall that we represent the
%integer numbers in binary form, and hence we have
%$\ket{0} = \ket{0^n} = \ket{0} \otimes \ket{0} \otimes \cdots
%\otimes \ket{0} = \ket{0}^{\otimes n}.$
Then, the quantum code $\rU_f$ acts on the density matrix $\vrh_0$ by
the relation $\rU_f \ket{0}\bra{0} \rU_f^\tpH$.  From
\feP{prop:rank-one-probability1} we have that $\rU_f \ket{0}\bra{0} \rU_f^\tpH$
is still a rank-one density matrix, and hence there exists a unit vector
$\ket{\Psi}$ such that $\rU_f \ket{0}\bra{0} \rU_f^\tpH = \ket{\Psi}\bra{\Psi}$.
To solve the problem, we will find an unitary matrix $\rU_f$ satisfying
$\rU_f^\tpH = \rU_f$ such that $\rU_f \ket{0} = \ket{\Psi}$. 

Assume that $a \in \D{Z}_{2^n}$ is the marked element.  Then we would like
to find a unit vector $\ket{\Psi}$ such that 
\[
\prob_{\ket{\Psi} \bra{\Psi}}(\rA = \lambda_a) = \svpi_{\ket{\Psi} \bra{\Psi}}(\ket{a}\bra{a})
= \ns{\bkt{\Psi}{a}}^2 = 1 - \varepsilon
\] 
holds for some $\varepsilon > 0.$
This unit vector $\ket{\Psi}$ is the solution of the Grover's algorithm.
It defines a quantum state $\svpi_{\ket{\Psi} \bra{\Psi}}$ such that law of the
random quantum variable $\rA$ is given by
\[
\prob_{\ket{\Psi} \bra{\Psi}}(\rA = \lambda_a) = 1 - \varepsilon \;
\text{ and } \; \prob_{\ket{\Psi} \bra{\Psi}}(\rA \neq \lambda_a) = \varepsilon.
\]
%In a real quantum computer, the output of the Grover's algorithm
%is the (empirical) law of the random quantum variable
%$(\rA,\svpi_{\ket{\Psi} \bra{\Psi}}).$

To proceed, we first take the unitary (and self-adjoint) matrix
$\rU_1 = \vH^{\otimes n}$, where
\[
\vH = \frac{1}{\sqrt{2}}\begin{bmatrix} 1 & \phantom{-}1 \\ 1 & -1 \end{bmatrix},
\]
is the Hadamard matrix.  Then, it is not difficult to see that
\[
  \ket{\Psi_1}:= \rU_1 \ket{0}^{\otimes n} = \vH^{\otimes n} \ket{0}^{\otimes n} =
  \left(\frac{1}{\sqrt{2}}(\ket{0} +\ket{1})\right)^{\otimes n} =
   \frac{1}{\sqrt{2^n}} \sum_{k=0}^{2^n-1} \ket{k}.
\]
Observe, that
\[
\prob_{\ket{\Psi_1}\bra{\Psi_1}}(\rA = \lambda_s)= \ns{\bkt{\Psi_1}{s}}^2 =
\left(\frac{1}{\sqrt{2^n}}\right)^2 = \frac{1}{2^{n}}
\; \text{ for  all } \; s \in \D{Z}_{2^n}.
\]
Now, the law of the quantum random variable $\rA$ under the state
$\svpi_{\ket{\Psi_1}\bra{\Psi_1}}$ with density matrix $\ket{\Psi_1}\bra{\Psi_1} =
\rU_1 \ket{0}^{\otimes n}\bra{0}^{\otimes n}\rU_1$
follows a discrete uniform distribution over the set $\D{Z}_{2^n}$.
This is a kind of state preparation.
Observe that
\[
   \ket{\Psi_1} = \frac{1}{\sqrt{2^n}} \sum_{k=0}^{2^n-1} \ket{k} =
   \frac{1}{\sqrt{2^n}} \ket{a} +  \frac{1}{\sqrt{2^n}}
   \sum_{\stackrel{k \in \D{Z}_{2^n}}{k\neq a}}\ket{k} =
   \frac{1}{\sqrt{2^n}} \ket{a} + \sqrt{\frac{2^n-1}{2^n}}\ket{a_{\perp}},
\]
where
\[
  \ket{a_{\perp}} = \frac{1}{\sqrt{2^n-1}} \sum_{\stackrel{k \in \D{Z}_{2^n}}{k\neq a}}\ket{k},
\]
is a unit vector orthogonal to $\ket{a}$.
Since $\theta_{\Psi_1,a}+\theta_{\Psi_1,a_{\perp}}= \frac{\pi}{2}$, we have
\[
   \prob_{\ket{\Psi_1}\bra{\Psi_1}}(\rA = \lambda_a)= \ns{\bkt{\Psi_1}{a}}^2 =
     \frac{1}{2^n} = \cos^2 \theta_{\Psi_1,a} = \sin^2\theta_{\Psi_1,a_{\perp}},
\]
and
\[
  \prob_{\ket{\Psi_1}\bra{\Psi_1}}(\rA \neq \lambda_a) =
  \ns{\bkt{\Psi_1}{a_{\perp}}}^2 = \left(\sqrt{\frac{2^n-1}{2^n}}\right)^2 =
  1-\frac{1}{2^n} = \cos^2 \theta_{\Psi_1,a_{\perp}}.
\]
Thus, we can write
\[
  \ket{\Psi_1} = \sin \theta_{\Psi_1,a_{\perp}} \ket{a} +
      \cos \theta_{\Psi_1,a_{\perp}} \ket{a_{\perp}}.
\]

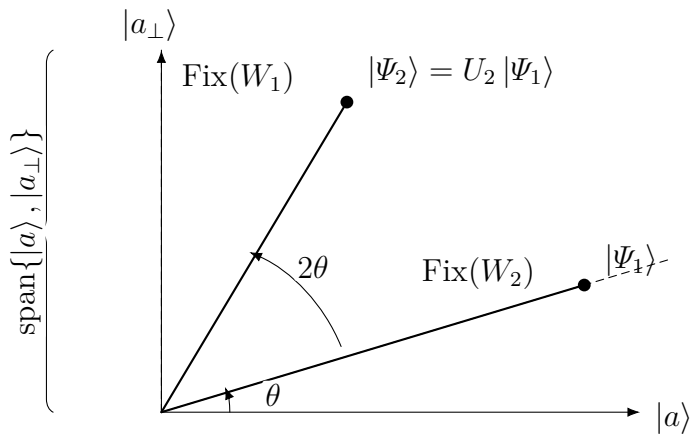
\begin{figure}[t]
\centering
\begin{tikzpicture}[>=Latex, line cap=round, line join=round, scale=1.5]

  % Parameters (numeric angles in degrees)
  \def\thetadeg{22} % numeric angle in degrees (do NOT use \theta here)
  \def\rA{4.2}      % x-axis length
  \def\rP{3.2}      % y-axis length

  % Coordinates
  \coordinate (O)  at (0,0);
  \coordinate (Ax) at (\rA,0);
  \coordinate (Ay) at (0,\rP);

  % State rays (in the invariant 2D plane span{|a>,|a_perp>})
  \coordinate (Psi1) at ({4.0*cos(\thetadeg)},{3.0*sin(\thetadeg)});
  \coordinate (Psi2) at ({4.0*cos(3*\thetadeg)},{3.0*sin(3*\thetadeg)}); % after one Grover iterate: +2*theta

  % Helper styles
  \tikzset{
    lbl/.style={fill=white, inner sep=1.2pt, outer sep=0pt, rounded corners=1pt},
    ray/.style={thick},
    dashray/.style={densely dashed}
  }

  % Axes / basis
  \draw[->] (O) -- (Ax);
  \node[lbl, anchor=west] at ($(Ax)+(0.10,-0.05)$) {$\ket{a}$};

  \draw[->] (O) -- (Ay);
  \node[lbl, anchor=south] at ($(Ay)+(-0.10,0.05)$) {$\ket{a_{\perp}}$};

  % Invariant subspace brace (left)
  \draw[decorate, decoration={brace, amplitude=5pt}]
    (-0.95,0) -- (-0.95,\rP)
    node[midway, lbl, xshift=-10pt, rotate=90]
    {$\mathrm{span}\{\ket{a},\ket{a_\perp}\}$};

  % Rays for states
  \draw[ray] (O) -- (Psi1);
  \node[lbl, anchor=south west] at ($(Psi1)+(0.15,0.05)$) {$\ket{\Psi_1}$};

  \draw[ray] (O) -- (Psi2);
  \node[lbl, anchor=south west] at ($(Psi2)+(0.15,0.10)$) {$\ket{\Psi_2}=U_2\ket{\Psi_1}$};

  % Mark the two states
  \fill (Psi1) circle (1.6pt);
  \fill (Psi2) circle (1.6pt);

  % Reflection fixed lines
  % Fix(W1): hyperplane fixed by W1 (here: the |a_perp|-axis; reflection flips along |a|)
  \draw[dashray] (O) -- (Ay);
  \node[lbl, anchor=west] at ($(Ay)+(0.15,-0.25)$) {$\mathrm{Fix}(W_1)$};

  % Fix(W2): line fixed by W2 (the ray spanned by |Psi1|)
  \draw[dashray] (O) -- ({4.8*cos(\thetadeg)},{3.6*sin(\thetadeg)});
  \node[lbl, anchor=south, yshift=10pt] at ({3.0*cos(\thetadeg)},{2.2*sin(\thetadeg)}) {$\mathrm{Fix}(W_2)$};
  % Angle theta between |a> and |Psi1|
  \draw[->] (0.60,0) arc (0:\thetadeg:0.60);
  \node[lbl] at ({1.0*cos(\thetadeg/2)},{0.8*sin(\thetadeg/2)}) {$\theta$};

  % Rotation angle 2theta from Psi1 to Psi2 (larger radius; label outside)
  \draw[->] ({1.70*cos(\thetadeg)},{1.55*sin(\thetadeg)}) arc (\thetadeg:3*\thetadeg:1.55);
  \node[lbl] at ({1.85*cos(2*\thetadeg)},{1.85*sin(2*\thetadeg)}) {$2\theta$};

  % Explanatory box on the right (fixed width)
  \node[anchor=west, align=left, text width=6.1cm, lbl] at (5.1,2.55) {%
    $W_1 = I - 2\ket{a}\bra{a}$ \hfill (oracle phase flip)\\[2pt]
    $W_2 = 2\ket{\Psi_1}\bra{\Psi_1}-I$ \hfill (inversion about mean)\\[2pt]
    $U_2 = W_2 W_1$ is a rotation in $\mathrm{span}\{\ket{a},\ket{a_\perp}\}$.%
  };

\end{tikzpicture}
\caption{Geometric strategy of Grover's algorithm in the invariant two-dimensional subspace
$\mathrm{span}\{\ket{a},\ket{a_\perp}\}$. The oracle $W_1$ and diffusion $W_2$ are reflections;
their composition $U_2=W_2W_1$ yields a rotation by $2\theta$ that amplifies the probability of
measuring the marked state.}
\label{fig:grover-geometry}
\end{figure}

Next we consider the unitary (and self-adjoint) matrix
$\rW_1 = \rI - 2\ket{a}\bra{a}$, where $\rI$ is the identity matrix.
Then we have
\[
   \rW_1 \ket{a} = -\ket{a}\; \text{ and } \;
   \rW_1 \ket{\Psi_1} = \ket{\Psi_1} - \frac{2}{\sqrt{2^n}}\ket{a}
       =  \ket{\Psi_1} - 2 \sin  \theta_{\Psi_1,a_{\perp}} \ket{a}.
\]
In a similar way we take the unitary (and self-adjoint) matrix
$\rW_2 = 2\ket{\Psi_1}\bra{\Psi_1}-\rI,$ to obtain
\[
   \rW_2 \ket{\Psi_1} = \ket{\Psi_1} \; \text{ and } \;
   \rW_2 \ket{a} = \frac{2}{\sqrt{2^n}}\ket{\Psi_1} - \ket{a} =
   2 \sin  \theta_{\Psi_1,a_{\perp}}\ket{\Psi_1}-\ket{a}.
\]
Geometrically, in the invariant two-dimensional subspace
$\mathrm{span}\{\ket{a},\ket{a_\perp}\}$ the oracle reflection $W_1$
and the diffusion reflection $W_2$ compose into a rotation $U_2=W_2W_1$,
which increases the amplitude on the marked state at each iteration;
see Figure~\ref{fig:grover-geometry}.

Finally, the unitary (and self-adjoint) matrix $\rU_2 = \rW_2\rW_1$
allows us to obtain
\begin{align*}
  \ket{\Psi_2} & := \rW_2 \rW_1 \ket{\Psi_1} =
     \rW_2 \left(\ket{\Psi_1} - 2 \sin  \theta_{\Psi_1,a_{\perp}}\ket{a}\right) \\ 
   & = \ket{\Psi_1} - 2 \sin  \theta_{\Psi_1,a_{\perp}}
        \left(2 \sin \theta_{\Psi_1,a_{\perp}}\ket{\Psi_1} - \ket{a}\right) \\ 
   & = \left( 1 - 4 \sin^2 \theta_{\Psi_1,a_{\perp}}\right) \ket{\Psi_1} -
        2 \sin \theta_{\Psi_1,a_{\perp}}\ket{a} \\ 
   & = \left(1 - 2 \sin^2 \theta_{\Psi_1,a_{\perp}} -
        2 \sin^2 \theta_{\Psi_1,a_{\perp}}\right)\ket{\Psi_1} -
        2 \sin \theta_{\Psi_1,a_{\perp}}\ket{a} \\ 
   & = \left( \cos 2 \theta_{\Psi_1,a_{\perp}} - 2 \sin^2 \theta_{\Psi_1,a_{\perp}}\right)
        \ket{\Psi_1} - 2 \sin \theta_{\Psi_1,a_{\perp}} \ket{a} \\ 
   & = \left( \cos 2 \theta_{\Psi_1,a_{\perp}} - (1 - \cos 2 \theta_{\Psi_1,a_{\perp}})\right)
        \ket{\Psi_1} - 2 \sin \theta_{\Psi_1,a_{\perp}}\ket{a}\\
   & = \left( 2 \cos 2 \theta_{\Psi_1,a_{\perp}} - 1 \right) \ket{\Psi_1} -
        2 \sin \theta_{\Psi_1,a_{\perp}} \ket{a}.
\end{align*}
Now,
\begin{align*}
  \bkt{\Psi_2}{a} & =  \left( 1 - 4 \sin^2 \theta_{\Psi_1,a_{\perp}}\right)
  \bra{\Psi}\ket{a} -2 \sin \theta_{\Psi_1,a_{\perp}} \\ 
  & = \left( 1 - 4 \sin^2 \theta_{\Psi_1,a_{\perp}}\right) \sin \theta_{\Psi_1,a_{\perp}}
  - 2 \sin \theta_{\Psi_1,a_{\perp}} \\ 
  & = -4 \sin^3 \theta_{\Psi_1,a_{\perp}} - \sin \theta_{\Psi_1,a_{\perp}} \\ 
  & = - 2 \sin \theta_{\Psi_1,a_{\perp}} (\sin^2 \theta_{\Psi_1,a_{\perp}} + 1).
\end{align*}
Thus,
\begin{align*}
  \prob_{\ket{\Psi_2}\bra{\Psi_2}}(\rA = \lambda_a) & = \ns{\bkt{\Psi_2}{a}}^2 \\ 
   & = 4 \sin^2 \theta_{\Psi_1,a_{\perp}}(\sin^2 \theta_{\Psi_1,a_{\perp}} + 1)^2 \\ 
  & = 4 \prob_{\ket{\Psi_1}\bra{\Psi_1}}(\rA = \lambda_a)
  (\prob_{\ket{\Psi_1}\bra{\Psi_1}}(\rA = \lambda_a)+1)^2,
\end{align*}
and hence the law of the quantum random variable $\rA$ under the state
$\svpi_{\ket{\Psi_2}\bra{\Psi_2}}$ satisfies
\[
   \prob_{\ket{\Psi_2}\bra{\Psi_2}}(\rA = \lambda_a) > 2^2
  \prob_{\ket{\Psi_1}\bra{\Psi_1}}(\rA = \lambda_a) = \frac{1}{2^{n-2}}.
\]
Clearly,
\[
\prob_{\ket{\Psi_1}\bra{\Psi_1}}(\rA \neq \lambda_a) < 1-\frac{1}{2^{n-2}}.
\]
In consequence, the quantum code $\rU_2$ acts on the matrix density
$\rU_1 \vrh_0 \rU_1$ by the relation $\rU_2\rU_1 \vrh_0 \rU_1\rU_2$,
providing a quantum state that has high probability of measuring
the marked element.

Since we can write
\[
  \ket{\Psi_2}   = \sin \theta_{\Psi_2,a_{\perp}}\ket{a} +
     \cos \theta_{\Psi_1,a_{\perp}}\ket{a_{\perp}},
\]
where
\[
  \sin^2 \theta_{\Psi_2,a_{\perp}} = \prob_{\ket{\Psi_2}\bra{\Psi_2}}(\rA = \lambda_a)\;
  \text{ and }\;
  \cos^2 \theta_{\Psi_2,a_{\perp}} = \prob_{\ket{\Psi_2}\bra{\Psi_2}}(\rA \neq \lambda_a),
\]
by taking $\ket{\Psi_3} = \rU_2 \ket{\Psi_2} = \rU_2^2\rU_1\ket{0}^{\otimes n}$
and proceeding in a similar way, we obtain a state $\svpi_{\ket{\Psi_3}\bra{\Psi_3}}$
where the law of $\rA$ satisfies
\[
  \prob_{\ket{\Psi_3}\bra{\Psi_3}}(\rA = \lambda_a) >
  2^2 \prob_{\ket{\Psi_2}\bra{\Psi_2}}(\rA = \lambda_a) >
  2^4 \prob_{\ket{\Psi_1}\bra{\Psi_1}}(\rA = \lambda_a) = \frac{1}{2^{n-4}}.
\]
Proceeding in this way, we obtain at step $k$ a unit vector
$\ket{\Psi_k} = \rU_2^k\rU_1\ket{0}$, such that it provides
a state $\svpi_{\ket{\Psi_k}\bra{\Psi_k}}$ where the law of $\rA$ satisfies
\[
   \prob_{\rU_2^k \rU_1\vrh_0 \rU_1 \rU_2^{k}}(\rA = \lambda_a) >
   2^{2k} \prob_{\rU_1\vrh_0 \rU_1}(\rA = \lambda_a) = \frac{1}{2^{n-2k}}.
\]

Thus, we can state the following proposition.

\begin{prop}  \label{prop:last-QC}
For $k \ge 1,$ consider the quantum code 
\begin{align}  \label{eq:quantum-code}
    \rU_0,\rU_1,\overbrace{\rU_2, \ldots, \rU_2}^{k-\text{times}},
\end{align}
where $\rU_0 = \rI, \rU_1= \vH^{\otimes n}$ and $\rU_2=\rW_2\rW_1$
are self-adjoint matrices in $\mathrm{U}(2^n)$.
Then the outcome of running the quantum code \feq{eq:quantum-code} in a
$n$-qubits digital quantum computer $(\rA,\svpi_{\ket{0^n}\bra{0^n}})$
is the law of the quantum random variable  $\rA$ in the final state
$\svpi_{\rU_2^k \rU_1\rho_0\rU_1 \rU_2^{k}}$ that satisfies
\[
   \prob_{\rU_2^k \rU_1\rho_0 \rU_1 \rU_2^{k}}(\rA = \lambda_a) >
   2^{2k} \prob_{\rU_1\rho_0 \rU_1}(\rA = \lambda_a) = \frac{1}{2^{n-2k}}.
\]
Furthermore, given $\varepsilon > 0$, the probability of measuring
a marked element $a \in \D{Z}_{2^n}$ is greater than $1-\varepsilon$
for $k = \lceil \frac{n}{2} \rceil$. This amplification mechanism is the planar rotation depicted in
Figure~\ref{fig:grover-geometry}.
\end{prop}

% --- Grover strategy: corrected (no \theta override) + clean labels (TikZ) ---
% Preamble:
% \usepackage{tikz}
% \usetikzlibrary{arrows.meta,calc,positioning,decorations.pathreplacing}

%%% END IGNORE

%  $Log: quant-comp-a_QC-alg.tex,v $
%  Revision 1.2  2026/01/15 21:04:52  hgm
%  updated version
%
%  Revision 1.1  2026/01/07 22:21:22  hgm
%  initial
%
%
%  Revision 1.0 2022/10/21 19:14:38  hgm
%  introduced content of quant-comp_QC-alg.tex
%
%
%
%
%
%

%%% Local Variables: 
%%% mode: latex
%%% TeX-master: "../23_QC-algebra"
%%% End: 

%  
\section{Conclusion} \label{S:concl}
% !TEX root = ../23_QC-algebra.tex
% !TEX encoding = UTF-8 Unicode
% RCSID:       $Id: conclusion_QC-alg.tex,v 1.9 2026/01/28 19:02:17 hgm Exp $
% Author:      $Author: hgm $
% Contact:     wire@tu-bs.de
% =================================

Here we have glanced a bit on algebraic probability, mainly keeping
to the finite dimensional case and thus avoiding many analytical difficulties.
It was shown that the classical connutative or Abelian case of probability
theory can be covered in the algebraic setting, so that one sees that this
is a more general theory (e.g.\ \citep{GhorbalSchuermann99, BenyRicht15, Plavala2021,
Ron2023}).

This algebraic point of view also allows
a natural approach to random matrices and tensor fields, where the random variables
do not necessarily have to commute, and the interesting object is the behaviour
of their spectra, a distinctly analytic and algebraic concept which is
much more complicated to treat with the usual measure-theoretic background.
Connected with this is one area which was treated only very shortly here,
and this is everything connected with \emph{free probability}.  We only
offer the pointers to
\citep{VoiculescuDykemaNica1992, HiaiPetz2000, Mitchener2005, MingoSpeicher2017}
for the interested reader.

While this material in \fsec{S:formal} is essentially mainly linear algebra,
although sometimes not so well known, it is surprising to see what happens
when random variables do not commute, and this gives the striking differences
to the classical case which were explained in  \fsec{S:op-represent}.
In particular, this section contains the important view on channels,
which are the formal descriptions of information transmission.
Equipped with all this background, in \fsec{S:basic-QC} a mini-introduction
to quantum computation is given, and it is shown how this relies on the
previous material.  As an example of a quantum algorithm, we explain the
Grover algorithm in some detail.

To each and any of the topics much more could be said.  We have been a bit more
detailed in the formulation of algebraic probability, and much briefer in the
subjects which build on it, and to continue on the foundational matters
\citep{Accardi00b, Accardi2018, Accardi2022} are probably good continuation
and starting points for further study.  In the applications, it is especially quantum
computation (e.g.\ \citep{NielsenChuang2011}) and quantum information
(e.g.\ \citep{Werner2001, NielsenChuang2011, Holevo2012, Wolf2012, Wilde2017, Quillen2025})
where many new developments occur.

We hope that these notes have given a glimpse at the algebraic treatment of probability,
and the differences which come with non-commutativity of random variables.  Likewise,
we hope to have sparked some interest among the readers with the small collection of
developments and applications of algebraic probability.  Thus they may have an idea now
of what it is, and where to continue in case they are interested in more.

\ignore{     %%% BEGIN IGNORE

\section{Old Text}
%

%\ignore{           %%%% BEGIN ignore
\subsection{Test of fonts} \label{SS:testf}
\paragraph{Slanted:} --- serifs
\[ a, \alpha, A, \Phi, \phi, \vphi, \qquad 
     \vek{a}, \vek{\alpha}, \vek{A}, \vek{\Phi}, \vek{\phi}, \vek{\vphi} \]

\paragraph{Slanted:} --- sans serif
\[ \tns{a}, \tns{\alpha}, \tns{A}, \tns{\Phi}, \tns{\phi}, \tns{\vphi}, \qquad
  \tnb{a}, \tnb{\alpha}, \tnb{A}, \tnb{\Phi}, \tnb{\phi}, \tnb{\vphi} \]
   
\paragraph{Other:}  --- no small letters
\[ \C{E}, \C{Q}, \C{R}; \D{E}, \D{Q}, \D{R}; \E{E}, \E{Q}, \E{R}\]

\paragraph{Fraktur:}
\[ \F{e}, \F{q}, \F{r}; \F{E}, \F{Q}, \F{R} \]

\paragraph{Upright:} --- serifs, no small greek letters
\[ \mrm{a}, \mrm{A}, \mrm{\Phi},\qquad \mat{a}, \mat{A}, \mat{\Phi} \]

\paragraph{Upright:} --- sans serif, no small greek letters
\[ \ops{a}, \ops{A}, \ops{\Phi},\qquad \opb{a}, \opb{A}, \opb{\Phi} \]
%}           %%%% END ignore

}    %%% END IGNORE

%  $Log: conclusion_QC-alg.tex,v $
%  Revision 1.9  2026/01/28 19:02:17  hgm
%  Written a conclusion
%
%  Revision 1.8  2026/01/15 21:05:24  hgm
%  updated version
%
%  Revision 1.7  2025/04/04 13:21:52  hgm
%  commented text taken out again
%
%  Revision 1.6  2025/04/03 21:23:22  hgm
%  commented text put here
%
%  Revision 1.5  2025/01/16 20:27:14  hgm
%  text saved inside
%
%  Revision 1.4  2024/05/09 09:50:13  hgm
%  some changes
%
%  Revision 1.3  2024/04/29 09:35:02  hgm
%  new text
%
%  Revision 1.2  2024/04/08 09:40:59  hgm
%  small changes
%
%  Revision 1.1  2023/12/07 16:32:47  hgm
%  from conclusion_RV-alg.tex
%
%  Revision 1.0  2022/05/16 20:08:37  hgm
%  inital check in, from conclusion_RV-alg.tex
%
%
%
%
%
%

%%% Local Variables: 
%%% mode: latex
%%% TeX-master: "../23_QC-algebra"
%%% End: 

%
%\appendix
%\section{Appendix} \label{S:appendix}
%\input{\thetext/appendix_RV-alg}
%\appendix

\section*{Acknowledgments}
The research reported in this publication was for HGM partly supported by funding 
from the Deutsche Forschungsgemeinschaft (DFG),
a Gay-Lussac Humboldt Research Award, as well as a stay at the
Erwin Schrödinger Institute in Vienna.

% ============================================================================
% Bibliography
% The BibTeX files come from my external common bibtex repository
% multiple .bib-files are simply concatenated (without whitespace!)

%\bibliographystyle{hgmplain-1}
%\bibliographystyle{spmpsci}
%\bibliography{\thebib/alg-prob}

\clearpage
\phantomsection
\addcontentsline{toc}{section}{References}
\bibliographystyle{hgmplain-1}
\bibliography{bib/alg-prob}
%\clearpage

   { %\color{gray9}
   \tiny
%   \begin{verbatim}
%    $Id: 23_QC-algebra.tex,v 1.16 2026/01/23 22:30:36 hgm Exp $
%         \roman{ \RCSId}
%   \end{verbatim}
    \texttt{\RCSId} 
   }

\end{document}